\documentclass[rmp,aps,twocolumn,preprintnumbers]{revtex4}
\usepackage{epsfig,graphicx}
\usepackage{amsmath}
\usepackage{xspace}
\usepackage{amssymb}
\usepackage{psfrag}

\usepackage{relsize}

\newcommand{\beqa}{\begin{eqnarray}}
\newcommand{\eeqa}{\end{eqnarray}}
\newcommand{\beq}{\begin{equation}}
\newcommand{\eeq}{\end{equation}}
\newcommand{\bsp}{\begin{split}}
\newcommand{\esp}{\end{split}}
\newcommand{\bal}{\begin{align}}
\newcommand{\eal}{\end{align}}
\renewcommand{\Re}{{\cal R}e}
\renewcommand{\Im}{{\cal I}m}
\def\be{\begin{eqnarray}}
\def\en{\end{eqnarray}}

\def\A{{\cal A}}

\def\3bar{{\bf \bar 3}}
\def\6bar{{\bf \bar 6}}
\def\10bar{{\bf \ov{10}}}
\def\ov{\overline}

\def\lsim{ {\
\lower-1.2pt\vbox{\hbox{\rlap{$<$}\lower5pt\vbox{\hbox{$\sim$}
}}}\ } }
\def\gsim{ {\
\lower-1.2pt\vbox{\hbox{\rlap{$>$}\lower5pt\vbox{\hbox{$\sim$}
}}}\ } }

\def\vev#1{\langle #1\rangle}

\def\babar{\mbox{\slshape B\kern-0.1em{\smaller A}\kern-0.1em
    B\kern-0.1em{\smaller A\kern-0.2em R}}\xspace}
\def\belle{\mbox{Belle}\xspace}
\def\superb{\mbox{Super$B$}\xspace}

\begin{document}
%\noindent
%BNL-HET-08/4
%\preprint{SLAC-PUB-12063}
\preprint{BNL-HET-08/4}
%\preprint{hep-ph/yymmnnn}

%\preprint{August 2006}

%\vspace*{10pt}
\title{New Physics at a Super Flavor Factory }
\def\addbnl{Physics Department, Brookhaven National Laboratory, Upton, New York 11973, USA}
\def\addIJS{J.~Stefan Institute, Jamova 39, 1000 Ljubljana, Slovenia}
\def\addFMF{Faculty of mathematics and physics, University of Ljubljana, Jadranska 19, 1000 Ljubljana, Slovenia}
\def\addmit{Center for Theoretical Physics, Massachusetts Institute 
for Technology,
Cambridge, MA 02139}
\def\addhawaii{Department of Physics, University of Hawaii, Honolulu,
Hawaii 968222, USA}  
\def\addwarwick{Department of Physics, University of Warwick, Coventry,
CV4 7AL, UK}
\def\addBucharest{National Institute for Physics and Nuclear Engineering, 
Department of Particle Physics, 077125 Bucharest, Romania}\
\def\addCERN{Theory Division, Department of Physics, CERN
CH-1211 Geneva 23, Switzerland}

\author{Thomas E. Browder}
\email{teb@phys.hawaii.edu}
\affiliation{\addhawaii}

\author{Tim Gershon}
\email{T.J.Gershon@warwick.ac.uk}
%\email{gershon@bmail.kek.jp}
\affiliation{\addwarwick}

\author{Dan Pirjol}
\email{pirjol@mac.com}
\affiliation{\addBucharest}

\author{Amarjit Soni}
\email{soni@quark.phy.bnl.gov}
\affiliation{\addbnl}

\author{Jure Zupan}
\email{jure.zupan@ijs.si}
\affiliation{\addCERN}
\affiliation{\addFMF}
\affiliation{\addIJS}

\begin{abstract} \vspace*{18pt}

Abstract

The potential of a Super Flavor Factory (SFF) for searches of New Physics
is reviewed. 
While very high luminosity $B$ physics is assumed to be at the core of 
the program, its scope
for extensive
charm and $\tau$ studies are also emphasized. 
The possibility to run at the $\Upsilon(5{\rm S})$
is also very briefly discussed; 
in principle, this could provide very clean measurements of $B_s$ decays. 
The strength and reach of a SFF is most notably due to
the possibility of examining an impressive array of very clean observables.
The angles and the sides of the unitarity triangle can be determined
with unprecedented accuracy. 
These serve as a reference for New Physics (NP) sensitive decays such 
as $B^+ \to \tau^+ \nu$ and penguin dominated hadronic decay modes,
providing tests of generic NP scenarios with an accuracy of a few percent. 
Besides, very precise studies of direct and time dependent CP
asymmetries in radiative $B$ decays and forward-backward asymmetry studies 
in $B \to X_s \ell^+ \ell^-$ and numerous null tests using $B$, charm and $\tau$ 
decays are also likely to provide powerful insights into NP. 
The dramatic increase in luminosity at a SFF will also open up 
entirely new avenues for probing NP observables,
{\it e.g.} by allowing sensitive studies using theoretically
clean processes such as $B \to X_s \nu \bar \nu$. 
The SFF is envisioned to be a crucial tool for essential studies of 
flavor in the LHC era,
and will extend the reach of the LHC in many important ways.

\end{abstract}

\maketitle

\tableofcontents
%%%%%%%%%%%%%%%%%%%%%%%%%%%%%%%%%%
%%%%%%%%%%%% I. Introduction %%%%%%%%%%%%%%%%%%
%%%%%%%%%%%%%%%%%%%%%%%%%%%%%%%%%%%

\section{Introduction}
\label{sec:intro}
The term {\it flavor} was first used in particle physics in the context of the quark model of hadrons. It was coined in 1971 by Murray Gell-Mann and his student at the time, Harald Fritzsch, at a Baskin-Robbins ice-cream 
store in Pasadena. Just as ice-cream has both color and flavor so do quarks \cite{Fritzsch:story}.

{\it Flavor physics} denotes physics of transitions between the three generations
of Standard Model (SM) fermions. With the LHC startup around the corner, why should
one pay attention to these low energy phenomena? For one thing, flavor
physics can probe new physics (NP)
through off-shell corrections, before the NP particles themselves are produced
in energy frontier experiments. As a historic example, the existence of the
charm quark was predicted from the suppression of $K_L\to \mu^+\mu^-$ 
before its discovery \cite{Glashow:1970gm}, while its mass was successfully predicted from $\Delta m_K$ \cite{Gaillard:1974hs}.  Flavor physics is also intimately 
connected with the origin of fermion masses. In the limit of vanishing masses the flavor physics is trivial -- no intergenerational transitions occur since weak and 
mass eigenbases trivially coincide. It is only the mismatch of weak and mass eigenbases (or the mismatch between the bases in which gauge and Yukawa terms are diagonal) that makes flavor physics interesting. In the quark sector of SM this mismatch is described by a single unitary matrix - the Cabibbo--Kobayashi--Maskawa (CKM) matrix. Finally, CP violation is closely related to flavor physics.  
A strong argument for the existence of new sources of CP violation 
is that the CKM mechanism is unable to account for the observed 
baryon asymmetry of the universe (BAU) through baryogenesis~\cite{Gavela:1994dt}.
This points at NP with new sources of CP violation in either the quark or
lepton sector (the latter potentially related to the BAU via
leptogenesis~\cite{Uhlig:2006xf}). 
It is therefore important to investigate the BAU by studying CP
violation in both quark and lepton sectors (see below).

\begin{figure}[tb!]
  \begin{center}
    \includegraphics[width=4.2cm]{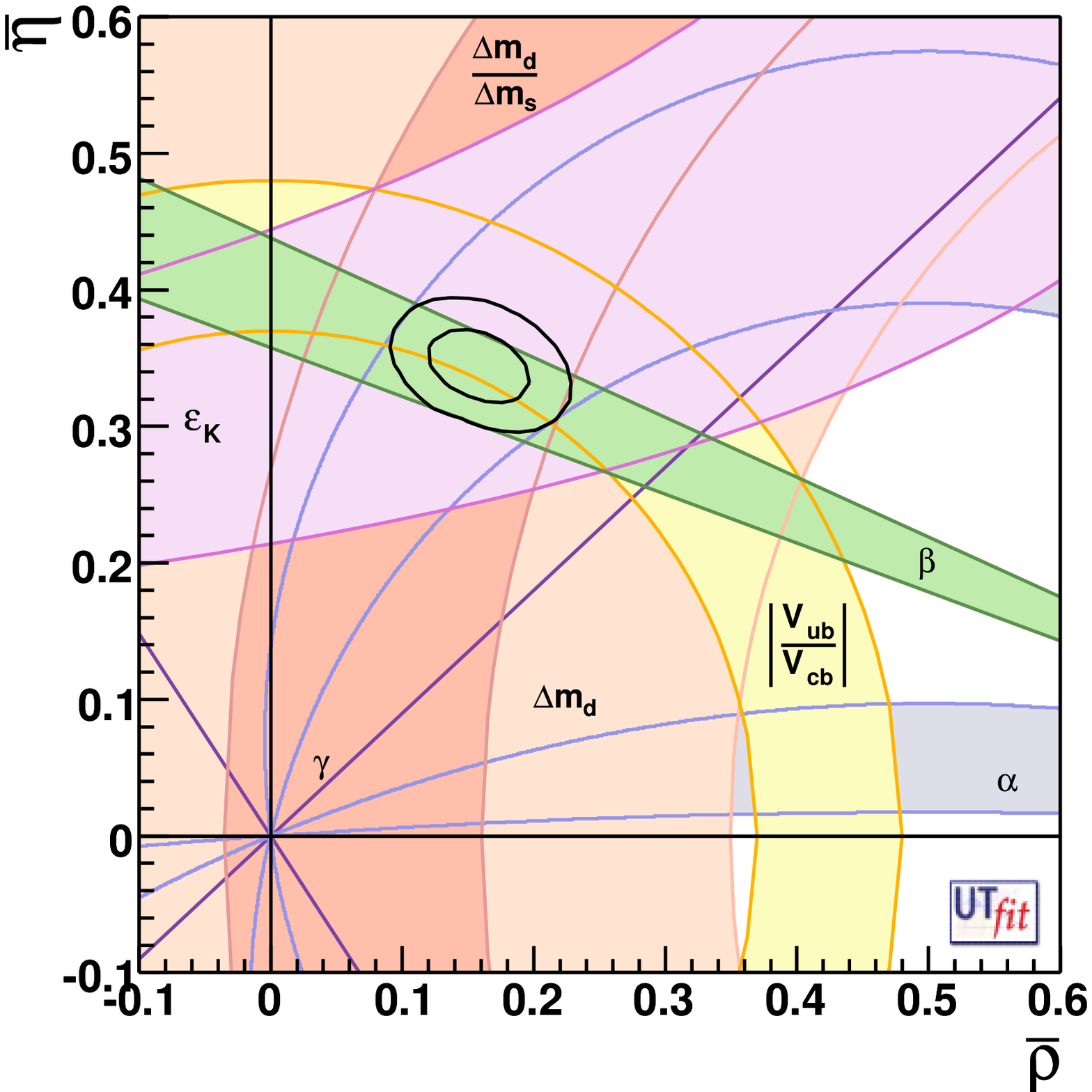}
    \includegraphics[width=4.2cm]{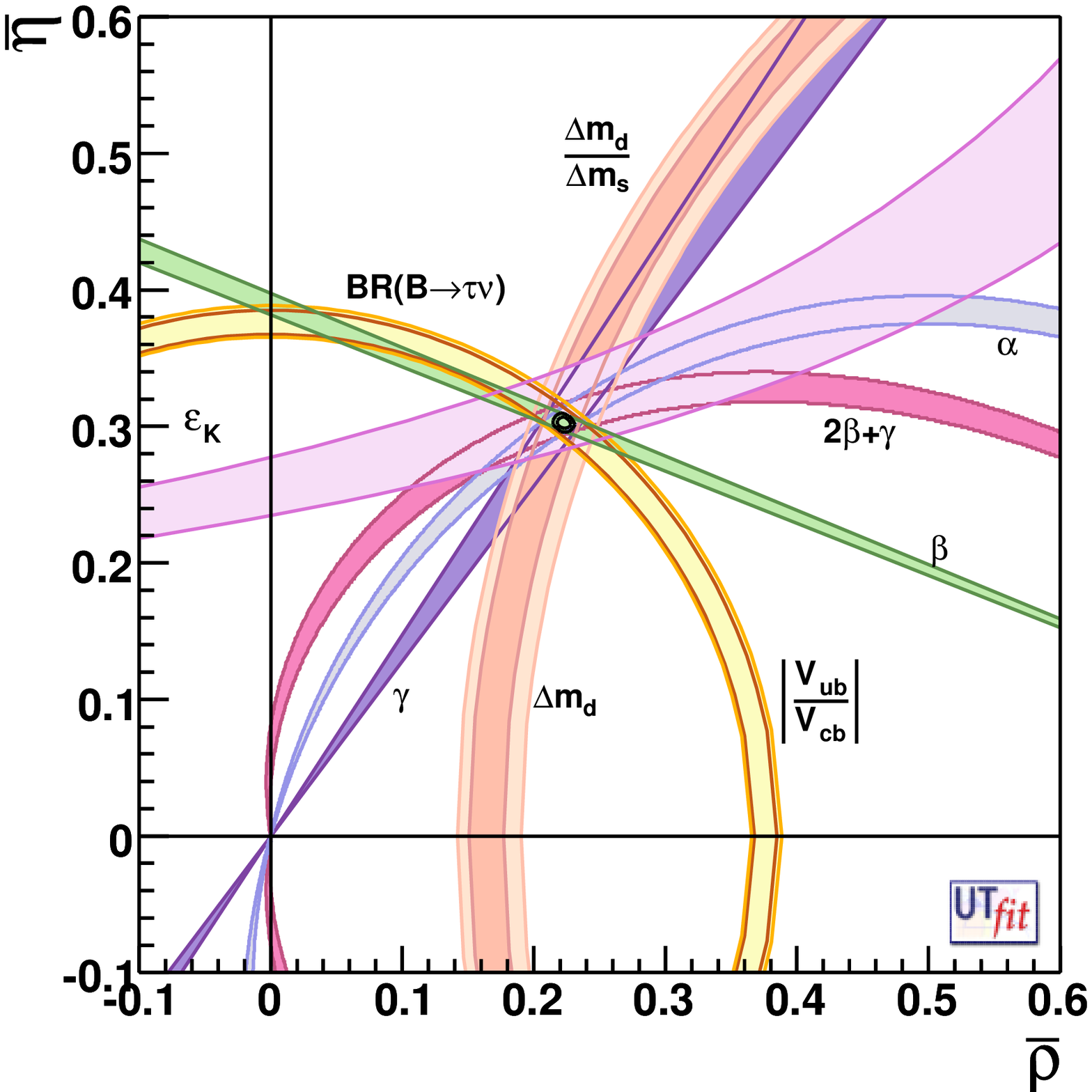}
    \caption{
      $95\%$ confidence level constraints on parameters $\bar\rho$ and
      $\bar\eta$ in the Wolfenstein parametrization of the CKM matrix. 
      Left: present constraints, right:
     with errors shrunk to the size expected at a SFF while tuning central values to have compatible constraints [from \cite{Browder:2007gg}].
    }
    \label{fig:ckm}
  \end{center}
\end{figure}

In the past ten years,
due to the spectacular performance of the two $B$-factories,
a milestone in our understanding of CP violation phenomena was reached.
For the first time, detailed experiments,  
\babar~\cite{Aubert:2001tu} and Belle~\cite{Abashian:2000cg},
provided a striking confirmation of the 
CKM-paradigm of CP violation~\cite{Cabibbo:1963yz,Kobayashi:1973fv}.  The
Kobayashi-Maskawa model of CP-violation, based on three families
and a single CP-odd phase, is able to account for
the observed CP violation in the $B$ system, 
as well as that in the $K$ system, to an accuracy of about 20\%,
as shown  in Fig.~\ref{fig:ckm}~\cite{Charles:2004jd, Bona:2006sa,Bona:2007vi,Lunghi:2007ak}. 
The impressive gain in precision on CKM constraints 
that is expected at a SFF is also shown in Fig. \ref{fig:ckm}.

While we celebrate this remarkable agreement  it is important to note that increasing the 
accuracy of CKM tests brings more than just an increased knowledge of fundamental CKM
parameters. Once NP particles are observed at LHC, flavor physics observables will provide a set of 
independent constraints on the NP Lagrangian. These constraints are complementary to the measurements
that are performed at high $p_T$ processes -- {\it i.e.} they provide a complementary constraint on the combination of 
couplings, mixing angles and NP masses and become much more powerful once NP mass spectra are already measured. 
However, to be relevant for TeV processes, high precision is needed. But, how precise is precise enough?
The answer depends on the NP flavor changing couplings. Taking as a conservative benchmark the case of minimally flavor violating
NP that has couplings to SM fermions comparable to weak gauge couplings, the present results from $B$ factories
allow for masses of NP particles below $\sim 100$ GeV. After completion of the
Super Flavor Factory (SFF) program this limit would be pushed to $\sim 600$
GeV \cite{Bona:2007vi,Browder:2007gg}, illustrating the complementarity of LHC
and SFF reach.\footnote{Note that the generic MFV scenario of weakly coupled NP is not the most
conservative scenario. The SFF constraint can be avoided, if couplings to SM
fermions are further suppressed 
(see, for instance, \textcite{Grossman:2007bd}.)}

Let us elaborate a bit more on this important point. The NP constraints
depend on both NP couplings to SM quarks and the NP masses and the two
cannot be disentangled. An important set of flavor physics observables
useful for NP searches are those from processes that proceed through flavor
changing neutral currents. These are loop suppressed in the SM, and hence NP contributions are easier to detect than in charged flavor changing transitions that occur at tree level in the SM.
Let us take as an explicit example corrections to the $\Delta F=2$ processes, 
{\it i.e.} to $K^0$--$\bar K^0$, $B^0_d$--$\bar B^0_d$ and $B^0_s$--$\bar B^0_s$ mixing. 
The corresponding SM weak Hamiltonian has a form
\beq\label{H_SM}
{\cal H}_{\rm eff}=\frac{1}{4}\dfrac{C_0}{\Lambda_0^2} \big(V_{t
i}^*V_{tj}\big)  \big[\bar d_{Li} \gamma_\mu d_{Lj}\big]^2,
\eeq
where $C_0$ is a Wilson coefficient that is of order ${\cal O}(1)$, 
$\Lambda_0 = 4\pi m_W/g^2\simeq 2.5 \ {\rm TeV}$ is the appropriate scale 
for a loop suppressed SM process, 
and $d_{i,j}$ are the down quark fields $d,s,b$. For simplicity let us also assume that NP leads to the effective operator with the same Dirac structure
as in the SM, so 
\beq\label{H_NP}
{\cal H}_{\rm eff}^{\rm NP}=\dfrac{C_{\rm NP}}{\Lambda_{\rm NP}^2}
\big[\bar d_{Li} \gamma_\mu d_{Lj}\big]^2.
\eeq
If NP couplings do not have any flavor structure, then $C_{\rm NP}\sim {\cal O}(1)$, while $\Lambda_{\rm NP}$ corresponds roughly
to the NP particles' masses, if these are exchanged at tree level. In this case the NP masses are well above the weak scale. For instance, present 
measurements exclude ${\cal O}(1)$ corrections to the $B^0_d - \bar B^0_d$ mixing, from which 
\beq
\begin{split}
{B^0_d-\bar B^0_d {\rm ~~mix.:}}&
{{\big(\underbrace{V_{t
b}^*}_{\sim1}\underbrace{V_{td}}_{\sim\lambda^3}\big)^2\dfrac{1}{4\Lambda_{0}^2}>\dfrac{C_{\rm NP}}{\Lambda_{\rm NP}^2}}}\\
 &\qquad\qquad\qquad \Rightarrow
 {\Lambda_{\rm NP} \gtrsim  500~{\rm TeV}},
\end{split}\label{Gen:NP}
\eeq
For $B^0_s-\bar B^0_s$ and $K^0 - \bar K^0$ mixings the corresponding 
 $\Lambda_{\rm NP}$ scales are 100 TeV and $10^4$ TeV, respectively. The fact that these scales are much larger than the weak scale
$\sim m_W$ is known as the {\it NP flavor problem}. 
 
If new physics particles with mass $M$ are exchanged at tree level with
${\cal O}(1)$ coupling constants, then $\Lambda_{\rm NP}\sim M$. This excludes
new physics with general flavor violation structure at the energies
accessible at the LHC. This conclusion holds even if new physics
particles are exchanged only at 1-loop order, where $\Lambda_{\rm NP}\sim 4\pi
M/g_{\rm NP}^2$. For $g_{\rm NP}\sim g$ even the weakest bound from
the $B^0_s - \bar B^0_s$ system still leads to 
new physics
particles
with masses $\gtrsim 7$ TeV.

In other words, if the hierarchy problem of the Standard Model is
resolved by adding more particles near the electroweak scale, this
extended sector must have a non-generic flavor structure. Having 
completely flavor blind new physics is unnatural since the SM
already contains flavor violation 
in the Yukawa couplings. 
The minimal possibility for the NP contribution
of Eq.~\eqref{H_NP} is that the NP flavor violation comes only from the SM
Yukawa couplings. This is the assumption underlying Minimal Flavor
Violation (MFV); see  
Section
\ref{Sec:MFV}. The NP contribution of Eq.~\eqref{H_NP} then obeys 
the same CKM hierarchy as the SM contribution of Eq.~\eqref{H_SM} and can be
rewritten
as
\beq
\tilde {\cal H}_{\rm eff}^{\rm NP}=\dfrac{\tilde C_{\rm
NP}}{\Lambda_{\rm NP}^2} \big(V_{t i}^*V_{tj}\big) \big[\bar d_{Li}
\gamma_\mu d_{Lj}\big]^2.
\eeq
In this case not observing ${\cal O}(1)$ effects from NP in the flavor transitions
translates to $\Lambda_{\rm NP}\gtrsim \Lambda_{0}\simeq 2.5$ TeV. If NP
contributions are loop suppressed (as those from the SM are), 
then this bound translates to a relatively
weak bound $M\gtrsim m_W$ (if $g_{\rm NP}\sim g$). 

We see that in this 
minimal scenario, where no new mechanisms of flavor violation beyond
those already present in the SM are introduced in the NP sector of the
theory, one requires precision measurements of $B$ physics
observables to have results that are complementary to the measurements of NP
spectrum at the LHC. In particular, as already mentioned, taking $g_{\rm NP}\sim g$ with NP contributing at 1-loop then SFF precision translates to a bound
on NP masses of around $600$ GeV \cite{Bona:2007vi,Browder:2007gg}. 

Another 
very powerful probe of NP effects are measurements of CP violating observables. Extensions of the SM generically lead
to new sources of CP-odd phases and/or new sources of
flavor breaking [for a review see, {\it e.g.}~\textcite{Atwood:2000tu}]. 
An elementary  example is provided by the SM itself. 
While a two-generation version of the SM does not exhibit  
CP violation, a single CP-odd phase in the CKM matrix occurs very
naturally as a consequence of the third quark family. 
Beyond the SM the existence of new CP odd phases 
can be seen explicitly in specific extensions such
as two Higgs doublet models~\cite{Lee:1973iz,Weinberg:1976hu}, 
the left-right symmetric model~\cite{Mohapatra:1974hk,Kiers:2002cz}, 
low energy SUSY~\cite{Grossman:1997pa} or models  
 with warped extra dimensions~\cite{Agashe:2004ay,Agashe:2004cp}.

Furthermore, while $B$-factory results have now established
that the CKM-paradigm works to good accuracy, 
as more data has been accumulated some possible indications of 
deviations from the SM have emerged.
These include the 
small ``tension'' between the direct and indirect
determinations of $\sin 2\beta$, 
as seen in 
Fig.~\ref{fig:ckm}~\cite{Charles:2004jd,Bona:2006sa,Bona:2007vi,Lunghi:2007ak}),
as well as the famous trend for 
$\sin 2\beta$
from hadronic $b \to s$ penguin dominated decays to be below 
that from $b \to c$ tree dominated decays.
While these measurements do not yet show compelling evidence for NP,
the results are quite intriguing -- it is also noteworthy that the
discrepancy between $\sin 2\beta$ from penguin dominated modes and 
from the indirect determination ({\it i.e.} from the SM fit) is  
larger~\cite{Lunghi:2007ak}.      
Several other measurements in penguin dominated decays 
show possible indications of NP that are, unfortunately, 
obscured by hadronic uncertainties.
Whether or not the currently observed effects are due to the intervention
of NP, this illustrates that these processes provide a 
sensitive tool to search for NP. 
Thus, it is all the more important to focus on theoretically clean observables,
for which hadronic uncertainties cannot cloud 
the interpretation of possible NP signals.
In most cases this requires a significant increase in statistics,
and therefore will only be possible at a 
SFF.

A key strength of a SFF is that it offers the opportunity
to examine a vast array of observables that allow 
a wide range of tests of the SM and sensitively probe 
many NP models.
In order to achieve this core physics program,
it will be necessary to accumulate  
$50-100 \ {\rm ab}^{-1}$ of integrated luminosity after a few years of running,
corresponding to an increase of nearly two orders of magnitude
over the final data samples available at the current $B$-factories.
It is important to stress that not only will a SFF
enable exciting $B$ physics, it will also 
provide over $5 \times 10^{10}$ charm hadron and $\tau$ lepton pairs,
enabling powerful studies of NP effects in the up-type quark and lepton
sectors.
The breadth of precision tests in a wide range of clean observables 
that are excellent probes of NP is an extremely important aspect of the SFF
proposal.

While expectations for the SFF performance are based on the successes of
the current $B$-factories, it is important to emphasise that
the huge increase in statistics will provide a step change in
the physics goals and in NP sensitivity.
The program will include not only much more precise studies of
NP-sensitive observables for which initial studies have already been
carried out
({\it e.g.} $b \to sg$, $b \to s\gamma$ and $b \to s \ell^+\ell^-$ penguin
dominated processes),
but will also include channels that have either barely been seen,
or which, at their SM expectations, are beyond the capabilities of
current experiments
({\it e.g.} $b \to d$ penguin dominated processes, $b \to s \nu \bar
\nu$ decays).
Clean studies of several interesting inclusive processes will become possible
for the first time.
Furthermore, for some channels with very small SM expectations,
positive searches would provide unambiguous NP signals
({\it e.g.} lepton flavor violating $\tau$ decays, CP violation in 
charm mixing and/or decays,
$b \to dd \bar s$ decays) etc.
These provide examples of 
numerous ``null
tests''~\cite{Gershon:2006mt} that are accessible to a SFF.
It is notable that much of the SFF program 
will use the recoil analysis technique,
that takes advantage of the $e^+e^- \to \Upsilon(4{\rm S}) \to B\bar{B}$
production chain to provide kinematic constraints on unreconstructed
particles.  This is of great importance since it allows measurement of
theoretically clean processes with typically low experimental backgrounds.

In Section II we begin with a very brief discussion of 
design issues for the new machine(s), Section III 
presents a review of 
NP effects
in FCNC processes. For illustration we discuss
three class of 
NP scenarios
that are very popular: Minimal Flavor Violation (MFV),
Minimal Supersymmetric Standard Model and models
of warped extra dimensions. We then discuss (Section IV) the prospects
for improved determinations of the angles of the 
UT by ``direct measurements'' through the cleanest methods
that have been devised so far. Section V briefly
reviews the determination of the sides of the UT.
We then discuss the time dependent CP asymmetry measurements in
penguin-dominated modes (Section VI) that have been the focus of much
attention in the past  
few years, followed by 
a section on null tests (Section VII).
Section VIII is devoted to the powerful
radiative $B$ decays; here we discuss both on-shell
photonic $b \to s \gamma$  as well as $b \to s \ell\ell$ in several
different manifestations. Sections IX is devoted to a very brief
presentation of highlights of $B_s$ physics possibilities at a 
SFF. Sections X and XI deal with
charm and $\tau$ physics potential of a SFF.
Section XII briefly discusses how the SFF and LHCb efforts
complement each other in important ways and Section
XIII is the Summary.

\section{Design issues}

\subsection{Machine design considerations}

Quite recently, two different designs for a Super Flavor Factory (SFF) have
emerged.  The SuperKEKB design~\cite{Hashimoto:2004sm} is an upgrade
of the existing KEKB accelerator with expected peak instantaneous luminosity
of $8 \times 10^{35} \ {\rm cm}^{-2} \, {\rm s}^{-1}$.
This is achieved by increasing the beam currents, while reducing
the beam sizes and 
improving the specific luminosity with crab cavities that provide
the benefits of effective head-on collisions with a nonzero crossing
angle~\cite{Akai:2004np,Oide:1989qz,Abe:2007iz}. 
While this is a conventional upgrade scenario,
it presents several challenges, particularly related to 
higher order mode heating, collimation and coherent synchrotron radiation.
A great deal of effort has gone into understanding and solving these
problems including prototypes 
(for a detailed discussion, see~\textcite{Hashimoto:2004sm}).

The \superb\ design~\cite{Bona:2007qt} uses a completely different approach to
achieve a peak instantaneous luminosity in excess of $10^{36} \ {\rm cm}^{-2}
\, {\rm s}^{-1}$.  The basic idea is that high luminosity is achieved through
reduction of the vertical beam size by more than
an order of magnitude, rather than by increasing the currents.
With such small emittance beams, a large crossing
angle~\cite{Piwinski:1977ts,Hirata:1994jn} is necessary to maintain beam
stability at the interaction point. Any degradation in luminosity due to the
crossing angle is recovered with a ``crab'' of the focal
plane~\cite{Raimondi:2007vi}. The \superb\ design could be built anywhere in
the world, though the most likely home for this facility is a green field site
on the Tor Vergata campus of the University of Rome.

Some of the key parameters of the SuperKEKB and \superb\ machines
are compared in Table~\ref{tab:machineParameters}.
One important number to compare is the wall power,
which dominates the operating costs of the machine.
The total costs are kept low by recycling as much hardware as possible --
from KEKB magnets and the \belle\ detector in the case of SuperKEKB,
and from PEP-II hardware and the \babar\ detector 
in the baseline design for \superb.

\begin{table}[tb]
  \begin{center}
    \begin{ruledtabular}
   \caption{
      Comparison of some of the key parameters of the 
      SuperKEKB~\cite{Hashimoto:2004sm} and \superb~\cite{Bona:2007qt} designs.
      \label{tab:machineParameters}
    }
    \begin{tabular}{lcc}
      Parameter & SuperKEKB & \superb \\
      \hline
      Beam energies ($e^+ \, / \, e^-$, ${\rm GeV}$) & 
      $3.5 \, / \, 8$ & $4.0 \, / \, 7.0$ \\
      Beam currents ($e^+ \, / \, e^-$, ${\rm A}$) & 
      $9.4 \, / \, 4.1$ & $2.3 \, / \, 1.3$ \\
      Bunch size ($\sigma_x^* \, / \, \sigma_y^*$, ${\rm nm}$) & 
      $42000 \, / \, 367$ & $ 5700 \, / \,  35$ \\
      Bunch length ($\sigma_z$, ${\rm mm}$) &
      3 & 6 \\
      Emittance ($\epsilon_x \, / \, \epsilon_y$,  ${\rm nm}$-${\rm rad}$) & 
      $9 \, / \, 0.045$ & $1.6 \, / \, 0.004$\\
      Beta function at IP ($\beta_x^* \, / \, \beta_y^*$, ${\rm mm}$) &
      $200 \, / \, 3$ & $20 \, / \, 0.3$ \\
      Peak luminosity ($10^{36} \ {\rm cm}^2 \ {\rm s}^{-1})$ & 0.8 & $>\,1$ \\
      Wall power (${\rm MW}$) & 83 & 17 \\
%      Cost (M\euro) & 300 & 340 \\
    \end{tabular}
\end{ruledtabular}
  \end{center}
\end{table}

Aside from high luminosity -- the higher the better --
there are several other desirable features for a SFF
to possess.
Although the physics goals appear to be best served by operation 
primarily at the $\Upsilon(4{\rm S})$ resonance,
the ability to change the centre-of-mass energy and run
at other $\Upsilon$ resonances, and even down to the tau-charm
threshold region (albeit with a significant luminosity penalty), 
enhances the physics capabilities of the machine.
The possibility to run with at least one beam polarized would add
further breadth to the physics program.

It is also important that the clean experimental environment enjoyed by the 
current $B$ factories must be achieved by a SFF.
How to achieve high luminosity while retaining low backgrounds
is a challenge for the design of the machine and the detector,
since the brute force approach to higher luminosity --
that of increasing the beam currents --
necessarily leads to higher backgrounds.
To some extent these can be compensated for by appropriate 
detector design choices,
but in such cases some compromise between luminosity and 
detector performance (and hence physics output) may be anticipated.

The background level in the detector depends on several factors.
One of these is the luminosity itself, and higher luminosity
unavoidably leads to larger numbers of physics processes
such as radiative Bhabha scattering and $e^+e^-$ pair production.
Other terms depend on the beam current.  
For example, synchrotron radiation is emitted wherever the 
beam is steered or bent,
some of which inevitably affects the detector in spite of 
careful design and shielding of the interaction region.
Another term that depends on the current arises from 
so-called beam gas interactions.
Although the interior of the beam pipe is maintained at high vacuum,
radiation from the beam will interact with material in the beampipe 
and cause particles to be emitted -- 
these in turn can be struck directly by the beam particles.
Consequently this term depends quadratically on the current.
The beam size is another consideration that has an impact on backgrounds.
As the beams become smaller the particles within them 
are more likely to undergo intrabeam scattering effects.
These include the Touschek effect, in which both particles 
involved in an intrabeam collision are ejected from the beam.
For very small emittance beams, the loss of particles can be severe,
leading to low beam lifetimes. 
The achievement of meeting the challenges of maintaining manageable
backgrounds and beam lifetimes represents a milestone for SFF
%(Super Flavor Factory)  
machine design~\cite{Hashimoto:2004sm,Bona:2007qt}. 

% \subsection{Beam Energy Asymmetry}

A related issue pertains to the asymmetry of the beam energies.
To obtain the optimal asymmetry, several factors must be taken into account.
From the accelerator design perspective, 
more symmetric beam energies lead to longer beam lifetimes
and potentially higher luminosities.
However, a certain degree of beam asymmetry is necessary in order 
to measure time-dependent CP asymmetries,
and these are an important part of the physics program of the SFF, 
%Super Flavor Factory, 
as discussed below.
An equally important part of the program, however,
relies on measurements that benefit from the hermeticity of the detector
in order to reconstruct decay modes with missing particles,
such as neutrinos.
Thus the physics considerations are subtly different from those 
that 
informed the design choices for the current $B$ factories,
and a somewhat smaller asymmetry than either
\babar ($9.0 \ {\rm GeV} \ e^-$ on $3.1 \ {\rm GeV} \ e^+$) or
\belle ($8.0 \ {\rm GeV} \ e^-$ on $3.5 \ {\rm GeV} \ e^+$),
may be optimal.
However, a change in the beam energies would require the 
design of the interaction region, and to a lesser extent the detector,
to be reconsidered.
In order to be able to reuse components of the existing detectors 
in the final SFF,
as discussed below,
it would be prudent to keep the asymmetry similar to those 
in successful operation today.
However, preliminary studies indicate that either \babar or \belle detectors
could quite easily be modified to operate with beam energies 
of $7 \ {\rm GeV}$ on $4 \ {\rm GeV}$.

\subsection{Detector design considerations}

The existing $B$ factory detectors~\cite{Aubert:2001tu,Abashian:2000cg}
provide a very useful baseline from which to design a SFF
detector that can provide excellent performance in the areas of 
vertex resolution, momentum resolution,
charged particle identification (particularly kaon-pion separation),
electromagnetic calorimetry and 
close to $4\pi$ solid angle coverage with high efficiency for
detection of neutral particles that may otherwise fake missing
energy signatures (particularly $K^0_L$ mesons).
However, some upgrades and additions are necessary.

As it is desirable to operate with reduced beam energy asymmetry
compared to the current $B$ factories,
improved vertex resolution is necessary in order to obtain 
the same performance in terms of $c \Delta t = \Delta z / (\beta\gamma)$,
where $(\beta\gamma)$ is the Lorentz boost factor of the 
$\Upsilon(4{\rm S})$ in the laboratory frame.\footnote{
  The use of the symbols $\beta$ and $\gamma$ here is unrelated to their
  use to represent angles of the Unitarity Triangle or,
  in the case of $\beta$, the ratio of Higgs vacuum expectation values.
}
In fact, it is highly desirable to improve the performance further,
since results from the current $B$ factories have demonstrated the 
utility of vertex separation as a powerful tool to reject backgrounds.
The ultimate resolution depends strongly on the proximity of the inner 
layer to the interaction point.
For reference, the radii of the innermost layers of the 
existing \babar\ and \belle\ vertex detectors are $30 \ {\rm mm}$
and $20 \ {\rm mm}$ respectively~\cite{Re:2006nk,Aihara:2006dh}.
To position silicon detectors close to the interaction region
requires careful integration with the beampipe design,
and a choice of technology that will not suffer from high occupancy.

While the inner radius of the vertex detector is of great importance
for almost all measurements that will be made by a SFF,
the outer radius has a large impact on a subset of channels,
namely those where the $B$ decay vertex position must be obtained 
from a $K^0_S$ meson (typically $B^0 \to K^0_S \pi^0$, 
$B^0 \to K^0_S \pi^0 \gamma$ and $B^0 \to K^0_S K^0_S K^0_S$).
The existing \babar\ and \belle\ vertex detectors have outer radii
of $144 \ {\rm mm}$ and $88 \ {\rm mm}$ respectively,
and the former appears to be a suitable choice for a SFF.
A larger outer radius for the silicon detector has 
a useful consequence in that the tracking chamber,
which can be based on a gaseous detector, 
does not have to extend too close to the interaction region
where the effect of high backgrounds would be most severe for this detector.
Therefore, assuming the same magnetic field ($1.5 \ {\rm T}$)
as \babar\ and \belle, similar momentum resolution would be expected~\cite{Bona:2007qt,Hashimoto:2004sm}.

The choice of particle identification technology
for a SFF
presents some challenges.
At present, \belle\ achieves good $K-\pi$
separation 
through a combination of measurements from 
time-of-flight and aerogel Cherenkov counters.
Some upgrades are necessary to cope with the 
SFF
physics demands and environment.
For an upgrade based on \babar, the existing technology using 
detection of internally reflected Cherenkov light
appears almost irreplaceable for the barrel, though this requires
a novel imaging and readout scheme.
Possibilities for particle identification capabilities in both
forward and backward regions are also being considered.

The high efficiency to reconstruct photons is one of the 
significant advantages of a SFF
compared 
to experiments in a hadronic environment.
The existing electromagnetic calorimeters of \babar\ and \belle
(and indeed of CLEO) are based on CsI(Tl) crystals;
studies show that technology can perform well at higher rates in the
barrel region.  However, in the endcaps where rates are highest alternative
solutions are necessary.
Various options, including pure CsI crystals or LYSO 
are under consideration~\cite{Hashimoto:2004sm,Bona:2007qt}.
Improvements to the calorimeter solid angle coverage and hence hermeticity
would benefit the physics output (especially for an upgrade based on the
\babar\ detector, which does not have a backward endcap calorimeter).

Another important consideration with respect to detector hermeticity
is the detection of $K^0_L$ mesons,
which if unreconstructed can fake missing energy signatures.
Both \babar\ and \belle\ have instrumentation in their magnetic
flux returns which allows the detection of showers that initiate in the yoke, 
that may be associated with tracks (as for muons) or with 
neutral particles ($K^0_L$ mesons).
The efficiency depends on the amount of material in the flux return,
while the background rates generally depend on radiation coming
from up- and down-stream bending magnets~\cite{Hashimoto:2004sm,Bona:2007qt}.
Both of these problems appear well under control for
operation.

Finally, it is important to note that the extremely high physics trigger
rate will present some serious challenges for data acquisition and computing.
However, in these areas one can expect to benefit from Moore's Law
and from the distributed computing tools that are under development
for the LHC.  
Thus there is no reason to believe that these challenges cannot be met.

To summarize, there exist two well-developed proposals
and approaches to achieving the luminosity and performance required
for the measurements of NP in flavor ~\cite{Hashimoto:2004sm,Bona:2007qt}.

\section{New Physics and Super Flavor Factory} 
\label{NP-sec}

A Super Flavor Factory offers a variety of observables sensitive 
to NP such as rare $B$ decays, CP asymmetries, lepton flavor violation, etc. 
To gauge their sensitivity to NP we review in this section
several examples of NP models whose imprint in flavor physics has 
been extensively discussed in the literature: the model independent 
approach of Minimal Flavor Violation, two Higgs doublet models, low 
energy SUSY models and extra dimensions. This list is by no means 
exhaustive. Other beyond the SM extensions not covered in this 
section have interesting flavor signals as well, for instance 
little Higgs models with conserved $T$ parity 
\cite{Cheng:2003ju,Blanke:2006eb,Blanke:2007db} or
the recent idea of ``Unparticle Physics''~\cite{Georgi:2007ek} -- 
a possible nontrivial scale invariant sector weakly coupled to the SM 
that could also have flavor violating signatures~\cite{Chen:2007pua,Lenz:2007nj,Huang:2007ax,Mohanta:2007ad,Zwicky:2007vv} [see, however the comments in \cite{Grinstein:2008qk}].

\subsection{Effective weak Hamiltonian}
The weak scale $\mu_{\rm weak}\sim m_W$ and the typical energy scale 
$\mu_{\rm low}$ of the low energy processes occurring at SFF are well separated. 
For instance, the typical energy scale in $B$ decays  is a few GeV, about a 
factor $\sim 50$ smaller than $m_W$. This means that using OPE the effects 
of weak scale physics can be described at low energies  by a set of local 
operators, where the expansion parameter is $\mu_{\rm low}/\mu_{\rm weak}$. 
The matching onto local operators is performed by integrating out the heavy 
fields - the top, the massive weak gauge bosons, the Higgs boson, and the 
possible new physics particles.  At low energies one then works only within 
the effective field theory (EFT). 

\begin{figure}[!t]
 \centerline{
  \mbox{\epsfxsize=8.0truecm \hbox{\epsfbox{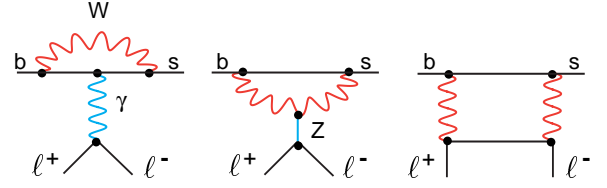}} }
  } 
%\vskip-0.3cm
\caption[1]{Sample diagrams contributing to the matching for $b\to s\ell^+\ell^-$ at one-loop order.}
\label{fig:bsllmatch} 
%\vskip-0.5cm
\end{figure}

For example, the SM
effective weak Hamiltonian for $\Delta S=1$ $B$ transitions is \cite{Buchalla:1995vs}
\beq\label{HW}
H_W=\frac{G_F}{\sqrt{2}}\sum_{p=u,c}\lambda_p^{(s)} \Big(C_1 O_1^p+C_2 O_2^p+\sum_{i=3}^{10, 7\gamma, 8g}C_iO_i\Big),
\eeq
where the CKM factors are $\lambda_p^{(s)}=V_{pb} V_{ps}^*$ and the standard basis of four-quark operators is
\beq\label{Oi}
\begin{split}
O_1^p=(\bar pb)_{} (\bar s p)_{-}, \quad &O_2^p=(\bar p_\beta b_\alpha)_{} (\bar s_\alpha p_\beta)_{-},
\\
O_{3,5}=(\bar s b)_{} (\bar q q)_{\mp },\quad &O_{4,6}=(\bar s_\alpha b_\beta)_{} (\bar q_\beta q_\alpha)_{\mp},
\\
O_{7,9}=\frac{3e_q}{2}(\bar s b)_{} (\bar q q)_{\pm },\quad
&O_{8,10}=\frac{3e_q}{2}(\bar s_\alpha b_\beta)_{} (\bar q_\beta q_\alpha)_{\pm},
\end{split}
\eeq
with the abbreviation $(\bar q_1\gamma^\mu(1-\gamma_5) q_2)(\bar q_3\gamma^\mu(1\mp\gamma_5) q_4)\equiv (\bar q_1 q_2)
(\bar q_3 q_4)_{\mp}$. The color indices $\alpha, \beta$  are displayed only when the sum is over fields in 
different brackets. In the definition of the penguin operators $O_{3-10}$ in Eq.~\eqref{Oi}
there is also an implicit sum over $q=\{u,d,s,c,b\}$. The electromagnetic and chromomagnetic operators are
\beq\label{magnetic}
O_{\{7\gamma,8g\}}=-\frac{m_b}{4\pi^2}\bar s \sigma^{\mu\nu}\{eF_{\mu\nu},g G_{\mu\nu}\}P_R b,
\eeq
with $P_{L,R}=1\mp\gamma_5$, while the effective Hamiltonian for $b\to s\ell^+\ell^-$ contains in addition \cite{Grinstein:1988me}
\beq
\begin{split}
Q_{\{9\ell,10\ell\}} =& \frac{e^2}{8\pi^2} (\bar \ell\gamma^\mu\big\{1,\gamma_5\big\}\ell)(\bar s \gamma_\mu P_L b)\,.
\end{split}
\eeq
 These two operators arise at 1-loop from matching the $W$ and $Z$ box and penguin 
diagrams shown in Fig.~\ref{fig:bsllmatch}. 
The operator $Q_{10\ell}$ is RG invariant to all orders in the strong coupling,
while the operator $Q_{9\ell}$ mixes with the four-quark operators $Q_{1,\dots,6}$
already at zeroth order in $\alpha_s$. 
Similarly, the operator for $b\to s\nu\bar\nu$ transition in SM is
\begin{eqnarray}
O_{11\nu}  = \frac{e^2}{4\pi^2\sin^2\theta_W}
(\bar\nu \gamma_\mu P_L \nu)
(\bar s\gamma_\mu P_L b)\,.
\end{eqnarray}

The weak Hamiltonian for $\Delta S=0$ $B$ decays is obtained 
from Eqs.~\eqref{HW}-\eqref{magnetic} through the replacement $s\to d$, 
while for $K$ decays another $b\to s$ replacement is needed. 
$B$--$\bar B$ mixing is governed in the SM by
%\beq
$Q_{\Delta B = 2} = (\bar b d)_-(\bar b d)_-$, 
%\eeq
with analogous operators for $B_{s}$--$\bar B_s$, $K$--$\bar K$ and $D$--$\bar D$ mixing.

The Wilson coefficients $C_i(\mu)$ are determined in a two-step procedure.
After matching at the high scale $\mu_h \sim m_W$, they are RG evolved down
to the low scale. For brevity we will discuss here only the case of $B$ decays, 
where the low scale is of the order 
$\mu \sim m_b$. 

The weak scale perturbative matching
is performed in a mass-independent scheme such as
$\overline{MS}$, giving the Wilson coefficients expanded in 
$\alpha_s(\mu_h)$ and $\alpha_{\rm em}(\mu_h)$
\beq
\begin{split}
C_i(\mu_h) &= C_i^{(0)} + \frac{\alpha_s(\mu_h)}{4\pi} C_i^{(1)}(\mu_h) \\
&+ 
\Big(\frac{\alpha_s(\mu_h)}{4\pi}\Big)^2 C_i^{(2)}(\mu_h) + 
\frac{\alpha_{\rm em}(\mu_h)}{4\pi} C_i^{(1e)}(\mu_h) +  \cdots
\end{split}
\eeq
At tree level all Wilson coefficients vanish apart from $C_2^{p(0)} = 1$. The
matching calculation includes both hard gluon and electroweak loop effects.

The Wilson coefficients are evolved from $\mu_h$ down
to a typical hadronic scale $\mu \sim m_b$ by solving  the Renormalization 
Group Equation (RGE)
\begin{eqnarray}
\mu \frac{d}{d\mu} \vec C(\mu) = (\hat \gamma)^T \vec C(\mu),
\end{eqnarray}
where the anomalous dimension matrix is also expanded 
\begin{eqnarray}\label{gamdef}
\hat \gamma = \frac{\alpha_s}{4\pi} \hat \gamma^{(0)}_s + 
\frac{\alpha_s^2}{(4\pi)^2} \hat \gamma^{(1)}_s + 
\frac{\alpha_{\rm em}}{4\pi} \hat \gamma^{(0)}_{\rm em} + \cdots \,.
\end{eqnarray}

The solutions of the RGE are renormalization-scheme and renormalization-scale invariant
to any given order only provided that the orders in matching and running are
chosen appropriately. 
Keeping the tree level matching $C_i^{(0)}$ and the one-loop order anomalous dimension
matrix $\hat \gamma^{(0)}$ yields the so-called leading-log approximation (LL) for the
Wilson coefficients. For instance the LL  values for tree and QCD penguin operators, $i=1,\dots, 6$,  are
$\bar C_i(\mu=4.8~{\rm GeV})=\{-0.248,1.107,0.011,-0.025,0.007,-0.031\}$.
The next-to-leading approximation (NLL) corresponds to keeping the
one-loop matching conditions $C_i^{(1)}$ and the two-loop anomalous dimension 
matrix $\hat\gamma^{(1)}$, and so on.  The NLL  values for  $i=1,\dots, 6$ are
$\bar C_i(\mu=4.8~{\rm GeV})=\{-0.144, 1.055 , 0.011 ,-0.034,0.010,-0.039\}$ .

Note that for higher loop calculations it has become
customary to use a different operator basis than that of Eq.~\eqref{Oi}. 
In the basis introduced by \textcite{Chetyrkin:1996vx}, 
$\gamma_5$ does not appear explicitly (except in the magnetic operators), 
which allows a use of dimensional regularization 
with fully anticommuting $\gamma_5$,  simplifying multiloop calculations. 
The present status of the coefficients entering the RGE 
is as follows.

%\item 
%$\bullet$ 
The two-loop matching corrections to the Wilson coefficients $C_i(\mu_h)$
were computed by \textcite{Bobeth:1999mk}. The three-loop matching correction
to the coefficient of the dipole operator
$C_{7}(\mu_h)$ was recently obtained by \textcite{Misiak:2004ew}.
%\item 
%$\bullet$ 
The leading 2-loop electroweak corrections to the Wilson coefficient of the
dipole operator $C_7$ were computed by \textcite{Czarnecki:1998tn}, 
while the leading electromagnetic logs  
$\alpha_{\rm em} \alpha_s^n \log^{n+1}(m_W/m_b)$ were 
resummed for this coefficient in \textcite{Kagan:1998ym,Baranowski:1999tq}. 
A complete two-loop matching 
of the electroweak corrections was performed by \textcite{Gambino:2000fz,Gambino:2001au}. 
%\item 
%$\bullet$ 
The three-loop anomalous dimension matrix of the four-quark operators was
computed in \textcite{Gorbahn:2004my,Gorbahn:2005sa}.

%\end{itemize}

The presence of new physics (NP) has several effects 
on the form of the effective Hamiltonian in Eq.~\eqref{HW}. 
First, it shifts the values of the Wilson coefficients away from the SM values
\beq
\lambda_p^{(q)}C_i= \lambda_p^{(q)}C_i^{\rm SM}+C_i^{\rm NP}.
\eeq
Note that the NP contribution to the Wilson coefficient 
may not obey the CKM hierarchy of the SM term, 
and can also depend on new weak phases. 
Second, NP contributions can also enlarge the basis of the operators, 
for instance by introducing operators of opposite chirality 
to those in Eq.~\eqref{HW}, 
or even introducing four quark operators with scalar interactions. 
We will discuss the two effects in more detail in the subsequent
subsections, where we focus on particular NP models.

\subsection{Minimal Flavor Violation}
\label{Sec:MFV}

In SM the global flavor symmetry group 
\beq\label{GF}
G_F=U(3)_Q\times U(3)_{U_R}\times U(3)_{D_R}\times U(3)_{L_L}\times 
U(3)_{E_R}
\eeq
 is broken only by the Yukawa couplings, $Y_U, Y_D$, and $Y_E$ (with $U(1)$'s also broken by anomalies). 
In a generic extension of SM, on the other hand, additional sources of flavor 
violation can appear.  If the extended particle spectrum is to solve the 
hierarchy problem (for instance by doubling of the spectrum as in MSSM) 
these new particles have to have masses comparable to the electroweak scale. 
This then leads to a clash with low energy flavor physics experimental data. 
Namely, virtual exchanges of particles with TeV masses and with completely 
generic flavor violating couplings lead to flavor changing neutral currents (FCNCs)
that are orders of magnitude larger than observed, cf. Eq. \eqref{Gen:NP}. 

TeV scale NP therefore cannot have a generic flavor structure. 
On the other hand, it cannot be completely flavor blind either
since 
the Yukawa couplings in SM already break flavor symmetry. 
This breaking will then translate to a NP sector through 
renormalization group running as long as the NP fields couple to the SM fields. 
Thus, the minimal choice for the flavor violation in the extended theory 
is that its flavour group is also broken {\it only} by the SM Yukawa couplings.
This is the Minimal Flavor Violation (MFV) hypothesis 
\cite{Chivukula:1987py,Hall:1990ac,Ciuchini:1998xy,Buras:2000dm,D'Ambrosio:2002ex,Buras:2003jf}.

The 
idea of MFV was formalized by \textcite{D'Ambrosio:2002ex} by 
promoting the Yukawa couplings to spurions that transform under flavor group $G_F$. 
Focusing only on the quark sector, the transformation properties under 
$SU(3)_Q\times SU(3)_{U_R}\times SU(3)_{D_R}$ are 
\begin{eqnarray}
  Y_U\sim (3,\bar 3,1), \qquad Y_D(3,1,\bar 3)
\end{eqnarray}
so that the Yukawa interactions 
\begin{eqnarray}
{\cal L}_{Y}=\bar Q_L Y_D d_R H+\bar Q_L Y_U u_R H^c+h.c,
\end{eqnarray}
are now formally invariant under $G_F$, Eq.~\eqref{GF}. 
Above we suppressed the generation indices on the left-handed quark isodoublet 
$Q_i = (u_L, d_L)_i$, on right-handed quark isosinglets $u_R,d_R$ and on Yukawa
matrices $Y_{U,D}$, while for the Higgs isodoublet the notation $H^c=i
\tau_2 H^*$ was used. 
Minimally flavor violating NP is also formally invariant under $G_F$ 
with the breaking coming only from insertions of spurion fields $Y_{U,D}$. 
Integrating out the heavy fields 
({\it i.e.} the NP fields, Higgs, top, $W$ and $Z$) 
one then obtains the low-energy EFT that is also invariant under $G_F$. 

A particularly convenient basis for discussing transitions between 
down-type quarks is the basis in which the Yukawa matrices take the following form
\begin{eqnarray}\label{basis}
Y_D=\lambda_D,  \qquad Y_U=V^\dagger \lambda_U.
\end{eqnarray}
Here $\lambda_{D,U}$ are diagonal matrices proportional to the quark masses and $V$ is the CKM matrix.  In a theory with a single Higgs (or in a small $\tan \beta$ regime
of the 2HDM or MSSM) one has $\lambda_D\ll 1$, $\lambda_U\sim {\rm diag}(0,0,1)$. The dominant non-diagonal structure for down-quark processes is thus provided by $Y_UY_U^\dagger$ transforming as $(3\times \bar 3, 1, 1)$. Its off-diagonal elements exhibit the CKM hierarchy $(Y_UY_U^\dagger)_{ij}\sim\lambda_t^2V_{ti}^*V_{tj}$. Furthermore, multiple insertions of $Y_UY_U^\dagger$ give $(Y_UY_U^\dagger)^n\sim \lambda_t^{2n} V_{ti}^*V_{tj}$ and are thus equivalent to a single $Y_UY_U^\dagger$ insertion, while multiple insertions of $Y_D$ beyond leading power can be neglected. 
This makes the MFV framework very predictive.

The particular realization of MFV outlined above is the so-called 
constrained minimal flavor violation (cMFV) framework 
\cite{Buras:2000dm,Blanke:2006ig}. 
The assumptions that underlie cMFV are (i) the SM fields are the only 
light degrees of freedom in the theory, (ii) there is only one light Higgs 
and (iii) the SM Yukawas are the only sources of flavor violation. 
The NP effective Hamiltonian following from these assumptions is
\begin{eqnarray}\label{weakMFV}
{\cal H}_{\rm eff}^{\rm NP}=\dfrac{C_i^{\rm NP}}{\Lambda_{\rm NP}^2} \big(V_{t i}^*V_{tj}\big) Q_i,
\end{eqnarray}
where $Q_i$ are exactly the same operators as in the SM effective weak Hamiltonian of Eq.~\eqref{HW}. 
[This is sometimes taken to be the definition of cMFV \cite{Blanke:2006ig,Buras:2000dm,Buras:2003jf}]. 
Note that Eq.~\eqref{weakMFV} provides a very nontrivial constraint.
 For instance already in two-Higgs doublet models 
or in MFV MSSM even with small $\tan \beta$, 
sizeable contributions from operators with non-SM chiral structures 
in addition to Eq.~\eqref{weakMFV} are possible (see next sections).

In cMFV the Wilson coefficients of the weak operators
deviate from the SM values, but remain real, 
so that no new sources of CP violation 
are introduced. 
In phenomenological analyses it is also useful to assume that NP contributions 
are most prominent in the EWP Wilson coefficients ($C_{8,\dots,10}$),
the dipole operators ($C_{7\gamma, 8g}$), 
and the four-fermion operators involving quarks and leptons 
($C_{9\ell}, C_{10\ell}, C_{11\nu}$). 
The rationale for this choice is that the Wilson coefficients of these 
operators are small in the SM, so that NP effects can be easier to spot. 
In contrast, NP effects are assumed to be negligible in the tree, $C_{1,2}$, 
and QCD penguin operators, $C_{3,\dots,6}$. 

Because cMFV is a very constrained modification of the weak Hamiltonian 
Eq.~\eqref{weakMFV}, one can experimentally distinguish it from other 
BSM scenarios by looking at the correlations between observables in $K$ and 
$B$ decays. A sign of cMFV would be a deviation from SM predictions that 
can be described without new CP violating phases and without enlarging the SM 
operator basis. A deviation in $\beta$ from 
$B^0 \to \phi K_S$ (see Section~\ref{time-dependent})
on the other hand would rule out the cMFV framework. 

\begin{figure}
\includegraphics[width=\columnwidth]{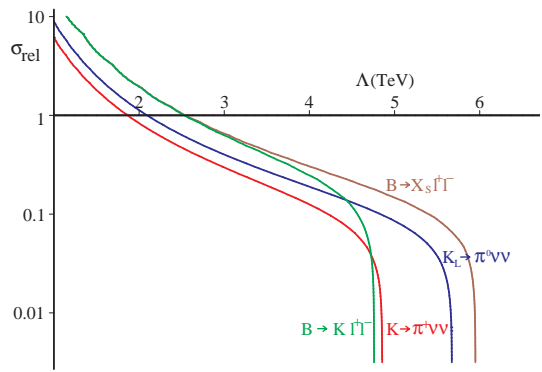}
\caption{
  Expectations for bounds on $\Lambda_{NP}$ for 
  $(\bar Q_L Y_U^\dagger Y_U \gamma_\mu Q_L)(\bar L_L\gamma_\mu L_L)$ 
  that would follow from relative experimental precision $\sigma_{\rm rel}$, 
  with currently expected theoretical uncertainties~\cite{D'Ambrosio:2002ex}.
}\label{Fig:MFV}
\end{figure}

How well one can bound NP contributions 
depends both on the experimental and theoretical errors. 
The observables in which theoretical errors are below $10\%$ 
have a potential to probe $\Lambda_{NP}\sim 10$ TeV (taking $C_i^{NP}=1$). 
The most constraining FCNC observable at present is the 
inclusive $B\to X_s\gamma$ rate  with the experimental and theoretical 
error both below $10\%$ after the recent (partially completed) NNLO
calculation \cite{Becher:2006pu,Misiak:2006zs,Misiak:2006ab}. 
Using older theoretical predictions and experimental data, 
the $99\%$ confidence level (CL) bound is $\Lambda_{NP}>6.4 (5.0)$ TeV in the case of 
constructive (destructive) interference with SM~\cite{D'Ambrosio:2002ex}.  
Constraints from other FCNC observables are weaker. 
As an illustrative example we show in Figure \ref{Fig:MFV} 
expected $\Lambda_{NP}$ bounds following from observables sensitive to 
the operator $(\bar Q_L Y_U^\dagger Y_U \gamma_\mu Q_L)(\bar L_L\gamma_\mu L_L)$
for improved experimental precisions [see also \cite{Bona:2005eu,Bona:2007vi}].

The MFV hypothesis has been extended to the leptonic sector (MLFV) in
\textcite{Cirigliano:2006su,Cirigliano:2005ck}. In MLFV the most sensitive 
FCNC probe in the leptonic sector is $\mu \to e\gamma$, while 
$\tau\to \mu\gamma$ could be suppressed below the SFF sensitivity. 
The MLFV scenario also predicts correlations between the rates of 
various LFV processes.
Studies of LFV in tau decays at a SFF are therefore crucial to test
the MLFV framework (see Section~\ref{sec:LFV}).

An extension of MFV to the Next-to-Minimal Flavor Violation (NMFV)
hypothesis was put forward in \textcite{Agashe:2005hk} by demanding that NP 
contributions  only roughly obey the CKM hierarchy, and in particular 
can have ${\cal O}(1)$ new weak phases.
This definition of NMFV is equivalent to having an additional spurion
$Y_S$ transforming as $Y_UY_U^\dagger$ or $Y_DY_D^\dagger$ under $G_F$, 
where the transformation between $Q_L$ weak basis and the $Y_S$ 
eigenbasis is demanded to be aligned with the CKM matrix. 
The consequences of $Y_S$ transforming differently under $G_F$ 
than the SM Yukawas have been worked out by \textcite{Feldmann:2006jk}.

\subsection{Two-Higgs Doublet Models}
\label{sec:2HDM}
The scalar sector of SM contains only a single scalar electroweak doublet.
This is no longer true (i) in low energy supersymmetry, 
where holomorphism of the superpotential requires at least two scalar doublets;
(ii) in many of the solutions to the strong CP problem~\cite{Peccei:1977hh,Peccei:1977ur}; 
(iii) in models of spontaneous CP breaking~\cite{Lee:1973iz}. 
Here we focus on the simplest extension, the two-Higgs doublet model (2HDM), 
where the scalar sector is composed of two Higgs fields, $H_U, H_D$,
transforming as doublets under $SU(2)_L$. 
More complicated versions with Higgs fields carrying higher weak isospins are possible, but are also more constrained by electroweak precision data, in particular that the $\rho$ parameter is equal to one up to radiative corrections.
The 2HDM model is also a simplified version of the MSSM Higgs sector, 
to be considered in the next subsection.

The Yukawa interactions of a generic 2HDM are
\beq\label{2HDML}
\begin{split}
{\cal L} =& \bar Q_L f^{D} H_D d_R + \bar Q_L f^{U}H_D^c u_R \\
&+ 
\bar Q_L g^{U} H_U u_R + \bar Q_L g^{D} H_U^c d_R + 
\mbox{ h.c.},
\end{split}
\eeq
where $H_{D,U}^c=i\tau_2 H_{D,U}^*$, and the generation indices are suppressed. 
If all the $3\times 3$ Yukawa matrices $f^{D,U}$ and $g^{D,U}$ are nonzero 
and take generic values, this leads to tree level FCNCs from neutral Higgs 
exchanges that are unacceptably large. 

Tree level FCNCs are not present, if up and down quarks
couple only to one Higgs doublet \cite{Glashow:1976nt}. 
This condition can be met in two ways, which also define two main classes of 2HDM.
In type-I 2HDM both up- and down-type quarks couple only to one of the two Higgses (as in SM), {\it i.e.} either $g^{U} = g^{D} = 0$ or $f^{U} = f^{D} = 0$.
 In type-II 2HDM up- and down-type quarks couple to two separate Higgs doublets, {\it i.e.} $f^{U} = g^{D} = 0$ \cite{Haber:1978jt}. 
 
The remaining option that all $f^{D,U}$ and $g^{D,U}$ are nonzero is known as type-III 2HDM \cite{Atwood:1996vj,Cheng:1987rs,Hou:1991un}. 
The tree level flavor violating couplings to neutral Higgs then need to be suppressed in some other way, for instance by postulating a functional dependence of the couplings $f_{U,D}, g_{U,D}$ on the quark masses \cite{Cheng:1987rs,Antaramian:1992ya}. 
A particular example of type-III 2HDM is also the so-called T2HDM \cite{Das:1995df,Kiers:1998ry}, 
which evades the problem of large FCNC effects in the first two generations
by coupling $H_D$ to all quarks and leptons except to the top quark, while $H_U$ couples only to the top quark. 

After electroweak symmetry breaking the fields $H_{U,D}$ acquire 
vacuum expectation values $v_{1,2}$ 
\begin{eqnarray}
\langle H_U \rangle = \Big(
\begin{array}{c}
\frac{1}{\sqrt2} v_2 \\
0 \\
\end{array}
\Big)\,,\qquad
\langle H_D \rangle = \Big(
\begin{array}{c}
0 \\
\frac{1}{\sqrt2} v_1 \\
\end{array}
\Big)\,,
\end{eqnarray}
where it is customary to define $\tan\beta = v_2/v_1$, while $v_1^2+v_2^2=v^2$, with $v = 246$ GeV.
In type-II 2HDM the up and down quark masses are $m_t \sim v_2, m_b \sim v_1$.
The large hierarchy $m_t/m_b \sim 35$ can thus be naturally explained in this model by a large ratio of the vevs $v_2/v_1 = \tan\beta \gg 1$.

The physical degrees of freedom in 2HDM scalar sector consist of one charged
Higgs boson $H^\pm$, two  CP-even neutral Higgs bosons $H_{1,2}$, and one CP-odd Higgs boson $A$. The phenomenology of the 2HDM of type-I, II is similar to that of the SM with the
addition of the charged Higgs flavor-changing interactions. 
 These $S\pm P$ couplings are for type-II 2HDM given by
\beq\label{H+-coupl}
\begin{split}
& \frac{H^+}{v}   \Big[\tan\beta
\bar u_{L} V M_{D} d_{R} +\frac{1}{\tan\beta}
\bar u_{R} M_{U} V d_{L} \Big] + \mbox{h.c.},
\end{split}
\eeq
while the type-I 2HDM interactions are obtained by replacing 
$\tan\beta\to - 1/\tan\beta$ in the first term. 
The matrix $V$ is the same CKM matrix as in the $W^\pm$ couplings, while
$M_{D(U)}$ are diagonal matrices of down (up) quark masses. As mentioned before, type-III
2HDM contains in addition also flavor violating neutral Higgs couplings.

\begin{figure}
\begin{center}
   \epsfxsize=5.5cm
   \centerline{\epsffile{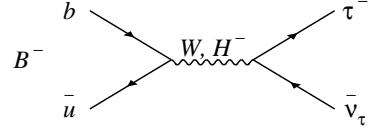}}
\vspace*{-0.7cm}
\end{center}
\caption{\label{fig:B2taunu} Contribution to the
$B\to \tau\bar\nu_\tau$ decay mediated by $W, H^\pm$ exchange in 2HDM.}
\end{figure}

The most sensitive probes of interactions in Eq.~\eqref{H+-coupl} 
are processes where $H^\pm$ can be exchanged at tree level: 
semileptonic $b \to c \tau \bar \nu_\tau$ decays
and the weak annihilation decay $B^-\to \tau\bar\nu_\tau$, 
see Fig.~\ref{fig:B2taunu}, 
giving a constraint on the ratio $m_{H^+}/\tan\beta$~\cite{Grossman:1994ax,Kiers:1997zt}. 

The inclusive semitauonic decays have been studied 
at LEP~\cite{Barate:2000rc,Abbiendi:2001fi}.
Assuming type-II 2HDM, these give a 90\% CL upper bound of 
$\tan\beta/M_{H^+} \leq 0.4 \ {\rm GeV}^{-1}$.
A comparable constraint on $\tan\beta /m_{H^+}$ can be obtained from exclusive
$B\to D^{(*)}\tau \bar \nu_\tau$ decays~\cite{Tanaka:1994ay,Chen:2006nua,Nierste:2008qe}.
First observations of these decays have recently been made at the 
$B$ factories~\cite{Matyja:2007kt,Aubert:2007ds}, 
with significant improvements in precision expected at a SFF. Furthermore, the study of $B\to D \tau \nu_\tau$ decay distributions can discriminate between $W^+$ and $H^+$
contributions \cite{Kiers:1997zt,Miki:2002nz,Grzadkowski:1992qj,Nierste:2008qe}. In particular, in the decay chain 
 $\bar B\to D  \bar \nu_\tau\tau^-[\to\pi^- \nu_\tau]$ the differential distribution with respect to the angle between
three-momenta $\vec p_D$ and $\vec p_\pi$ can be used to measure both the magnitude and the weak phase of the 
charged Higgs scalar coupling to quarks \cite{Nierste:2008qe}.

\begin{figure}[b]
\begin{center}
\includegraphics[width=6cm]{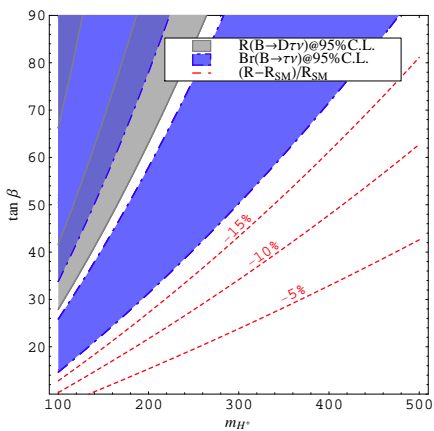}
\end{center}
\caption{\label{fig:MHtanb} 
  Exclusion region in $(M_{H^+}, \tan\beta)$ due to present data on 
$B\to \tau\nu$ (blue) and $R={\cal B}(B\to D\tau\nu)/{\cal B}(B\to De\nu)$ (gray).
Red dashed lines represent percentage deviation from the SM prediction of
$R$ in the presently allowed region \cite{Kamenik:2008tj}.
}
\end{figure}

In the annihilation decay $B^-\to \tau\bar\nu_\tau$,
$H^+$ exchange may dominate over helicity suppressed $W^+$ exchange contribution. 
The two contributions interfere destructively~\cite{Hou:1992sy}.
Recent measurements~\cite{Ikado:2006un,Aubert:2007xj,Aubert:2007bx} give
\begin{eqnarray}
R_{B\tau\nu} = \frac{{\cal B}^{\rm exp}(B^- \to \tau \nu)}
{{\cal B}^{SM}(B^- \to \tau \nu)} = 0.93 \pm 0.41 \, ,
\end{eqnarray}
compatible with the presence of $H^+$ contribution.
The present status of the constraints on
$(M_{H^+},\tan\beta)$ from the tree level processes 
$B\to \tau\nu$ and $B\to D\ell\bar\nu_\ell$, $\ell=e,\tau$ is shown 
in Fig.~\ref{fig:MHtanb}. 
More precise measurements of these mode, 
and of the complementary leptonic decay $B^- \to \mu^-\nu_\mu$,
will be possible at a SFF.

Loop mediated FCNC such as $B_s$--$\bar B_s$ mixing
and $b\to s\gamma$ decays can also constrain the parameters of 2HDM models. 
In $b\to s\gamma$ the charged Higgs boson contribution comes from penguin 
diagrams with top and $H^+$ running in the loop, which are known at 
NLO ~\cite{Ciuchini:1997xe,Borzumati:1998tg} [LO calculations were done 
by~\textcite{Ellis:1985xy,Hou:1987kf,Grinstein:1988me}]. In type-I 2HDM the 
$W^+$ and $H^+$ contributions to the electromagnetic dipole Wilson coefficient 
$C_{7\gamma}(\mu)$ can interfere with either sign, while in type-II
2HDM they always interfere constructively. 
The present WA of the branching fraction 
${\cal B}(B\to X_s\gamma) = (3.55 \pm 0.24 ^{+0.09}_{-0.10} \pm 0.03) \times 10^{-4}$ 
implies the lower bound $M_{H^+}>300$ GeV~\cite{Misiak:2006zs}.

Type-III models have a richer flavor violating structure with FCNC transitions 
generally allowed at tree level. Here we will focus on type-III models where
the Peccei-Quinn symmetry violating terms $g^D$ and $f^U$  in Eq.~(\ref{2HDML}) are only a small peturbation. 
These models are close to  a type-II 2HDM and correspond to 
the situation encountered in the MSSM.  
We further restrict ourselves to the conservative case of MFV. 
The matrices $g^D$ and $f^U$ are functions of large Yukawa matrices 
$Y^U\equiv g^U$ and $Y^D\equiv f^D$ in accordance with spurion analysis 
using flavor group Eq.~\eqref{GF}. 
The most general Yukawa term involving down-type quarks in a type-III 2HDM 
with MFV is then
\cite{D'Ambrosio:2002ex}
\beq\label{HdtypeIII}
\begin{split}
{\cal L}_{\epsilon Y_D} =&
\bar Q_{L} \Big[ H_D + \big(\epsilon_0 + \epsilon_1 \Delta + 
\epsilon_2 Y_U Y_U^\dagger +\epsilon_3 Y_U Y_U^\dagger\Delta \\
&+\epsilon_4 \Delta Y_U Y_U^\dagger\big) H_U^c \Big] Y_D d_R + \mbox{ h.c.}
\end{split}
\eeq
with $\epsilon_i$ some unknown coefficients, 
where we have used the mass eigenstate basis in which 
$Y_U$ and $Y_D$ have the form of Eq.~\eqref{basis}. 
In particular  $Y_D$ is diagonal, 
so that  $Y_DY_D^\dagger\propto {\rm diag}(0,0,1)\equiv \Delta$. 
The additional couplings to $H_U^c$ in Eq.~\eqref{HdtypeIII} 
introduce new
flavor changing vertices both in the charged currents $W^\pm qq$ and charged
Higgs vertices $H^\pm qq$. 
In addition, new FCNC couplings to the neutral Higgses 
$H^0, h^0, A^0$ are  introduced. 
Integrating out the heavy Higgs fields
gives new scalar operators mediating FCNC transitions. 
These can be especially important in the large $\tan\beta$ regime, 
where $\epsilon_i \tan\beta$ can be ${\cal O}(1)$.

The large $\tan\beta$ limit of the MFV hypothesis 
has two important consequences for the low energy 
effective weak Hamiltonian of Eq.~\eqref{weakMFV}: 
(i) the basis of FCNC operators is larger than in the SM and includes 
scalar operators arising from tree level FCNC neutral Higgs exchanges, 
and (ii) the $\Delta$ insertions Eq.~\eqref{HdtypeIII} 
decouple the third generation decays from the first two. 
The correlation between $B$ and $K$ meson observables present in the 
low $\tan\beta$ MFV scenario (cMFV) discussed in subsection \ref{Sec:MFV}, 
is thus relaxed. For instance, the new contributions in Eq.~\eqref{HdtypeIII} 
allow us to modify separately $\Delta M_{B_d}$ and $\epsilon_K$.

The effect of flavor violation in the large $\tan\beta$
limit is particularly dramatic for 
$b \to s\ell^+\ell^-$ transitions and $B_{(s)} \to \ell^+\ell^-$ decays.
These are helicity suppressed in SM, but now receive tree level
contributions from neutral Higgs exchange. 
An enhancement of $B\to \ell^+ \ell^-$ by two orders of magnitude is then, 
in general, possible.
Conversely, experimental data on these processes translate into constraints in 
the $(M_{H^+}/\tan\beta, \epsilon_i \tan\beta)$ plane~\cite{D'Ambrosio:2002ex}. 
These in turn impose useful constraints on the underlying 
physics producing the couplings $\epsilon_i$. This program is especially
powerful in the context of a specific model, for instance in the
case of a supersymmetric theory like the MSSM discussed in the next section

While $B\to \ell^+ \ell^-$ has already been searched for 
at the Tevatron~\cite{Aaltonen:2007kv,Abazov:2007iy}
and will be searched for at LHCb~\cite{Buchalla:2008jp}, 
a SFF has an important role in pinning down the large $\tan \beta$ scenario 
by (i) precisely measuring also non-helicity suppressed decays 
({\it e.g.} $B\to (K,K^*) \ell^+\ell^-$ where ${\cal O}(10\%)$ 
breakings of flavor universality would be expected~\cite{Hiller:2003js}), 
and (ii) by measuring $B\to X_s \tau^+\tau^-$ and $B\to \tau^+\tau^-$~\cite{Isidori:2001fv}. 
In a completely general large $\tan \beta$ MFV analysis using EFT there are no 
correlations between $B\to \ell\nu$, $B\to \ell^+\ell^-$, 
$\Delta M_{B_s}$ and $B\to X_s\gamma$, 
but these do exist in a more specific theory, 
for instance in MFV MSSM with large $\tan\beta$~\cite{D'Ambrosio:2002ex,Isidori:2007jw,Isidori:2006pk,Lunghi:2006uf}. 
In this scenario one gets $\sim (10\%-40\%)$ suppression of 
${\cal B}(B^+\to \tau^+\nu)$, enhancement of $(g-2)_\mu$, 
SM-like Higgs boson with $m_{h^0}\sim 120$ GeV and small 
effects in $\Delta M_{B_s}$ and ${\cal B}(B\to X_s\gamma)$ 
quite remarkably in agreement with the present tendencies in the data~\cite{Isidori:2007jw,Isidori:2006pk}.

\subsection{Minimal Supersymmetric Standard Model}
Low energy supersymmetry (SUSY) offers a possible solution to the hierarchy problem. 
In SUSY the quadratically divergent quantum corrections to the
scalar masses (in SM to the Higgs boson mass) are cancelled by introducing 
superpartners with opposite spin-statistics for each of the particles. 
The simplest supersymmetrization of the Standard Model is the so-called 
Minimal Supersymmetric Standard Model (MSSM), to which we 
restrict most of the discussion in the following.
(For more extended reviews see, {\it e.g.},
\textcite{Nilles:1983ge,Haber:1984rc,Martin:1997ns,Misiak:1997ei}). 

The matter content of MSSM is shown in Table \ref{table:field}. 
The structure of SUSY demands two Higgs doublets $H_{U,D}$ 
that appear together with their superpartners, Higgsinos $\tilde h_{U,D}$. 
These mix with the fermionic partners 
of the $W$ and $Z,\gamma$ gauge bosons into
the chargino $\tilde \chi^\pm$ and the neutralinos $\tilde\chi^0$. 
The superpartner of the gluon is the gluino, $\tilde g$.
In addition, there are also the scalar partners of the fermion fields with 
either chirality, the squarks $\tilde q_R, \tilde q_L$, 
and the sleptons and sneutrinos $\tilde e_L, \tilde e_R, \tilde \nu$.  

\begin{table}[tb]
\begin{center}
\begin{ruledtabular}
\caption{Field content of the Minimal Supersymmetric Standard Model.
The spin-$0$ fields are complex scalars, and the spin-$1/2$ fields are 
left-handed two-component Weyl fermions. Last column gives gauge representations in a
$(SU(3)_C ,\, SU(2)_L ,\, U(1)_Y)$ vector. In addition there are also fermionic superpartners
of gauge bosons: gluino, wino and bino. \label{table:field}}
\begin{tabular}{ccccc}
\multicolumn{2}{c}{Superfield notation} 
& spin 0 & spin 1/2 & gauge repr.
\\  \hline
squarks, quarks & $ {\cal Q}$ & $({\widetilde u}_L\>\>\>{\widetilde d}_L )$&
 $(u_L\>\>\>d_L)$ & $(\>{\bf 3},\>{\bf 2}\>,\>{\frac{1}{6}})$
\\
($\times 3$ families) & $\bar {\cal U}$
&${\widetilde u}^*_R$ & $u^\dagger_R$ & 
$(\>{\bf \overline 3},\> {\bf 1},\> -{\frac{2}{3}})$
\\ & $\bar {\cal D}$ &${\widetilde d}^*_R$ & $d^\dagger_R$ & 
$(\>{\bf \overline 3},\> {\bf 1},\> {\frac{1}{3}})$
\\  \hline
sleptons, leptons & ${\cal L}$ &$({\widetilde \nu}\>\>{\widetilde e}_L )$&
 $(\nu\>\>\>e_L)$ & $(\>{\bf 1},\>{\bf 2}\>,\>-{\frac{1}{2}})$
\\
($\times 3$ families) & $\bar {\cal E}$
&${\widetilde e}^*_R$ & $e^\dagger_R$ & $(\>{\bf 1},\> {\bf 1},\>1)$
\\  \hline
Higgs, Higgsinos &${\cal H}_U$ &$(H_u^+\>\>\>H_u^0 )$&
$(\widetilde h_u^+ \>\>\> \widetilde h_u^0)$& 
$(\>{\bf 1},\>{\bf 2}\>,\>+{\frac{1}{2}})$
\\ &${\cal H}_D$ & $(H_d^0 \>\>\> H_d^-)$ & $(\widetilde h_d^0 \>\>\> \widetilde h_d^-)$& 
$(\>{\bf 1},\>{\bf 2}\>,\>-{\frac{1}{2}})$
\end{tabular}
\end{ruledtabular}
\end{center}
\end{table}

The superpotential describing the Yukawa couplings of the two Higgs fields to 
the quark and lepton chiral superfields is 
\beq\label{MSSMsuperpotential} 
\begin{split}
{\cal W} = &Y_U^{ij} {\cal H}_U {\cal Q}_i \bar {\cal U}_j + Y_D^{ij} {\cal H}_D {\cal Q}_i \bar {\cal D}_j\\
&+ Y_L^{ij} {\cal H}_D {\cal L}_i \bar {\cal E}_j + \mu {\cal H}_U {\cal H}_D.
\end{split}
\eeq
The Yukawa matrices $Y_U, Y_D, Y_L$ act on the family indices $i,j$. 
The last term is the so-called $\mu$ term coupling the two Higgs fields. 
The above superpotential is the most general one that conserves $R-$parity 
under which SM particles are even, while the superpartners are odd. 
$R$-parity ensures $B$ and $L$ quantum numbers conservation at a renormalizable level.
Comparing the superpotential of Eq.~\eqref{MSSMsuperpotential} 
with the 2HDM Yukawa interactions in Eq.~\eqref{2HDML}, 
we see that at tree level this gives quark-Higgs couplings of a type-II 2HDM.
Loop corrections induced by the $\mu$ term, however, introduce also the
Higgs-quark couplings of the ``wrong-type'', 
effectively changing the interaction into a type-III 2HDM
(see Fig.~\ref{fig:YdH2}).

SUSY predicts fermion-boson mass degeneracy, 
which is not observed in Nature, so SUSY must be broken. 
The required breaking needs to be soft, 
{\it i.e.} only from super renormalizable terms, 
in order not to introduce back quadratic divergences 
and sensitivity to the high scale. 
The general soft SUSY breaking Lagrangian in the squark sector of MSSM 
is then (for a review see, {\it e.g.} \textcite{Chung:2003fi})
\beq
\begin{split}
{\cal L}_{\rm soft} &= (M_{\tilde Q}^2)_{ij} (\tilde u_{Li}^\dagger \tilde u_{Lj} +
\tilde d_{Li}^\dagger \tilde d_{Lj} ) \\
&+ (M_{\tilde U}^2)_{ij} \tilde u_{Ri}^\dagger \tilde u_{Rj}  + (M_{\tilde D}^2)_{ij} 
\tilde d_{Ri}^\dagger \tilde d_{Rj} \\
&+  (A_U)_{ij} \widetilde Q_i  H_U \tilde u_{Rj}^*
+ (A_D)_{ij} \widetilde Q_i  H_D \tilde d_{Rj}^*, 
\end{split}
\eeq
with $\widetilde Q_i=(\tilde u_L,\tilde d_L)$ and $H_{U,D}$ Higgs doublets. 
The precise form of the soft squark masses 
$M_{\tilde Q}, M_{\tilde U}, M_{\tilde D}$ and the trilinear terms 
$A_U, A_D$ depends on the specific mechanism which breaks SUSY. 
In its most general form the soft SUSY breaking introduces a large number of
unknown parameters which can induce large observable FCNC effects. 
A detailed counting gives that the flavor sector of the MSSM contains 69
real parameters and 41 phases~\cite{Dimopoulos:1995ju,Haber:1997if}, 
compared with nine quark and lepton masses, 
three real CKM angles and one phase in the SM.
The generically large FCNCs from soft SUSY breaking 
is known as the SUSY flavor problem, and to solve it any realistic SUSY model 
must explain the observed FCNC suppression. 
We address this issue next.

\subsubsection{Flavor violation in SUSY}
In MSSM there are two main sources of flavor violation beyond the SM: 
i) if the squark and slepton mass matrices are neither flavor universal 
nor are they aligned with the quark or the lepton mass matrices, 
and ii) the flavor violation that is induced by the wrong-Higgs 
couplings to quarks and leptons. 

The first effect is most transparent in the super-CKM basis, 
in which the quark mass matrices are diagonal, while the squark fields 
are rotated by the same matrices that diagonalize the quark masses. 
The squark mass matrices, however, need not be diagonal in this basis 
\beq
{\cal M}_U^2 =
\begin{pmatrix}
M_{U_{LL}}^2 & M_{U_{LR}}^2 \\
M_{U_{LR}}^{2\dagger} & M_{U_{RR}}^2 \\
\end{pmatrix}
,\quad
{\cal M}_D^2 = 
\begin{pmatrix}
M_{D_{LL}}^2 & M_{D_{LR}}^2 \\
M_{D_{LR}}^{2\dagger} & M_{D_{RR}}^2 \\
\end{pmatrix}.
\eeq
Explicitly, the $3\times 3$ submatrices are
\begin{align}
M_{U_{LL}}^2 =& M_{\tilde Q}^2 + M_U^2 + \frac16 M_Z^2 \cos 2\beta 
(3-4\sin^2\theta_W), \label{MULL}\\
M_{U_{LR}}^2 =& M_U (A_U - \mu\cot\beta ), \\
M_{U_{RR}}^2 =& M_{\tilde U}^2 + M_U^2 + \frac23 M_Z^2 \cos 2\beta 
\sin^2\theta_W,\label{MURR}
\end{align}
and similarly for the down squarks. 
While the quark mass matrices $M_{U,D}$ are diagonal in the super-CKM basis, 
the soft breaking terms $M_{\tilde Q}^2$, $M_{\tilde U, \tilde D}^2$ 
and $A_{U,D}$ are not, in general. 
The flavor violation, that in the super-CKM basis resides in the squark sector, 
then translates into flavor violation in the quark processes
through loop effects -- 
in particular, squark-gluino loops 
since the $q\tilde q \tilde g$ coupling is proportional to $g_s$. 

\begin{figure}
\includegraphics[width=4.0cm]{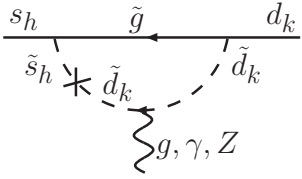}
\caption{Example of squark-gluino $\Delta S=1$ penguin diagram with $h,k={L,R}$.}\label{Fig:squarkgluino}
\end{figure}

In order to suppress FCNC transitions, 
the squark mass matrices $M_{\tilde Q}^2$ and $M_{\tilde U, \tilde D}^2$ must be
either very close to the unit matrix (flavor universality),
or proportional to the quark mass matrices (alignment). 
These properties can arise from the assumed SUSY breaking mechanism, 
for instance in gauge mediated SUSY breaking, 
if the hidden sector scale is below the flavor breaking scale~\cite{Giudice:1998bp}, 
in anomaly mediated SUSY breaking \cite{Randall:1998uk} 
or from assumed universality in SUGRA \cite{Girardello:1981wz,Kaplunovsky:1993rd,Brignole:1993dj}. 
Alternatively, alignment can follow from a symmetry, for instance from 
horizontal symmetries \cite{Nir:1993mx,Dine:1993np,Leurer:1993gy,Barbieri:1995uv}.

The minimal source of flavor violation that is necessarily present 
is due to the Yukawa matrices $Y_U, Y_D$. 
The Minimal Flavor Violation assumption, discussed in section \ref{Sec:MFV}, 
means that these are also {\it the only} sources of flavor violation,  
a scenario that is natural in, for instance, 
models with gauge mediated SUSY breaking. 
The most general structure of soft squark mass terms allowed by MFV is \cite{D'Ambrosio:2002ex}
\beq
\begin{split}
\label{softMFV}
M_{\tilde Q}^2 &= \tilde M^2 \big(a_1 + b_1 Y_U Y_U^\dagger + \cdots \big), \\
M_{\tilde U}^2 &= \tilde M^2 \big( a_2 + b_2 Y_U^\dagger Y_U \big), \\
M_{\tilde D}^2 &= \tilde M^2 \big( a_3 + b_3 Y_D^\dagger Y_D \big), \\
A_U &= A \big(a_4 + b_4 Y_D Y_D^\dagger\big) Y_U, \\
 A_D &= A \big(a_5 + b_5 Y_U Y_U^\dagger\big) Y_D,
\end{split}
\eeq
with $\tilde M^2$ a common mass scale, and $a_i, b_i$ undetermined parameters. 
These can be completely uncorrelated, but are fixed in more constrained scenarios, 
such as the constrained MSSM to be discussed below.

\begin{figure}[b]
\begin{center}
\includegraphics[width=7.cm]{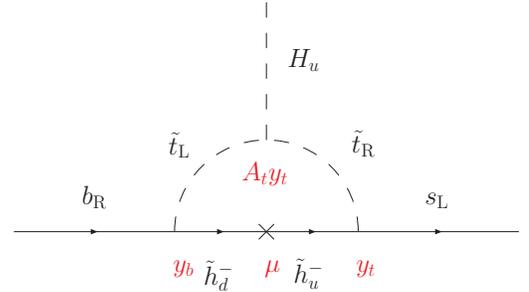}
\end{center}
\caption{\label{fig:YdH2} Flavor changing coupling of the up Higgs-boson $H_u$
to the down type quarks (from \textcite{Lunghi:2006uf})}
\end{figure}

The second source of flavor violation in the MSSM 
is due to the wrong-Higgs couplings, 
{\it e.g.} the $H_U$ coupling to down quarks. 
These are introduced by loop corrections to the $H\bar qq$ vertex. 
There are two such contributions in the MSSM:
the gluino-$\tilde d$ graph, and the Higgsino-$\tilde u$ graph 
(see Figure \ref{fig:YdH2}). 
These induce a type-III 2HDM quark-Higgs interaction Lagrangian of the
form given in Eq.~(\ref{HdtypeIII}). 
The loop induced effects are proportional to $\tan\beta$, 
and thus become important for large $\tan\beta$.

\subsubsection{Constraints on the MSSM parameter space}

The MSSM has 124 free parameters making a direct study of its parameter space intractable. 
Due to the complexity of the problem, 
it is convenient to divide the discussion into two parts. 
We start by first considering a flavor blind MSSM, 
keeping only the SM flavor violation in  the quark sector, 
but neglecting any other sources of flavor violation. 
In the second step we include the two new flavor violating effects 
of the MSSM discussed above.

A particularly simple version of a flavor blind MSSM 
is the so-called constrained MSSM (cMSSM) \cite{Kane:1993td}. 
The soft SUSY breaking masses and trilinear terms are assumed 
to be universal at some high scale, 
for instance at the GUT scale $M_{\rm GUT} \sim 10^{16}$ GeV 
\beq
\begin{split}
& (M_Q^2)_{ij} = (M_U^2)_{ij} = (M_D^2)_{ij} = (M_L^2)_{ij} = 
M_0^2 \delta_{ij},\\
& (A_{U,D})_{ij} = A_0 e^{i\phi_0} (Y_{U,D})_{ij}\,.
\end{split}
\eeq
The gaugino masses are also assumed to be universal at $M_{\rm GUT}$ and equal to $M_{1/2}$.
The cMSSM has only six unknown parameters that can be taken to be: 
the universal gaugino mass $M_{1/2}$, 
the squark and slepton soft breaking mass scale $M_0$, 
the trilinear coupling $|A_0|$, 
the ratio of Higgs vevs $\tan\beta$, and two phases $\phi_\mu = \arg(\mu)$ and
$\phi_A = \arg(A)$. 
In minimal supergravity (mSUGRA), 
an additional constraint $B_0(\tan\beta)=A_0-M_0$ is imposed, 
but the terms cMSSM and mSUGRA are often used interchangeably in the literature. 
The masses and couplings at the electroweak scale are found by
RG running in the MSSM. 
In particular this introduces a flavor structure of the form shown in Eq.~\eqref{softMFV}.  

We consider here only the cMSSM with conserved $R-$parity, 
for which the lightest neutralino (the lightest supersymmetric particle) 
is identified as the dark matter particle.
The experimental constraints on cMSSM parameters are then:
\begin{itemize}

\item The lower bound on light neutral Higgs boson mass, 
$M_{h_0} \geq 120 $ GeV, rules out very low values of $\tan\beta$ 
and constrains a combination of $A_0$ and $M_0^2$ parameters.

\item The anomalous magnetic moment of the muon $a_\mu = \frac12(g-2)_\mu$
appears to differ from the SM prediction at about $3\sigma$ level, 
($a_\mu^{\rm exp} - a_\mu^{SM}) \simeq (27.5\pm 8.3) \times 10^{-10}$ 
\cite{Bennett:2006fi,Miller:2007kk}.
The sign of the difference suggests that $\mu >0$ is strongly favored.
 
\item The radiative decays $b\to s\gamma$. 
The $H^\pm-$top and $W^\pm-$top penguin loops interfere constructively, 
while the chargino diagram has a relative sign given by $-\mbox{sgn}(A_t\mu)$ 
and can thus interfere either constructively or destructively. 
To preserve the good agreement with the SM prediction for $C_7$,
the $H^\pm$ and chargino contributions must cancel to a good approximation,
which requires $\mu >0$. 
An alternative possibility would be a large destructive chargino contribution,
finely tuned to give $C_7=-(C_7)_{\rm SM}$, but this possibility is ruled out 
by the measurement of ${\cal B}(b\to s\ell^+\ell^-)$~\cite{Lunghi:2006uf,Gambino:2004mv}.

\item Electroweak precision observables~\cite{Heinemeyer:2004gx}. 
The good agreement with the SM predictions constrains the mass splitting
of the superpartners, especially in the third generation.

\end{itemize}

Recent detailed cMSSM analyses with special emphasis on $B$ meson phenomenology
were done in ~\cite{Barenboim:2007sk,Goto:2007ee,Ciuchini:2007ha,Carena:2006ai,Ellis:2007fu} [see also earlier works referenced therein]
Here we mention a few implications of these studies that are valid in cMSSM.

The gluino dominance of the RG evolution leads to strong correlations between
gaugino and squark masses at the weak scale. The lower bound on chargino mass
from direct searches then translates to a lower bound of about $250$ GeV 
on the mass of the ligtest squark, the stop. The constraint from $b\to s\gamma$
implies heavy charged Higgs in most of the parameter space, $m_{H^+}\gtrsim 400$ GeV \cite{Bartl:2001wc}.
For large values of $\tan\beta$ smaller masses are possible, if the charged Higgs contribution to $b\to s\gamma$ is cancelled
by the chargino contribution. This simultaneously requires large squark masses
above TeV, while ${\cal B}(B^-\to\tau^-\bar \nu_\tau)$ then
puts a constraint $m_{H^+}\geq 180$ GeV \cite{Barenboim:2007sk}.

The cMSSM contains new sources of CP violation, 
the phases $\phi_\mu$ and $\phi_A$. 
These are constrained by the experimental upper bound on the 
electron electric dipole moment (EDM) $|d^e| \leq 4.0 \times 10^{-27}$~\cite{Regan:2002ta}.
In the MSSM one-loop chargino and neutralino contributions 
lead to a nonzero electron EDM. 
Although each of these two contributions restricts $\phi_\mu, \phi_A$ 
to be very small, cancellations can occur so that $\phi_\mu \leq 0.1$ 
and unrestricted $\phi_A$ are still allowed. In this case  
$A_{\rm CP}(b\to s\gamma)$ can be of order a few percent \cite{Bartl:2001wc}, while if $\phi_\mu$ is set to zero 
the resulting $A_{\rm CP}(b\to s\gamma)$ is hard to distinguish from SM \cite{Goto:2007ee}. 
Measurements of this asymmetry can thus give important information 
about the structure of CP violation beyond the SM.

\subsubsection{Flavor violation in the generic $\tan\beta$ scenario}

For moderate values of $\tan\beta \sim 5$--$15$, 
the only new flavor violating effects are from the off-diagonal terms 
in the squark mixing matrices (in the super-CKM basis). 
It is convenient to parameterize this matrix in a 
way which is simply related to FCNC data. 
Using data to bound the off-diagonal squark mixing matrix elements, 
one would then gain insight into the 
flavor structure of the soft breaking terms.

A convenient way to formulate such constraints makes use of 
the mass insertion approximation in terms of the $\delta_{ij}$ parameters 
\cite{Hall:1985dx,Gabbiani:1996hi}
\begin{eqnarray}
(\delta_{AB}^d)_{ij} = \frac{(M_{D_{AB}}^2)_{ij}}{M_{\tilde q}^2}\,,\quad A,B\in \{L,R\},
\end{eqnarray}
where $M_{\tilde q}$ is an average squark mass. 
Often this is chosen to be the generation dependent quantity,  
$M^2_{\tilde q} =  M_{\tilde q_{Ai}} M_{\tilde q_{Bj}}$.
Analogous parameters can be defined in the up squark sector. 

The most recent constraints on $\delta_{AB}^d$ from \textcite{Ciuchini:2007ha} 
are summarized in Table \ref{table:delta}. 
These bounds are derived in the mass insertion approximation, 
keeping only the dominant gluino diagrams. 
The best constrained parameters are the off-diagonal $\delta^d_{LL}$, 
which contribute to FCNC processes in the down quark sector. 

The $(\delta^d_{AB})_{12}$ parameters (see Table \ref{table:delta}) are 
constrained by measurements in the kaon sector 
of $\Delta M_K, \varepsilon, \varepsilon^\prime/\varepsilon$.
Data on $B_d$--$\bar B_d$ mixing constrain $(\delta_{AB}^d)_{13}$. 
Finally, in the 2--3 sector there are several constraints: 
from rare radiative decays $b\to s\gamma$, $b\to s\ell^+\ell^-$, 
and the recently measured $B_s$--$\bar B_s$ mixing. 
Constraints on the mass insertions in the up sector 
can be derived from recent $D$--$\bar D$ mixing data~\cite{Ciuchini:2007cw}.

\begin{table}[tb]
\begin{center}
\begin{ruledtabular}
\caption{Upper bounds (90\% CL) on the $(\delta_{AB}^d)_{ij}$ squark mixing
parameters obtained from experimental data~\cite{Ciuchini:2007ha}.
\label{table:delta}}
\begin{tabular}{ccccc}
$ij/AB$ & LL & LR & RL & RR \\
\hline
12 & $1.4 \times 10^{-2}$ & $9.0 \times 10^{-5}$ & $9.0 \times 10^{-5}$ &
$9.0 \times 10^{-3}$ \\
13 & $9.0 \times 10^{-2}$ & $1.7 \times 10^{-2}$ & $1.7 \times 10^{-2}$ &
$7.0 \times 10^{-2}$ \\
23 & $1.6 \times 10^{-1}$ & $4.5 \times 10^{-3}$ & $6.0 \times 10^{-3}$ &
$2.2 \times 10^{-1}$ \\
\end{tabular}
\end{ruledtabular}
\end{center}
\end{table}

\subsubsection{Large $\tan\beta$ regime}
\label{Large_tan_beta}

The loop induced couplings of $H_u$ to down-type
quarks render the Yukawa interactions equivalent to a type-III
2HDM, cf. Fig. \ref{fig:YdH2} and Eq.~(\ref{HdtypeIII}). 
These new flavor violating effects are enhanced by $\tan\beta$.
Assuming MFV the new interactions are restricted 
to the form in Eq.~(\ref{HdtypeIII}). 
The $\epsilon_i$ coefficients are calculable from SUSY loop diagrams: 
$\epsilon_0$ contains the effect of the gluino diagram, while 
$\epsilon_{1,2}$ are induced by the Higgsino diagrams of Fig.~\ref{fig:YdH2}.
The induced low energy EFT operators give enhanced contributions 
to several $B$ physics processes. 
We discuss here $B_s \to \ell^+\ell^-$, $B_s$ mixing and $b\to s\gamma$, which
have a distinctive phenomenology in the large $\tan\beta$ scenario with MFV.

The $B_s \to \ell^+\ell^-$ decay receives an enhanced contribution
from tree level exchange of neutral Higgs bosons, which induce scalar 
operators of the form $m_b (\bar b_R s_L) (\bar \ell \ell)$ and 
$m_b (\bar b_R s_L) (\bar \ell\gamma_5 \ell)$. 
The branching fraction of this mode scales as 
${\cal B}(B_s\to \ell^+\ell^-) \sim \tan^6\beta/M_A^4$,
and can thus be easily enhanced by several orders of magnitude 
compared to the SM prediction~\cite{Babu:1999hn,Chankowski:2000ng,Bobeth:2001sq,Bobeth:2002ch}.

Tree level exchange of neutral Higgs bosons induces also the double penguin
operators $(\bar b_R s_L)(\bar b_L s_R)$, 
which contribute to $B_s$--$\bar B_s$ mixing. 
The contributions are enhanced by a factor of $\tan^4\beta$ and decrease 
the $\Delta M_{B_s}$ mass difference compared with the SM \cite{Buras:2001mb}.

The radiative decay $b\to s\gamma$ receives contributions from 
neutral Higgs loops in the large $\tan\beta$ limit. 
An important effect is the presence of corrections of order 
$(\alpha_s\tan\beta)^n$, which can be resummed to all 
orders~\cite{Carena:2000uj,Dedes:2002er,Ellis:2007kb}. 
The effect of the resummation can be appreciable for 
sufficiently large values of $\tan\beta$.

\begin{figure}%[b]
\includegraphics[width=7cm,angle=-90]{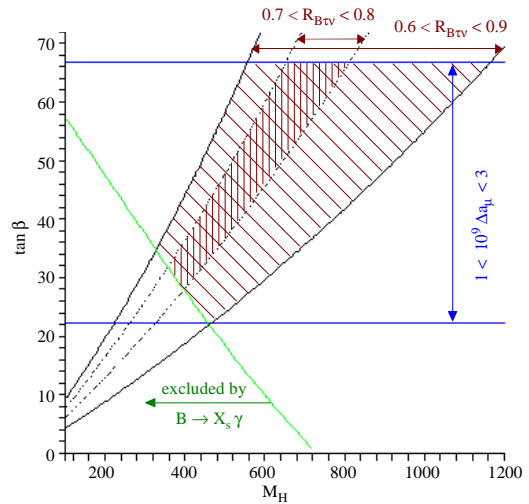}
\caption{Constraints from $B$ physics observables and 
$(g-2)_\mu$ in the $(M_{H^\pm}, \tan\beta)$ plane, with fixed
$\mu = 0.5$ TeV and $A_U=0$~\cite{Isidori:2006pk}} \label{fig:TanbMH} 
\end{figure}

The correlation of these observables can be studied in the $(M_{H^+},\tan\beta)$
plane, as shown in Fig.~\ref{fig:TanbMH}, for fixed values of $A_U,\mu$. 
The tree mediated decay $B_u \to \tau \nu$ is included in these constraints. 
In the MSSM this is given by the same expression as in the 2HDM, 
up to a gluino correction which becomes important 
in the large $\tan\beta$ limit. 

\subsection{Models of Warped Extra Dimensions}
\label{sec:warped}
One of the most interesting models of New Physics
is based on the idea of a warped extra dimension~\cite{Randall:1999ee}. 
This notion has great appeal as it can lead to a simultaneous resolution 
to the hierarchy problem as well as the flavor problem of the SM
by accomodating rather naturally
the observed large disparity 
of fermion masses~\cite{Grossman:1999ra,Gherghetta:2000qt,Davoudiasl:2007zx}.
For lack of space we do not discuss the implications of 
universal extra dimensions, for which we refer the reader 
to the recent review by \textcite{Hooper:2007qk}. 

In RS setup the 5-dimensional space-time has anti-de Sitter geometry (AdS$_5$).
A slice of AdS$_5$ (bulk) is truncated by flat 4D boundaries, 
the Planck (UV) and the TeV (IR) branes.  
This setup gives a warped metric in the bulk~\cite{Randall:1999ee}  
\beq
ds^2 = e^{-2 k r_c |\phi|} \eta_{\mu\nu} dx^\mu dx^\nu - r_c^2 d\phi^2,
\label{metric}
\eeq
where $k$ is the 5D curvature scale, $r_c$ the radius of compactification and  
$\phi\in [-\pi,\pi]$ the coordinate along the $5^{\rm th}$ dimension. 
The warp factor $e^{-2 k r_c |\phi|}$ leads to different length scales 
in different 4D slices along the $\phi$ direction, 
which provides a solution to the hierarchy problem.  
In particular, the Higgs field is assumed to be localized
near the TeV-brane so that the metric ``warps'' 
$\vev{H}_5 \sim M_5\sim M_P \sim 10^{19}$~GeV down to the weak scale, 
$\vev{H}_4 = e^{-kr_c \pi} \vev{H}_5$.  
For $kr_c \approx 12$ then $\vev{H}_{\tt SM} \equiv \vev{H}_4 \sim 1$~TeV.  

Originally all the remaining SM fields were 
assumed to also reside at the IR-brane~\cite{Davoudiasl:1999jd}.  
However, the cutoff of the effective 4D theory is then also 
red-shifted to the weak scale. 
This in turn leads to unsuppressed higher dimensional operators 
and thus large violations of EWP data and unacceptably large FCNCs.

This problem can be solved by realizing that the points along the warped 
$5^{\rm th}$ dimension correspond to different effective 4D cut-off scales. 
In particular, by localizing the first and second generation fermions 
close to the UV-brane the higher dimensional operators get suppressed 
by effectively larger scales \cite{Gherghetta:2000qt}. 
Note that this explains why first and second generation fermions are light: 
the Yukawa interactions are small because of small overlap between
IR localized Higgs and UV localized light fermion zero modes. 
The top quark on the other hand is localized near the TeV brane 
to obtain a large top Yukawa coupling.

This configuration
suppresses FCNCs substantially
(however, see below) 
and reproduces the fermion mass hierarchies 
without invoking large disparities in the Yukawa couplings of the fundamental
5D action~\cite{Grossman:1999ra,Gherghetta:2000qt}. 
It thus has a built in analog of the SM GIM mechanism (the RS GIM) 
and reproduces the approximate flavor symmetry among the light fermions.

Similarly to the SM GIM, the RS GIM is violated by the large top quark mass. 
In particular, $(t,b)_L$ needs to be localized near the TeV brane 
otherwise the 5D Yukawa coupling becomes too large and makes the theory
strongly coupled at the scale of the first KK excitation.
This has two consequences:
(1) in the interaction basis, the coupling of $b_L$ to gauge KK modes
(say the gluons),
$g_{ G^{ \rm KK } }^b$, is
large compared to the couplings of the lighter quarks.
This is a source of flavor violation 
leading to FCNCs. 
(2) The Higgs vev mixes the zero mode of $Z$ and its KK modes, 
leading to a non-universal shift 
$\delta g_Z^{ b}  \sim  g_{Z^{\rm KK }}^{b}
\sqrt{ \log \left( M_{\rm Pl} / \hbox{TeV} \right) }
{ m_Z^2 }/{ m_{\rm KK }^2 }$
in the coupling of
$b_L$ to the physical $Z$~\cite{Agashe:2003zs,Burdman:2003ya}. Here $g_{Z^{\rm KK }}^{b}$ is the  
coupling
between $b_L$ and a KK $Z$ state before EWSB.
The factor $\sqrt{ \log \left( M_{\rm Pl} / \hbox{TeV} \right) }$ comes from
enhanced Higgs coupling to gauge KK modes, 
which are also localized near the TeV brane.
Electroweak precision measurements of $Z\to b_L\bar b_L$ require that
this shift is smaller than $\sim 1 \%$.
Using $g^b_{Z^{\rm KK} } \sim g_Z$ this is satisfied for $m_{ KK } \sim 3 \ \hbox{TeV}$. 
In passing we also note that with enhanced bulk electroweak gauge symmetry, 
$SU(2)_L \times SU(2)_R \times U(1)_{B-L}$, and KK masses of $\approx$ 3 TeV,
consistency with constraints from electroweak precision measurements 
are achieved~\cite{Agashe:2003zs}.  

The tension between obtaining a large top Yukawa coupling 
and not introducing too large flavor violation and disagreement with EWP data~\cite{Agashe:2003zs,Burdman:2003ya} 
is solved in all models by assuming
(1) a close to maximal 5D Yukawa coupling, $\lambda_{\rm 5D} \sim 4$, 
so that the weakly coupled effective theory contains 3-4 KK modes, 
and (2) by localizing $(t,b)_L$ as close
to the TeV brane as allowed by  $\delta g_Z^{ b}\sim 1 \%$. 
This almost unavoidable setup leads to sizeable NP contributions 
in the following three types of FCNC processes that are top quark dominated:
(i) $\Delta F = 2$ transitions, 
(ii) $\Delta F = 1$ decays governed by box and EW penguin diagrams; 
(iii) radiative decays. 

Sizeable modifications of $\Delta F=2$ processes are possible 
from tree-level KK gluon exchanges.
The $\Delta F=1$ processes receive contributions from 
tree level exchange of KK $Z$ modes. 
These tend to give smaller effects than KK gluon exchanges. 
Nevertheless it can lead to appreciable effects in the branching ratio,
direct CP asymmetry and the spectrum of $b \to s \ell^+ \ell^-$~\cite{Burdman:2003ya,Agashe:2004ay,Agashe:2004cp}. 
In $b \rightarrow s \bar{q} q$ QCD penguin dominated 
$B\to (\phi, \eta^\prime, \pi^0, \omega, \rho^0) K_s$ decays 
on the other hand the RS contributions from flavor-violating $Z$ vertex
are at least $\sim g_Z^2 / g_s^2 \sim 20 \%$ suppressed and thus
subleading~\cite{Agashe:2004ay,Agashe:2004cp}. 
Consequently, RS models can accommodate only mild deviations
from the SM in the corresponding time dependent CP asymmetries.

We should emphasise that these models are not fully developed yet so 
there can be appreciable uncertainties in the specific predictions. 
For instance, the particular framework outlined above runs into at least 
two problems unless the relevant KK-masses are much larger than $~3$ TeV:
(i) the presence of right-handed couplings can cause enhanced contributions 
to $\Delta S=2$ processes, $K$--$\bar K$ mixing and $\epsilon_K$~\cite{Beall:1981ze,Bona:2007vi}, 
and (ii) the simple framework with ${\cal O}(1)$ complex phases tends to give 
an electron electric dipole moment about a factor $\sim 20$ above the 
experimental bound~\cite{Agashe:2004ay,Agashe:2004cp}. 
An interesting proposal for the flavor dynamics in the
RS setup was recently put forward by~\textcite{Fitzpatrick:2007sa} who
introduced 5D anarchic minimal flavor violation in the quark sector 
(see also ~\textcite{Cacciapaglia:2007fw}). 
This gives a low energy effective theory that falls in the NMFV class, 
consistent with both FCNC and dipole moment constraints 
(see section~\ref{Sec:MFV}).
In this picture new flavor and CP violation phases are present, 
however, their dominant effect occurs only in the up type quark sector.

\subsection{Light Higgs searches}  

Existing LEP constraints on the Higgs mass do not rule out the existence
of a very light Higgs boson $h$ with a mass well below the present limit
 of $114.4$ GeV, if the SM is extended either in the gauge or Higgs sector~\cite{Dermisek:2006py,Fullana:2007uq}.
Such states for instance appear naturally in 
extensions of the MSSM motivated by the $\mu-$problem. 
The most popular models are nonminimal supersymmetric models, 
where one or more gauge singlets are added to the two Higgs doublets of the MSSM 
\cite{Han:2004yd,Barger:2006dh,Dermisek:2006py}. 
The simplest case of one gauge singlet is the 
next-to-minimal supersymmetric standard model (NMSSM), which contains
seven physical Higgs bosons, two of which are neutral pseudoscalars.

A light Higgs boson would be difficult to observe at the LHC because of 
significant backgrounds, and a SFF could play a complementary
role in this respect. 
The main detection mode is $\Upsilon \to h(\to \ell^+\ell^-)\gamma$~\cite{Wilczek:1977zn}. 
The presence of a light Higgs may manifest itself as an enhancement of the $\Upsilon(1S)\to \tau^+ \tau^-$ channel relative
to other dilepton modes $(e, \mu)$.
In NMSSM at large $\tan\beta$, the $b\to sh$ vertex with $h$ a light Higgs 
produces observable effects in
rare $B,K$ decays. It can be search for in $\Upsilon$ or $B\to K$ decays with missing energy. 
The presence of new pseudoscalar in NMSSM  also breaks the correlation between $B_s\to \mu^+\mu^-$ decay and
$B_s$--$\bar B_s$ mixing that is present in MSSM~\cite{Hiller:2004ii}. 

In passing, we mention a related topic.
Invisible decays of quarkonia can be used to search for 
light dark matter candidates~\cite{Gunion:2005rw,McElrath:2005bp}.
An initial analysis of this type has been carried out at Belle~\cite{Tajima:2006nc},
illustrating the potential for this physics at a SFF.

\subsection{Flavor signals and correlations}

\begin{table}
  \begin{center}
%    \begin{ruledtabular}
      \caption{
        Summary of expected flavor signals in selected observables considered
        by \textcite{Goto:2007ee}.
        After imposing present experimental constraints, observables
        denoted by $\surd$ typically have a non-negligible deviation from the
        SM; those marked $\bullet$ have deviations which could become
        measurable at future experimental facilities such as LHCb, SFF, MEG;
        empty space indicates that deviations smaller than the expected
        sensitivities are anticipated.
        Lepton decay processes were not considered in the $U(2)$ model.}
      \label{table_obs_model}
      \begin{tabular}{c|c|c|c|c}
        \hline \hline
        Process & cMSSM & \multicolumn{2}{c|}{SU(5) SUSY GUT} & U(2) \\
        \cline{3-4}
        & & degen. & non-degen. & \\
        \hline
        $A^{\text{dir}}_{CP}(X_s\gamma)$ &  &  &  & $\surd$ \\
        $S(K^*\gamma)$& &$\bullet$ & $\surd$ & $\surd$ \\
        $A^{\text{dir}}_{CP}(X_d\gamma)$ & & &  &   \\
        $S(\rho\gamma)$ & &$\bullet$ & $\surd$ & $\surd$ \\
        $\Delta S(\phi K_S)$& &$\bullet$ & $\surd$ & $\surd$ \\ 
        $S(J/\psi \phi)$& &$\bullet$ & $\surd$ & $\surd$ \\ 
        \hline
        $\frac{\Delta M_s}{\Delta M_d} \mbox{ vs } \gamma$ 
        & & & $\bullet$ & $\bullet$ \\ 
        \hline
        $\mu \to e\gamma$  & & $\surd$ & & $-$ \\ 
        $\tau\to\mu\gamma$ & & & $\surd$ & $-$ \\ 
        $\tau\to e\gamma$& & & $\surd$ & $-$ \\
        \hline \hline
      \end{tabular}
%    \end{ruledtabular}
  \end{center}
\end{table}

How well can one distinguish various NP models from flavor data? 
This can be achieved by studying correlations among different 
flavor violating observables. 
As mentioned in previous subsections such correlations appear 
in models of flavor violation motivated by simple symmetry arguments, 
{\it e.g.} in MFV scenarios. 
An example of how flavor observables can distinguish among a restricted set 
of models is given in \textcite{Goto:2002xt,Goto:2003iu,Goto:2007ee}.  
The authors considered four classes of SUSY models, 
which are typical solutions of the SUSY flavor problem (restricted to the
low $\tan\beta$ regime): 
(i) cMSSM (which for this analysis is equivalent to mSUGRA), 
(ii) cMSSM with right-handed neutrinos, 
(iii) SU(5) SUSY GUT with right-handed neutrinos, 
and (iv) MSSM with U(2) flavor symmetry. 
The right-handed neutrinos were taken to be degenerate or nondegenerate, 
the latter with two specific neutrino matrix ans\" atze.  
Constraints from direct searches, $b\to s\gamma$, $B_{(s)}$--$\bar B_{(s)}$ 
and $K$--$\bar K$ mixing, and upper bounds on $l_i\to l_j\gamma$ 
and on EDMs were imposed on the models. 
Table \ref{table_obs_model} lists typical deviations from SM 
for each of the models that are then still allowed.

In addition to the patterns in Table \ref{table_obs_model}, 
certain correlations are expected between subsets of observables. 
For example, $\Delta M_{B_s}/\Delta M_{B_d}$ and $\gamma$ 
are correlated in all considered models, 
but to constrain the NP parameters this requires improved lattice QCD 
determination of the $\xi$ parameter at a few percent level. 
In Table \ref{table_obs_model}
 we do not list results for cMSSM with right-handed neutrinos, 
where the only observable deviations are expected in $\mu\to e\gamma$ 
for degenerate and in $\tau\to \mu\gamma, e\gamma$ 
for nondegenerate right-handed neutrinos. 

%%%%%%%%%%%%%%%%%%%%%%%%%%%%%%%%%%%%%%%%%
%%%%%%%%%%%%%%%%% II. Direct measurements of angles  %%%%%%%%%%%%%%%%%%%%
%%%%%%%%%%%%%%%%%%%%%%%%%%%%%%%%%%%%%%%%%
\section{Direct measurements of unitarity triangle angles}
\label{sec:ut}

\begin{figure}
  \includegraphics[width=5cm]{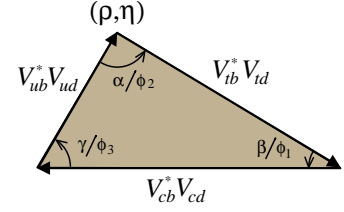}
  \caption{The standard CKM unitarity triangle.}
  \label{CKMtriangle}
\end{figure}

We now discuss methods for direct determination of the angles in the standard CKM unitarity triangle.
They test the CKM unitarity
requirement for the first and the third column of the CKM matrix (see
Fig. \ref{CKMtriangle}). We focus on methods that use little or no theoretical 
assumptions: the determinations of 
(i)  $\beta$ from $B^0 \to J/\psi K_{S,L}$ and $B^0 \to D h^0$, (ii)  $\gamma$ from $B \to DK$ and  $2\beta+\gamma$ from $B \to D^{(*)}\pi/\rho$, 
$D^{0(*)}K^{0(*)}$ and (iii)  $\alpha$ from $B \to \pi\pi$, $\pi\rho$, $\rho\rho$.
These decays are tree dominated so new physics effects are expected to be small.  
Together with measurements of the sides discussed in Section \ref{6sides}, a
determination of the ``standard model CKM unitarity triangle'' is possible
either using tree-level processes alone, or by also including
 $\Delta F = 2$ (mixing) processes~\cite{Buras:2000dm,Charles:2004jd,Bona:2006sa}. This should be compared with the determinations using methods sensitive to new physics discussed in the later sections.

Let us set up the notation. Assuming CPT invariance
the time dependent decay of an initially tagged $B^0$ is 
given by 
\beq
\label{time-dependence}
\begin{split}
  \Gamma&(B^0(t)\to f)\propto 
  e^{-\Gamma t}\; 
  \Big[
  \cosh \Big(\frac{\Delta \Gamma t}{2}\Big)+\\
  +&H_f \sinh \Big(\frac{\Delta \Gamma t}{2}\Big)- {\cal A}^{\mathrm{CP}}_f\cos\Delta m t- S_f \sin\Delta m t 
  \Big],
\end{split}
\eeq
where $\Gamma$ is the average neutral $B$ meson decay width, while $\Delta \Gamma=\Gamma_H-\Gamma_L$ is the difference of decay widths 
between heavier and lighter $B_q^0$ mass eigenstates, so that the mass difference $\Delta m=m_H-m_L>0$.
In this section  we focus on $B^0_d$ mesons, but 
Eq.~\eqref{time-dependence} applies also to the $B^0_s$ system discussed in Section \ref{sec:Bs}.
Using shorthand notation $\bar{A_f}=A(\bar B^0\to f)$, ${A_f}=A(B^0\to f)$, 
the coefficient of $\cos\Delta m t$ is
\beq\label{directCP}
{\cal A}_f^{\rm CP}=\frac{|\bar{A_f}|^2-|A_f|^2}{|\bar{A_f}|^2+|A_f|^2},
\eeq
and is equal to direct CP asymmetry in the case of a CP eigenstate 
$f$ (in
  the literature $C_f = - {\cal A}_f^{\rm CP}$ is also used). The coefficient of $\sin\Delta m t$ describes CP violation in interference between mixing and decay and is
\beq\label{Sf}
S_f= -2 \frac{\Im\lambda_f}{1+|\lambda_f|^2}, \qquad  
\lambda_f=\left(\frac{q}{p}\right)_{B}\frac{\bar{ A_f}}{A_f}, 
\eeq
where parameters $q_B, p_{B}$ describe 
the flavor composition of the $B^0$ mass eigenstates. 
In Eq.~\eqref{directCP} we neglected CP violation in mixing taking $|({q}/{p})_{B}|=1$, which we assume to be the case.
The time dependent decay width $\Gamma(\bar{B}^0(t)\to f)$ 
is then obtained from Eq.~\eqref{time-dependence} by flipping the signs 
of the $\cos(\Delta m t)$ and $\sin(\Delta m t)$ terms.
The time dependent CP asymmetry is thus
\beq
\begin{split}\label{aCPt}
  a_{CP}(B(t)\to f) &=
  \frac{\Gamma(\bar B(t)\to f)-\Gamma(B(t)\to f)}{\Gamma(\bar B(t)\to f)+\Gamma( B(t)\to f)}\\
  &=
  {\cal A}^{\mathrm{CP}}_f\cos( \Delta m t)+ S_f \sin (\Delta m t). 
\end{split}
\eeq

In the $B_d^0$ system the observable $H_f$ is  negligible
since $(\Delta \Gamma/\Gamma)_{B^0_d} \ll 1$.
For the $B^0_s$ system, on the other hand, 
a much larger decay width difference is predicted within the Standard Model
$(\Delta \Gamma/\Gamma)_{B^0_s}=-0.147\pm0.060$~\cite{Lenz:2006hd}.
Experimentally, the current world average from an angular analysis of $B_s^0 \to J/\psi \phi$ decays is  
$(\Delta \Gamma/\Gamma)_{B^0_s}=-0.206 \, ^{+0.111}_{-0.106}$~\cite{Abazov:2005sa,Acosta:2004gt,HFAG}
[a more precise value of $-0.104 \, ^{+0.084}_{-0.076}$~\cite{HFAG}
is obtained by including the $B^0_s$ lifetime measurements from flavor specific decays].
Thus, in the $B^0_s$ system both $S_f$ and
\beq\label{Hf}
H_f = -2{\Re\lambda_f}/{(1+|\lambda_f|^2)},
\eeq
are experimentally accessible~\cite{Dunietz:1995cp}.
While sensitivity to the $S_f$ term requires the ability to resolve the fast
$B^0_s$ oscillations, for which the large boost of a hadronic machine is
preferable, the $H_f$ term is measured from the coefficient of the 
$\sinh (\Delta \Gamma t/2)$ dependence, which can be achieved at a 
SFF operating at the $\Upsilon(5{\rm S})$.

\begin{table}[!tb] 
  \begin{center}
    \begin{ruledtabular}
  \caption{ 
    Precision on the parameters of the 
    standard CKM unitarity triangle expected from direct determinations.
    For each observable discussed in the text both the theoretical uncertainty
    and the estimated precision that can be obtained by a
    Super Flavor Factory~\cite{Akeroyd:2004mj,Bona:2007qt} are given.}
  \label{tab:angles}
    \begin{tabular}{l@{\hspace{-5mm}}cc}
%      \hline \hline
      Observable    & Theoretical error & Estimated precision \\
      & &  at a Super Flavor Factory \\
      \hline
      $\sin(2\beta)$ ($J/\psi K^0$)    &  0.002  &  0.01  \\ 
      $\cos(2\beta)$ ($J/\psi K^{*0}$) &  0.002  &  0.05  \\ 
      $\sin(2\beta)$ ($Dh^0$)          &  0.001  &  0.02  \\
      $\cos(2\beta)$ ($Dh^0$)          &  0.001  &  0.04  \\ 

      $\gamma$ ($DK$)                  & $\ll 1^\circ$ & $1$--$2^\circ$ \\ 
      $2\beta+\gamma$ ($DK^0$)         & $< 1^\circ$ & $1$--$2^\circ$ \\
      $\alpha$ ($\pi\pi$)              & $2$--$4^\circ$ & $3^\circ$  \\
      $\alpha$ ($\rho\pi$)             & $1$--$2^\circ$ & $1$--$2^\circ$ \\
      $\alpha$ ($\rho\rho$)            & $2$--$4^\circ$ & $1$--$2^\circ$ \\      
      $\alpha$ (combined)              & $\approx 1^\circ$ & $1^\circ$ \\ 
%      \hline \hline
    \end{tabular}
\end{ruledtabular}
  \end{center}
\end{table}

\subsection{Measuring $\beta$}
\label{sec:ut:beta}

The measurement of 
$\beta$ 
is the primary benchmark of the current $B$-factories. 
The present experimental world average from 
decays into charmonia-kaon 
final states, 
$\sin2\beta=0.680 \pm 0.025$~\cite{Chen:2006nk,Aubert:2007hm,HFAG},
disagrees slightly with an indirect 
extraction that is obtained 
using all other constraints on the unitarity triangle. CKMFitter group for instance obtains $\sin2\beta = 0.799^{+0.044}_{-0.094}$~\cite{Charles:2004jd}, while a similar 
small inconsistency is found in~\cite{
Bona:2006sa,Bona:2007vi,Lunghi:2007ak}. Improved accuracy in experiment and in theory are needed
to settle this  important issue. The theoretical error in the direct determination 
is negligible as discussed below. 
The theoretical error in the indirect determination, on the other hand, is a combination 
of theoretical errors in all of the constraints used in the fit,
and comes 
appreciably from the lattice inputs.

That the extraction of the weak phase $\beta$ 
from $B^0\to J/\psi K_S$ is theoretically 
very clean was realized long ago~\cite{Bigi:1981qs,Carter:1980tk,Hagelin:1981zk}.
The  decay is dominated by a $\bar b\to \bar c c \bar s$ tree level transition.
The complex parameter describing the mixing induced CP violation in $B\to J/\psi K_S$
is 
\beq
\begin{split}
\lambda_{J/\psi K_S}&=
-\left(\frac{q}{p}\right)_{B^0_d} \left(\frac{p}{q}\right)_{K^0}
\frac{A(\bar B^0 \to J/\psi \bar K^0)}{A(B^0 \to J/\psi K^0)}\\
&\simeq
\left(\frac{q}{p}\right)_{B^0_d}\left(\frac{p}{q}\right)_{K^0}
\frac{V_{cb}V_{cs}^*}{V_{cb}^* V_{cs}}.
\end{split}\label{lambdaJpsiKS}
\eeq
The $({p}/{q})_{K^0}$ factor is due to $K-\bar K$ mixing, cf. Eq.~\eqref{Sf}.
In going to the second line we have used CP symmetry to relate the two matrix elements, keeping only the tree-level operator 
$V_{cb} V_{cs}^* (\bar c b)_{V-A} (\bar s c)_{V-A} + h.c.$ 
in the effective weak Hamiltonian 
(the relative minus sign arises since the $J/\psi K$ final state has $L=1$).  The remaining pieces are highly suppressed in the SM.
In the standard phase convention for the CKM matrix~\cite{Aleksan:1994if}, 
$V_{cb} V_{cs}^*$ is real, while $(q/p)_{B^0_d}=-e^{-2i\beta}$
and $(q/p)_{K^0}=-1$ up to small 
corrections to be discussed below, so that
$S_{J/\psi K_S}=\sin 2 \beta$, 
${\cal A}_{J/\psi K_S}^{\rm CP} = 0$.
The time dependent CP asymmetry of Eq.~\eqref{aCPt} is then 
\beq\label{simpleS}
a_{CP}(B(t)\to J/\psi K_S) =\sin(2\beta) \sin(\Delta m t),
\eeq
with a vanishingly small $\cos (\Delta m t)$ coefficient. 
Corrections to this simple relation arise from 
subleading corrections to the $B^0_d - \bar B^0_d$ mixing,  
the $K^0 - \bar K^0$ mixing and the $B\to J/\psi K$ decay amplitude 
that have been neglected in the derivation of Eq.~\eqref{simpleS}. 
Including these corrections
\beq
\begin{split}\label{fullS} 
  a_{CP}(&B(t)\to J/\psi K_S)  
  =\big[\sin(2\beta)+\Delta S^{B\rm{mix}}\\
  +&\Delta S^{K\rm{mix}}+\Delta S^{{\rm decay}}+\frac{\Delta \Gamma_B t}{4}\sin4\beta \big] \sin\Delta m t\\
  +&\big[\Delta {\cal A}^{B\rm{mix}}+\Delta {\cal A}^{K\rm{mix}}+\Delta {\cal A}^{{\rm decay}}\big] \cos\Delta m t.
\end{split}
\eeq
Here~\cite{Boos:2004xp}
\beq \label{DeltaSBmix}
\begin{split}
\Delta S^{B\rm{mix}}&
=-\Im\frac{\Delta M_{12}}{|M_{12}|}
=(2.08\pm1.23)\cdot 10^{-4},\\
\end{split}
\eeq
is the correction due to $u$ and $c$ quarks 
in the box diagram which
mixes neutral $B$ mesons.
These contributions have 
a different weak phase than the leading $t$ quark 
box diagram and thus modify the relation $\arg(q/p)_{B^0_d}=2\beta$.

The correction~\cite{Grossman:2002bu}
\beq
\Delta S^{K\rm{mix}}=-2 \cos (2\beta)\, \Im (\epsilon_K)\simeq -2.3 \cdot 10^{-3},
\eeq
arises from the deviation of  
$(q/p)_{K^0}$ from $-1$, 
and from the fact that the experimental identification through $K_S \to \pi\pi$ decay
includes a small admixture of $K_L$.

The correction due to the penguin contributions in the 
$B\to J/\psi K$ decay is~\cite{Grossman:2002bu}
\beq\label{Sdecay}
\Delta S^{{\rm decay}}=
-2 \cos (2\beta)\, \Im \frac{\lambda_u^{(s)}}{\lambda_c^{(s)}} \,r \, \cos\delta_r,
\eeq
where $\lambda_q^{(s)}=V_{qb}V_{qs}^*$, 
$r$ is the ratio of penguin to tree amplitudes and 
$\delta_r$ the strong phase difference. 
Because of the strong CKM suppression ($|\lambda_u^{(s)}/\lambda_c^{(s)}|\sim 1/50$) 
these effects are small, of the order of the other two $\Delta S$ corrections. 
The calculation of $\Delta S^{{\rm decay}}$ is highly uncertain. The factorization theorems 
for  two-body decays into two light mesons are not applicable due to the large $J/\psi$ mass.
Even so, calculations have been attempted.
Using a combination of QCD factorization and pQCD \textcite{Li:2006vq} obtain
$\Delta S^{{\rm decay}}=(7.2^{+2.4}_{-3.4})\cdot 10^{-4}$. \textcite{Boos:2004xp}
find $\Delta S^{{\rm decay}}=-(4.24\pm1.94)\cdot 10^{-4}$ using 
a combination of the 
BSS mechanism~\cite{Bander:1979jx} 
and naive factorization and keeping only the $u\bar u$ loop contribution. An alternative approach uses SU(3) flavor symmetry to relate the 
$B^0\to J/\psi K^0$ amplitude to the $B^0\to J/\psi \pi^0$ amplitude,
neglecting annihilation-like contributions~\cite{Ciuchini:2005mg}. 
In $B^0\to J/\psi \pi^0$ decay the penguin contributions are CKM-enhanced,
increasing the sensitivity to $r$ and $\delta_r$. 
Using the experimental information available in 2005 \textcite{Ciuchini:2005mg} obtained 
$\Delta S^{{\rm decay}} = 0.000\pm0.017$. The error is dominated by the experimental errors
and is not indicative of the intrinsic $\Delta S^{{\rm decay}}$ size. 

In summary, $\Delta S_{J/\psi K_S}$ is expected to be 
$\Delta S_{J/\psi K_S}\simeq -1.4 \cdot 10^{-3}$. 
This is also the typical size of the term 
due to a
nonzero decay width difference, 
$\sin 4\beta(\Delta \Gamma \tau_{B^0})/4 \simeq -1 \cdot 10^{-3}$~\cite{Boos:2004xp}. 
Thus, any discrepancy significatly above permil level between $S_{J/\psi K_S}$ measurement
and $\sin2 \beta$ obtained from the CKM fits 
would be a clear signal of new physics~\cite{Hou:2006du}.
The theoretical uncertainty in the 
measurement of $\sin2\beta$ from $S_{J/\psi K_S}$ is likely to remain smaller
than the experimental error even at a 
SFF.
Extrapolations of the current analyses suggest that 
imperfect knowledge of the vertex detector alignment and beam spot position
will provide a limiting systematic uncertainty,
with the ultimate sensitivity of $0.5$--$1.2\%$~\cite{Akeroyd:2004mj,Bona:2007qt}.

Digressing briefly from the determination of the unitarity triangle,
the situation for the direct CP asymmetries in $B \to J/\psi K$ is rather
similar~\cite{Boos:2004xp,Grossman:2002bu,Li:2006vq} 
\begin{align}
-\Delta {\cal A}^{B\rm{mix}}&=
\Im\frac{\Gamma_{12}}{2 M_{12}}=
-(2.59\pm1.48)\cdot 10^{-4},\\
-\Delta {\cal A}^{K\rm{mix}}&=2 \Re (\epsilon_K)\simeq 3.2 \cdot 10^{-3},\\
-\Delta {\cal A}^{{\rm decay}}&=2 \Im \frac{\lambda_u^{(s)}}{\lambda_c^{(s)}} r \sin\delta_r=(16.7^{+6.6}_{-8.7})\cdot 10^{-4}, \label{Adecay}
\end{align}
giving a combined CP asymmetry
${\cal A}_{J/\psi K_S}\simeq -4.6 \cdot 10^{-3}$.

This is nearly an order of magnitude smaller than the current experimental
uncertainty on this quantity~\cite{Chen:2006nk,Aubert:2007hm},
and comparable to the likely size of the limiting systematic uncertainty
at a SFF~\cite{Akeroyd:2004mj,Bona:2007qt}.
New physics contributions to this quantity 
could enhance the CP asymmetry to the $1\%$ level or even higher,
 while obeying all other constraints 
from flavor physics~\cite{Hou:2006du,Bergmann:2001pm}.

A complementary measurement of $\beta$ is provided by 
a time dependent $B^0\to [K_S\pi^+\pi^-]_D
h^0$ Dalitz plot analysis~\cite{Bondar:2005gk}. Here  $h^0=\pi^0, \eta,
\omega, \dots$, while also $D^{*0}$ can be used in place of $D^0$. 
This channel provides measurements
of both $\sin 2\beta$ and $\cos 2\beta$ resolving the $\beta\to \pi/2-\beta$ discrete ambiguity. The resulting mesurement of $\beta$ is theoretically extremely clean
since it does not suffer from penguin pollution. The only theoretical uncertainty is due
to the $D^0$ decay model, which at present gives an error of $\sim 0.2$ on
$\cos 2\beta$~\cite{Aubert:2007rp,Krokovny:2006sv}, and can be reduced in
future using the same methods as for the $B \to DK$ analysis (see the
discussion in Section~\ref{sec:ut:gamma}).  $D$ decays to CP
eigenstates can also be used. However, these are only sensitive to $\sin 2\beta$~\cite{Fleischer:2003ai}.

\subsection{Measuring $\gamma$}
\label{sec:ut:gamma}
\subsubsection{$\gamma$ from $B\to D K$}
The most powerful method to measure $\gamma$ uses the interference between
$b\to c \bar{u} s$ and $b\to u \bar{c} s$ amplitudes in 
$B\to DK$ decays~\cite{Gronau:1991dp,Gronau:1990ra} [for a recent review see, {\it e.g.}, \cite{Zupan:2007zz}].
In the case of charged $B$ decays the interference is between 
$B^-\to D K^-$ amplitude, $A_B$, followed by $D\to f$ decay,  and 
$B^-\to \bar{D} K^-$ amplitude, $A_B r_B e^{i(\delta_B-\gamma)}$, followed by $\bar{D}\to f$ decay,
where $f$ is any common final state of $D$ and $\bar{D}$. The $B^+\to D(\bar D) K^+$ decay amplitudes are obtained by $\gamma\to -\gamma$ sign-flip. Neglecting CP violation in the $D$ decays we further have
\beq
\label{A_B}
\begin{array}{ccl}
  A(     D^0\to f)= & A(\bar D^0\to \bar f) = & A_f, \\
  A(\bar D^0\to f)= & A(     D^0\to \bar f) = & A_f r_f e^{i\delta_f}.
\end{array}
\eeq
The parameters $\delta_B$ and $\delta_f$ above are strong phase differences
in $B$ and $D$ decays respectively, while $A_B, r_B$, $A_f,r_f$ are real. The sensitivity to $\gamma$ is strongly dependent on the ratio
$r_B \sim 0.1$. Since there are no penguin contributions in this class of modes,
there is almost no theoretical uncertainty in the resulting
measurements of $\gamma$; all hadronic unknowns can in principle be
obtained from experiment.

Various choices for the final state $f$ are possible:
(i) CP eigenstates ({\it e.g.} $K_S \pi^0$)~\cite{Gronau:1991dp}, 
(ii) quasi-flavor specific states ({\it e.g.} $K^+\pi^-$)~\cite{Atwood:1996ci,Atwood:2000ck},
(iii) singly Cabibbo suppressed decays ({\it e.g.} $K^{*+} K^-$)~\cite{Grossman:2002aq} or 
(iv) many-body final states ({\it e.g.} $K_S\pi^+\pi^-$)~\cite{Giri:2003ty,Atwood:2000ck,Poluektov:2004mf}. 
There are also other extensions,
using many body $B$ decays 
({\it e.g.} $B^+\to D K^+\pi^0$)~\cite{Aleksan:2002mh,Gronau:2002mu}, 
using $D^{*0}$ in both $D^{*0} \to D\pi^0$ and 
$D^{*0} \to D\gamma$ decay modes~\cite{Bondar:2004bi}, 
using self tagging $D^{**}$ decays~\cite{Sinha:2004ct}.
Neutral $B$ decays (both time dependent and time integrated) 
can also be used~\cite{Gronau:2004gt,Kayser:1999bu,Atwood:2002vw,Fleischer:2003ai}.

For different $D$ decays in $B\to (f)_D K$,
the parameters $A_B, r_B,\delta_B,\gamma$ related to the $B$ decay are common,
so that there is significant gain in combining results from different $D$ 
decay channels~\cite{Atwood:2003jb}. It is therefore not suprising  that three body $D$ decays, {\it e.g.} $B^\pm\to [K_S\pi^+\pi^-]_D K^\pm$, 
provide the most sensitivity in the extraction of $\gamma$
as they represent an essentially continuous set of final states $f$. 
Also, for $D\to f$ multibody decays both the magnitude of $A_f$ and the strong phase variation
over the Dalitz plot can be determined using a decay model where $A_f$ is described by a sum of resonant (typically Breit-Wigner) terms~\cite{Giri:2003ty,Poluektov:2004mf}. The decay model can be determined from flavor tagged data 
[for details, see~\cite{Cavoto:2007fp,Poluektov:2006ia,Aubert:2006am}].

Flavor tagged $D$ decays do not provide direct information on the 
strong phase differences between $D^0$ and $\bar{D}^0$ amplitudes. In multibody decays the information comes from the interferences of the resonances, where the phase variation across the Dalitz plot is completely described 
by the chosen decay model.
The question is then what is the modelling error introduced through this
approach and how can it be reliably estimated? 
At present the modelling error on $\gamma$ is estimated to be $\sim 10^\circ$, 
which is obtained 
through an apparently conservative approach of including or excluding 
various contributions to the model.
In future it will be possible to reduce this error by using entangled $\psi(3770)\to D\bar D$ decays at a tau-charm
factory to arrive at a direct information
on the strong phases~\cite{Giri:2003ty,Atwood:2003mj}.

\begin{figure}
  \includegraphics[width=5cm]{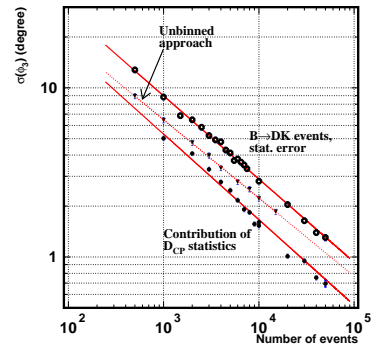}
  \caption{
    Statistical error on $\gamma$ ($\phi_3$) as a function of the number of
    reconstructed $B^\pm\to DK^\pm$ decays and $D_{CP}$ decays as given by toy
    MC study with $r_B=0.2$, $\gamma=70^\circ$, $\delta_B=180^\circ$ and $4
    \cdot 10^5$ $D_{CP}$ decays~\cite{Bondar:2005ki}. 
    Dotted line shows the error on $\gamma$ from model-dependent unbinned 
    Dalitz plot fit with the same input parameters.
    \label{Bondar_study}
  }
\end{figure}

Alternatively, the modelling error can be avoided entirely by using a model
independent approach~\cite{Giri:2003ty,Atwood:2000ck}.
After partitioning the $D\to K_S\pi^+\pi^-$ Dalitz plot into bins,
variables $c_i,s_i$ are introduced that are the cosine and sine of the strong phase
difference averaged over the $i$-th bin.
Optimally, these are determined from charm factory running at $\psi(3770)$~\cite{Soffer:1998un,Gronau:2001nr,Atwood:2003mj,Giri:2003ty}. Recent
studies~\cite{Bondar:2005ki,Bondar:2008hh} show that if measurements of $c_i$
from CP-tagged $D$ decays are included in the analysis, the resulting error
on $\gamma$ using rectangular Dalitz plot binning is only $30\%$ worse  than the unbinned model dependent
approach~\cite{Bondar:2005ki}, or even only $4\%$ worse for optimal binning~\cite{Bondar:2008hh}. Studies of charm
factory events in which both $D$ mesons decay to multibody final states such
as $K_S\pi^+\pi^-$ can also provide information on the $s_i$ terms~\cite{Bondar:2008hh}. As shown
in Fig.~\ref{Bondar_study}, approximately $10^4$ CP tagged $D$ decays are
required to keep the contribution to the uncertainty on $\gamma$ below the
$2^\circ$ statistical accuracy expected from a SFF. 

To reduce the statistical uncertainty, one can also include
additional $B$ decay modes.  For each, the hadronic factors $A_B$, $r_B$ and
$\delta_B$ can be different, so additional unknown parameters are introduced.
To date, $B^\pm\to DK^{\pm}$, $B^\pm\to DK^{*\pm}$ and $B^\pm\to D^*K^{\pm}$
(with 
$D^* \to D\pi^0(\gamma)$ 
~\cite{Bondar:2004bi}) have
been used.

Another useful approach 
is to include neutral $B^0$ decays. 
These have smaller decay rates, however the statistical error on $\gamma$ does not scale with the rate but roughly as
the smaller of the two interfering amplitudes.  Using isospin one sees that these
differ only by a factor of $\sqrt 2$~\cite{Gronau:2004gt}
\beq
A_B r_B\simeq \sqrt{2}A_B^n r_B^n.
\eeq
Here we have introduced $A_B^n$ and $r_B^n$ parameters in the same way as for
the charged decays above Eq.~\eqref{A_B}.  Although time dependent measurements are
needed to extract the full information in the $B^0 \to D K_S$
system~\cite{Kayser:1999bu,Atwood:2002vw,Gronau:1990ra,Fleischer:2003ai,Gronau:2007bh}, untagged time integrated rates 
alone provide sufficient information to
determine $\gamma$~\cite{Gronau:2004gt,Gronau:2007bh}, while $B^0 \to D K^{*0}$ decays are
self-tagging. Therefore, we expect these modes to make a significant
contribution to the measurement of $\gamma$ 
at a SFF~\cite{Akeroyd:2004mj,Bona:2007qt}.

We now discuss the theoretical errors. The determination
of $\gamma$ from $B\to DK$ decays is theoretically extremely clean since these
are pure tree decays. The largest uncertainty is due to $D^0 - \bar D^0$
mixing, assumed to be absent so far. The SM $D^0 - \bar D^0$ mixing
parameters are  $x_D \equiv \frac{\Delta m_D }{ \Gamma_D} \sim y_D  \equiv
\frac{\Delta \Gamma_D}{ 2\Gamma_D} \sim {\cal O}(10^{-2})$, with a negligible
CP violating phase, $\theta_D\sim {\cal O}(10^{-4})$ (see
Section~\ref{sec:charm}).

The effect of CP conserving $D^0 - \bar D^0$ mixing is to change the effective
relative strong phase (irrelevant for $\gamma$ extraction) and to dilute the
interference term, resulting in a shift 
$\Delta\gamma \propto (x_D^2+y_D^2)/r_f^2$~\cite{Grossman:2005rp}. 
Thus the shift is larger for the cases where $r_f$ is smaller,
but even for doubly Cabibbo suppressed decays $\Delta \gamma\lesssim 1^\circ$.
Furthermore, this bias can be removed by explicitly
including $D^0 - \bar D^0$ mixing
into the analysis once $x_D$ and $y_D$ are well 
measured~\cite{Silva:1999bd,Atwood:2003jb}.
Moreover in the model independent Dalitz plot analysis no changes are needed,
since there the method already includes the averaging (dilution) of the
interference terms.

The remaining possible sources of theoretical error are from higher
order electroweak corrections or from CP violation in the $D$ system. 
The latter would lead to  
$\Delta\gamma\sim {\cal O}(x_D\theta_D, y_D\theta_D)$. 
In the SM the error is conservatively  $\Delta \gamma<10^{-5}$, 
while even with large NP in the charm sector
one finds $\Delta\gamma \sim {\cal O}(10^{-2})$.

In summary, a precise measurement of $\gamma$ can be
achieved at a SFF  from a combination of $B \to DK$ type decays with multiple $D$
decay final states.  The precision can be improved using charm factory data on strong phases.  Although extrapolations of the
current data are difficult, studies suggest that an error on $\gamma$
of ${\cal O}(1^\circ)$ can be achieved~\cite{Akeroyd:2004mj,Bona:2007qt}.
This would represent a significant improvement on the constraints from any
other experiment, and yet the 
experimental uncertainty on $\gamma$ would still be far above the
irreducible theory error.

\subsubsection{$\sin(2\beta + \gamma)$}

The combination $\sin(2\beta +\gamma)$ can in principle be extracted 
from $B\to D^{(*)\pm}\pi^\mp$ time dependent analysis~\cite{Dunietz:1997in,Suprun:2001ms}. However, the ratio of the two interfering amplitudes
$r=|A(B^0\to D^{(*)+}\pi^-)/A(\bar B^0\to D^{(*)+}\pi^-)|$ is too small to be determined experimentally from 
${\cal O}(r^2)$ terms and significant input from theory is required. Related methods use $B^0 \to D^{*+}\rho^-, D^{*+}a_1^-$, where $r$ can be determined from the interference of different helicity
amplitudes \cite{London:2000zi,Gronau:2002nk}.  These modes are difficult experimentally because of $\pi^0$ reconstruction and no measurements exist to date.
Another option are rare decays such as $B \to D^{(*) \pm} X^\mp$,
$X=a_0,a_2,b_1, \pi(1300)$, where $r$ is $O(1)$  as pointed out by \textcite{Diehl:2001ey}.

Time dependent $B^0 \to D^{0(*)}
K^{0(*)}$ analyses are perhaps the most promising~\cite{Kayser:1999bu,Atwood:2002vw}.  The theoretical error is 
expected to be similar to that in $\gamma$ extraction from $B \to
DK$,   
and thus well below SFF sensitivity.  Another good candidate, $B_s\to D_s^\pm K^\mp$, is better suited for
experiments in an hadronic environment \cite{Fleischer:2003rx}.
Other alternatives, including three body modes such as 
$B \to D^\pm K_S \pi^\mp$~\cite{Charles:1998vf,Aleksan:2002mh,Polci:2006aw} 
could also lead to a precise measurement of $2\beta+\gamma$.

\subsection{Measuring $\alpha$}
\label{sec:ut:alpha}
\begin{figure}
\includegraphics[width=5cm]{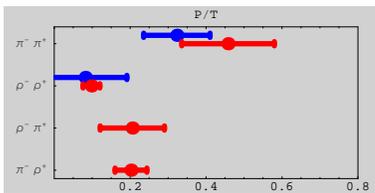}
\caption{Summary of the present constraints from isospin (blue/dark grey) and SU(3)
  flavor symmetry (red/light grey) on the $P/T$ ratio in the ``c-convention''. Only statistical errors are shown.}\label{fig:r_list} 
\end{figure}

Although in the SM $\alpha$ is not independent from $\gamma$ and $\beta$, it
is customary to separate the methods for the determination of the angle
$\gamma$ that involve $B^0_d - \bar B^0_d$ mixing from those that do not. 
In this subsection we will therefore briefly discuss the determination of $\alpha$ from
the decays $B \to \pi\pi$, $\rho\pi$ and $\rho\rho$ [for a longer review see {\it e.g.} \cite{Zupan:2007fq}]. 
The angle $\alpha$ is determined from the $S_f$ parameter of Eq.~\eqref{Sf}. For example in 
 $B \to \pi\pi$ this is
\beq
S_{\pi^+\pi^-}=\sin 2\alpha +2 r \cos \delta \sin(\beta+\alpha)\cos 2\alpha
+{\cal O}(r^2),  
\eeq
where the expansion is in penguin--to--tree ratio $r=P/T$. The ``tree'' (``penguin'') is a
term that carries a weak phase (or not),  $
A(B^0 \to \pi^+ \pi^-) =  T e^{i \gamma} + P e^{i \delta},  
$ while $\delta$ is a strong phase difference.\footnote{This is the so called ``c-convention'' where ``penguin'' is proportional to $V_{cb}^* V_{cd}$. The other option is a ``t-convention'', where ``penguin'' is proportional to $V_{tb}^* V_{td}$ and
carries weak phase $-\beta$.} 
In the $r=0$ limit  one has $S_{\pi^+\pi^-}=\sin
2\alpha$. If ${\cal O}(r)$ ``penguin pollution'' term is known, $\alpha$ can be extracted from $S_{\pi^+\pi^-}$. 
This is
achieved by using symmetries of QCD, isospin or flavor SU(3), or by the
$1/m_b$ expansion in frameworks such as QCD factorization, pQCD, and SCET.
The  
theoretical error on extracted $\alpha$ depends crucially on
the size of $r$. Using isospin and/or SU(3) flavor
symmetry one finds (see also Fig. \ref{fig:r_list})
\beq \label{r-hierarchy}
r(\pi^+\pi^-)>r(\rho^+\pi^-)\sim r(\pi^+\rho^-)>r(\rho^+\rho^-).
\eeq
We can 
expect a similar hierarchy 
for the theoretical errors on 
 $\alpha$ in the different channels. This simple rule, however,
does not apply for methods based on isospin symmetry as discussed in more
detail below.

\subsubsection{$B\to \pi\pi$}
\begin{figure}[t] 
  \begin{center}
    \begin{tabular}{cc}
      \includegraphics[width=3.7cm]{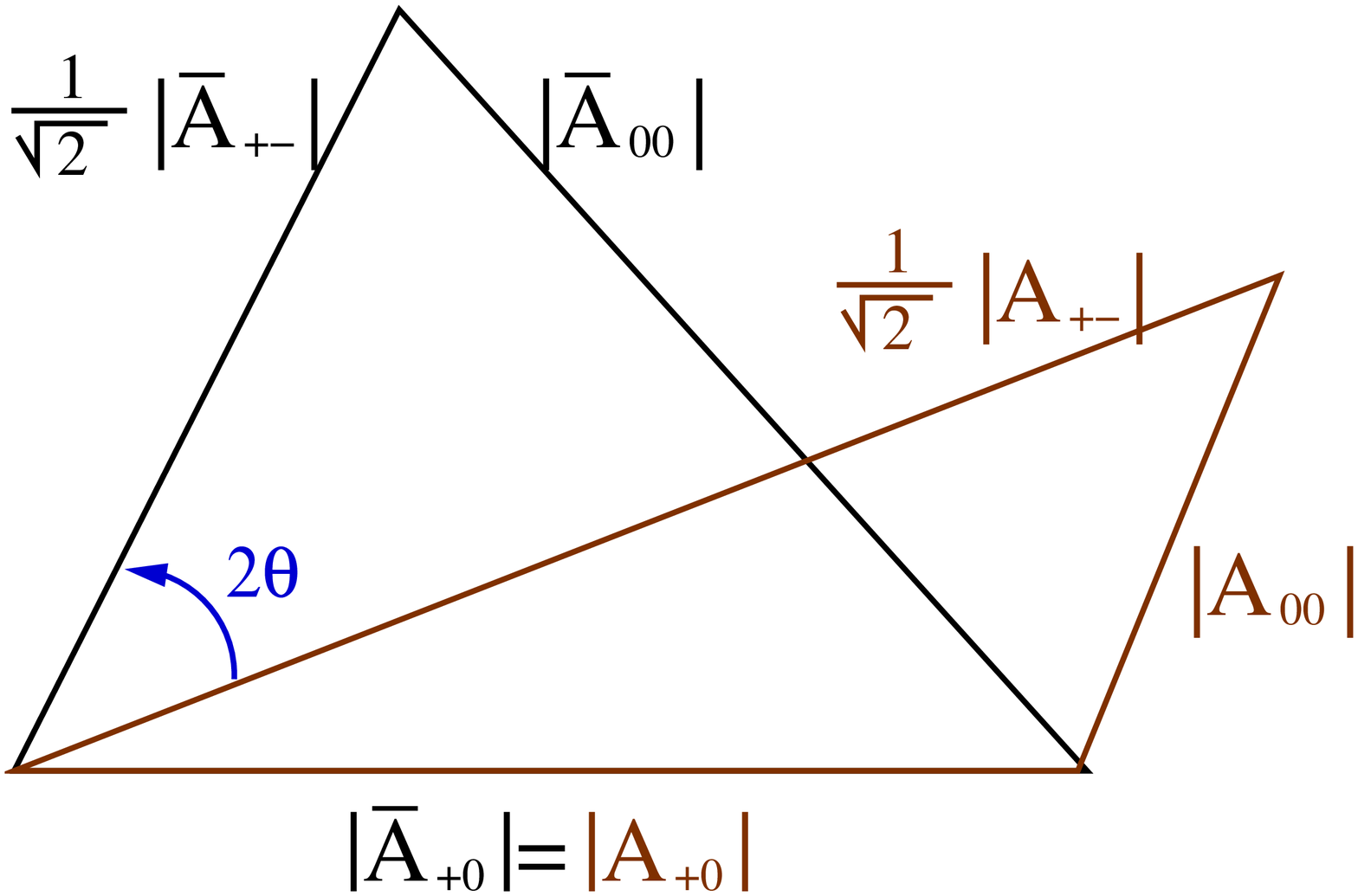} &
      \includegraphics[width=3.9cm]{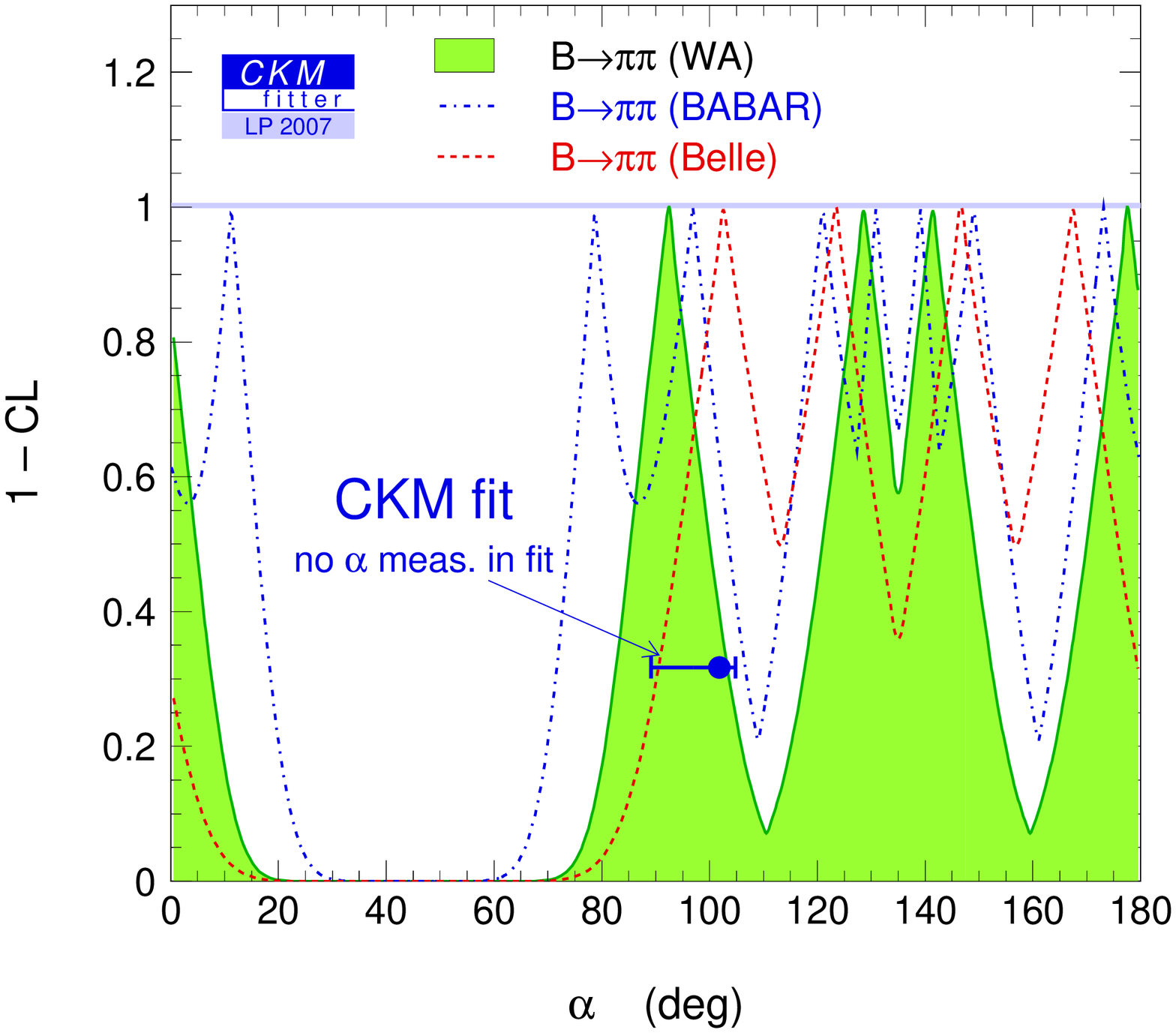}
    \end{tabular}
  \end{center}
  \caption{
    Left: the isospin triangle relations due to~\textcite{Gronau:1990ka}, with the notation $A_{ij}\equiv A(B^0\to \pi^i\pi^j)$. 
    Only one of four possible triangle orientations is shown. 
    Right: constraints on $\alpha$ from isospin analysis of
    $B\to\pi\pi$~\cite{Charles:2004jd}.
    Note that solutions at $\alpha \approx 0$ need very large values of $T,P$ with fine-tuned cancellation and are thus excluded \cite{Bona:2007qta}.
  }
  \label{ZupanGLtriangle-fig}
\end{figure}

Let us first review the extraction of $\alpha$ from $B\to \pi\pi$ using
isospin decomposition~\cite{Gronau:1990ka}. In isospin limit $\pi$ forms a triplet and $B$
a doublet of isospin. In general $B\to \pi\pi$ transition could be mediated by
$\Delta I=1/2,3/2$ and $5/2$ interactions. However, $\Delta I=5/2$ operators
do not appear in the effective weak Hamiltonian of Eq.~\eqref{HW},
so that $B\to \pi^0\pi^0, \pi^+\pi^-, \pi^+\pi^0$ amplitudes are 
related as shown in Fig. \ref{ZupanGLtriangle-fig}. 

Another important input  
is that 
aside from possible electroweak penguin (EWP) contributions, $A_{+0}$ is a
pure tree 
(notation is as in Fig.~\ref{ZupanGLtriangle-fig}). Neglecting EWP the weak phase of $A_{+0}$ is fixed, so that for instance $e^{i\gamma} A_{+0} = e^{-i\gamma} \bar A_{+0}$.
Then the observable $\sin(2\alpha_{\rm eff}) = 
S_{\pi\pi}/\sqrt{1 - ({\cal A}^{\rm CP}_{\pi\pi})^2}$ is
directly related to $\alpha$ through $2\alpha=2 \alpha_{\rm eff}-2 \theta$,
where $\theta$ is defined in Fig.~\ref{ZupanGLtriangle-fig}, left. The present
constraints on $\alpha$ following from the isospin analysis with the most
recent experimental results~\cite{Aubert:2007mj,Ishino:2006if} are shown in
Fig.~\ref{ZupanGLtriangle-fig}, right. Note that in the determination of $\alpha$ the contribution of $\Delta I=1/2$ terms cancel. This implies that the isospin analysis is insensitive to NP in
QCD penguin operators,
and would still return the SM
value of $\alpha$ even if such NP contributions were large. 

Let us now turn to the question of theoretical uncertainties in the isospin
analysis which come from isospin breaking. 
This has several effects: (i) different $d$ and $u$ charges lead to EWP operators 
$Q_{7,\dots,10}$ in ${\cal H}_{\rm eff}$ of Eq.~\eqref{HW}, (ii) the $\pi^0$ mass and isospin eigenstates no longer coincide, leading to  $\pi^0-\eta-\eta'$ mixing, (iii) reduced matrix elements for states in the same isospin multiplet may no longer be related simply
by SU(2) Clebsch-Gordan coefficients, and (iv) $\Delta I=5/2$ operators may be induced, {\it e.g.} from electromagnetic rescattering. 

In the literature only the first two effects have been analyzed in some
detail. The effect of EWP is known quite precisely since the $\Delta I=3/2$ part of the EWP Hamiltonian is related to the tree part of the weak Hamiltonian~\cite{Neubert:1998pt,Neubert:1998jq,Neubert:1998re,Buras:1998rb,Gronau:1998fn}. The relation between the bases of triangles in Fig. \ref{ZupanGLtriangle-fig} is now modified to $  e^{i\gamma} A_{+0} = e^{-i(\gamma+{ \delta})} \bar A_{+0}$, where  $\delta = ( 1.5 \pm 0.3\pm 0.3 )^\circ$~\cite{Gronau:1998fn,Gronau:2005pq}. 
The $\pi^0-\eta-\eta'$ mixing modifies also the Gronau-London 
triangle relations of Fig.~\ref{ZupanGLtriangle-fig}~\cite{Gardner:1998gz}. Since $\pi^0-\eta-\eta'$ is small, the resulting shift in the extracted value of $\alpha$ is small as well, $|\Delta\alpha_{\pi-\eta-\eta'}|  < 1.6^\circ $~\cite{Gronau:2005pq}. 

These two examples of isospin breaking effects show that
while not all of the isospin breaking effects can be calculated or
constrained at present, the ones that can are of the expected size, 
$ \delta\alpha \sim (m_u-m_d)/\Lambda_{\rm QCD} \sim 1\%$.

Experimentally, the isospin triangle approach is limited by the need to
measure $|A_{00}|$ and $|\bar A_{00}|$, {\it i.e.} to measure direct CP
violation in $B^0\to \pi^0 \pi^0$ decays.  In addition, the method suffers
from ambiguities in the solutions for $\alpha$ (as can be seen in
Fig.~\ref{ZupanGLtriangle-fig}, right).  A SFF will enable
both problems to be overcome, since the large statistics will allow a precise
measurement of ${\cal A}^{\rm CP}_{00}$, while the sample of events 
with photon conversions
will allow $S_{00}$ to be measured, removing 
one ambiguity~\cite{Ishino:2007pt}.
Including these effects, we expect a SFF to reach a precision
of $\sim 3^\circ$ on $\alpha$ from $B \to \pi\pi$~\cite{Akeroyd:2004mj,Bona:2007qt}.

\subsubsection{$B\to \rho \rho$}
The isospin analysis in $B\to \rho\rho$ follows the same lines as for $B\to
\pi \pi$, but with separate isospin triangles, Fig.  \ref{ZupanGLtriangle-fig}, for
each polarization. The longitudinally polarized final state is found
to dominate the other two, 
simplifying the analysis considerably.
Another difference from the $\pi\pi$ system is that $\rho$ resonances have a
non-negligible decay width.  In addition to experimental complications, this
allows the two $\rho$ resonances in the final state to form an $I=1$ state, if
the respective invariant masses are different~\cite{Falk:2003uq}, leading to
${\cal O}(\Gamma_\rho^2/m_\rho^2)$ effects.  This effect can in principle be
constrained experimentally by making different fits to the mass
distributions~\cite{Falk:2003uq}, though very high statistics would be
necessary for such a procedure to be effective.

The remaining theoretical errors are due to isospin
breaking effects. While the shift due to EWP is exactly the same as in
$\pi\pi$,  $\rho-\omega$ mixing is expected to cause a relatively large,
${\cal O}(1)$, effect near the $\omega$ mass in the $\pi^+\pi^-$ invariant mass
spectrum.  However, integrated over all phase space, the effect is of the
expected size for isospin breaking, as indeed are all effects that can
currently be estimated~\cite{Gronau:2005pq}.

An ingredient that makes the $\rho\rho$ system favourable over $\pi\pi$ is the
small penguin pollution, cf. Fig.~\ref{fig:r_list}.  Moreover, the fact that $B^0 \to \rho^0\rho^0$
results in an all charged final state means that $S_{00}$ can be
determined~\cite{Aubert:2007qs}.  
Consequently, 
$\alpha$ determination from isospin analysis of $B \to
\rho\rho$ at the SFF is expected to remain more precise than
that from 
$B \to \pi\pi$, {\it i.e.} $1^\circ$--$2^\circ$~\cite{Akeroyd:2004mj,Bona:2007qt}.

Somewhat surprisingly, the small penguin
pollution makes the method based on the SU(3) symmetry as theoretically clean
as the isospin analysis~\cite{Beneke:2006rb}. This is because SU(3) symmetry is used to
directly 
constrain $P/T$, while the isospin construction involves also relations
between the tree amplitudes, so that isospin breaking on the larger
amplitudes translate to the corrections. 
The basic idea is to relate $\Delta S=0$ 
decays in which tree and penguin terms have CKM elements of similar size 
to $\Delta S=1$ decays in which the $P/T$ ratio has a relative enhancement of
$\sim 1/\lambda^2$.  The $\Delta S=1$ decays can then be used to 
constrain  $P/T$.
For example, ${\cal B}(B^+ \to K^{*0}\rho^+)$ can be used to bound 
the penguin contribution to  $B^0\to \rho^+\rho^-$~\cite{Beneke:2006rb}:
\beq
|A_L(K^{*0}\rho^+)|^2 = 
F\left(\frac{|V_{cs}|f_{K^*}}{|V_{cd}|f_{\rho}}\right )P^2,
\eeq
where the $F$ parameterises SU(3) breaking effects ($F=1$ in the limit of
exact SU(3)). Using a conservative range of $0.3 \le F \le 1.5$ results in 
theoretical error of $\sim 4^\circ$ on $\alpha$, comparable to the
theoretical error in the isospin analysis.

\subsubsection{$B\to \rho\pi$}
Since $\rho^\pm \pi^\mp$ are not CP eigenstates, extracting $\alpha$ from this
system is more complicated. Isospin analysis similar to the one for $B\to \rho\rho, \pi\pi$ leads to an isospin pentagon contruction~\cite{Lipkin:1991st} that is not competitive. It requires a large amount of experimental
data and suffers from multiple solutions.
Two more useful approaches are: (i) to exploit the full
time-dependence of the $B^0 \to \pi^+\pi^-\pi^0$ Dalitz plot together with
isospin~\cite{Snyder:1993mx}, or (ii) to use only the $\rho^\pm\pi^\mp$ region
with SU(3) related modes~\cite{Gronau:2004tm}.

For the Snyder-Quinn isospin analysis two important differences compared to
the isospin analysis of $B\to \pi\pi$ and $B\to \rho\rho$ are (i) that in
$B\to \rho\pi$ only the isospin relation between
penguin amplitudes is needed, and (ii) that from the full time-dependent $B^0 \to \pi^+\pi^-\pi^0$ Dalitz
plot the magnitudes and relative phases of $A(B^0\to \rho^+ \pi^-), A(B^0\to
\rho^- \pi^+), A(B^0\to \rho^0 \pi^0)$ and the CP conjugated amplitudes are obtained. As a result 
the Snyder-Quinn approach does not suffer from multiple ambiguities, giving a
single (and highly competitive) value for $\alpha$ in $\left[0,\pi\right]$. This approach has been implemented by both $B$
factories~\cite{Aubert:2007jn,Kusaka:2007dv,Kusaka:2007mj}.

A potential problem is
that the peaks of $\rho$ resonance
bands do not fully overlap in the Dalitz plot, but are separated by
approximately one decay width, so one is sensitive to the precise lineshape of
the $\rho$ resonance. Isospin breaking effects on the other hand are expected to be
$P/T$ suppressed, since only the isospin relation between penguins was used.
The largest shift is expected to be due to EWP and is known precisely, as in $B\to\pi\pi, \rho\rho$ case~\cite{Gronau:2005pq}. 
Other isospin breaking effects are expected to be small. 
For instance,  the shift due to
$\pi^0-\eta-\eta^\prime$ mixing was estimated to be $|\Delta\alpha_{\pi-\eta-\eta'}| \le  0.1^\circ$~\cite{Gronau:2005pq}, showing
that the expected $P/T$ suppression exists.

An alternative use of the same data is provided by the SU(3) flavor
symmetry. In this way the potential sensitivity of the Snyder-Quinn method on the form of $\rho$ resonance tails  can be avoided. The required information on $P/T$ is obtained from the SU(3) related $\Delta S=1$
modes, $B^0\to K^{*+}\pi^-, K^+\rho^-$ and $B^+\to K^{*0} \pi^+,
K^0 \rho^+$. Since penguin pollution is
relatively small, the error on the extracted value of $\alpha$ due to SU(3)
breaking is expected to be small as well, of a few degrees~\cite{Gronau:2004tm}. 
Unlike the Snyder-Quinn approach this method does suffer from discrete
ambiguities.

In summary, theory errors in the above direct measurements of $\alpha$ are difficult to determine completely. Our best estimates for the error on $\alpha$
from isospin analysis of the $\pi \pi$ and $\rho \rho$ systems are around a
few degrees. The uncertainty is expected to
be smaller for the Snyder-Quinn analysis of $\rho\pi$ which relies on an
isospin relation between only penguin amplitudes.  Since a SFF can make determinations of $\alpha$ in all of the
above modes, we can be cautiously optimistic that most sources
of theoretical uncertainty can be controlled with data.  Therefore, there is a good
chance that the final error on $\alpha$ from a SFF will be around $1^\circ$.

Finally, Table \ref{tab:angles} summarizes the estimates  on the theory
error and also the expected accuracy at the SFF for each angle
through the use of these direct methods.

\section{Sides of the triangle}
\label{6sides}

In this section we review briefly the strategies 
for measurements of the magnitudes of CKM matrix elements.
For a more extensive review see \textcite{VcbYao:2006px}.

While the determinations of 
$|V_{ub}|$, $|V_{cb}|$, $|V_{td}|$ and $|V_{ts}|$ mainly rely 
on CP conserving observables -- 
the CP averaged $B$ decay branching ratios -- their values 
do constitute an independent check of the CKM mechanism.
The information on $|V_{ub}|/|V_{cb}|$ for instance 
determines the length of the unitarity 
triangle side opposite to the well measured 
angle $\beta$, cf. Fig. \ref{CKMtriangle}. 
Together with the direct determination 
of $\gamma$ it provides a consistency check between 
the constraints from $b\to u$ tree transitions 
and the constraint from the loop induced 
$B$--$\bar B$ mixing. \\

\subsection{Determination of $|V_{cb}|$}
\label{Vcb}

Both exclusive and inclusive $b\to c$ decays are used,
giving consistent determinations \cite{VcbYao:2006px}
\beq
\begin{split}
|V_{cb}|_{\rm excl.}&=(40.9\pm1.8)\times 10^{-3},\\
|V_{cb}|_{\rm incl.}&=(41.7\pm0.7)\times 10^{-3}.
\end{split}
\eeq
The value of $|V_{cb}|$ from the exclusive decay $\bar B\to D^{*} l
\bar \nu_l$ ($\bar B\to D^{} l \bar \nu_l$) is at present determined
with a $4\%$ ($12\%$) relative error, where the 
theoretical and experimental contributions to the errors are comparable. 
In the heavy quark limit the properly normalized form 
factors are equal to $1$ at zero recoil, $v_B\cdot v_{D^{(*)}}=1$. This
prediction has perturbative and nonperturbative corrections
\beq
\begin{split}
{\cal F}_{D^*}(1)&=1+c_A(\alpha_s)+\frac{0}{m_Q}+\frac{c_{\rm nonp.}^*}{m_Q^2},\\
{\cal F}_{D}(1)&=1+c_V(\alpha_s)+\frac{c_{\rm nonp.}}{m_Q^2}.
\end{split}
\eeq
The absence of $1/m_Q$ corrections in ${\cal F}_{D^*}(1)$ 
is due to Luke's theorem \cite{Luke:1990eg}. 
The perturbative corrections $c_{A,V}$ are known to 
$\alpha_S^2$ order \cite{Czarnecki:1996gu,Czarnecki:1997cf}, 
while the first nonperturbative corrections $c_{\rm nonp.}^{(*)}$ 
are known only from 
quenched lattice QCD \cite{Hashimoto:2001nb,Hashimoto:1999yp} 
or from phenomenological models. 
Improvement can be expected in the near future when unquenched lattice QCD
results become available. 
The projected uncertainty is $2$-$3\%$ \cite{VcbYao:2006px,Laiho:2007pn}, 
which is comparable to presently quoted errors 
in quenched calculations  \cite{Hashimoto:2001nb,Hashimoto:1999yp}, 
but the results will be  more reliable. 
Further improvements in precision will be needed, however, 
to reach the $1\%$ uncertainty projected 
for the inclusive $|V_{cb}|$ determination discussed below. 
To achieve this goal analytical work is also needed: 
the calculation of higher order matching of latticized HQET 
to continuum QCD is already in progress
\cite{Nobes:2003nc,Oktay:2003gk}, 
while other ingredients such as the radiative corrections 
to the $1/m_Q$ and $1/m_Q^2$ suppressed terms in the currents 
are not yet being calculated. 
The difficulty of this task is comparable or even greater 
than the same order calculation needed 
for the inclusive determination of $|V_{cb}|$ \cite{VcbYao:2006px}.
On the experimental side, reduction of the uncertainty with larger statistics
is not guaranteed, since systematic errors already limit the precision
\cite{Aubert:2004bw,Aubert:2006cx}.

The inclusive determination of $|V_{cb}|$ is 
based on the operator product expansion leading to 
a systematic expansion 
in $1/m_b$ \cite{Manohar:1993qn,Bigi:1993fe,Bigi:1993fm}. 
Present fits to $\bar B\to X_c l \bar \nu_l$ include terms 
up to order $1/m_b^3$ and $\alpha_s^2 \beta_0$. The same 
nonperturbative elements also appear 
in the predictions of $B\to X_s\gamma$ so that global fits to 
electron and photon energy moments from data are performed, 
giving  $|V_{cb}|$ with a relative error of about 
$1.7\%$ \cite{VcbYao:2006px}. 
Improvements on the theoretical side can be made by calculating higher 
order perturbative corrections \cite{Neubert:2005nt} 
and by calculating the perturbative corrections 
to the matrix elements that define the heavy quark expansion parameters.
Experimentally, systematic errors are already limiting the most recent 
results in these analyses~\cite{Schwanda:2006nf,Urquijo:2006wd}.
However, some improvement is certainly possible 
with the large statistics of a SFF,
so that a precision on $|V_{cb}|$ 
around $1\%$ may be possible. \\

\subsection{Determination of $|V_{ub}|$}
\label{Vub}

Both exclusive and inclusive determinations are being pursued. 
At present there is some slight tension (at the $1\sigma$ level)
between the two types of determinations; 
as discussed below.

The theoretical and experimental difficulty with 
the inclusive extraction of 
$|V_{ub}|$ from $\bar B\to X_u l\bar \nu_l$ is 
due to the large charm background from 
$\bar B\to X_c l\bar \nu_l$. As a result 
one cannot obtain the full inclusive rate experimentally. 
The region of phase space without 
charm contamination is typically a region where 
the inclusive hadronic state forms a jet, so that the OPE is not valid. 
Still, one can find a $\Lambda_{\rm QCD}/m_b$ expansion, 
and using SCET one can show that there is a 
factorization of the structure functions 
(in terms of which the branching ratio is 
expressed) into hard, jet and shape functions, see Eq.~\eqref{factincl} below. 
Each of these factors encode physics at scales of 
the order $m_b$, $\sqrt{\Lambda_{\rm QCD} m_b}$ and $\Lambda_{\rm QCD}$. 
The jet and shape functions are currently known at ${\cal O}(\alpha_s(m_b))$ 
\cite{Bauer:2003pi,Bosch:2004th} 
and ${\cal O}(\alpha_s^2(\sqrt{\Lambda_{\rm QCD} m_b}))$ 
\cite{Becher:2006qw} respectively, 
while the power corrections have been included 
only at ${\cal O}(\alpha_s^0)$
\cite{Lee:2004ja,Bosch:2004cb,Beneke:2004in}. 
In the BLNP approach the parameters for the models 
of the LO shape function are extracted from the $\bar B\to X_s\gamma$ spectrum
\cite{Lange:2005yw}, 
while subleading shape functions are modeled. 
The HFAG average using this approach is 
$|V_{ub}|_{\rm incl. (BLNP)}=(4.49 \pm 0.19 \pm0.27)\times 10^{-3}$~\cite{Bizjak:2005hn,Aubert:2007rb,HFAG},
where the first error is experimental and the second theoretical. 
Alternatively, as discussed in Section~\ref{constraintsCKM},
the ratio of $\bar B\to X_u l\bar \nu_l$ to $\bar B\to X_s\gamma$ decay rates 
can be used to reduce the dependence on the LO shape function 
\cite{Neubert:1993um,Leibovich:2000ey,Lange:2005qn,Lange:2005xz}.  
This approach has been used to obtain the value
$|V_{ub}|=(4.43 \pm 0.45\pm 0.29)\times 10^{-3}$~\cite{Aubert:2006qi},
where the first error is experimental and the second theoretical.
The combined theoretical error 
from using 2-loop corrections to jet functions, 
the subleading shape function corrections and the known $\alpha_s/m_b$ 
corrections has been estimated to be $5\%$~\cite{Lange:2005qn}. 
This error could be further reduced 
by using the $B\to X_s\gamma$ hard kernels 
at ${\cal O}(\alpha_s^2)$ a calculation of 
which is almost complete \cite{Becher:2006pu}, 
but a similarly demanding calculation of the hard kernel 
in $\bar B\to X_u l\bar \nu_l$ at the same order would be needed. 
Another hurdle is the estimation of the subleading shape functions 
-- to gain in precision one would need to go beyond modeling. 

A different approach that can reduce the dependence on shape functions 
is a combined cut on the leptonic momentum transfer $q^2$ 
and the hadronic invariant mass $M_X$~\cite{Bauer:2001rc,Bauer:2000xf}, 
so that a larger portion of phase space is used. 
Furthermore, it has been suggested~\cite{Bigi:1993bh,Voloshin:2001xi} 
that uncertainties from weak annihilation 
can be reduced by making a cut on the high $q^2$ region. 
Another theoretical approach, Dressed Gluon Exponentiation, 
that uses a renormalon inspired model for the leading shape function
has been advocated~\cite{Andersen:2005mj}.
Following these approaches, and taking advantage of the large statistics
at a SFF, a precision on $|V_{ub}|$ of 3--5\% from inclusive modes may
be possible.

For the exclusive $|V_{ub}|$ determination,
the decay $\bar B\to \pi l\bar \nu_l$ is primarily used,
although decays such as $\bar B\to \rho l\bar \nu_l$ also provide useful
information, and, as discussed in Section~\ref{sec:2HDM},
leptonic decays $\bar B\to l \bar \nu_l$ can be used to obtain a tree-level
determination of $|V_{ub}|$ that is sensitive to NP effects.
Nonperturbative information on $\bar B\to \pi l\bar \nu_l$ 
form factors comes from lattice QCD for $q^2>16 \ {\rm GeV}^2$, 
while light cone sum rules can be used for $q^2\to 0$. 
Using current lattice QCD results 
in their range of applicability $q^2>16 \ {\rm GeV}^2$, 
HFAG finds
$|V_{ub}|=(3.33 \pm 0.21^{+0.58}_{-0.38})\times 10^{-3}$~\cite{Hokuue:2006nr,Athar:2003yg,Aubert:2006px,HFAG}
using the unquenched HPQCD calculation \cite{Dalgic:2006dt}, 
and $|V_{ub}|=(3.55 \pm 0.22 ^{+0.61}_{-0.40})\times 10^{-3}$ 
for the unquenched calculation from the
FNAL collaboration \cite{Okamoto:2004xg}. 
A number of extrapolation ansaetze have been proposed so 
that the whole $q^2$ region can be used for $|V_{ub}|$ determination  
\cite{Becirevic:1999kt,Hill:2005ju,Becher:2005bg,Arnesen:2005ez,Boyd:1994tt,Boyd:1997qw}.  
A recent discussion of their use is given in \textcite{Ball:2006yj}.  

The current status is somewhat problematic: inclusive methods give $|V_{ub}|$ values
systematically larger than the exclusive methods,
and are also in disagreement with direct $\sin 2 \beta$ determination 
at $\sim 2 \sigma$ level~\cite{Charles:2004jd,Bona:2006sa,Bona:2007vi,Lunghi:2007ak}. 
\textcite{Neubert:2008cp} argued recently that, 
due to model dependence introduced by the shape function and 
contributions other than those from the $Q_{7\gamma}$ operator, 
the $b\to s \gamma$ data should not be used in the $|V_{ub}|$ determination. 
Using $m_b$ determined only from $b\to c l\nu$ 
and the theoretically cleanest $M_X$ cut, 
Neubert finds $|V_{ub}| = (3.70 \pm 0.15 \pm 0.28) \times 10^{-3}$, 
resolving the disagreement. 

The SFF will give much improved determinations of $|V_{ub}|$ using
the exclusive approach, where the statistical errors currently
control the precision of the measurements. Here one requires precise
determinations of the $q^2$ spectrum, in the low recoil region where the rate
is very small. The large data sample at a SFF will allow measurements of
binned spectra with precision of a few percent. Assuming that lattice QCD can 
reach a comparable level of precision, an error of 3--5\% on $|V_{ub}|$ from 
the exclusive approach appears attainable at a SFF.

\subsection{Determination of $|V_{td}|$ and $|V_{ts}|$ from loop processes}
\label{VtdVts}

The values of the CKM matrix elements $|V_{td}|$ and $|V_{ts}|$ 
can only be studied in loop processes at a SFF.
These include both mixing ($\Delta F = 2$) and decay ($\Delta F = 1$) processes.
Specifically, the ratio $|V_{td}|/|V_{ts}|$ can be obtained 
by comparing the $B_d$--$\bar B_d$ and $B_s$--$\bar B_s$ mass differences,
or from the ratio of, for example, $b\to d\gamma$ and $b\to s\gamma$ 
radiative decays. 
Since both are loop mediated processes they are sensitive to NP.

The oscillation frequencies in $B_{d,s}$--$\bar B_{d,s}$ mixing determine the
mass differences. These are short distance dominated and depend on 
the CKM matrix elements as
\beq
\begin{split}
\Delta M_d &\!=\! M_H^d-M_L^d = \\
&= \frac{G_F^2 M_{B_d}}{6 \pi^2} m_W^2 |V_{tb}V_{td}^*|^2  
\eta_B S_0(x_t) f_{B_d}^2 B_{B_d}, 
\end{split}
\eeq
and similarly for the $B_s$ system with the substitution $d\to s$. 
Here $\eta_B S_0(x_t)$ encodes the short-distance information in the Inami-Lim function $S_0(x_t)$
that depends on the top mass through $x_t=m_t^2/m_W^2$, 
while $\eta_B = 0.55$ is a numerical factor containing NLO QCD 
corrections due to running from $m_W$ to $\mu\sim m_b$ \cite{Buras:1990fn}. 

The mass difference is precisely measured 
in the $B_d$--$\bar B_d$ system with the
present WA $\Delta M_d = 0.505 \pm 0.005 \ {\rm ps}^{-1}$~\cite{Aubert:2005kf,Abe:2004mz,HFAG}.
Further improvement of this measurement at a SFF is not likely to reduce the
error on $|V_{td}|$, 
which is dominated at present by theory (lattice) errors. 
The $B_s$--$\bar B_s$ mixing parameter $\Delta M_{s}$ 
has recently been measured at the Tevatron
to be $\Delta M_s = 17.77 \pm 0.10 \pm 0.07 \ {\rm ps}^{-1}$~\cite{Abulencia:2006ze}.
Again, lattice errors limit the direct extraction of $|V_{ts}|$ from this result.

The parameters $f_{B_{d,s}}$ and $B_{B_{d,s}}$ 
have been computed in lattice QCD using a variety 
of methods (see \textcite{Okamoto:2005zg,Tantalo:2007ai} for recent reviews). 
Both quenched and unquenched determinations of the decay constants are available.
For the bag parameters the quenching effect is not very important. 
For instance, the analogous quantity $B_K$ of the kaon system  
has been computed in unquenched simulations using
domain wall quarks, and is now known to about $5-6\%$
error~\cite{Antonio:2007pb}.
In fact, separating out the decay constants 
from $f_{B_{d,s}} \sqrt{B_{B_{d,s}}}$ 
is a notational artefact remaining from the days of 
vacuum saturation approximation~\cite{Bernard:1998dg,Dalgic:2006gp}.
Calculating the product instead can lead to reduced errors.

The best constraint comes at present from the ratio of the mass differences 
\beq
\label{DmdDmsRatio}
\frac{\Delta M_s}{\Delta M_d} = 
\frac{M_{B_s}}{M_{B_d}} \xi^2 \left| \frac{V_{td}}{V_{ts}}\right|^2,
\eeq
where $\xi=f_{B_s} \sqrt{B_{B_s}}/f_{B_d} \sqrt{B_{B_d}}$. 
Several theoretical uncertainties cancel out in this ratio.
From Eq.\eqref{DmdDmsRatio} and the experimental values of 
$\Delta M_d$ and $\Delta M_s$ given above, one obtains
$|V_{td}/V_{ts}|=0.2060 \pm 0.0007 \, ^{+0.0081}_{-0.0060}$~\cite{Abulencia:2006ze}
where the first error is experimental and the second theoretical,
from the input value $\xi=1.21^{+0.047}_{-0.035}$ which is obtained from an 
average of $n_f=2$ partially quenched simulations \cite{Okamoto:2005zg}. 
Thus, the lattice uncertainty also dominates 
this constraint; indeed the stated errors here may well be an
underestimate. 
However, unquenched precision calculations of $\xi$ are 
underway; see {\it e.g.} \textcite{Dalgic:2006gp}
and certainly by the time of SFF the stated error on $\xi$ should
be confirmed.   

An alternative determination  of $|V_{td}/V_{ts}|$ can be obtained 
from the ratio of $b\to d\gamma$ and $b\to s\gamma$ rare radiative decays. 
This is discussed in more detail in Section~\ref{constraintsCKM}, 
and we give here only a brief account. 
Taking the ratio of $B\to \rho\gamma$ and $B\to K^* \gamma$ exclusive decays, 
the hadronic matrix elements cancel
to a good approximation, giving
\beq
\frac{{\cal B}(B\to \rho\gamma)}{{\cal B}(B\to K^*\gamma)} =
\left| \frac{V_{td}}{V_{ts}}\right|^2 \Big(\frac{{M_B^2}-{m_\rho^2}}
{{M_B^2}-{m_{K^*}^2}}\Big)^3
\zeta^2 (1 + \Delta R).
\eeq
Here $\zeta $ is the ratio of the $B\to\rho/K^*$ tensor form
factors and equals 1 in the SU(3) limit, and $\Delta R$
describes the effect of the weak annihilation in $B^\pm \to \rho^\pm\gamma$. 
As discussed in Section~\ref{constraintsCKM}, 
this gives results in good agreement with the 
determination from neutral $B_{d,s}$ meson mixing, albeit with larger errors 
that, for now, are predominatly experimental in origin. 
We note that the corresponding inclusive radiative modes can be used as well, 
provided that the $s\bar s$ background in $b\to d\gamma$ modes 
can be reliably taken into account.

Theoretically, an extremely clean determination of $|V_{td}/V_{ts}|$ 
is possible using the ratio \cite{Buras:2000dm}
\beq\label{VtdVtsfromneutrinos}
\frac{{\cal B}(B\to X_d\nu\bar \nu)}{{\cal B}(B\to X_s\nu\bar \nu)}=
\left|\frac{V_{td}}{V_{ts}}\right|^2,
\eeq
which is predicted in the SM with essentially no hadronic uncertainties. 
However, the inclusive modes in Eq.~\eqref{VtdVtsfromneutrinos} 
are very challenging experimentally because of the presence of 
the two undetected neutrinos. 
Nevertheless, studies of these decays,
in particular in exclusive final states, can be started at a SFF,
as we discuss in Section~\ref{ExclXllTheory}. 
We mention here that since the exclusive modes are subject to SU(3) breaking,
an extraction of $V_{td}/V_{ts}$ without theory uncertainty
can only be obtained from inclusive measurements.

Table \ref{table:sides} summarizes the current versus the estimated
error in the SFF era.  

\begin{table}[!tb]
  \begin{center}
    \begin{ruledtabular}
\caption{
Precision on sides determination, current
versus projected in the SFF era. Since in some cases the error is
dominated by
theory the projected improvements are based on expectations for theory.}
\label{table:sides}
 \begin{tabular}{ccc}  
Side & Current accuracy & Projected accuracy \\
\hline
$V_{cb}$ excl. & 4--5\% & 2--3 \% \\
$V_{cb}$ incl. & 1.5--2\% & 0.7--1\% \\
\hline
$V_{ub}$ excl. & $\sim$ 18\% & 3--5\% \\
$V_{ub}$ incl. & $\sim 8\%$  & 3--5\% \\
\hline
$V_{td}/V_{ts}$ & 5--6\% & 3--4\% \\
\end{tabular}
\end{ruledtabular}
\end{center}
\end{table}

\section{Time-dependent CP asymmetry in penguin-dominated modes}
\label{time-dependent}

Penguin dominated hadronic $B$ decays offer one of the most promising sets of observables to search for new sources
of CP violation. The time dependent CP asymmetry in channels such as $B^0 \to \phi K_S$ and $B^0 \to \eta^\prime K_S$
gives in the SM the value of $\sin 2\beta$ that should be the same (up to suppressed terms) as the one determined from the tree dominated ``golden'' mode $B^0 \to J/\psi K_S$ (cf. Section \ref{sec:ut:beta}). However, since $B^0 \to \phi K_S$ and $B^0 \to \eta^\prime K_S$ are loop dominated, NP contributions can modify this
prediction. 

\begin{figure}
  \begin{center}
    \includegraphics[width=0.8\columnwidth]{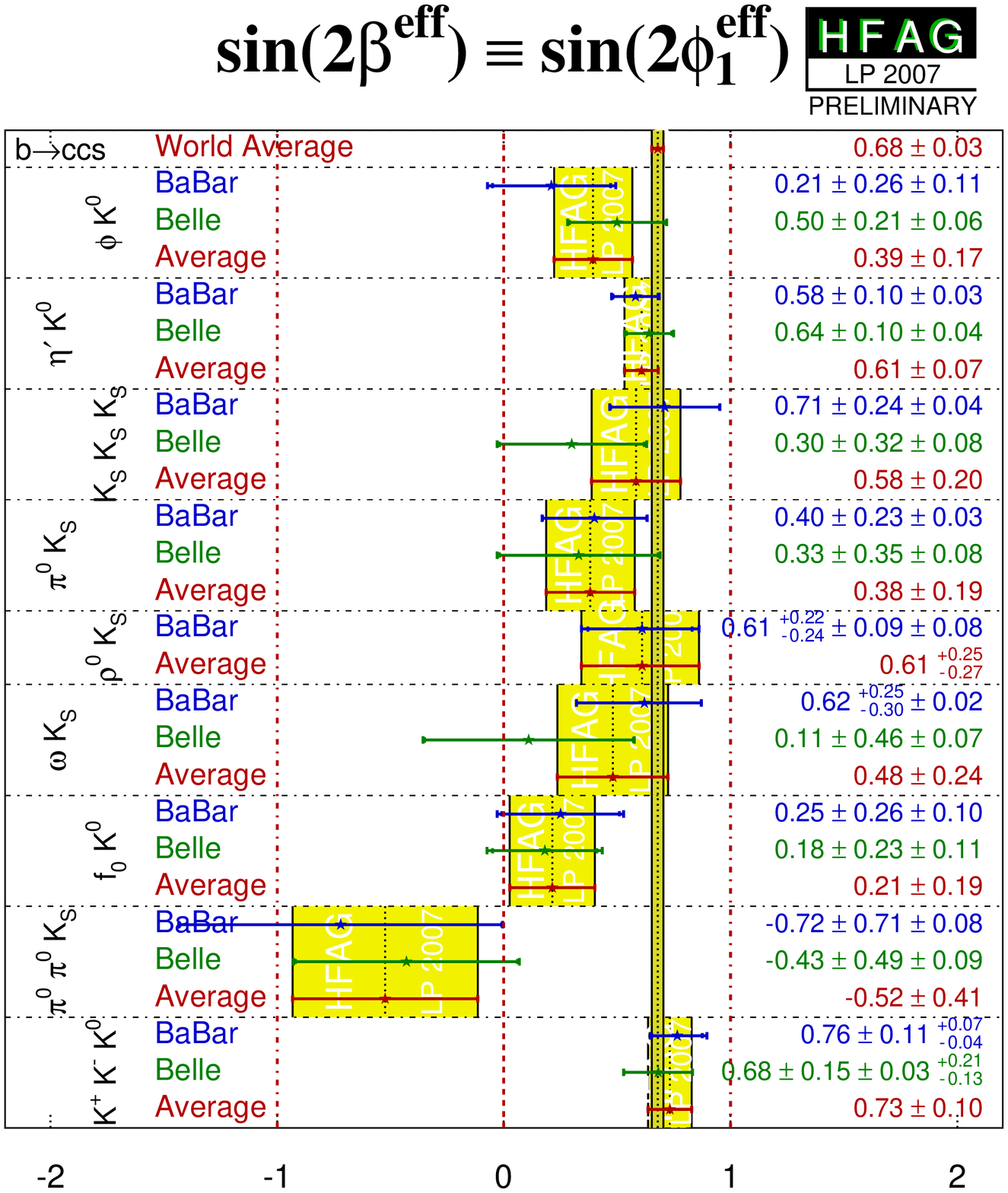}
    \caption{
      HFAG compilation of $\sin(2\beta^{\rm eff})\equiv -\eta_f S_f $ measurements
      in $b \to s$ penguin dominated decays~\cite{HFAG} compared to $\sin(2\beta)$ from $b \to c\bar{c}s$ decays
      to charmonia such as $B^0 \to J/\psi K^0$. 
      The figure does not include the recent \babar\ result on 
      $B^0 \to f_0 K_S^0$ from the time-dependent Dalitz plot analysis of 
      $B^0 \to \pi^+\pi^-K_S^0$~\cite{Aubert:2007vi}, which has highly non-Gaussian uncertainties.
      \label{fig:hfag_btos}
    }
  \end{center}
\end{figure}

The decay amplitude for the penguin dominated $\Delta S=1$ charmless $B$ decay can be written as
\beq  
\label{unitarity}
\begin{split}
  M(\ov B^0\to f) 
  & = \lambda_u^{(s)} A_f^u + \lambda_c^{(s)}A_f^c,
\end{split}
\eeq   
where the ``tree'' amplitude, $A_f^u$, and ``penguin'' amplitude, $A_f^c $, are multiplied by 
different CKM elements $\lambda_q^{(s)}=V_{qb}V_{qs}^*$. This is a
general decomposition. Using CKM unitarity, $\lambda_t^{(s)} = -\lambda_u^{(s)} -\lambda_c^{(s)}$, any
SM contribution can be cast in the form of Eq. \eqref{unitarity}. The ``tree'' contribution is suppressed by 
a factor $|\lambda_u^{(s)}/\lambda_c^{(s)}|\sim 1/50$ and can be neglected to first approximation. Following the same steps
as for the ``golden'', tree-dominated mode $B^0 \to J/\psi K_S$ in Eq.~\eqref{lambdaJpsiKS}, this then gives $\lambda_f\simeq \eta_f e^{-2i\beta}$
with $\eta_f=+1$ ($-1$) for CP-even (CP-odd) final states. Therefore, the SM expectation is that
\beq\label{expectation}
-\eta_f S_f \simeq \sin 2\beta, \qquad \A_f\simeq 0.
\eeq

The same is expected for mixing-induced CP violation in  $B^0\to J/\psi K^0$ as described in Section~\ref{sec:ut:beta}.
Here the measurements are quite mature, with the latest world average (including both $J/\psi K_S$ and $J/\psi K_L$ final states)~\cite{HFAG}
\beq  
\sin 2\beta \approx S_{J/\psi K^0}= 0.668 \pm 0.026 \, .
\eeq   

The $B$ factories 
have measured in the past few years time-dependent CP violation parameters 
for a number of $b \to s$ modes, including
$B^0 \to \phi K^0$, $B^0 \to \eta^\prime K^0$, $B^0 \to K_S K_S K_S$, 
$B^0 \to \pi^0 K_S$, $B^0 \to \rho^0 K_S$, $B^0 \to \omega^0 K_S$,
$B^0 \to f_0 K^0$, $B^0 \to \pi^0\pi^0K_S$ and $B^0 \to K^+K^-K^0$~\cite{Chen:2006nk,Aubert:2007sd,Aubert:2006wv,Aubert:2007me,Aubert:2007mgb,Abe:2006gy,Aubert:2007vi,Aubert:2006ai,Aubert:2007ub,Abe:2007xd}.
A recent compilation of these results is shown in Figure~\ref{fig:hfag_btos}.
To make the test of SM more transparent it is convenient to introduce
\beq \label{DeltaSf}
\Delta S_f\equiv -\eta_f S_f - S_{J/\psi K^0}.
\eeq  
where $f$ is a penguin-dominated final state.  Up to small corrections to be discussed below, one has $\Delta S_f=0$ in the SM.
A summary of the current experimental world averages for $\Delta S_f$
is given in Table~\ref{tab:deltaS_exp}.

\begin{table}
  \begin{center}
    \begin{ruledtabular}
      \caption{
        Current experimental world averages for $\Delta S_f$ and $\A_f$~\cite{HFAG}.
        The recent \babar\ result from  on 
        $B^0 \to f_0 K_S^0$ from time-dependent $B^0 \to \pi^+\pi^-K_S^0$ Dalitz plot analysis~\cite{Aubert:2007vi} is not included,
        since it has highly non-Gaussian uncertainties.
      }
      \label{tab:deltaS_exp}
      \begin{tabular}{ccc}
        Mode & $\Delta S_f$ & $\A_f$ \\
        \hline
        $\phi K^0$     &  $-0.28 \pm 0.17$ & $0.01 \pm 0.12$ \\
        $\eta^\prime K^0$ & $-0.06 \pm 0.08$ & $0.09 \pm 0.06$ \\
        $K_S K_S K_S$  &  $-0.09 \pm 0.20$ & $0.14 \pm 0.15$ \\ 
        $\pi^0 K_S$    &  $-0.29 \pm 0.19$ & $-0.14 \pm 0.11$ \\
        $\rho^0 K_S$   &  $-0.06^{+0.25}_{-0.27}$ & $-0.02 \pm 0.29$ \\
        $\omega^0 K_S$ &  $-0.19 \pm 0.24$ & $0.21 \pm 0.19$ \\
        $f_0 K^0$      &  $-0.46 \pm 0.18$ & $-0.08 \pm 0.12$ \\
        $\pi^0\pi^0K_S$&  $-1.19 \pm 0.41$ & $-0.18 \pm 0.22$ \\
        $K^+K^-K^0$    &  $0.06 \pm 0.10$ & $-0.07 \pm 0.08$ \\
      \end{tabular}
    \end{ruledtabular}
  \end{center}
\end{table}

So far we have neglected the ``tree'' amplitude $A_f^u$ of Eq.~\eqref{unitarity}. In many of the penguin dominated modes, 
{\it e.g.} $\omega K_S, \rho^0K_S, \pi^0K_S$, the amplitude
$A_f^u$ receives contributions from the $b\to u\bar u s$ tree operators which can partially lift the large CKM suppression. 
To first order in 
$r_f \equiv (\lambda_u^{(s)} A_f^u)/(\lambda_c^{(s)} A_f^c)$ one has~\cite{Gronau:1989ia,Grossman:2003qp,Cheng:2005ug}
\beq 
\label{eq:CfSf}
\begin{split}
  \Delta S_f = & 2|r_f|\cos 2\beta\sin\gamma\cos\delta_f,\\
  \A_f = & 2|r_f|\sin\gamma\sin\delta_f,
\end{split}
\eeq
with a strong phase $\delta_f={\rm arg}(A_f^u/A_f^c)$. Both $\Delta S_f$ 
and $\A_f$ can thus deviate appreciably from zero, if the ratio $A_f^u/A_f^c$ is large. Most importantly, the size of this ratio is channel dependent and 
will give different $\Delta S_f$ for different modes. 
We thus turn next to the theoretical estimates of $\Delta S_f$. 

\subsection{Theoretical estimates for $\Delta S_f$}

The original papers~\cite{Grossman:1996ke,Fleischer:1996bv,London:1997zk,Ciuchini:1997zp,Gronau:1989ia} that suggested $\Delta S_f$ (Eq.~\eqref{DeltaSf}) as a powerful tool for new physics searches used naive factorization. 
In recent years several theoretical reappraisals have been performed 
using several different approaches to calculate $\Delta S_f$  
(for detailed reviews, see {\it e.g.}~\cite{Silvestrini:2007yf,Zupan:2007ca}).
The methods used are either based on SU(3) symmetry relations~\cite{Grossman:2003qp,Gronau:2003kx,Gronau:2004hp,Engelhard:2005hu,Engelhard:2005ky,Gronau:2006qh,Fleischer:2007mq,Buras:2005cv,Buras:2004th,Buras:2004ub,Buras:2003dj,Buras:2003yc}; 
or use the $1/m_b$ expansion -- QCD factorization (QCDF)~\cite{Beneke:2005pu,Buchalla:2005us,Cheng:2005bg,Cheng:2005ug}
perturbative QCD (pQCD)~\cite{Li:2006jv,Ali:2007ff}, and
Soft-Collinear Effective Theory (SCET)~\cite{Williamson:2006hb}.
Table~\ref{tab:penguin} summarizes some of the findings.

The SU(3) relations typically give only loose constraints on $\Delta S_f$ 
since the bounds involve sums of amplitudes, where relative phases are unknown.
Furthermore, SU(3) breaking is hard to estimate and all the analyses are done 
only at leading order in the breaking. 
The $1/m_b$ expansion on the other hand provides a systematic framework 
where higher order corrections can in principle be included. 
The three approaches: QCDF, pQCD and SCET, 
while all using the $1/m_b$ expansion, 
differ in details such as the treatment of higher order corrections, 
charming penguins~\cite{Ciuchini:1997hb,Ciuchini:2001gv} 
and the scale at which the treatment is still deemed perturbative~\cite{Beneke:2004bn,Bauer:2005wb}. 

Experimental observations of large direct CP asymmetries 
in several exclusive $B$ decay modes, 
such as $K^+ \pi^-$~\cite{Chao:2004mn,Aubert:2007mj} and 
$\pi^+ \pi^-$~\cite{Ishino:2006if} require large strong phases. 
In different theoretical approaches these are seen to come from 
different sources. 
In pQCD~\cite{Keum:2000wi} they arise from annihilation diagrams 
and are deemed calculable using a phenomenological parameter $k_T$ 
as an endpoint divergence regulator. 
In QCDF the large strong phase is deemed nonperturbative and 
comes from endpoint divergent weak annihilation diagrams and the
chirally-enhanced power corrections to hard spectator scattering. 
It is then modeled using nonperturbative parameters. 
In SCET the strong phase is assigned to nonperturbative charming penguins, 
while annihilation diagrams are found to be real~\cite{Arnesen:2006vb}. 
The nonperturbative terms are fit from data. 
In the approach of~\textcite{Cheng:2004ru,Cheng:2005bg,Cheng:2005ug} 
the strong phases are assumed to come from final state interactions. 
These are then calculated from on-shell rescattering of 2-body modes, 
while QCDF is used for the short-distance part.

\subsection {Theoretically cleanest modes} 

The deviations $\Delta S_f$ are expected to be the smallest in 
$\eta^\prime K^0$, $\phi K^0$ and $K_S K_S K_S$~\cite{Gershon:2004tk} channels, 
making them the theoretically cleanest probes of NP, 
see Table \ref{tab:penguin}.
The tree pollution in the decays $B\to\phi K^0, K_S K_S K_S$ is small 
since the tree operators $Q_{1,2}$ do not contribute at all
(taking $\phi$ to be a pure $s\bar{s}$ state). 
Thus $\Delta S_f\ne 0$ arises only from EWP contributions. 
In $B\to \eta^\prime K^0$, on the other hand, tree operators do contribute.
However, the penguin contribution is enhanced,
as signaled by the large $B\to \eta^\prime K$ branching ratios~\cite{Schumann:2006bg,Aubert:2007si,HFAG}, 
giving again a small tree--to--penguin ratio $r_f$. 
The differences in the predicted values of $\Delta S_{\eta^\prime K_S}$
seen in Table \ref{tab:penguin}
can be attributed to different determinations of strong phases 
and nonperturbative parameters. 
While only the SCET prediction of $\Delta S_{\eta^\prime K_S}$ is negative 
(going in the direction of the experimental central value), 
all the calculations find $|\Delta S_{\eta^\prime K_S}|$ to be small. 
To establish clear evidence of NP effects in these decays, 
a deviation of $\Delta S_f$ from zero that is 
much larger than the estimated theoretical uncertainty is needed.

\begin{table}
  \begin{center}
    \begin{ruledtabular}
      \caption{Expectations for $\Delta S_f$ in three cleanest modes.}
      \label{tab:penguin}    
      \begin{tabular}{ccccc}
        Model & $\phi K^0$  & $\eta^\prime K^0$ & $K_S K_S K^0$ \\
        QCDF+FSI${}^{\rm a}$ &  
        $0.03^{+0.01}_{-0.04}$ & $0.00^{+0.00}_{-0.04}$ & $0.02^{+0.00}_{-0.04}$ \\
        QCDF${}^{\rm b}$ & 
        $0.02 \pm 0.01$ & $0.01 \pm 0.01$ \\
        QCDF${}^{\rm c}$ & 
        $0.02 \pm 0.01$ & $0.01 \pm 0.02$ \\
        SCET${}^{\rm d}$ &
        & $\begin{array}{c} -0.019 \pm 0.009 \\ -0.010 \pm 0.010 \end{array}$ \\
        pQCD${}^{\rm e}$ &
        $0.02 \pm 0.01$ \\
      \end{tabular}
    \end{ruledtabular}
  \end{center}
  \vspace{-7mm}
  \begin{flushleft}
    \begin{tabular}{l@{\hspace{5mm}}l}
      ${}^{\rm a}$\textcite{Cheng:2005bg,Cheng:2005ug} &
      ${}^{\rm b}$\textcite{Beneke:2005pu} \\
      ${}^{\rm c}$\textcite{Buchalla:2005us} &
      ${}^{\rm d}$\textcite{Williamson:2006hb} \\
      ${}^{\rm e}$\textcite{Li:2006jv}
    \end{tabular}
  \end{flushleft}
\end{table}

\subsection{Comparison with SM value of $\sin 2 \beta$}

As experimental errors reduce, for a number of modes the deviations
of $\Delta S_f$ from zero may become significant. 
The translation of the measured values of $\Delta S_f$ into a 
deviation from the SM then becomes nontrivial. 
However, forgetting about this issue and just averaging 
over the experimental data given in Table~\ref{tab:deltaS_exp} gives
a value of $\left< \Delta S_f \right> = -0.11 \pm 0.06$~\cite{HFAG}
(using only the theoretically cleanest modes 
$\eta^\prime K^0$, $\phi K^0$ and $K_S K_S K^0$,
one obtains instead $\left< \Delta S_f \right> = -0.09 \pm 0.07$). 

Different approaches that take into account theoretical predictions 
are possible~\cite{Zupan:2007ca}.
Correcting for the SM value of $\Delta S_f$ by defining 
$(\Delta {S_f})_{\rm corr}=(\Delta S_f)_{\rm exp}-(\Delta S_f)_{\rm th}$,
one has several choices that can be taken for $(\Delta S_f)_{\rm th}$,
including: 
(i) to use all available theoretical predictions in a particular framework 
({\it e.g.} QCDF), and to discard remaining experimental data, 
(ii) to use the theoretical prediction for each channel that is closest 
to the experimental data 
(and neglecting three-body decays where only one group has made predictions). 
The first prescription gives 
$\left< (\Delta {S_f})_{\rm corr}\right> = -0.133\pm0.063$~\cite{Zupan:2007ca}. 
Interestingly enough the second prescription gives almost exactly the same result.

\subsection{Experimental prospects}

Several previous studies have considered the potential of a 
SFF to improve the measurements of $\Delta S_f$ 
to at least the level of the current theoretical uncertainty
in a wide range of channels, including all the theoretically cleanest modes~\cite{Gershon:2006mt,Akeroyd:2004mj,Hewett:2004tv,Bona:2007qt,Hashimoto:2004sm}.
By extrapolating the current experimental measurements,
these studies show that data samples of at least $50 \ {\rm ab}^{-1}$
(containing at least $50 \times 10^9$ $B\bar{B}$ pairs) will be necessary.
This roughly corresponds to five years of operation for a facility
with peak luminosity of $10^{36} {\rm cm}^{-2} {\rm s}^{-1}$
and data taking efficiency comparable to the current $B$ factories.
These studies also indicate the systematic uncertainties are unlikely to 
cause any unsurmountable problems at the few percent precision level
that will be reached
(although the Dalitz plot structure of the $B^0 \to K^+K^-K^0$ decay~\cite{Aubert:2007sd}
will need to be clarified to obtain high precision on $S_{\phi K^0}$).

One may consider the potential of a hadronic machine to address these modes.
At present, it appears that $\phi K_S$ is difficult, but not impossible
to trigger and reconstruct in the hadronic environment, 
due to the small opening angle in $\phi \to K^+K^-$;
$\eta^\prime K_S$ is challenging since neutral particles are involved in the 
$\eta^\prime$ decay chain; for $K_SK_SK_S$ meanwhile, 
there are no charged tracks originating from the $B$ vertex,
and so both triggering and reconstruction seem highly complicated.
Modes containing $K_L$ mesons in the final state 
may be considered impossible to study at a hadron machine.
Furthermore, due to the theoretical uncertainties discussed above,
there is a clear advantage provided by the ability to study multiple channels
and to make complementary measurements that check that the theory errors 
are under control.
Thus, these modes point to a Super Flavor Factory,
with integrated luminosity of at least $50 \ {\rm ab}^{-1}$.

\section{Null tests of the SM}
\label{sec:null}
An important tool in searching for new flavor physics effects are the 
observables that vanish or are very small in the SM, have small calculable
corrections and potentially large new physics effects. Several examples
of such null tests of the SM are discussed at length in separate sections of 
this review:

\begin{itemize}
\item
As discussed in Section~\ref{InclTheory},
the untagged direct CP asymmetry $A_{\rm CP}(B\to X_{s+d}\gamma)$ 
vanishes in the U-spin limit~\cite{Soares:1991te,Hurth:2001yb}.\footnote{
  For neutral $B$ decays potential nonzero contributions, 
  such as  annihilation, start at $\alpha_s(m_b)/m_b^3$ order.
} 
The leading SU(3) breaking corrections are of order 
$(m_s/m_b)^2 \sim 5 \cdot 10^{-4}$ giving
$A_{\rm CP}(B\to X_{s+d}\gamma) \sim 3 \cdot 10^{-6}$~\cite{Hurth:2001yb}. 
This can be easily modified by new physics contributions. 
For instance, in the MSSM with nonvanishing flavor blind phases 
$A_{\rm CP}(B\to X_{s+d}\gamma)$ can be a few percent, 
while more general flavor violation can saturate the 
present experimental bounds~\cite{Hurth:2003dk}.  

\item
Photon polarization in $B\to V\gamma$ decays. As discussed in Section \ref{photonpol}, the time dependent
CP asymmetry, $S$,  in $B(t) \to \gamma K^* (K_S \pi^0, \rho,...)$ can be used as quasi-null tests of
the SM.

\item
Lepton flavor violating $\tau$ decays such as $\tau\to \mu \gamma$, 
$\tau \to 3\mu$, etc., would be a clear signal of new physics. 
The theoretical expectations and SFF reach are discussed in Section~\ref{sec:LFV}. 

\item
CP asymmetry from interference of decay and mixing in 
$\Delta S=1$ penguin dominated decays, $S_f$, 
is equal to $\sin 2\beta$ up to CKM suppressed hadronic corrections. 
As shown in Section \ref{time-dependent}, 
the precision of this test is at the few percent level or below 
for several modes such as 
$B\to \eta^\prime K_S, \phi K_S, K_S K_S K_S$ decays.
New physics contributions can easily accommodate much larger deviations. 

\end{itemize}  
In this section we give some further examples of null tests. 

\subsection{Isospin sum-rules in $B\to K\pi$} 
As first discussed  by \textcite{Lipkin:1998ie} and by \textcite{Gronau:1998ep}
the following sum of CP averaged $B\to K\pi$ decay widths 
\beq
\label{LipkinSumRule} 
\begin{array}{lcr}
  \multicolumn{2}{l}{
    \Delta L \equiv \frac{1}{\Gamma({\bar K^0\pi^-})}
    \Big[ 2  \Gamma({\bar K^0\pi^0}) - \Gamma({K^-\pi^+}) +
  } \\
  \hspace{30mm} &
  \multicolumn{2}{r}{
    2 \Gamma({K^-\pi^0}) - \Gamma({\bar K^0\pi^-})\Big],
  }
\end{array}
\eeq
vanishes in the SM up to second order in two small parameters: 
the EWP-to-penguin ratio and the doubly CKM suppressed tree-to-penguin ratio.
Assuming isospin symmetry, the LO SCET theory prediction is 
$\Delta L\stackrel{\rm Th.}{=}(2.0\pm0.9\pm0.7\pm0.4)\times 10^{-2}$~\cite{Williamson:2006hb}, 
which is compatible with and more precise than a QCDF prediction~\cite{Beneke:2003zv}.
Remaining isospin breaking contributions are small~\cite{Gronau:2006eb}.  
The experimental value has at present much larger errors, 
$\Delta L\stackrel{\rm Exp.}{=}0.13\pm0.09$~\cite{Aubert:2006gm,Aubert:2007hh,Aubert:2006fha,Aubert:2007mgb,Abe:2006qx,HFAG}.
The precision of the branching fraction measurements of all input modes
would need to be improved to make a significant reduction in this experimental
uncertainty at a SFF.
The measurements currently have comparable statistical and 
systematic uncertainties, so this is not straightforward.
However, some modest reduction of uncertainties due to 
$K_S$ and $\pi^0$ reconstruction efficiencies can be expected,
so that this test may become at least a factor two more stringent.

A quantity that is even further suppressed in SM is a similar sum of 
partial decay width differences 
$\Delta \Gamma=\Gamma(\bar B\to f)-\Gamma(B\to \bar f)$
\beq
\label{DiffSumRule}
\begin{array}{lcr}
  \multicolumn{2}{l}{
    \Delta_{\sum} = \frac{1}{\Gamma({\bar K^0\pi^-})}
    \Big[ 2  \Delta\Gamma({\bar K^0\pi^0}) - \Delta\Gamma({K^-\pi^+}) + 
  } \\
  \hspace{30mm} & 
  \multicolumn{2}{r}{
    2 \Delta\Gamma({K^-\pi^0}) - \Delta\Gamma({\bar K^0\pi^-})\Big].
  }
\end{array}
\eeq
In the limit of exact isospin and no EWP $\Delta_{\sum}$ vanishes~\cite{Atwood:1997iw,Gronau:2005gz}. 
Furthermore, the corrections due to EWP are subleading 
in the $1/m_b$ expansion~\cite{Gronau:2005kz}, 
so that $\Delta_{\sum}$ is expected to be below $1\%$. 
Experimentally, $\Delta_{\sum} \stackrel{\rm Exp.}{=} 0.01 \pm 0.10$~\cite{Aubert:2007mgb,Aubert:2007mj,Aubert:2006gm,Aubert:2007hh,Abe:2006gy,Abe:2006xs,HFAG}, 
where the uncertainty is dominated by the ${\cal A}_{\rm CP}(\pi^0 K^0)$ experimental error. 
This is large because the reconstructed final state for this mode ($\pi^0 K_S$)
is a CP eigenstate containing no information on the initial $B$ meson flavor.
The required flavour tagging comes at a statistical cost that is, however,
less severe at an $e^+e^-$ $B$ factory than at a hadron collider.
Therefore, this SM test is unique to a SFF,  where a significant improvement 
compared to the current precision can be expected.

The above sum rules given in Eq.~\eqref{LipkinSumRule} and Eq.~\eqref{DiffSumRule} 
can be violated by NP that breaks isospin symmetry. 
An example is given by NP contributions to EWP, 
extensively discussed in the literature (see \textcite{Buras:2004ub,Baek:2004rp} and references therein).

\subsection{$b\to ss\bar d$ and $b\to dd\bar s$ decays}
In the SM $b\to ss\bar d$ and $b\to dd\bar s$ transitions are highly 
suppressed, proceeding through a $W$--up-type-quark box diagram~\cite{Huitu:1998vn}. 
Compared to the penguin transitions  $b\to q\bar q s$ and $b\to q\bar q d$ 
they are additionally suppressed by the CKM factor 
$V_{td}V_{ts}^*\sim \lambda^5 \simeq 3 \cdot 10^{-5}$ and
are thus exceedingly small in the SM, 
with inclusive decay rates at the level of $10^{-12}$ and $10^{-14}$ 
for $b\to ss\bar d$ and $b\to dd\bar s$, respectively~\cite{Fajfer:2006av}. 

These amplitudes can be significantly enhanced in SM extensions, 
for instance in MSSM with or without conserved $R$ parity, 
or in the models containing extra $U(1)$ gauge bosons. 
For example,  the $b\to ss\bar d$ decays 
$B^-\to K^{*-}\bar K^{*0}$ and $B^-\to K^{-}\bar K^{*0}$ can reach 
$\sim 6\cdot 10^{-9}$ in the MSSM, while they are 
$\sim 7 \cdot 10^{-14}$ in the SM \cite{Fajfer:2000ny}. 
Note that the flavor of $\bar K^{*0}$ is tagged using the decay into 
the $K^-\pi^+$ final state. 
The $b\to dd\bar s$ transitions 
$B^-\to \pi^{-}K^{*0}$ and $B^-\to \rho^{-} K^{*0}$ can be enhanced 
from $\sim 10^{-16}$ in the SM to $\sim 10^{-6}$ 
in the presence of an extra $Z^\prime$ boson~\cite{Fajfer:2006av}.
The relevant experimental upper limits are 
${\cal B}(B^- \to K^-K^-\pi^+) < 1.3 \times 10^{-6}$ and
${\cal B}(B^- \to K^+\pi^-\pi^-) < 1.8 \times 10^{-6}$~\cite{Aubert:2003xz}.
Although these decays are background limited, improvements in these limits by almost two orders of
magnitude can be expected from a SFF.

Although the observation of highly suppressed SM decays would provide
the clearest signal for NP in these decay amplitudes,
there are a number of other possible signals for such wrong sign kaons~\cite{Chun:2003rg}.
For example, these amplitudes could invalidate the isospin relations given above,
cause a non-zero CP asymmetry in $B^- \to K_S\pi^-$,
induce a difference in rates between $B^0 \to K_S\pi^0$ and $B^0 \to K_L\pi^0$
or a difference in rates between $B^0 \to K_SK_S$ and $B^0 \to K_LK_L$,
as well as resulting in a non-zero rate for $B^0 \to K_SK_L$.

\subsection{CP asymmetry in $\pi^+ \pi^0$}
Since $\pi^+ \pi^0$ is an $I=2$ final state, only tree and EWP operators 
contribute to the $B^+\to \pi^+\pi^0$ decay amplitude. 
Therefore, the direct CP asymmetry ${\cal A}_{\pi^+\pi^0}$
is expected to be very small. 
Theoretical estimates range between 
$\lsim 0.1\%$~\cite{Gronau:1998fn,Beneke:2003zv} to ${\cal O}(1\%)$~\cite{Cheng:2004ru}.
The current average of the $B$ factory results is
${\cal A}_{\rm CP}(B^+ \to \pi^+ \pi^0) = 0.06 \pm 0.05$~\cite{Aubert:2007hh,HFAG}.
Further theoretical studies of this observable would be desired 
to match the precision attainable at a SFF.

\subsection{Semi-inclusive hadronic $B$ decays}
Several semi-inclusive hadronic decays can be used to test the SM. 
For instance, the decays $B\to D^0 X_{s,d}$ and $B\to \bar D^0 X_{s,d}$ 
have zero CP asymmetry in the SM, 
because they proceed through a single diagram, 
and provide a check for non-SM corrections to the value of $\gamma$ 
extracted from $B\to DK$  decays (Section~\ref{sec:ut:gamma}).
Another test is provided by flavor untagged semi-inclusive
$B^{\pm} \to M^0 (\bar{M}^0) X^{\pm}_{s+d}$ decays, where $M^0$  is either
an eigenstate of $s \leftrightarrow d$ switching symmetry,
{\it e.g.} $K_S$, $K_L$, $\eta^\prime$ or any charmonium state, or 
$M^0$ and $\bar M^0$ are related by the $s \leftrightarrow d$ transformation,
{\it e.g.} $K^{*0}$, $\bar{K}^{*0}$, and one sums over the two states. 
In the SM the CP asymmetry of such semi-inclusive decays
vanish in the SU(3) flavor limit~\cite{Soni:2005jj,Gronau:2000zy} 
(this follows from the same considerations as for the
direct CP asymmetry in $B\to X_{s+d}\gamma$ in Section~\ref{InclTheory}). 
The CP asymmetries are thus both doubly CKM $(\sim \lambda^2)$ 
and $m_s/\Lambda_{\rm QCD}$ suppressed. 

If the tagged meson $M^0$ is light the CP asymmetries can be 
reliably calculated using SCET in the end-point region, 
where $M^0$ has energy close to $m_b/2$~\cite{Chay:2006ve,Chay:2007ej}. 
This gives CP asymmetries for $B^{\pm} \to M^0 X^{\pm}_{s+d}$
below $1\%$ for each of $M^0 = (K_S, \eta^\prime, (K^{*0}+\bar K^{*0})$~\cite{Soni:2005jj,Atwood:1997de,Atwood:1998ib,Hou:1997wy}.

These modes can be studied at a SFF using inclusive reconstruction 
of the $X$ system by taking advantage of the recoil analysis technique
that is possible due to the $e^+e^- \to \Upsilon(4{\rm S}) \to B^+B^-$ 
production chain. 
The method has been implemented for measurement of inclusive charmless 
$B \to K^+(K^0) X$ decays~\cite{Aubert:2006ana},
as well as having multiple applications for studies of {\it e.g.}
$b \to s\gamma$ and $b \to s\ell^+\ell^-$.
With SFF data samples, this class of important null tests
can be probed to ${\cal O}(1\%)$ precision.

\subsection{Transverse $\tau$ polarization in semileptonic decays}
The transverse polarization of tau leptons produces in
$b \to c \tau \nu$ decays,  defined as
$p^T_{\tau} \equiv \vec S_{\tau} \cdot \vec p_{\tau} \times \vec p_X / 
|\vec p_{\tau} \times \vec p_X|$,
where $\vec S_{\tau}$ is the spin of the $\tau$,
is a very clean observable since it vanishes in the SM.
On the other hand it is very sensitive to the presence of 
a CP-odd phase in scalar interactions. 
It is thus well suited as a probe of $CP$ violating multi-Higgs doublet 
models~\cite{Atwood:1993ka,Grossman:1994eb,Garisto:1994vz}. 

Since $p^T_{\tau}$ is a naive $T_N$-odd observable
it does not require a non-zero strong phase. The fact that 
$p^T_{\tau}$ arises from an underlying CP-odd phase can be
verified experimentally by comparing the asymmetry in $B$ with $\bar B$
decays whence it should change sign reflecting a change
in the sign of the CP-odd phase.

In principle any charged lepton could be used for such searches.
Indeed, the transverse muon polarization in kaon decays has been of interest
for a very long time~\cite{Abe:2004px,Abe:2006de}. 
The advantage of using the tau lepton is that $\tau$ decays serve as 
self-analyzers of the polarization. 
This propery has already been exploited at the $B$ factories~\cite{Inami:2002ah}. 
On the other hand, any semitauonic $B$ decay contains at least two neutrinos, 
so that kinematic constraints from the reconstruction 
of the recoiling $B$ are essential. 

In passing we mention that, as mentioned in Section~\ref{sec:2HDM},
the rates and differential distributions in $B \to D^{(*)}\tau\nu$ decays
are sensitive to contributions from 
charged Higgs exchanges~\cite{Kiers:1997zt}. 
The first studies of these are being carried out 
at the $B$ factories~\cite{Matyja:2007kt,Aubert:2007ds},
though much larger data samples are needed for precise measurements. 
On the other hand, $a_{CP}^\tau$ is theoretically extremely clean, 
so that experimental issues are the only limiting factor. 
Thus, transverse polarization studies in these semitauonic
decays will be a unique new possibilty for exploration at a SFF.

\section{Rare $b\to s\gamma$ and $b\to s\ell^+\ell^-$ decays}

The decays $b\to s\gamma$ and $b\to s\ell^+\ell^-$ are forbidden  at tree level in the Standard Model. 
They do  proceed at loop level, through diagrams with internal $W$ bosons
and 
charge +2/3 quarks,
which has several important implications. 
First, the $b\to s/d\gamma$ amplitudes are particularly sensitive to the weak
couplings of the 
top quark -- the CKM matrix elements $V_{tb}$, $V_{ts}$ 
and $V_{td}$. Along 
with $B-\bar B$ mixing, these processes are the only 
(low energy) 
experimental probes of $V_{td}$, one of the least well-known 
CKM matrix elements. 
Second, the loop suppression of SM contributions makes them an important probe
of possible contributions from new physics particles. As a consequence a great deal of theoretical and experimental work is dedicated to these decays.

In this Section we review the implications of the rare radiative decays
for constraining the Standard Model parameters, and their relevance
in new physics searches. We start by briefly reviewing the present theory status and then proceed to
describe the observables of interest.

\subsection{$B\to X_{s/d}\gamma$ decays}

\subsubsection{Inclusive $B\to X_{s/d}\gamma$ decays}
\label{InclTheory}

The application of the effective Hamiltonian (\ref{HW}) to actual hadronic
radiative decays requires knowledge of the matrix elements for the 
operators $O_i^p$ acting on hadronic states. This 
difficult problem
can be
addressed in a model independent way only in a limited number of cases.

In inclusive radiative decays $b \to s\gamma$,
the operator product expansion 
(OPE) and quark-hadron duality can be used to make
clean predictions for sufficiently inclusive observables: the inclusive rate, the photon energy spectrum or the hadronic invariant mass
spectrum 
\cite{Chay:1990da,Blok:1993va,Manohar:1993qn,Falk:1993dh}. 
These observables can be computed using the heavy quark expansion
in $\Lambda_{\rm QCD}/m_b$, where 
$\Lambda_{\rm QCD} \sim 500$ MeV is the scale of strong interactions.

The starting point 
is the optical theorem, which relates the imaginary part of the forward scattering amplitude $T(E_\gamma) = i\int d^4 x T\{{\cal H}_W\,, {\cal H}_W \}$ to the inclusive rate 
\beq
\begin{split}
  \Gamma(B \to X_s\gamma) =
  & \frac{1}{2M_{B}} \Big(-\frac{1}{\pi}\Big) 
  \mbox{Im }
  \langle B |T(E_\gamma)| B\rangle.
\end{split}
\eeq
Here $E_\gamma$ is the photon energy. 
In the heavy quark limit the energy release into hadronic final states is very
large, so that the forward scattering amplitude $T(E_\gamma)$ is dominated by short
distances $x\sim 1/m_b\to 0 $. This implies that $T(E_\gamma)$, and thus the total $B\to X_s\gamma$
rate, can be expanded 
in powers of
$\Lambda_{\rm QCD}/m_b$ using OPE
\begin{eqnarray} \label{OPE}
-\frac{1}{\pi} \mbox{Im } T = {\cal O}_0 + \frac{1}{m_b} {\cal O}_1 + 
\frac{1}{m_b^2} {\cal O}_2 + \cdots.
\end{eqnarray}
Here ${\cal O}_j$ are the most general local operators of dimension $3+j$ which can
mediate the $b \to b$ transition. 
At leading order there is only one such operator ${\cal O}_0 = \bar b b$. Its matrix element 
is known exactly from $b$ quark number conservation. 
The dimension 4 operators ${\cal O}_1$ vanish by 
the equations of motion~\cite{Chay:1990da},
while the matrix elements of the dimension-5 operators ${\cal O}_2$ can be
expressed in terms of two nonperturbative parameters
\beq
\begin{split}
\lambda_1 &= \frac{1}{2M_B} \langle \bar B |\bar b_v (iD)^2 b_v | \bar B
\rangle \,, \\
3\lambda_2 &= \frac{1}{2M_B} \langle \bar B |\bar b_v \frac{g}{2}
\sigma_{\mu\nu} G^{a\mu \nu } T^a b_v | \bar B \rangle \,,
\end{split}
\eeq
where $b_v$ is the static heavy quark field. 
The 
$B\to X_s\gamma$ decay rate  following from the OPE (\ref{OPE})
is thus 
\beq\label{B2Xsincl}
\begin{split}
\Gamma(B\to X_s\gamma) = &\frac{\alpha G_F^2}{16\pi^4} m_b^5 |\lambda_t^{(s)}|^2 \times \\
 &
    \times |C_{7\gamma}(m_b)|^2
    \Big[1  + \frac{\lambda_1 - 9 \lambda_2}{2m_b^2} \Big]\,.
\end{split} 
\eeq 
The leading term represents the parton level $b\to s\gamma$ decay width, which is 
thus recovered
as a model-independent prediction 
in the heavy quark limit. 
The nonperturbative corrections to the LO result are 
doubly suppressed, by $\Lambda_{\rm QCD}^2/m_b^2$. 
In a physical picture they arise from the so-called Fermi motion 
of the heavy quark inside the hadron,
and from its interaction with the color gluon field inside the hadron. 
At each order in the $\Lambda_{\rm QCD}/m_b$ expansion, these effects are 
parameterized in terms of a small number of nonperturbative parameters.

In the endpoint region of the photon spectrum, where $M_B - 2 E_\gamma \sim \Lambda_{\rm QCD}$, the
heavy quark expansion in $\Lambda_{\rm QCD}/m_b$ breaks down. It is replaced 
with a simultanous expansion in powers of $\Lambda_{\rm QCD}/m_b$ and $1-x$, 
where $x = 2E_\gamma/M_B$~\cite{Neubert:1993um,Mannel:1994pm,Bigi:1993ex}.
In this region the invariant mass of the hadronic state is 
$M_X^2 \sim M_B\Lambda_{\rm QCD}$. The photon spectrum is given 
by a factorization relation~\cite{Korchemsky:1994jb,Bauer:2001yt}
\beq\label{factincl}
\begin{split}
    \frac{1}{\Gamma_0} &\frac{d\Gamma(E_\gamma)}{dE_\gamma} = H(E_\gamma, \mu) 
S(k_+) \star J(k_+ \negmedspace+ m_b-2E_\gamma),
\end{split}
\eeq
where $H(E_\gamma,\mu)$ contains the effects of hard loop momenta, 
$J$ is the jet function describing the physics of the hard-collinear loops with $M_B\Lambda_{\rm QCD}$ off-shellness, 
$S(k_+)$ is the shape function parameterizing 
bound-state effects in the $B$ meson, while the star denotes a convolution over soft momentum $k_+$. 
The nonperturbative shape function has to be either extracted 
from data or modelled [commonly used shape function
parameterizations can be found in \cite{Bosch:2004th}].

The present world average for the inclusive branching fraction
is \cite{Aubert:2005cua,Koppenburg:2004fz,Chen:2001fja,HFAG}
\beq
\begin{split}
    &{\cal B}^{\rm exp}(B\to X_s\gamma)|_{E_\gamma > 1.6 \ {\rm GeV}} =\\
&~~(3.55^{+0.09}_{-0.10}|_{\rm shape} \pm 0.24|_{\rm stat/sys} \pm 0.03|_{\rm d\gamma}) \times 10^{-4}.
\end{split}
\eeq
The errors shown are due to the 
shape function,
experimental (statistical and systematic combined), 
and the contamination from $b\to d\gamma$ events, respectively.

On the theory side, the SM prediction for the inclusive branching fraction
has recently been advanced to NNLO~\cite{Misiak:2006zs}, with the result
\beq
{\cal B}(B\to X_s\gamma)|^{\rm NNLO}_{E_\gamma > 1.6 \ {\rm GeV}} =
(3.15 \pm 0.23) \times 10^{-4} \,,
\eeq
where the error combines in quadrature several types of uncertainties: 
nonperturbative (5\%),
parametric (3\%), higher-order (3\%) and $m_c-$interpolation ambiguity (3\%).
The leading unknown nonperturbative corrections to this prediction 
arise from spectator contributions with one hard gluon exchange. 
They scale like ${\cal O}(\alpha_s \Lambda_{\rm QCD}/m_b)$ 
in the limit $m_c \ll m_b/2$ and like 
${\cal O}(\alpha_s \Lambda_{\rm QCD}^2/m_c^2)$ in the limit $m_c \gg m_b/2$.
An alternative estimate, with the photon energy cut dependence 
resummed using an effective theory formalism, gives~\cite{Becher:2006pu}
\beq
\begin{split}
    {\cal B}(B\to X_s\gamma)|^{\rm NNLO}_{E_\gamma > 1.6 \ {\rm GeV}} &= (2.98^{+0.13}_{-0.17}|_{\rm pert}\pm\\
  \pm 0.16|_{\rm hadr}   \pm 0.11|_{\rm pars}& \pm 0.09|_{m_c}) \times 10^{-4} .
\end{split}
\eeq
This result is about $1.4\sigma$ below the central value of the 
experimental measurement. 

The $B\to X_s\gamma$ branching ratio is an important constraint on new physics models as discussed in Section \ref{NP-sec}.
At present the largest error limiting the precision of the test 
arises from experimental uncertainties. 
Furthermore, using the statistics that would be available at 
a Super Flavor Factory, it would be possible to reduce the photon energy cut,
which can help improve the theoretical understanding.
Theoretical uncertainties will, however, ultimately limit the precision,
to about the $5\%$ level.

Another important observable in weak radiative decays is the 
direct CP asymmetry, often called the partial rate asymmetry (PRA) 
\begin{eqnarray}
  A_{\rm CP} = \frac{\Gamma(\bar B\to X\gamma) - \Gamma(B\to \bar X\gamma)}
  {\Gamma(\bar B\to X\gamma) + \Gamma(B\to \bar X\gamma)},
\end{eqnarray}
where $\bar X$ is the CP conjugate of the $X$ state.

In general, decay amplitudes can be written as 
the sum of two terms with different weak phases
(see also Eq.~\eqref{unitarity})
\begin{eqnarray}
  A(\bar B\to X\gamma) = P + e^{i\psi} A = 
  P\, (1 +  \varepsilon_A e^{i(\delta+\psi)}) \,,
\end{eqnarray}
where $\varepsilon_A e^{i\delta} = A/P$, and $\delta$ and $\psi$ are the strong and weak phase differences.
One finds for the direct CP asymmetry
\begin{eqnarray}
  A_{\rm CP} = \frac{2 \varepsilon_A \sin\delta\sin\psi}
  {1 + 2\varepsilon_A\cos\delta\cos\psi + \varepsilon_A^2} \,,
\end{eqnarray}
in agreement with the well-known result that for $A_{CP}\ne 0$ both strong and weak phase differences need to be nonzero  
[see, {\it e.g.}~\cite{Bander:1979px}]. 
The direct CP asymmetry in $b\to s\gamma$ is then suppressed by three concuring
small factors: 
i) CKM suppression by 
$\varepsilon_A \propto |\lambda_u^{(s)}/\lambda_t^{(s)}| \sim \lambda^2$, 
ii) a factor of $\alpha_s(m_b)$ required in order to generate the strong phase,
and iii) a GIM suppression factor $(m_c/m_b)^2$, reflecting the fact that in the limit $m_c = m_u$ the charm
and up quark penguin loop contributions cancel in the CP asymmetry.

The OPE approach discussed above 
can be used to compute also the $B\to X_s\gamma$ direct CP
asymmetry~\cite{Soares:1991te,Kagan:1998bh,Kiers:2000xy}. 
The most recent update by \textcite{Hurth:2003dk} gives
\beq
\begin{split}
    A_{\rm CP}&(B\to X_s\gamma)|_{E_\gamma > 1.6 \ {\rm GeV}} =\\
 &   (0.44 ^{+0.15}_{-0.10}|_{m_c/m_b} \pm 0.03|_{\rm CKM} \pm^{+0.19}_{-0.09} |_{\rm RG}) \% \, .
\end{split}
\eeq
This can be compared to the current world average~\cite{HFAG,Coan:2000pu,Aubert:2004hq,Nishida:2003yw}
\begin{eqnarray}
  A_{\rm CP}(b\to s\gamma) = 0.004 \pm 0.036 \,,
\end{eqnarray}
which is compatible with a vanishing or very small direct CP 
asymmetry as expected in the SM. 
The experimental uncertainty is still an order of 
magnitude greater than the theory error, so that a dramatic improvement in the 
precision of this SM test can be achieved with a SFF.
The ultimate precision is expected to be limited by experimental systematics
at about the same level as the current theory error.

The theoretical error can be further reduced if one considers an even more inclusive $B\to X_{s+d}\gamma$ decay. 
In the U-spin symmetry limit, the inclusive partial rate asymmetries in 
$B^\pm\to X_s\gamma$ and $B^\pm\to X_d \gamma$ 
are equal and of opposite signs, $  \Delta \Gamma(B^\pm \to X_s\gamma) = - \Delta \Gamma(B^\pm \to X_d\gamma)$~\cite{Hurth:2001yb}.
A similar relation holds also for neutral $B^0$ meson decays, but with corrections due to annihilation and other $1/m_b$ suppressed terms.
In the SU(3) limit $(m_d=m_s)$ therefore the inclusive untagged CP asymmetry 
$A_{\rm CP}(B\to X_{s+d}\gamma)$ vanishes in the SM, while the leading SU(3) breaking correction is 
of order $(m_s/m_b)^2 \sim
10^{-4}$~\cite{Hurth:2003dk}. The inclusive untagged CP asymmetry thus
provides a clean test of the SM, with very little 
uncertainty.
Any measurement of a nonzero value would be a clean signal for NP.

A first measurement of the 
untagged CP asymmetry has been made by \babar~\cite{Aubert:2006gg}, 
\begin{equation}
  A_{\rm CP}(B\to X_{s+d}\gamma)\negmedspace = 
  -0.110 \pm 0.115|_{\rm stat} \pm 0.017|_{\rm {sys}}.
\end{equation}
A significant reduction of the uncertainity is necessary to 
provide a stringent test of the SM prediction.
A SFF will be able to measure this quantity to about $1\%$ precision.

%%%%%%%%%%%%%%%%%%%%%%%%%%%%%%%%%%%%%%%%%%%%%%%%%%%%%%%%%%%%%%%%%%%%%%%%%%%%%%%%%%%%%%%%%%%%%%%%%%%%%
%%%%%%%%%%%%%%%%%%%%%%%%%%%%%%%%%%%%%%%%%%%%%%%%%%%%%%%%%%%%%%%%%%%%%%%%%%%%%%%%%%%%%%%%%%%%%%%%%%%%%
%%%%%%%%%%%%%%%%%%%%% Exclusive %%%%%%%%%%%%%%%%%%%%%%%%%%%%%%%%%%%%%%%%%%%%%%%%%%%%%%%%%%%%%%%%%%%
%%%%%%%%%%%%%%%%%%%%%%%%%%%%%%%%%%%%%%%%%%%%%%%%%%%%%%%%%%%%%%%%%%%%%%%%%%%%%%%%%%%%%%%%%%%%%%%%%%%%%
%%%%%%%%%%%%%%%%%%%%%%%%%%%%%%%%%%%%%%%%%%%%%%%%%%%%%%%%%%%%%%%%%%%%%%%%%%%%%%%%%%%%%%%%%%%%%%%%%%%%%
%%%%%%%%%%%%%%%%%%%%%%%%%%%%%%%%%%%%%%%%%%%%%%%%%%%%%%%%%%%%%%%%%%%%%%%%%%%%%%%%%%%%%%%%%%%%%%%%%%%%%

\subsubsection{Exclusive $B\to V_{s,d}\gamma$ decays}
\label{ExclTheory}

The exclusive decays such as $B\to K^*\gamma $ or  $B\to K\pi\gamma , K\pi\pi\gamma$ are experimentally much cleaner than the inclusive channels due to simpler event identification criteria and background elimination. 
They are, however, more theoretically 
challenging which limits 
their usefulness for NP searches. In this subsection we 
review the theoretical progress on $B\to V_{s,d}\gamma$ branching ratios and direct CP asymmetries. Theoretically clean observables related to photon polarization are then 
covered in the next subsection.
The extraction of CKM parameters from  
$B\to V_{s,d}\gamma$ decays is 
reviewed in Section \ref{constraintsCKM}.

The $B\to V\gamma$ decays  are dominated by the electromagnetic dipole
operator $O_{7\gamma}$, Eq.~(\ref{HW}). Neglecting for the moment the remaining smaller contributions,
this gives
\beq\label{B2KstBr}
{\cal B}(B \to K^*\gamma) = \tau_B \frac{\alpha G_F^2 |\lambda_t^{(s)}| }{16\pi^4}
|C_{7\gamma}|^2 m_b^2 E_\gamma^3 |T_1(0)|^2,
\eeq
where $T_1(q^2)$ is a tensor current form factor. Its nonperturbative nature is at the heart of theoretical uncertainties in $B\to V\gamma$ decay. In 
principle it can be obtained model independently from 
lattice QCD~\cite{Bernard:1993yt}, with first unquenched studies presented in \cite{Becirevic:2006nm}. 
Lattice QCD results are obtained only at large values of the momentum transfer $q^2 \sim m_b^2$.
Extrapolation to low $q^2$ then
introduces  
some model
dependence. Using the BK parametrization \cite{Becirevic:1999kt}, \textcite{Becirevic:2006nm} find $T_1^{BK^*}(0) = 0.24 \pm 0.03^{+0.04}_{-0.01}$.

Another nonperturbative approach is based on QCD sum rules, where 
OPE is applied to correlators of appropriate interpolating operators. Relying on quark-hadron duality the OPE result is related 
to properties of the hadronic states. The heavy-to-light form factors in the large energy release region can be computed from a modification
of this approach, called light-cone QCD sum rules. Using this framework \textcite{Ball:2004rg}
find $T_1^{(\rho)}(0)=0.267\pm 0.021$ and $T_1^{(K^*)}(0)=0.333\pm 0.028$.

Relations to other form factors 
follow in the large energy limit $E_M \gg \Lambda_{\rm QCD}$. In this limit the heavy-to-light $B\to V$ form factors have been studied in QCDF \cite{Beneke:2000wa}  and in SCET \cite{Bauer:2002aj,Beneke:2003pa,Hill:2004if}
at leading order in $\Lambda_{\rm QCD}/E_M$.
The main result 
is a factorization formula for
heavy-to-light form factors 
consisting of perturbatively calculable factorizable terms and a nonfactorizable soft term common to several form factors. The analysis can be systematically extended to higher orders. 

\begin{figure}
%\vspace*{0.5cm}
%\hspace*{-2.8cm}
\includegraphics[width=8cm]{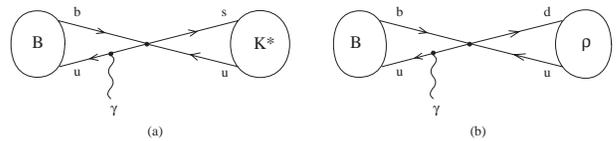}
%\vskip0.6cm
\caption{Typical contributions to the weak annihilation amplitude
in $B\to K^*\gamma$ (a) and $B\to \rho\gamma$ (b) weak radiative decays.
Additional diagrams with the photon attaching to the final state quarks
are not shown.
}\label{wafig}
\end{figure}

Eq.~\eqref{B2KstBr} neglects the contributions from the four-quark operators 
$O_{1-6}$ and the 
gluonic dipole operator $O_{8g}$ in the weak hamiltonian, Eq.~\eqref{HW}.
These contributions are of two types: i) 
short-distance dominated loop corrections 
absorbed into effective 
Wilson coefficients in factorization formula 
and  ii) weak annihilation (WA) type contributions,
Fig.~\ref{wafig} 
\cite{Beneke:2001at,Ali:2001ez,Bosch:2001gv,Descotes-Genon:2004hd}. 
The 
WA amplitude is power suppressed, ${\cal O}(\Lambda_{\rm QCD}/m_b)$, 
but occurs at tree level and is thus 
also 
relatively enhanced.
It is proportional to $\lambda_u^{(q)}$ and is 
CKM suppressed in $b\to s\gamma$ transitions, but not in
$b\to d\gamma$ decays, for instance 
in $B\to \rho\gamma$~\cite{Atwood:1994rw}. 
At LO in $\alpha_s$ and $\Lambda_{\rm QCD}/m_b$
the WA amplitude factorizes as shown in \cite{Grinstein:2000pc,Beneke:2001at,Bosch:2001gv}. 

Direct CP asymmetries in exclusive modes such as $B\to K^*\gamma$ can 
be 
estimated using the factorization formula.
This gives 
$A_{\rm CP}(B\to K^*\gamma) = -0.5\%$~\cite{Bosch:2001gv}, 
in agreement with the experimental 
world average $A_{\rm CP}(B\to K^*\gamma) = -0.010 \pm 0.028$~\cite{HFAG,Aubert:2004te,Nakao:2004th}.
Since the theory prediction 
depends on 
poorly known light-cone wave functions and unknown power corrections, 
this observable does not offer a precision test of the SM.
Some theoretical uncertainties can be overcome by exploiting 
the cancellation of partial rate asymmetries in the U-spin limit~\cite{Hurth:2001yb}, 
but symmetry breaking corrections are difficult to compute in a clean way.
Other possible uses of exclusive radiative decays to test the SM are discussed 
below.

%%%%%%%%%%%%%%%%%%%%%%%%%%%%%%%%%%%%%%%%%%%%%%%%%%%%%%%%%%%%%%%%%%%%%%%%%%%%%%%%%%%%%%%%%%%%%%%%%%%%%
%%%%%%%%%%%%%%%%%%%%%%%%%%%%%%%%%%%%%%%%%%%%%%%%%%%%%%%%%%%%%%%%%%%%%%%%%%%%%%%%%%%%%%%%%%%%%%%%%%%%%
%%%%%%%%%%%%%%%%%%%%%                              %%%%%%%%%%%%%%%%%%%%%%%%%%%%%%%%%%%%%%%%%%%%%%%%%%%
%%%%%%%%%%%%%%%%%%%%%   Photon polarization        %%%%%%%%%%%%%%%%%%%%%%%%%%%%%%%%%%%%%%%%%%%%%%%%%%%
%%%%%%%%%%%%%%%%%%%%%                              %%%%%%%%%%%%%%%%%%%%%%%%%%%%%%%%%%%%%%%%%%%%%%%%%%%
%%%%%%%%%%%%%%%%%%%%%%%%%%%%%%%%%%%%%%%%%%%%%%%%%%%%%%%%%%%%%%%%%%%%%%%%%%%%%%%%%%%%%%%%%%%%%%%%%%%%%
%%%%%%%%%%%%%%%%%%%%%%%%%%%%%%%%%%%%%%%%%%%%%%%%%%%%%%%%%%%%%%%%%%%%%%%%%%%%%%%%%%%%%%%%%%%%%%%%%%%%%

\subsubsection{Photon polarization in $b\to s\gamma$ }
\label{photonpol}

\begin{figure}
\includegraphics[width=2.6cm]{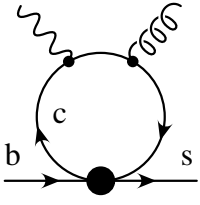}
\caption{Diagram with insertion of the operator $O_2^c$ which contributes
to right-handed photon emission. The wavy line denotes a photon and the curly line
a gluon. }\label{charmloop}
\end{figure}

In the SM the photons emitted in $b\to s\gamma$ are predominantly left-handed
polarized, and those emitted in $\bar b\to \bar s\gamma$ are predominantly
right-handed, in accordance with the form of electromagnetic operator $O_{7\gamma}$, 
Eq. \eqref{magnetic}. In the 
presence of NP the decay into photons of opposite chirality 
can be enhanced 
by a chirality flip on the internal heavy NP lines. 
This observation underlies the proposal 
to use the mixing-induced asymmetry in $B^0(t)\to f\gamma$ decays
as a null test of the SM~\cite{Atwood:1997zr}. 
The value of $S_{f\gamma}$ parameter significantly away from zero would signal the presence of NP.
The precision of the test depends on the 
SM
ratio of the wrong polarization decay amplitude $A(\bar B\to f_q \gamma_R)$ and the 
right polarization decay amplitude $A(\bar B \to f_q \gamma_L)$ 
for given $f_q$
($q=d,s$)
\begin{eqnarray}\label{rfdef}
r_f e^{i(\phi_q + \delta_f)}\equiv
\frac{A(\bar B\to f_q \gamma_R)}{A(\bar B \to f_q \gamma_L)}\,.
\end{eqnarray}
Here $\phi_q$ is a weak phase, and $\delta_f$ a strong phase. 
For  a CP eigenstate $f$ the resulting $B^0(t)\to f\gamma$ in terms of $r_f,\delta_f$ 
is given in Eq.~(\ref{SCPeigen}).
Keeping only the 
dominant electromagnetic penguin contribution one finds a very small ratio 
$r_f =m_q/m_b$ and $\phi_q=\delta_f=0$, independent of the final state $f_q$. 
This estimate can be changed, however, by hadronic effects \cite{Grinstein:2004uu}. 
The right-handed photon amplitude receives contributions from charm- and up-quark 
loop graphs in Fig.~\ref{charmloop} with the four-quark operators $O_{1-6}$ in the 
weak vertex. The largest contributions come from the operator $O_2^c$.

For inclusive $B\to X_s\gamma_R$ decays one finds $r\simeq 0.11$ when integrating over the partonic phase space with
$E_\gamma > 1.8$ GeV~\cite{Grinstein:2004uu}. This estimate includes the numerically important 
${\cal O}(\alpha_s^2\beta_0)$ correction. Note that the obtained $r$ is much larger than the estimate from electromagnetic penguins only $r\sim m_s/m_b\sim 0.02$.

An effect of similar size is found for $B\to V_q\gamma$ decays using SCET,
following from a nonfactorizable contribution suppressed by $\Lambda_{\rm QCD}/m_b$ \cite{Ligeti:1997tc}. 
By dimensional arguments 
the estimate for the 
$r_{K^*/\rho}$ ratio is 
\beq\label{rdim}
r_f = \frac{m_q}{m_b} + c_f \frac{C_2}{3C_{7\gamma}} \frac{\Lambda_{\rm QCD}}{m_b}.
\eeq
Here $|c_f|$ is a dimensionless parameter of order one that 
depends on the final hadronic state $f$. The second term remains in the limit of a massless light quark $m_q\to 0$. 
Although power suppressed, 
it is enhanced by the large 
ratio $C_2/C_{7\gamma} \sim 3$. A dimensional estimate is thus $r_f\sim 0.1$, 
which would translate into an asymmetry ($S$) of about 10\%, 
much larger than the LO estimate of $m_q/m_b$ that gives an asymmetry in $b \to s$ transitions of around 3\%.
A more reliable 
estimate requires a challenging dynamical computation of the 
nonlocal nonfactorizable 
matrix element. However, these theoretical difficulties need not stand
in the way of experimental progress as there is data driven method to separate
the SM contamination by studying the dependence of the asymmetry on the final
state~\cite{Atwood:2004jj} as we discuss below.

First steps in the direction of explicit model calculations find $S(B \to K^* \gamma) = -0.022 \pm 0.015$ using
QCD sum rules \cite{Ball:2006cva}, consistent with the leading order estimate. 
In particular, expanding the relevant nonlocal operator 
in powers of $\Lambda_{\rm QCD} m_b/m_c^2\sim 0.6$
and then keeping only the first term,
they obtain  
 for $f=K^*,\rho$ 
[see also~\cite{Khodjamirian:1997tg}]
\beq\label{rffinal}
r_f = \frac{m_q}{m_b} - 
\frac{C_2}{C_7}\frac{L-\tilde L}{36 m_b m_c^2 T_1^{BV}(0)} = 
\frac{m_q}{m_b} - (0.004 \pm 0.007),
\eeq
with $L,\tilde L$ parametrizing $B\to K^*$ matrix elements of the nonlocal operator.
Another calculation using pQCD obtained a
very similar result for the asymmetry,
$S ( B \to K_s \pi^0 \gamma) = -0.035 \pm 0.017$ \cite{Matsumori:2005ax}.

Experimentally, the photon polarization can be measured from time dependent $B^0(t) \to f\gamma$ decay utilizing
the interference of $B-\bar B$ mixing with the right- and left-handed
photon amplitudes~\cite{Atwood:1997zr,Atwood:2004jj}.
In particular, taking the time-dependent asymmetry summed over the unobserved photon polarization 
\beq
\begin{split}
  A_{CP}(t) & =\frac{\Gamma(\bar B^0(t) \to f \gamma_{L+R}) - 
    \Gamma(B^0(t) \to f \gamma_{L+R})}{\Gamma(\bar B^0(t) \to f \gamma_{L+R}) +
    \Gamma(B^0(t) \to f \gamma_{L+R})} \\
  & = S_{f\gamma} 
  \sin (\Delta m t) - C_{f\gamma} \cos (\Delta m t),
\end{split}
\eeq
the two coefficients are
\beq\label{SCgen}
\begin{split}
S_{f\gamma} &= \frac{2\mbox{Im }
\Big[\Big(\frac{q}{p}\Big)_B (\bar A_L A_L^* + \bar A_R A_R^*)\Big]}
{|\bar A_L|^2 + |A_L|^2 + |\bar A_R|^2 + |A_R|^2}, \\
C_{f\gamma} &= \frac{|\bar A_L|^2 - |A_L|^2 + |\bar A_R|^2 - |A_R|^2}
{|\bar A_L|^2 + |A_L|^2 + |\bar A_R|^2 + |A_R|^2}, 
\end{split}
\eeq
where $\bar A_{L,R} \equiv A(\bar B \to f\gamma_{L,R})$ and $A_{L,R} \equiv A( B \to f\gamma_{L,R})$. Note that $S_{f\gamma}=0$ when the ``wrong'' polarization 
amplitudes $\bar A_{R}$ and $A_{L}$ vanish. This can be made more transparent in a simplified case where $f$ is a CP eigenstate with eigenvalue $\eta_{CP}(f)$, while also assuming that the $B\to f\gamma$ transitions are dominated by a single weak phase $\phi_q$, so that $\bar A_{L,R}  = e^{i\phi_q} \bar a_{L,R} $ and
$A_{L,R} = e^{-i\phi_q} \eta_{CP}(f) \bar a_{R,L}$, where $a_{L,R}$ and $\bar a_{L,R}$ are strong amplitudes. Then
\beq\label{SCPeigen}
S_{f\gamma} = \eta_{CP}(f) \frac{2r_f\cos\delta_f}{1+r_f^2} \mbox{Im }
\Big[ \Big(\frac{q}{p}\Big)_{B} e^{2i\phi_q}\Big]\,,
\eeq
and $C_{f\gamma} = 0$. Here $r_f \exp(i\delta_f) = \bar A_R/\bar A_L$ as in
Eq.~\eqref{rfdef}. 
The asymmetry $S_{f\gamma}$ vanishes in the
limit of 100\% left-handed photon polarization ($r_f=0$).

The value of $S_{f\gamma}$ depends crucially also on the mismatch between the
weak phase $\phi_q$ of the decay amplitude and the $B_{d,s}$ mixing phases,
$(q/p)_{B_d}= -\exp(-2i\beta)$ and $(q/p)_{B_s}=- 1$.  There are two distinct
categories. For $B_d \to f_s\gamma$ and $B_s \to f_d\gamma$ decays this phase
difference is large ($2\beta$) and 
$S_{f\gamma} = \frac{\pm 2 r_f}{1+r_f^2}\cos\delta_f\sin 2\beta$ 
is suppressed only by $r_f$. For $B_d \to f_d\gamma$ and $B_s \to f_s\gamma$
decays, on the other hand, the weak phase difference vanishes so that in SM
$S_{f\gamma} = 0$ with negligible theoretical uncertainty. For NP to modify
these predictions it has to induce large right-handed photon polarization
amplitude, while for $B_d \to f_d\gamma$ and $B_s \to f_s\gamma$ decays also a
new weak phase is needed to have $S_{f\gamma} \ne 0$. 

Current results give a world average $S_{K^*\gamma} = -0.19 \pm 0.23$~\cite{HFAG,Ushiroda:2006fi,Aubert:2007qj},
and the first measurement of time-dependent asymmetries in $b \to d\gamma$ decays has recently been reported, $S_{\rho\gamma} = -0.83 \pm 0.65 \pm 0.18$~\cite{Ushiroda:2007jf}.
These are compatible, within experimental errors, with the SM predictions.
A SFF could reduce the uncertainty on the former to about $2$--$3\%$ 
and on the latter to about $10\%$ (see Table~\ref{tab:sff_sensivite}).

Measurements have also been made over an extended range of 
$K\pi$ invariant mass in $B\to K_S\pi^0\gamma$. In multibody exclusive radiative decays, 
a nonvanishing right-handed photon amplitude can be present 
at leading order in the $1/m_b$ expansion. However, using a combination of SCET and chiral perturbation theory (ChPT) methods applicable in 
kinematical region with one energetic kaon and a 
soft pion  
$r_{K\pi}$ was found to be numerically less than 1\% due to kinematical suppression 
\cite{Grinstein:2005ry,Grinstein:2005nu}. With the SFF data, the multibody radiative 
decays will be most useful to search for 
SM corrections to the photon polarization. These effects depend on the Dalitz plot position,
in contrast to NP effects, which should be universal~\cite{Atwood:2004jj}. 
The LO dipole moment operator (as well as NP) would give rise to an asymmetry
that is independent of the energy of the photon whereas the 
soft gluon effects will give rise to an asymmetry that depends on photon energy.  
Thus there is a model independent, completely data driven
method to search for NP effects by studies of time dependent asymmetries. 
In addition, further decay modes, such as $B\to K_S\eta\gamma$,
$B\to K_S\pi^+\pi^-\gamma$ and $B\to K_S\phi\gamma$ can also
be used~\cite{Atwood:2007qh} in a very similar fashion.

Other approaches for probing the photon polarization in 
$b\to s\gamma$ decays have been suggested and can be employed at a SFF.
One powerful idea is to relate the photon polarization information to angular 
distributions of the final state hadrons. Examples relevant for a SFF are 
$B\to K\pi\pi\gamma$~\cite{Gronau:2001ng,Gronau:2002rz}, 
and $B^\pm \to K^\pm \phi\gamma$ \cite{Atwood:2007qh,Orlovsky:2007hv}. Similar tests have been suggested also using $\Lambda_b$ decays, such as 
$\Lambda_b \to \Lambda\gamma$ \cite{Mannel:1997pc,Gremm:1995nx,Hiller:2001zj,Hiller:2007ur} and 
$\Lambda_b\to pK\gamma$ \cite{Legger:2006cq}.

We consider $B\to X_s\gamma$ decays,
 where the final hadronic state $X_s = K\pi\pi,K\bar KK$
originates from the strong decay of resonance $K_{\rm res}$, 
produced in the weak decay $B\to K_{\rm res} \gamma$. 
The lowest lying vector state, the $K^*$, cannot be used  for this purpose, since 
the $K^*$ polarization is not observable in its two-body decay $K^*\to K\pi$.
This is due to the fact that it is impossible to form a T-odd quantity from 
only two vectors, the photon momentum 
and the $K$ momentum, in the $K^*$ rest frame.

The photon polarization can then be measured through higher resonance $K_{\rm res}\to K\pi\pi$ decays. The angular
distribution of the decay width in $K_{\rm res}$ rest frame is~\cite{Gronau:2002rz}
\beq\label{3res}
\begin{array}{lcl}
  \multicolumn{2}{l}{    
    \frac{\mbox{d}^2\Gamma}{\mbox{d}s \mbox{d}\cos\tilde\theta} = 
  } &
  |c_1|^2  \left\{
    1 + \cos^2\tilde \theta + 4 P_\gamma R_1 \cos\tilde \theta \right\} \\
  \hspace{3mm} &
  \multicolumn{2}{l}{
    +  |c_2|^2  \left\{
      \cos^2\tilde\theta + \cos^2 2\tilde\theta + 
      12 P_\gamma R_2 \cos\tilde\theta \cos 2\tilde\theta\right\}
  } \\
  \hspace{3mm} &
  \multicolumn{2}{l}{
    + |c_3|^2 B_{K^*_1}(s) \sin^2\tilde\theta
  } \\
  \hspace{3mm} &
  \multicolumn{2}{l}{
    + \left\{ c_{12} \frac12(3\cos^2\tilde\theta -1) + 
      P_\gamma c^\prime_{12} \cos^3\tilde\theta \right\} \,.
  }
\end{array}
\eeq
Here $\tilde \theta$ is the angle between the direction opposite to the photon momentum ($-\vec q$) and the vector
$\vec p_{\pi_{\rm slow}}\times \vec p_{\pi_{\rm fast}}$(the pions
are ordered in terms of their momenta). The first three terms in Eq.~\eqref{3res} correspond respectively to decays 
through $K_{\rm res}$ resonances with $J^P = 1^+, 2^+$ and $1^-$,
while the last terms come from $1^+$--$2^+$ interference.
The hadronic parameters $R_{1,2}$ can be computed from the Breit-Wigner resonant model \cite{Gronau:2001ng,Gronau:2002rz}.
The $K_1(1400)$ resonance decays predominantly to $K^*\pi$. The relevant parameters in $R_1$ are then fixed by isospin, leading to a precise determination  $R_1 = 0.22 \pm 0.03$.
Thus, measurements of the angular distribution Eq.~(\ref{3res}) restricted to the $K_1(1400)$ mass range
can be used to extract the photon polarization parameter $P_\gamma$. So far
only an upper bound on ${\cal B}(B\to K_1(1400)\gamma)<1.5 \times 10^{-5}$
exists \cite{Yang:2004as}. 
The use of the narrow resonance $K_1(1270)$, with a larger branching ratio 
${\cal B}(B\to K_1(1270)\gamma)=(4.3\pm1.2) \times 10^{-5}$~\cite{Yang:2004as,Aubert:2005xk},
may be more advantageous experimentally.
A drawback is the estimate of $R_1$ in which a strong phase between
$K_1(1270)\to K^*\pi$ and $K_1(1270)\to K\rho$ decay amplitudes needs to be obtained 
from an independent measurement.

The  method outlined above works only for certain charge states, for which two $K^*\pi$ channels interfere to produce
the up-down asymmetry in $\cos\tilde\theta$. These channels are $K^0\pi^+\pi^0$, where the interfering channels are $K^{*+}\pi^0$ and $K^{*0}\pi^+$, and 
$K^+\pi^-\pi^0$, where the interfering channels are $K^{*+}\pi^-$ and $K^{*0}\pi^0$.

\subsection{$B\to X_{s/d}\ell^+\ell^-$ and $B\to X_{s/d}\nu\bar \nu$ decays}
The rare $B\to X_s\ell^+\ell^-$ decays form another class of FCNC processes, 
which proceed in the SM only through loop effects. 
The richer structure of the final state allows tests 
complementary to those performed in 
weak radiative $B\to X_s\gamma$ decays. 
In addition to the total branching fraction, 
one can study also the dilepton invariant mass, the forward-backward asymmetry,
and various polarization observables. 
We discuss these predictions, 
considering in turn the exclusive and inclusive channels.

\subsubsection{Inclusive $B\to X_s\ell^+\ell^-$ decays}
\label{InclXllTheory}

In inclusive $B\to X_{s/d}\ell^+\ell^-$ decays there are three distinct regions of dilepton invariant mass 
$q^2=(p_{\ell^+} + p_{\ell^-})^2$: 
(i) the low $q^2$ region, $q^2< 6$ GeV$^2$, 
(ii) the high $q^2$ region $q^2 > 12$ GeV$^2$, 
and (iii) the charm resonance region $q^2\sim (6-12)$ GeV$^2$. 
In the intermediate region (iii) $c\bar c$ resonances couple to the dilepton pair 
through a virtual photon, leading to nonperturbative strong interaction
effects which are difficult to compute in a model independent way.

In the low$-q^2$ and high$-q^2$ regions, 
a model independent computation of the decay rate is possible 
using an 
OPE and heavy quark expansion, 
similar to that used for the rare radiative decays discussed in Section~\ref{InclTheory}. 
QCD corrections have been evaluated at NNLO including the complete three-loop mixing of the four quark operators 
$O_{1,2}$ into $O_9$ necessary for a complete solution of 
the RGE to NNLL order~\cite{Asatryan:2001zw,Ghinculov:2002pe,Ghinculov:2003qd,Asatrian:2002va,Gambino:2003zm,Bobeth:2003at,Huber:2007vv}. This calculation has been further improved by including electromagnetic log enhanced contributions ${\cal O}(\alpha_{\rm e.m.} \log(m_W^2/m_b^2))$ that appear only if the integration 
over dilepton mass is restricted to a range but vanish for the full rate \cite{Huber:2007vv,Huber:2005ig,Bobeth:2003at}.
Nonperturbative power suppressed effects have been considered in~\cite{Falk:1993dh,Ali:1996bm}.
Effects of the $c\bar c$ intermediate states in the resonance region 
can be modeled assuming factorization of the four-quark operator 
$(\bar s c)(\bar cb)$~\cite{Kruger:1996cv}.

Integrating over the low dilepton invariant mass range $q^2=(1,6)$ GeV$^2$,
the partial branching fractions corresponding to the low$-q^2$ region
are~\cite{Huber:2005ig}
\begin{eqnarray}
  {\cal B}(B\to X_s\mu^+\mu^-) & = & (1.59 \pm 0.11) \times 10^{-6}\,,\\
  {\cal B}(B\to X_s e^+ e^-) & = & (1.64 \pm 0.11) \times 10^{-6}
\end{eqnarray}
where the dominant theoretical uncertainty $(\pm 0.08)$ arises from 
scale dependence, along with smaller uncertainties from the quark masses, 
CKM matrix elements, and nonperturbative 
${\cal O}(1/m_b^2, \alpha_s \Lambda_{\rm QCD}/m_b)$ corrections. 
The predictions agree well with the present average of the 
\babar\ and Belle experimental measurements of this quantity~\cite{Aubert:2004it,Iwasaki:2005sy,Huber:2005ig} ${\cal B}(B\to X_s\ell^+\ell^-) = (1.60 \pm 0.51) \times 10^{-6}$.
The present (SM) theory error for 
the branching fraction is below the total experimental uncertainty. 
At a SFF the situation would be reversed. 

Additional uncertainty in these predictions is introduced if a cut on the 
hadronic mass $M_{X_s} < M_D$ is imposed to eliminate charm backgrounds.
This introduces sensitivity to the shape function, which however can be 
eliminated using $B\to X_s\gamma$ data \cite{Lee:2005pwa}.
In the high $q^2$ region, an improvement in theory 
is possible, if the integrated decay rate 
is normalized to the semileptonic $b\to u l\nu$ rate with the same $q^2$ cut \cite{Ligeti:2007sn}. 
This drastically reduces the size of $1/m_b^2$ and $1/m_b^3$ power corrections.

Besides the dilepton invariant mass spectrum the observable most often discussed is the 
forward-backward asymmetry. However, recently \textcite{Lee:2006gs} pointed out that a third 
constraint can be obtained from $B\to X_{s}\ell^+\ell^-$ double differential decay width
\beq
\begin{split}
\frac{d^2\Gamma}{dq^2dz}=\frac{3}{8}\big[&(1+z^2)H_T(q^2)+2 z H_A(q^2)\\
&+2(1-z^2)H_L(q^2)],
\end{split}
\eeq
where $z=\cos\theta$, with $\theta$ the angle between $\ell^+$ and the $B$ meson three-momentum in the $\ell^+\ell^-$ center-of-mass frame. The functions
$H_i$ do not depend on $z$. The sum $H_T(q^2)+H_L(q^2)$ gives the dilepton invariant mass spectrum $d\Gamma/d q^2$, while the forward-backward
asymmetry (FBA) is conventionaly defined as $dA_{FB}(q^2)/dq^2=3H_A(q^2)/4 $. 
The importance of the $H_i$ functions is that they are calculable in the low$-q^2$
and high$-q^2$ regions, and also depend differently on the Wilson coefficients of the 
effective weak Hamiltonian of Eq.~\eqref{HW}. 
This suffices to determine the sizes and signs of 
all the relevant coefficients, probing in this way 
NP effects. At leading order
they have a general structure \cite{Lee:2006gs}
\beq
\begin{split}\label{Hi}
H_T(q^2)&\propto 2(1-s)^2 s\Big[\Big({\cal C}_9+\frac{2}{s}{\cal C}_{7\gamma}\Big)^2+{\cal C}_{10}^2\Big],\\
H_A(q^2)&\propto -4(1-s)^2 s{\cal C}_{10}\Big({\cal C}_9+\frac{2}{s}{\cal C}_{7\gamma}\Big),\\
H_L(q^2)&\propto (1-s)^2 \Big[\Big({\cal C}_9+2{\cal C}_{7\gamma}\Big)^2+{\cal C}_{10}^2\Big],
\end{split}
\eeq
where $s=q^2/m_b^2$. 
 The modified Wilson coefficients ${\cal C}_{7\gamma, 9,10}$ 
are 
$\mu$ independent 
linear combinations of the Wilson coefficients $C_{7\gamma, 9,10}$  and $C_{1,\dots,6,8g}$ 
in weak Hamiltonian of Eq.~\eqref{HW}.
They are related to
the NNLO ``effective'' Wilson coefficients $C_{7,8}^{\rm eff}$ calculated in 
\cite{Asatryan:2001zw,Beneke:2001at,Ghinculov:2003qd}.

\begin{figure}
%\vspace*{0.5cm}
%\hspace*{-2.8cm}
\includegraphics[width=4.3cm]{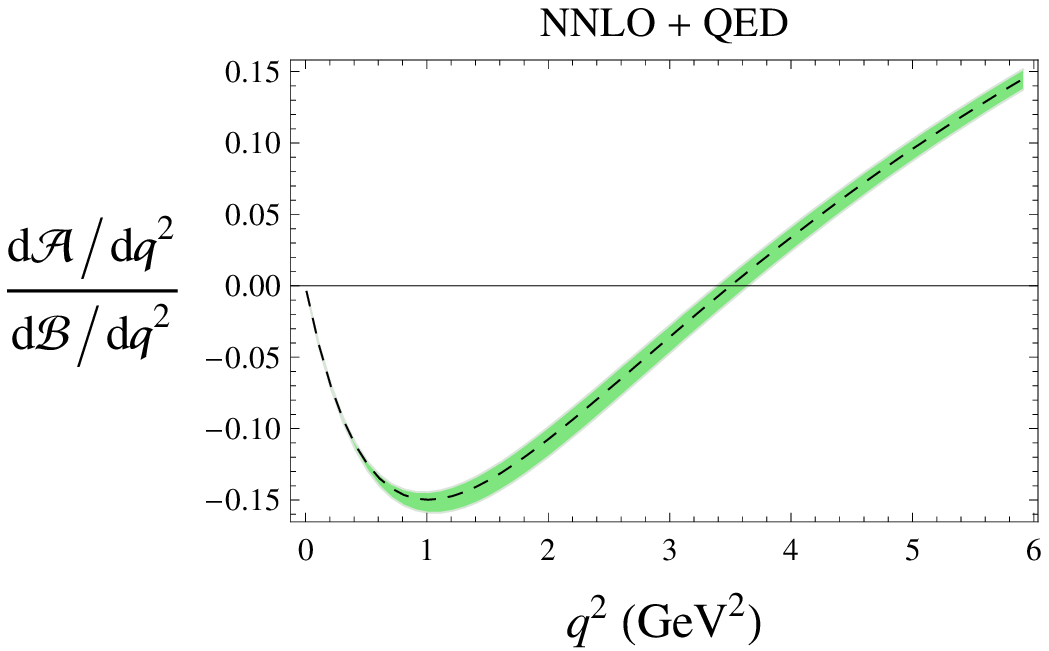}
%\hspace*{0.2cm}
\includegraphics[width=4.0cm]{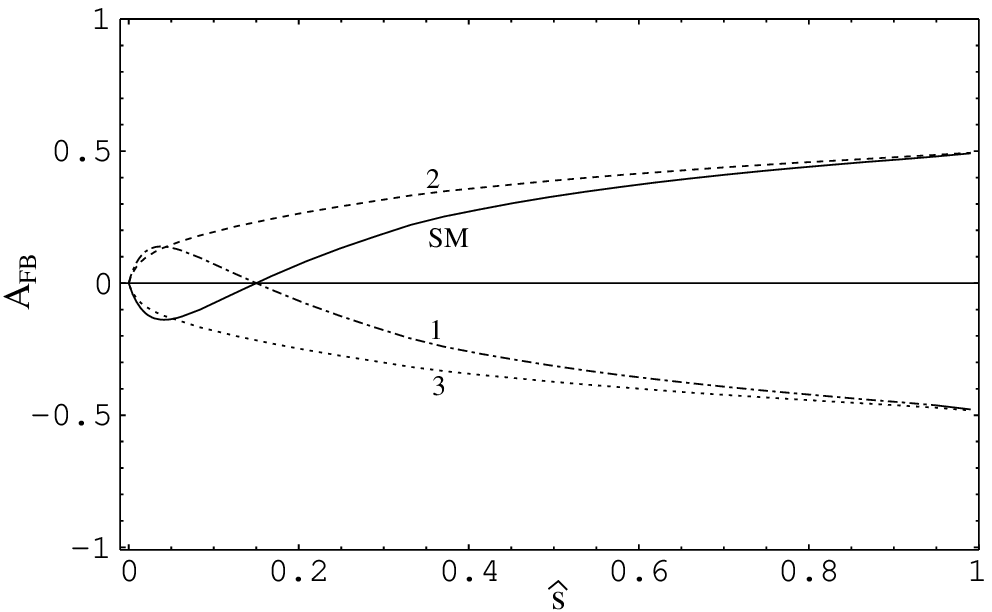}
%\vskip0.6cm
\caption{ Left: the full NNLO prediction for $B\to X_s\ell^+\ell^-$ forward-backward asymmetry normalized to the dilepton mass distribution (dashed line) and the total - parametric and perturbative - error band (shaded area) [from \cite{Huber:2007vv}]. Right: $dA_{FB}/dq^2$ in SM (solid line), with sign of $C_{10}$ opposite to SM (line 1), with reversed $C_{7 \gamma}$ sign (line 2), both $C_7\gamma$ and $C_{10}$ signs reversed (line 3) [from \cite{Ali:2002jg}].
}\label{fig:b2sll}
\end{figure}

Note 
that in $H_T$ and $H_A$ the coefficient ${\cal C}_7$ is enhanced by a $1/s$ pole 
so that measuring the dilepton mass dependence gives further information. 
Also, $H_A(q^2)$ has a zero at $q^2_0$. The existence of a zero of the FBA and the relative insensitivity to 
hadronic physics effects was first pointed out for exclusive channels~\cite{Burdman:1998mk}, 
and subsequently extended also to the
inclusive channels \cite{Ali:2002jg,Ghinculov:2002pe}. 
In the SM the zero appears in the low $q^2$ region, sufficiently away
from the charm resonance region to allow a precise computation 
of its position in perturbation theory. The value of the zero of the FBA is one of the most 
precisely calculated observables in flavor physics with a theoretical error at the order 
of $5\%$. For $B\to X_s\mu^+\mu^-$, for instance, the improved NNLO prediction is 
$(q^2_0)_{\mu\mu}=(3.50\pm0.12) {\rm ~GeV}^2$ \cite{Huber:2007vv},
where the largest uncertainty is due to the remaining scale dependence (0.10). The position 
of the zero is directly related to the relative size and sign of 
the Wilson coefficients $C_7$ and $C_9$. Thus it is very sensitive to new physics effects 
in these parameters. This quantity has not yet been measured, but estimates show that 
a precision of about $5\%$ could be obtained at a SFF~\cite{Bona:2007qt,Hashimoto:2004sm}.

\subsubsection{Exclusive $B\to X_s\ell^+\ell^-$ and $B\to X_s\nu\bar\nu$ decays}
\label{ExclXllTheory}

The channels $B\to M \ell^+\ell^-$ are
experimentally cleaner than inclusive decays, but more 
complicated theoretically. The $B\to M$ transition amplitude  depends on 
hadronic physics through form factors. The theoretical formalism described in Sec.~\ref{ExclTheory} for exclusive radiative decays 
 can be applied to this case as well.

The simplest are the decays with one pseudoscalar meson, 
such as $B\to K\ell^+\ell^-$ or $B\to \pi\ell^+\ell^-$. 
Unlike $B\to K/\pi\gamma$ decays that are not possible due to angular momentum conservation, 
the dilepton decays are allowed since the dilepton can carry zero helicity.
Especially interesting for NP searches is the angular dependence on $\theta_+$, the angle 
between the $\ell^- (\ell^+)$ and the $B (\bar B)$ momenta in the dilepton rest frame. 
In the SM the dependence is simply 
$d\Gamma \sim \sin^2\theta_+$. 
Allowing for scalar and pseudoscalar couplings to the leptons, 
which are possible in extensions of the SM, the general 
angular distribution is~\cite{Bobeth:2001sq}
\beq
\frac{1}{\Gamma} \frac{d\Gamma}{d\cos\theta_+} = \frac34 (1 - F_S) \sin^2\theta_+ + \frac12 F_S + 
A_{\rm FB} \cos\theta_+.
\eeq
The coefficient $F_S$ receives contributions 
from the scalar and pseudoscalar couplings to the leptons, 
while $A_{\rm FB}$ depends on the interference 
between the vector and scalar couplings. 
As these terms vanish in the SM, their measurement is a null test sensitive
to new physics from scalar and pseudoscalar penguins - see
\cite{Bobeth:2007dw} for a detailed study.
The first measurement of these parameters has been carried out 
in $B^+\to K^+\ell^+\ell^-$ decays by \babar~\cite{Aubert:2006vb}. 
The results are compatible with zero:
$A_{\rm FB} = 0.15 ^{+0.21}_{-0.23} \pm 0.08$ and 
$F_S=0.81^{+0.58}_{-0.61} \pm 0.46$,
where the first error is statistical and the second systematic.
These measurements could become an order of magnitude more precise,
and measure or set tight bounds on coefficients of NP operators 
which can produce these asymmetries.

\begin{figure}
  \includegraphics[width=8cm]{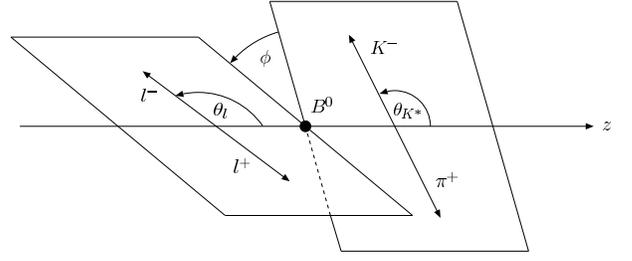}
  \caption{Parameterization of the final state in the rare decay $B\to K^*(\to K\pi) \ell^+\ell^-$.
  }\label{fig:angles}
\end{figure}

\begin{figure}
\vspace*{3.6cm}
\hspace*{-7cm}
%\epsffile{figs3/rateKst.eps}
%\epsffile{figs3/AFBKst.eps}
\includegraphics[width=1.9cm]{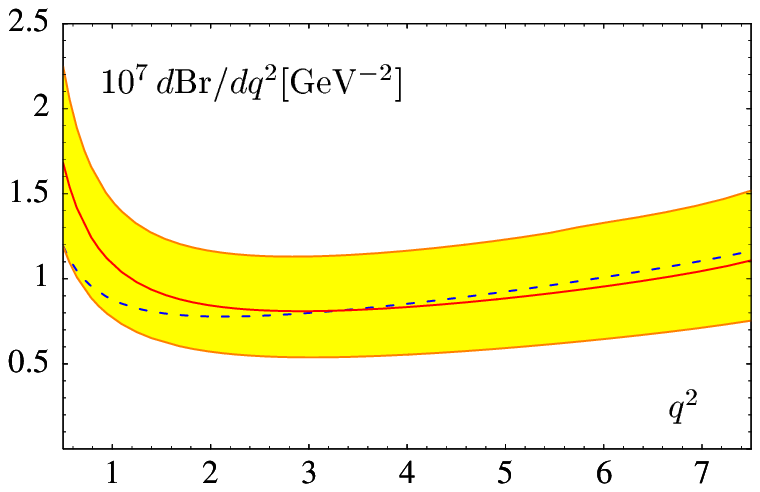}
\hspace*{2.4cm}
\includegraphics[width=1.9cm]{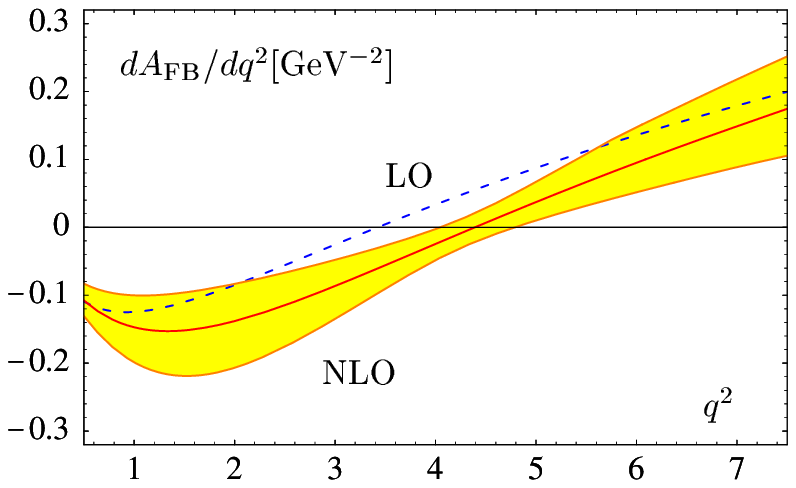}
\vskip0.3cm
\caption{
Differential decay rate $d{\cal B}(B^-\to {K^*}^-\ell^+\ell^-)/dq^2$ 
and the forward-backward asymmetry $A_{\rm FB}(B^-\to K^{*-}\ell^+\ell^-)$ 
\cite{Beneke:2001at}. The solid center line shows the
next-to-leading order result, and the dashed line shows the leading order 
result. The band reflects all theoretical uncertainties from 
parameters and scale dependence combined, with most of the 
uncertainty due to the form factors. }\label{fig:AFBKst}
\end{figure}

We turn next to the decays with a vector meson in the final state, 
such as $B \to K^* \ell^+\ell^-$ and $B\to \rho \ell^+\ell^-$. 
Since vector mesons carry a polarization, the final state has a more complex structure.
The $K^*$ decays to $K\pi$, and 
the final state
is 
specified by three angles defined as in Fig.~\ref{fig:angles}. After integrating over 
$(\phi,\theta_{K^*})$ the rate is described by three functions of $q^2$
as in the inclusive case, Eq. \eqref{Hi}, with the difference that the Wilson coefficients 
$C_{7\gamma,9,10}$ are also multiplied by
$B\to K^*$ form factors. As in inclusive case, the transverse helicity amplitudes 
are dominated by the photon pole
in the low $q^2$ region. 
In the high $q^2$ region, 
the $C_{9,10}$ 
terms dominate the amplitudes. Fig.~\ref{fig:AFBKst} shows results for the decay rate and 
the  
FBA in the exclusive mode 
$B\to K^*\ell^+\ell^-$~\cite{Beneke:2001at}. Due to form factor uncertainties the 
determination of the Wilson coefficients $C_{7\gamma}, C_9, C_{10}$ and the resulting NP 
constraints  have substantially larger theoretical errors than the ones following from the 
inclusive decays (compare for instance Fig. \ref{fig:b2sll} with Fig. \ref{fig:AFBKst}).

In the large recoil limit the  $B \to K^*/\rho \ell^+\ell^-$ amplitudes satisfy factorization
relations at leading order in $\Lambda/m_b$ 
\cite{Beneke:2001at,Bauer:2002aj,Beneke:2003pa,Hill:2004if}. These factorization 
relations 
reduce the number of unknowns by expressing the amplitudes as combinations of soft overlap factors $\zeta_\perp^{BV},
\zeta_\parallel^{BV}$ and factorizable contributions, multiplied with hard coefficients.
The factorization relations predict
that in the SM the right(left)-handed helicity amplitudes for $\bar B(B) \to K^*\ell^+\ell^-$
are power suppressed.
Any non-standard chirality structure could change this. 
A second prediction in the large recoil limit is that the left-handed helicity amplitude
$H_-^{(V)}(q^2)$ has a zero at dilepton invariant mass $q_0^2$. In the SM this is predicted 
to be~\cite{Beneke:2001at,Beneke:2004dp}
\beq
\begin{split}
q_0^2[K^{*0}] &= (4.36^{+0.33}_{-0.31}) {\rm~GeV}^2\,,\\
q_0^2[K^{*+}] &= (4.15\pm 0.27) {\rm~GeV}^2. 
\end{split}
\eeq
This result was improved recently by including the resummation of the Sudakov logs in 
SCET~\cite{Ali:2006ew}, reducing the scale dependence uncertainty.
The measurement of  $q_0^2$ can be translated into a measurement of 
$\mbox{Re}\big( {C_7}/{C_9}\big)$, up to a correction depending on the ratio of two
form factors $V(q^2)/T_1(q^2)$, which has been computed in factorization 
\cite{Beneke:2001at,Beneke:2005gs}.
Whether the soft overlap and the factorizable contributions in these form factors are 
comparable or not is still a subject of discussion, and may lead to larger errors than 
usually quoted in the literature \cite{Lee:2006gs}. 
Additional uncertainty can be introduced by the $\Lambda/m_b$ power corrections.

Various other observables are accessible in $b \to s \ell^+\ell^-$ decays,
including time-dependent 
~\cite{Kim:2007fx}
and transverse polarization asymmetries~\cite{Kruger:2005ep,Lunghi:2006hc}.
These provide additional possibilities to probe the suppression 
of right-handed amplitudes and to search for NP operators with non-standard chirality at 
a SFF. We note the presence of possible SM contamination to these observables
due to ${\cal O}(1)$ 
contributions to the right-handed amplitude in the multibody channel 
$\bar B \to K\pi \ell^+\ell^-$ in the soft pion region
\cite{Grinstein:2005ud}\footnote{These contributions also introduce a shift in 
the position of the FBA zero in $B\to K^*(\to K\pi)\ell^+\ell^-$, as the $K^*$ is 
always observed through the $K\pi$ final state.}.
This is similar to the effect discussed above for $B\to K\pi\gamma$,
and could be reduced by applying phase space cuts on the pion energy.

Further observables are accessible in the case with massive leptons,
$b\to s\tau^+\tau^-$. 
The $\tau$ polarization asymmetry 
\begin{eqnarray}
  P_\tau(q^2) \equiv \frac{d{\cal B}_{\lambda=-1} -d{\cal B}_{\lambda=+1}}
  {d{\cal B}_{\lambda=-1} +d{\cal B}_{\lambda=+1}},
\end{eqnarray}
integrated over the region $q^2/m_b^2 \geq 0.6$, is about $-48\%$ in the SM, 
but NP effects can change this prediction~\cite{Hewett:1995dk, Dai:1996vg}. 
No experimental studies of $b\to s\tau^+\tau^-$ decays exist,
making predictions of the SFF sensitivity unreliable. 
However, it appears that exclusive modes could be measured.

Another related mode is $b\to s\nu\bar\nu$, mediated in the SM
through the box and $Z$ penguin diagrams, which are matched onto the operator 
$O_{11\nu}$.
In extensions of the SM, additional diagrams can contribute, 
such as Higgs-mediated penguins in models with an extended Higgs sector, 
and models with modified $bsZ$ couplings~\cite{Grossman:1995gt,Bird:2004ts}. 
The SM expectation for the branching fractions of these modes is
${\cal B}(B\to X_s\nu\bar \nu) \sim 4\times 10^{-5}$~\cite{Buchalla:1995vs}, and 
${\cal B}(B\to X_d\nu\bar \nu) \sim 2\times 10^{-6}$. 
The dominant exclusive modes are $B\to K^{(*)}\nu\bar\nu$, 
which are expected to occur with branching fractions of about $10^{-6}$. 
Present data give only an upper bound for ${\cal B}(B^+\to K^+\nu\bar\nu)$ 
at the level of $40\times 10^{-6}$~\cite{Aubert:2004ws,Chen:2007zk},
which is one order of magnitude above the SM prediction.
These modes are very challenging experimentally because of the presence 
of two undetected neutrinos.
Nonetheless, the expected precision of the measurement of 
${\cal B} (B^+\to K^+\nu\bar\nu)$ at a SFF is 20\%, 
while the $B^+\to \pi^+\nu\bar\nu$ mode should be at the limit of 
observability~\cite{Bona:2007qt}. 

\subsection{Constraints on CKM parameters}
\label{constraintsCKM}

The radiative $b\to s(d)\gamma$ are sensitive to the CKM elements involving the
third generation quarks. In the following we briefly review the methods proposed for precision 
determination of the CKM parameters, and indicate the types of
constraints which can be obtained.

$\bullet$ {\sl $|V_{ub}|/|V_{tb} V_{ts}^*|$ from inclusive $b\to s\gamma$ and $b\to u \ell\nu$}: The inclusive radiative decays $B\to X_s\gamma$ were discussed in 
Section~\ref{InclTheory} and the inclusive semileptonic decays $B\to X_u \ell^-\bar\nu_\ell$ in Section \ref{6sides}. For both types of the decays only part
of the phase space is accessible experimentally. In semileptonic decays a cut on lepton energy or hadronic invariant mass needs to be made to avoid charm background, while in $B\to X_s\gamma$ the photon needs to be energetic enough to reduce background. Experimentally accessible is the so called shape function region of the phase space, where the inclusive state forms an energetic jet with mass $M_X^2 \sim \Lambda_{\rm QCD} Q$. Restricted to this region the OPE breaks down, while instead SCET is applicable. The decay widths factorize in a form shown in Eq.~(\ref{factincl}) for $B\to X_s\gamma$. Both radiative and semileptonic decays
depend, at LO in $1/m_b$, on the the same shape function $S(k_+)$ 
describing the
nonperturbative dynamics of the $B$ meson. The dependence on the shape function can be eliminated by combining the radiative and semileptonic rates. This then determines $|V_{ub}|/|V_{tb} V_{ts}^*|$, with different methods of implementing the basic idea discussed in detail in Sec.~\ref{6sides} (see also a review by \textcite{Paz:2006me} and recent developments in \textcite{Lee:2008vs}).

$\bullet$ {\sl $|V_{td}/V_{ts}|$ from $B\to (\rho/K^*)\gamma$ }: 
The radiative $B\to \rho\gamma$ and $B\to K^*\gamma$ amplitudes are dominated by electromagnetic penguin contributions proportional 
to $V_{td}^* V_{t b}$ and $V_{ts}^* V_{t b}$ CKM elements respectively. The ratio of the charge-averaged rates is then
\beq\label{rho2K}
\begin{split}
&\frac{\overline{\cal B}(B_q\to \rho\gamma)}{\overline{\cal B}(B_q\to K^{*}\gamma)} =\\ 
& \hspace{5mm} \kappa_q^2 \left| \frac{V_{td}}{V_{ts}} \right|^2 R_{\rm SU(3)} 
\Big( \frac{m_B^2-m_\rho^2}{m_B^2-m_{K^*}^2} \Big)^{3/2} \big(1+ r_{\rm WA}\big),
\end{split}
\eeq
where $B_q = (B^-, \bar B_d)$, $\rho_q = (\rho^-, \rho^0)$ and $\kappa_q = (1, 1/\sqrt2)$ for $q=(u,d)$ spectator quark flavors.  
The coefficient $r_{\rm WA}$ denotes the WA contribution in 
$B\to \rho\gamma$, while it is negligible for $B\to K^*\gamma$. 
The coefficient $R_{\rm SU(3)}$ parameterizes the
SU(3) breaking in the ratio of tensor form factors. 
The theory error 
in the determination of $|V_{td}/V_{ts}|$ 
is thus due to these two coefficients. 
The coefficient $r_{\rm WA}$ can be calculated using factorization. 
Writing
\beq
r_{\rm WA} =  2 \Re( \delta a) \cos \alpha |\lambda_u^{(d)}/\lambda_t^{(d)}|+
{\cal O}(\delta a ^2)\,,
\eeq
\textcite{Bosch:2004nd} find  $\Re (\delta a)=0.002^{+0.124}_{-0.061}$ for $B^0\to \rho^0\gamma$, and  $\Re(\delta a)=-0.4\pm0.4$ for $B^+\to \rho^+\gamma$.
(For an alternative treatment, see \textcite{Ball:2006eu}.)
The WA amplitude is larger for charged $B$ decays, where it is color allowed, in contrast to neutral $B$ decays, where it is color suppressed. Along with $|\lambda_u^{(d)}/\lambda_t^{(d)}|\sim 0.5$
the above values of $\delta a$ show that the uncertainty introduced by the WA contribution is minimal 
in neutral $B$ radiative decays.

\begin{figure}
%\vspace*{0.5cm}
%\hspace*{-2.8cm}
\includegraphics[width=5cm]{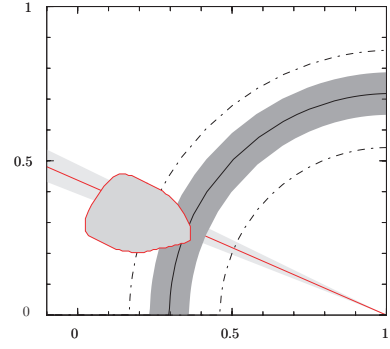}
%\vskip0.6cm
\caption{Typical constraint from $B^0\to \rho^0\gamma$ in the 
$(\bar\rho,\bar\eta)$ plane~\cite{Bosch:2004nd}. 
The constraint assumes ${\cal B}(B^0 \to \rho^0\gamma) = (0.30 \pm 0.12)
\times 10^{-6}$. The dark band corresponds to varying the SU(3) breaking ratio 
$R_{\rho/K^*}^{-1} = 1.31\pm 0.13$ at fixed $R_0$. The allowed region from the standard
CKM fit (grey area) and the constraint from $\sin 2\beta$ (angular area) are also shown.
}\label{rhoetaB2Vg}
\end{figure}

The second source of theoretical uncertainty is given by SU(3) breaking.
The parameter $R_{\rm SU(3)}$ was estimated
using QCD sum rules with the most recent result 
%\begin{eqnarray}
$R_{SU(3)} = 1.17 \pm 0.09$ \cite{Ball:2006nr}.
%\end{eqnarray}
It seems rather difficult to improve on this calculation in a model independent
way.

This method for determining $|V_{td}/V_{ts}|$ has been used to obtain 
$| {V_{td}}/{V_{ts}} | = 0.199 \, ^{+0.026}_{-0.025} \, ^{+0.018}_{-0.015}$~\cite{Abe:2005rj}
and 
$| {V_{td}}/{V_{ts}} | = 0.200 ^{+0.021}_{-0.020} \pm 0.015$~\cite{Aubert:2006pu},
where the first errors are experimental and the second theoretical,
and in both cases the average over the $B \to (\rho/\omega) \gamma$ channels is used.
A dramatic improvement in experimental error can be expected at a SFF,
and while the theoretical error can be reduced by using only the cleaner 
$B^0 \to \rho^0 \gamma$, the precision is likely to be limited at about
$4\%$ due to the SU(3) breaking correction discussed above.
This could possibly be improved using data collected at the $\Upsilon(5S)$,
as discussed in Section~\ref{sec:Bs:ckm}.

$\bullet$ {\sl $|V_{ub}/V_{td}|$ from $B\to \rho\gamma$ and 
$B\to \rho \ell\bar\nu_\ell$}: 
The ratio of CKM matrix elements $|V_{ub}/V_{td}|$ can be constrained by
combining the semileptonic mode $B\to \rho\ell\bar\nu$ with the radiative
decay $B\to\rho\gamma$~\cite{Bosch:2004nd,Beneke:2005gs}. 
In the large recoil limit the relevant form factors
satisfy factorization relations. 

The doubly differential semileptonic rate expressed in terms of the helicity 
amplitudes is
\beq
\begin{split}
\frac{d^2\Gamma(\bar B\to \rho \ell\bar\nu)}{dq^2 d\cos\theta} =& 
\frac{G_F^2 |V_{ub}|^2}{96\pi^3 m_B^2} q^2 |\vec q\,| 
\big( (1 + \cos\theta)^2 H_-^2 \\
&+ (1 - \cos\theta)^2 (H_+^2 + 2 H_0^2) \big),
\end{split}
\eeq
where $\theta$ is the angle between the $\bar \nu$ and the $B$ meson 
momentum in the $\ell\bar\nu$ center of mass frame. At $\theta=0$
only the left-handed helicity amplitude $H_-$ contributes. 
The $q^2\to 0$ limit of
the ratio of the $\bar B\to \rho_L \ell\bar\nu$ partial rate to the $\bar B\to \rho \gamma$
rate depends only on 
\beq
\begin{split}
\Big(\frac{H_-(0)}{T_1(0)}\Big)^2 \longrightarrow 2(m_B + m_V) \frac{1}{{\cal R}_2^2(0)},
\end{split}
\eeq
where $T_1(q^2)$ is a tensor current form factor Eq.~\eqref{B2KstBr}, 
while ${\cal R}_2(0)$ is calculable in a perturbative expansion 
in $\alpha_S(m_b)$ and $\alpha_S(\sqrt{\Lambda_{\rm QCD} m_b})$. 
This ratio has been computed to be $1/{\cal R}_2^2 = 0.82 \pm 0.12$~\cite{Beneke:2005gs},
allowing for a 60\% uncertainty in the spectator-scattering contribution.
This amounts to a 10\% uncertainty on this determination of $|V_{ub}/V_{td}|$, 
which however does not include uncertainties from power suppressed contributions.

$\bullet$ {\sl $|V_{ub}|^2/|V_{tb} V_{ts}^*|^2$ from $B\to K^*\ell^+\ell^-$ and
$B\to \rho \ell\bar\nu_\ell$:} In the low recoil region $q^2 \sim (M_B-M_{K^*})^2$, the
$B\to K^*\ell^+\ell^-$ amplitude can be computed in an expansion in $\Lambda/m_b, 4m_c^2/Q^2,
\alpha_s(Q)$ \cite{Grinstein:2004vb}, relating it to the semileptonic decay $B\to \rho\ell\nu$,
up to SU(3) breaking correction in the form factors. These can be eliminated using semileptonic
$D$ decay rates by forming the Grinstein double ratio \cite{Ligeti:1995yz}
\begin{eqnarray}
\frac{d\Gamma(B\to \rho\ell\nu)/dq^2}{d\Gamma(B\to K^*\ell^+\ell^-)/dq^2 }
\cdot \frac{d\Gamma(D\to K^*\ell\nu)/dq^2}{d\Gamma(D\to \rho\ell\nu)/dq^2} 
\end{eqnarray}
which is proportional to $|V_{ub}|^2/|V_{tb} V_{ts}^*|^2$. The theory error on $|V_{ub}|$
of this method is about 5\%, but measurements of the required branching fractions in
the region $q^2 = (15,19)$ GeV$^2$ require SFF statistics. 

$\bullet$ {\sl Constraints from the isospin asymmetry in $B\to \rho\gamma$}: Assuming dominance by the penguin amplitude in $B\to \rho\gamma$, 
isospin symmetry relates the charged and neutral modes to be 
$\Gamma(B^\pm \to \rho^\pm\gamma)= 2\Gamma(B^0\to \rho^0\gamma)$. 
The present experimental data point to a possible isospin asymmetry. 
The most recent world averages give~\cite{HFAG,Aubert:2006pu,Abe:2005rj}
(using CP-conjugate modes)
\begin{eqnarray}
  \Delta(\rho\gamma) & = & 
  \frac{\Gamma(B^+ \to \rho^+\gamma)}{2\Gamma(B^0\to\rho^0\gamma)}-1   \\ 
  & = & 
  \frac{(0.88\,^{+0.28}_{-0.26})\times10^{-6}}{2\times(0.93\,^{+0.19}_{-0.18})\times10^{-6}}-1 = 
  -0.53\,^{+0.18}_{-0.17} \,.\nonumber
% -0.35 \pm 0.27\,.
 %   (-4.6 \, {}^{+5.4}_{-4.2} \big|_{\rm CKM} 
   %   {}^{+5.8}_{-5.6} \big|_{\rm had})\%.
\end{eqnarray}

Several mechanisms can introduce a nonzero isospin asymmetry:
i) the $m_u-m_d$ quark mass difference leading to isospin asymmetry in the
tensor form factor $T_1$; ii) contributions from operators other than $O_{7\gamma}$
where the photon attaches to the spectator quark in the $B$ meson; 
iii) spectator diagrams such as those in Fig.~\ref{wafig}, which 
depend on the spectator quark $q$ through its electric charge, and the
hard scattering amplitude. 

\begin{figure}[t]
\begin{center}
   \epsfxsize=6cm
   \centerline{\epsffile{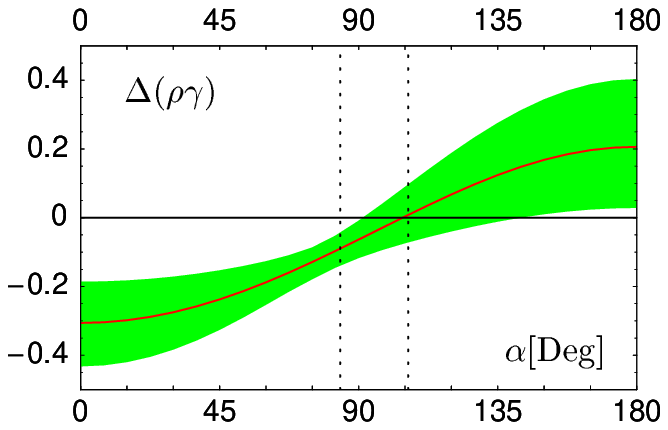}}
\vspace*{-0.7cm}
\end{center}
\caption{\label{fig:isorhogamma} Isospin asymmetry $\Delta(\rho\gamma)$ as a
  function of the CKM angle $\alpha$. 
  The band displays the total theoretical uncertainty which is 
  mainly due to weak annihilation. The vertical dashed lines 
  limit the range of $\alpha$ obtained from the CKM unitarity 
  triangle fit.}
\end{figure}

The dominant contribution to the isospin asymmetry in the SM 
is given by the last mechanism (iii), 
mediated by the four-quark operators $O_{1-6}$. 
The matrix elements of these operators  can be computed using factorization
and the heavy quark expansion
\cite{Khodjamirian:1995uc,Ali:1995uy,Grinstein:2000pc,Beneke:2001at,Bosch:2001gv,Ali:2001ez}.  
Since the four-quark operators contribute with a different weak phase 
to the penguin amplitude, the result is sensitive to CKM parameters, in
particular to the weak phase $\alpha$. Using as inputs the parameters from the
CKM fit, an isospin asymmetry of a few percent is possible, 
with significant uncertainty from hadronic parameters~\cite{Beneke:2004dp}
\begin{eqnarray}
  \Delta(\rho\gamma) =
  (-4.6 \,^{+5.4}_{-4.2} \big|_{\rm CKM} \,^{+5.8}_{-5.6} \big|_{\rm had})\%.
\end{eqnarray}

This prediction can be turned around to obtain constraints on the CKM
parameters $(\bar\rho,\bar\eta)$, using the $\rho\gamma$ asymmetries. 
As discussed in \cite{Beneke:2004dp}, measurements of the direct CP asymmetry and of the
isospin asymmetry in $B\to \rho\gamma$ give complementary constraints, which
in principle allow a complete determination of the CKM parameters. However,
the precision of such a determination is ultimately going to be limited by 
hadronic uncertainties and power corrections.

\section{$B_s$ physics at $\Upsilon(5{\rm S})$}
\label{sec:Bs} 
The $\Upsilon(5{\rm S})$ resonance is heavy enough 
that it decays both to $B_{u,d}^{(*)}$ and $B_{s}^{(*)}$ mesons. 
So far, several $e^+ e^-$ machines have operated
at the $\Upsilon(5{\rm S})$ resonance resulting in  
$0.42 \ {\rm fb}^{-1}$ of data collected 
by the CLEO collaboration~\cite{Besson:1984bd,Lovelock:1985nb}, 
followed by $1.86 \ {\rm fb}^{-1}$ of data collected by Belle
collaboration during an $\Upsilon(5{\rm S})$ 
engineering run~\cite{Drutskoy:2006dw} and a sample of 
about $21 \ {\rm fb}^{-1}$ collected by 
Belle during a one month long run in June 2006. \textcite{Baracchini:2007ei} 
performed a comprehensive analysis of the physics opportunities 
that would be offered by much larger data samples 
of $1 \ {\rm ab}^{-1}$ ($30 \ {\rm ab}^{-1}$) from a short (long) run of 
a SFF at the $\Upsilon(5{\rm S})$, where the data sample is recorded in
special purpose runs. Collecting $1 \ {\rm ab}^{-1}$ should
require less than one month at a peak luminosity 
of $10^{36} {\rm cm}^{-2} {\rm s}^{-1}$. As a result, a SFF can give 
information on the $B_s$ system 
that is complementary to that from hadronic experiments. 
In Table~\ref{tab:Bs} 
we give the expected precision from a SFF and LHCb 
for a sample of observables, clearly showing complementarity. 
In particular, the SFF can measure inclusive decays and modes with neutrals, 
which are inherently difficult in hadronic environment while LHCb
provides superior time-dependent measurements of all-charged final states. 

Physical processes involving $B_s$ mesons add to the 
wealth of information already available from the $B_{d,u}$ 
systems because the initial light quark is an $s-$quark. 
As a result, $B_s$ decays are sensitive to a different set 
of NP operators transforming between $3^{\rm rd}$ 
and $2^{\rm nd}$ generations than are $b\to s$ decays of $B_{d,u}$. 
The prime examples are $B_s\to \mu^+\mu^-$ 
where semileptonic $b\to s$ operators are probed 
and $B_s$--$\bar B_s$ mixing where $\Delta B=2$ NP operators are
probed. In addition, $B_s$ can improve knowledge of hadronic processes 
since $B_s$ and $B_d$ are related by U-spin. 
In the application of flavor SU(3) to hadronic matrix elements 
then the commonly used dynamical assumption of 
small annihilation-like amplitudes may no longer be needed.

\begin{table}
% {\renewcommand{\arraystretch}{1.5}
  \begin{center}
    \begin{ruledtabular}
      \caption{Expected precision on a subset of important observables that
        can be measured at SFF running at the $\Upsilon(5{\rm S})$ and/or
        LHCb. The first two columns give expected errors after short (less
        than a month) and long SFF runs~\cite{Bona:2007qt,Baracchini:2007ei},
        while the third lists expected {\it statistical} errors after 1 year
        of LHCb running at design luminosity~\cite{Buchalla:2008jp}.}
      \label{tab:Bs}
      \begin{tabular}{l@{\hspace{-6.5pt}}ccc}
% \hline\hline
        Observable  & SFF ($1 {\rm ab}^{-1}$)  & SFF ($30 {\rm ab}^{-1}$) &
        LHCb ($2 {\rm fb}^{-1}$)\\
        \hline
        $\Delta \Gamma_s/\Gamma_s$ & $0.11$  & $0.02$ &$0.0092$  \\
        $\beta_s$ $(J/\Psi\phi)$ & $20^\circ$    &$8^\circ$ & $1.3^\circ$ \\
        % $\beta_s$ ($J/\Psi\phi$) & $10^\circ$     &$3^\circ$ & $-$ \\
        $\beta_s$ ($B_s\to K^0\bar K^0$)  & $24^\circ$    &$11^\circ$ & $-$\\
        $A_{\rm SL}^s$     & $0.006$  &$0.004$ & $0.002{}$\\
        % $A_{\rm CH}$   & $0.004$    &$0.004$ & $-$ \\
        ${\cal B}(B_s\to \mu^+\mu^-)$  & $-$     & $<8\cdot 10^{-9}$ & $3 \sigma$ evidence \\
        $|V_{td}/V_{ts}|$ from $R_s$     & $0.08$ & $0.017$ & $-$  \\
        ${\cal B}(B_s\to \gamma\gamma)$  & $38\%$     & $7\%$ & $-$\\
        % $\Delta \Gamma_s$ & $0.16 \ {\rm ~ps}^{-1}$  & $0.03 \ {\rm ~ps}^{-1}$ &$?$  \\
        % $\Gamma_s$   	& $0.07{\rm ~ps}^{-1}$  & $0.01{\rm ~ps}^{-1}$ & $?$\\ 
% \hline\hline
      \end{tabular}
    \end{ruledtabular}
  \end{center}
\end{table}

\subsection{$B_s$--$\bar B_s$ mixing parameters }
\label{sec:Bs:mix} 
$B_s$--$\bar B_s$ mixing is described by the 
mass difference $\Delta m_s$ of the two eigenstates, 
the average of two decay widths $\Gamma_s$ 
and their difference $\Delta \Gamma_s$, by $|q/p|$ 
and by the weak mixing phase $\beta_s=-1/2 \arg(q/p)$, which 
is very small in the SM, 
$\beta_s=\arg(-V_{tb}{V_{ts}^{*}}/V_{cb}{V_{cs}^{*}})=(-1.05\pm0.05)^\circ$
\cite{Charles:2004jd}, see also Eq.~\eqref{time-dependence}. 
All these parameters can be modified 
by NP contributions and are, for instance, 
very sensitive to the large $\tan \beta$ regime of the MSSM as discussed
in Section~\ref{Large_tan_beta}. 

The oscillation frequency $\Delta m_s$ has been measured 
recently~\cite{Abulencia:2006ze}, and is found to be
consistent with SM predictions, within somewhat large theory errors.  
These oscillations are too fast to be resolved at a SFF, 
which thus cannot measure $\Delta m_s$. However, measurements of the other
parameters, $\Gamma_s$, $\Delta \Gamma_s$ and $\beta_s$ 
are possible through time dependent {\sl untagged} decay rates. 
Explicitly, for a $B_s, \bar B_s$ pair 
produced from $B_s^*, \bar B_s^*$ at the $\Upsilon (5{\rm S})$ 
this is given by~\cite{Dunietz:2000cr}
\beq
\begin{split}
\Gamma({B_s(t)\to f})&+\Gamma({\bar B_s(t)\to f}) =  \\
{\cal N}
\Gamma_s e^{-\Gamma_s |t|}
 &\Big[\cosh\Big(\frac{\Delta \Gamma_s t}{2}\Big)+H_f \sinh \Big(\frac{\Delta \Gamma_s t}{2}\Big)\Big],
\end{split}
\eeq
where $f$ is a CP-eigenstate and $H_f$ is given in Eq.~\eqref{Hf}. 
The normalization factor is given by 
${\cal N} = \frac{1}{2}(1 - (\frac{\Delta \Gamma_s}{2\Gamma_s})^2)$,
neglecting possible effects due to CP violation in mixing.

At the $\Upsilon(5{\rm S})$,
CP-tagged initial states can also be used to extract
the untarity angle $\gamma$ rather
cleanly~\cite{Falk:2000ga,Atwood:2001js},
and to constrain lifetime difference
$\Delta \Gamma_s$ 
through time independent measurements~\cite{Atwood:2002ak}.

The most promising channel for measuring  $B_s$--$\bar B_s$ mixing parameters 
at a hadronic machine is $B_s\to J/\psi \phi$, 
where angular analysis is needed to separate CP-even and CP-odd components. 
Recent measurements at D0 and CDF favor large $|\beta_s|$ 
making further studies highly interesting \cite{Aaltonen:2007he,Abazov:2008fj}.
As shown in Table \ref{tab:Bs} a SFF cannot compete with LHCb 
in this analysis, either for $\beta_s$ or for 
$\Delta \Gamma_s/\Gamma_s$ measurements, 
assuming systematic errors at LHCb are negligible. 
However, LHCb and a SFF can study complementary channels. 
For example, $B_s\to J/\psi \eta^{(\prime)}$
or $\beta_s$ from the $\Delta S=1$ penguin dominated $B_s\to K^0\bar K^0$, 
are difficult measurements at hadronic machines as shown in
Table \ref{tab:Bs}. 
The latter mode would be complementary to $B_s\to \phi \phi$, where
a precision of $0.11$ is expected after 
$2 \ {\rm fb}^{-1}$ of data at LHCb (1 year of nominal luminosity running). 
Other interesting modes that can be studied at a SFF 
include $B_s\to D_s^{(*)+}D_s^{(*)-}$, $B_s\to D_s^{(*)}K_S$,  
$B_s\to D_s^{(*)}\phi$, $B_s\to J/\psi K_S$, $B_s\to \phi \eta'$ and 
$B_s\to K_S \pi^0$ \cite{Bona:2007qt}.

Another important observable is the 
semileptonic asymmetry $A_{SL}^s$, 
which is a measure of CP violation in mixing
\beq
\begin{split}
A_{SL}^s&=\frac{{\cal B}(B_s\to D_s^{(*)-}\ell^+\nu_l)-{\cal B}(\bar B_s\to D_s^{(*)+}\ell^-\bar \nu_l)}{{\cal B}(B_s\to D_s^{(*)-}\ell^+\nu_l)+{\cal B}(\bar B_s\to D_s^{(*)+}\ell^-\bar \nu_l)}\\
&=\frac{1-|q/p|^4}{1+|q/p|^4}.
\end{split}
\eeq
The error on  $A_{SL}^s$ will become systematic dominated 
relatively soon. Taking as a guide the systematic error  
$\sigma_{\rm syst.}(A_{SL}^d)=0.004$ from current measurements
at the $\Upsilon(4{\rm S})$, this will happen at an integrated luminosity
of about $3 \ {\rm ab}^{-1}$ at the $\Upsilon(5{\rm S})$. 
Thus systematics will saturate the error quoted in 
Table \ref{tab:Bs} for $30 \ {\rm ab}^{-1}$ (where
the statistical error is only $0.001$) \cite{Bona:2007qt}. 
Note that the LHCb estimate in 
Table \ref{tab:Bs} gives only the statistical error on $A_{SL}^s$, 
while systematic errors could be substantial due to the hadronic environment.

\subsection{Rare decays}
\label{sec:Bs:rare} 
One of the most important $B_s$ decays 
for NP searches is $B_s\to \mu^+\mu^-$. 
In the SM this decay is chirally and loop suppressed with a
branching fraction of
${\cal B}(B_s\to \mu^+\mu^-)=(3.35\pm0.32)\times 10^{-9}$ 
\cite{Blanke:2006ig}. 
Exchanges of new scalar particles can lift this suppression, significantly
enhancing the rate. For instance, in the MSSM it is $\tan\beta^6$ 
enhanced in the large $\tan \beta$ regime 
(cf. Section  \ref{Large_tan_beta}). After one year of nominal 
LHCb data taking $3 \sigma$ evidence at the SM rate will be possible, 
while the SFF sensitivity to 
this channel is not competitive as indicated in Table \ref{tab:Bs}. 

A SFF can make a significant impact in radiative $B_s$ decays
and decay modes with neutrals. 
One example is $B_s\to \gamma\gamma$. 
Here the SM expectation is 
${\cal B}(B_s\to \gamma\gamma)\simeq (2-8)\cdot 10^{-7}$
\cite{Reina:1997my}, while NP effects can significantly enhance the
rate; for instance, the rate is enhanced 
by an order of magnitude in 
the $R$ parity violating MSSM \cite{Gemintern:2004bw}. 
The Belle $\Upsilon(5{\rm S})$ sample of $23.6 \ {\rm fb}^-1$ 
has already been used to demonstrate the potential of the SFF approach; 
the first observation of the penguin decay mode $B_s\to \phi\gamma$ 
has recently been reported, along with a statistics limited upper limit on 
$B_s\to \gamma\gamma$ a factor of ten above the SM level~\cite{Wicht:2007ni}.

\subsection{Improved determinations of $V_{td}/V_{ts}$ and of $V_{ub}$}
\label{sec:Bs:ckm} 

As described in Section~\ref{constraintsCKM},
exclusive radiative decays mediated by $b \to d$ and $b \to s$
penguins can be used to obtain constraints on the CKM ratio $V_{td}/V_{ts}$. 
An analogous treatment to that for $B^0 \to \rho^0 (K^{0*}) \gamma$ 
can be applied to $B_s \to K^{0*} (\phi) \gamma$,
where the theoretical error is expected to be reduced.
This is due to the simple observation that the final states 
$K^{0*}$ and $\phi$ are close in mass and are related by U-spin, 
which should help studies on the lattice. 
Moreover, a comparison of $B_s \to K^{0*} \gamma$ to 
$B^0 \to K^{0*} \gamma$ offers a determination of $V_{td}/V_{ts}$ 
that is free from SU(3) breaking corrections in the form factors~\cite{Baracchini:2007ei,Bona:2007qt}.
An improved determination of $V_{td}/V_{ts}$ from $\Delta B = 1$ 
radiative decays will be very helpful to compare to that from $B$ mixing,
and with the SM $\rho-\eta$ fit.

Study of the inclusive $B_s \to X_{us} l\nu$ and 
exclusive $B_s \to K^{(*)} l \nu$ charmless semileptonic decays can  
play a very important role in an improved $V_{ub}$ determination.
For the lattice calculation of $B_s\to K, K^*$ form factors 
a smaller extrapolation in valence light quark masses is needed 
than for $B\to \pi, \rho$ form factors, reducing the errors. 
Since $B_s \to K^{(*)} l \nu$ modes 
have significant branching ratios of ${\cal O}(10^{-4})$,
this can be an important early application of $B_s$ studies.

\section{Charm physics }
\label{sec:charm}

There are many reasons for vigorously pursuing charm physics at a SFF.
Perhaps most important is the intimate relation of charm to the top quark. Because of its
large mass top quark is sensitive to NP effects in many models. New interactions involving the top quark quite naturally also imply modified interactions of the charm quark. For example, models of warped extra-dimensions, discussed in Section~\ref{sec:warped}, inevitably
lead to flavor-changing interactions for the 
charm quark~\cite{Agashe:2004cp,Agashe:2005hk,Fitzpatrick:2007sa}.
The same is true of two Higgs doublet models, in which the top quark has
a special role~\cite{Das:1995df,Wu:1999nc}.

Charm also provides a unique handle on mixing effects in the up-type (charge
$+\tfrac{2}{3}$) sector.  The top quark does not form bound states,
which makes $D-\bar D$ the only system where this study is possible. Importance of these studies is nicely illustrated by the constraint that they provide on the MSSM squark spectrum and mixing \cite{Nir:2007ac}. The squark-quark-gluino flavor violating coupling that mixes the 
first two generations is given by $g_s \sin\theta_q$ with $q=u(d)$ for up (down) squarks. The difference of the two mixing angles needs to reproduce the Cabibbo angle  
\beq
\sin\theta_u-\sin\theta_d=\sin\theta_C\simeq 0.23. \label{relation_Cabibbo}
\eeq
 Small enough $\sin\theta_d$ can sufficiently suppresses SUSY corrections to $K-\bar K$ mixing even for nondegenerate squarks with TeV masses. This is possible in the absence of information on $D-\bar D$ mixing. The smallness of $D-\bar D$ mixing, however, requires that 
also $\sin\theta_u$ is small, which violates the  relation to the Cabibbo
angle in Eq.~\eqref{relation_Cabibbo}. The squarks with masses light enough to be observable at LHC thus need to be degenerate \cite{Nir:2007ac}.

We next summarize the salient aspects of charm physics -- 
detailed reviews can be 
found in \cite{Bianco:2003vb,Burdman:2003rs,Artuso:2008vf}. 
Within the SM, some aspects of the charm system are under excellent 
theoretical control.  In particular, one expects negligible CP asymmetry in
charm decays since the weak phase comes in CKM suppressed.  
The strong phases on the other hand are expected to be large in the charm 
region as it is rich with resonances.  
This means that a NP weak phase is likely to lead to observable CP violation. 
Moreover, although the absolute size of $D$ mixing cannot be reliably
calculated in the SM because of long distance contamination, the rate of
mixing can be used to put bounds on NP parameters in many
scenarios~\cite{Golowich:2007ka}.  
Furthermore, the indirect CP violation is negligibly small in the SM. 
It arises from a short distance contribution that is subleading in 
$D$--$\bar D$ mixing compared to the long distance piece 
and is furthermore CKM suppressed by $V_{cb}V_{ub}^*/V_{cb} V_{cs}^*$. 
It therefore provides a possibility for a very clear NP signal.

The most promising modes to search for direct CP violation in charm decays are singly
Cabibbo suppressed channels, such as $D^+ \to K^+ \bar K^0$, $\phi \pi^+$,
$D_s \to \pi^+K^0$, $K^+ \pi^0$, 
which in the SM receive contributions from two weak amplitudes, 
tree and penguin \cite{Grossman:2006jg}. 
As already mentioned indirect CP violation is very small, 
while direct CP violation is both loop and CKM suppressed making it negligible as well. Supersymmetric squark-gluino loops on the other hand
can saturate the present experimental sensitivity of ${\cal O}(10^{-2})$ \cite{Grossman:2006jg}.
Doubly Cabibbo suppressed modes may also be useful in the search
for NP effects since the SM cannot give rise to any direct CP violation and
thus the SM ``background'' contribution is small. 

The prospects for finding a BSM CP-odd
phase via $D^0$ oscillations dramatically improved in 2007.  
Using time-dependent measurements from their large charm
data samples,  Belle and \babar\ reported the first evidence for
$D^0$--$\bar D^0$ mixing~\cite{Staric:2007dt,Aubert:2007wf}. As
discussed above the existence of mixing makes it possible to search for new
physics (CP-odd) phases in the charm sector via CP-violating asymmetries. 
      
The phase of $D^0$ mixing, $\phi_D = \Im(q/p)_{D^0}$ is the analogue of the
phases of $B^0_d$ mixing or $B^0_s$ mixing discussed in
Section~\ref{sec:ut}.\footnote{ 
  Here we assume that any large phase is due to new physics. In this case, the
  quantity that is measured is the phase of $D^0$ mixing via $M_{12}$. In the SM,
  it is possible that $M_{12}\sim\Gamma_{12}$ in which case the relation
  between the experimental phase and the phase of $D^0$ mixing is more
  complicated.
}
While the phase of $B^0_d$ mixing is large in the SM, the phases of $D^0$
mixing and $B_s$ mixing are small in the SM; both are examples of
null tests, with the phase of $D^0$ mixing particularly clean since it is
expected to be of order $10^{-3}$ in the SM.  We emphasise that new physics
that appears in the $D$ sector (involving up-type quarks) may be completely
different from that in the $B$ sector.

Currently, the best sensitivity on $\phi_D$, of ${\cal O}(20^\circ)$,  is
obtained from time-dependent $D(t)\to K_S \pi^+\pi^-$ Dalitz plot analysis~\cite{Abe:2007rd}. Assuming that there are no fundamental systematic limitations
in the understanding of this Dalitz plot structure, the sensitivity to $\phi_D$ at a SFF will be about $1^\circ$--$2^\circ$. 
The use of other modes such as $D^0\to K^- K^+$ and $D^0\to K^{*0}\pi^0$ can
improve the overall sensitivity and help to eliminate ambiguous solutions for
the phase~\cite{Sinha:2007ck}.   

Searches for CP-violation via triple correlations are also very
powerful. These searches require final states that contain several linearly
independent 4-momenta and/or spins.  
A crucial advantage is
that this class of somewhat complicated final states
does not require the presence of a CP-conserving (rescattering) phase;
in $T_N$ odd-observables the CP asymmetry
is proportional to the real part of the Feynman 
amplitude~\cite{Atwood:2000tu}.
Many final states such as $K K\pi \pi$, $K\pi\pi\pi$, $\pi\pi ll$ and $KK ll$
can be used; initial studies of some of these have been carried out~\cite{Link:2005th}.
Semi-leptonic rare decays are of special interest as their small
branching fractions can translate into large CP-asymmetries. In practice, the
search for triple correlations requires the presence of a term in the angular
distribution that is proportional to $\sin \phi$, where $\phi$ is the angle
between the planes of the two pseudoscalars and the two leptons.
It has recently been pointed out by~\textcite{Bigi:2007vs} that this asymmetry could be enhanced using
data taken by a SFF in the charm energy region ({\it i.e.} at the $\psi(3770)$
resonance).
In this scenario, one uses the process 
$e^+ e^- \to \gamma^* \to D_{\rm short}D_{\rm long}$
followed by tagging of the short-lived state via, {\it e.g.}, 
$D_{\rm short} \to K^+ K^-$.
This then allows analysis of the $D_{\rm long} \to K^+ K^- \ell^+ \ell^-$ decay.
The operation of a SFF at the $\psi(3770)$ resonance would also
provide important input to the determination of $\gamma$ from $B
\to DK$ decays, as discussed in
Section~\ref{sec:ut:gamma}
~\cite{Soffer:1998un,Gronau:2001nr,Atwood:2003mj,Bondar:2005ki,Bondar:2008hh}.

CP violation in mixing can be probed using inclusive semileptonic CP asymmetry of ``wrong sign''
leptons~\cite{Bigi:2007vs}: 
\begin{eqnarray}
  a_{SL}(D^0) & \equiv &
  \frac{\Gamma (D^0(t) \to \ell^-X) - \Gamma (\bar D^0(t) \to \ell^+X)}
  {\Gamma (D^0(t) \to \ell^-X) + \Gamma (\bar D^0(t) \to \ell^+X)} \nonumber \\
  & = & \frac{|q|^4 - |p|^4}{|q|^4 + |p|^4} \, .
\end{eqnarray}
A nonnegligible value requires a BSM CP violating phase in $\Delta C=2$ dynamics and depends on both
$\sin \phi_D$ and $\Delta\Gamma/\Delta M$.  In the $D^0$ system, while $\Delta
\Gamma$ and $\Delta M$ are both small, the ratio $\Delta \Gamma/\Delta M$ need
not be.  In fact the central values in the present data are consistent with
unity or even a somewhat bigger value. The asymmetry $a_{SL}(D^0)$ is driven
by this ratio or its inverse, whichever is smaller.  Thus although the
rate for ``wrong sign'' leptons is small, their CP asymmetry might not be if
there is a significant NP phase $\phi_D$~\cite{Bigi:2007vs}.
Due to the smallness of the rate for ``wrong sign'' leptonic decays,
NP constraints from this measurement would still be statistics limited at a SFF.

Finally, although we have focused on CP violation phenomena in this section,
there is also a number of rare decays that can be useful probes of new physics effects.  
For example, searches for lepton flavor violating charm decays 
such as $D^0 \to \mu e$ or $D_{(s)} \to M \mu e$, 
where $M$ is a light meson such as $K$ or $\pi$, 
can clearly help improve the bounds on exotica.
In addition, studies of $D^+_{(s)} \to l \nu$ decays provide complementary 
information to leptonic $B^+$ decays (discussed in Section~\ref{sec:2HDM}),
and are useful to bound charged Higgs contributions in the large $\tan\beta$
limit~\cite{Rosner:2008yu,Akeroyd:2003jb,Akeroyd:2007eh}.

\section{NP tests in the tau lepton sector}
\label{sec:LFV}

\subsection{Searches for Lepton Flavor Violation}
\label{subsec:phys:LFV}

The discovery of neutrino oscillations~\cite{Davis:1968cp,Fukuda:1998mi,Ahmad:2002jz,Eguchi:2002dm,Aliu:2004sq,Kajita:2006cy}
provides direct experimental evidence that the accidental lepton flavor 
symmetries of the renormalizable Standard Model are broken in nature. 
It is therefore compelling to search for lepton flavor violation (LFV)
also in the decays of charged leptons. 
LFV decays of tau leptons can be searched for at a Super Flavor Factory. 
The list of interesting LFV modes includes $\tau\to l\gamma$, 
$\tau\to l_1 l_2l_3$ and $\tau \to l h$, where $l_i$ stands for $\mu$ or $e$, 
while the hadronic final state $h$ can be, for example, 
$\pi^0, \eta^{(\prime)}$, $K_S$, or a multihadronic state.  
These searches will complement studies of LFV in the muon sector.
The decay $\mu \to e\gamma$ will be searched for at 
MEG~\cite{Grassi:2005ac,Ritt:2006cg}, 
while $\mu \to e$ conversion will be searched for at 
PRISM/PRIME~\cite{Kuno:2005mm,Sato:2006zza}. 
Another interesting way to search for NP effects is to test lepton flavor universality in 
$ B\to K e^+e^-$ vs. $B\to K\mu^+\mu^-$ decays. The decays into muons can be well measured in hadronic environment, while the electron decays are easier to measure at a SFF. 
The current  and expected future sensitivities of 
several LFV modes of interest are summarized in Table~\ref{tab:lfv}
(for more details, see~\textcite{Raidal:2008jk}).

\begin{table}[tb]
  \begin{center}
    \begin{ruledtabular}
      \caption{
        Current and expected future $90\%$ CL upper limits on the branching
        fractions and conversion probabilities of several lepton flavor
        violating processes. 
        % All limits are at $90\%$ confidence level.
        The expectations given for $\mu^- \to e^- \gamma$ and 
        $\mu^- {\rm Ti} \to e^- {\rm Ti}$ conversion are 
        single event sensitivities (SES).
      }
      \label{tab:lfv}
      \begin{tabular}{l@{\hspace{3mm}}c@{\hspace{3mm}}c}
        Mode & Current UL & Future UL/SES \\
        \hline
        $\mu^- \to e^- \gamma$ & $1.2 \times 10^{-11}$${\ }^{\rm (a)}$ & $(1-10)\times 10^{-13}$${\ }^{\rm (b)}$ \\
        $\mu^- \to e^- e^+ e^-$ & $1.0 \times 10^{-12}$${\ }^{\rm (c)}$ & --- \\
        $\mu^- {\rm Ti} \to e^- {\rm Ti}$ & $6.1 \times 10^{-13}$${\ }^{\rm (d)}$ & $5 \times 10^{-19}$${\ }^{\rm (e)}$ \\
        $\tau^- \to \mu^- \gamma$ & $5.0\times 10^{-8}$${\ }^{\rm (f)}$ & $(2-8)\times 10^{-9}$${\ }^{\rm (g)}$ \\
        $\tau^- \to e^- \gamma$   & $5.0\times 10^{-8}$${\ }^{\rm (h)}$ & $(2-8)\times 10^{-9}$${\ }^{\rm (g)}$ \\
        $\tau^- \to \mu^- \mu^+ \mu^-$ & $3.2\times 10^{-8}$${\ }^{\rm (i)}$ & $(0.2-1)\times 10^{-9}$${\ }^{\rm (g)}$ \\
        $\tau^- \to \mu^- \eta$ & $6.5 \times 10^{-8}$${\ }^{\rm (j)}$ & $(0.4-4)\times 10^{-9}$${\ }^{\rm (g)}$ \\
      \end{tabular}
    \end{ruledtabular}
  \end{center}
  \vspace{-7mm}
  \begin{flushleft}
    \begin{tabular}{ll}
      \multicolumn{2}{l}{
        ${}^{\rm (a)}$\textcite{Brooks:1999pu,Ahmed:2001eh}
      } \\
      ${}^{\rm (b)}$\textcite{Grassi:2005ac,Ritt:2006cg} & 
      ${}^{\rm (c)}$\textcite{Bellgardt:1987du} \\
      ${}^{\rm (d)}$\textcite{Dohmen:1993mp} &
      ${}^{\rm (f)}$\textcite{Hayasaka:2007vc} \\
      ${}^{\rm (e)}$\textcite{Kuno:2005mm,Sato:2006zza} \\
      \multicolumn{2}{l}{
        ${}^{\rm (g)}$\textcite{Akeroyd:2004mj,Bona:2007qt}
      } \\
      ${}^{\rm (h)}$\textcite{Aubert:2005wa} &
      ${}^{\rm (j)}$\textcite{Miyazaki:2007jp} \\
      \multicolumn{2}{l}{
        ${}^{\rm (i)}$\textcite{Miyazaki:2007zw,Aubert:2007pw}
      } \\
    \end{tabular}
  \end{flushleft}
\end{table}

Extending to the leptonic sector the concept of minimal flavor violation,
described in Section~\ref{Sec:MFV},  
provides an effective field theory estimate of LFV~\cite{Cirigliano:2005ck,Grinstein:2006cg,Davidson:2006bd}. 
The minimal lepton flavor violation (MLFV) hypothesis supposes that
the scale $\Lambda_{LN}$ at which the total lepton number gets broken is
much larger than the mass scale $\Lambda_{LF}$  of the lightest new
particles extending the SM leptonic sector~\cite{Cirigliano:2005ck}.
These new particles could, for instance, be the sleptons of MSSM. 
The assumption of MLFV is that the new particles break flavor minimally, 
{\it i.e.} only through charged lepton and neutrino Yukawa matrices. 

MLFV predictions have several sources of theoretical uncertainties. 
First, unlike the quark sector the MFV prescription is not unique for 
the leptons because of the ambiguity in the neutrino sector. 
The minimal choice for the SM neutrino mass term is
\beq
{\cal L}_{\rm dim 5}=-\frac{1}{2\Lambda_{LN}}g_\nu^{ij}(\bar L_L^{ci}\tau _2 H)(H^T\tau_2 L_L^j)+h.c.~,
\eeq
with $g_\nu$ a spurion of MLFV. 
This mass term could arise from integrating out heavy right-handed neutrinos. 
In this case there is an additional spurion $y_\nu$ from heavy neutrino-light 
neutrino Yukawa terms with $g_\nu\sim \lambda_\nu^T\lambda_\nu$. 
This then changes the spurion analysis, 
giving different predictions on the size of LFV processes.  
Further ambiguities are due to unknown absolute size of neutrino masses, 
{\it i.e.} whether neutrinos have normal or inverted mass hierarchy, 
and from the size of CP violation in the leptonic sector. 
Most importantly, the minimal size of LFV effects is not fixed. 
Rescaling simultaneously the coupling matrix $g_\nu\to k^2 g_\nu$ 
and the lepton number violation scale $\Lambda_{LN}\to k^2 \Lambda_{LN}$ 
does not change the neutrino mass matrix, while it changes 
${\cal B}(e_i\to e_j\gamma) \to k^4 \log k {\cal B}(e_i \to e_j\gamma)$ 
(keeping $\Lambda_{LF}$ fixed at the same time). 
The rates of the lepton flavor violating processes therefore increase 
as the masses of the heavy neutrinos are raised\footnote{They do decrease with increased $\Lambda_{LF}$, the mass scale of low energy NP particles (such as slepton), as for the most NP sensitive measurements.}. 
This dependence cancels in the ratio
${\cal B}(\tau\to\mu\gamma)/{\cal B}(\mu\to e\gamma)$. 
Normalizing to the charged-current decay
\beq
{\cal B}(l_i\to l_j\gamma) \mapsto \frac{{\cal B}(l_i\to l_j\gamma)}{{\cal B}(l_i\to l_j\nu\nu)},
\eeq
\textcite{Cirigliano:2005ck} obtain that 
${\cal B}(\mu\to e\gamma)\sim (0.1-10^{-4})\times {\cal B}(\tau\to \mu\gamma)$ 
depending on the value of $\sin \theta_{13}$ angle, with smaller values of 
${\cal B}(\mu\to e\gamma)$ obtained for smaller values of $\sin \theta_{13}$. 
Saturating the present experimental bound on ${\cal B}(\mu\to e\gamma)$ at 
$\sin \theta_{13}\sim 0.05$ gives ${\cal B}(\tau\to \mu\gamma)\sim 10^{-8}$,
within the reach of a SFF. 

\begin{figure}[tb]
  \begin{center}
    \includegraphics[width=0.99\columnwidth]{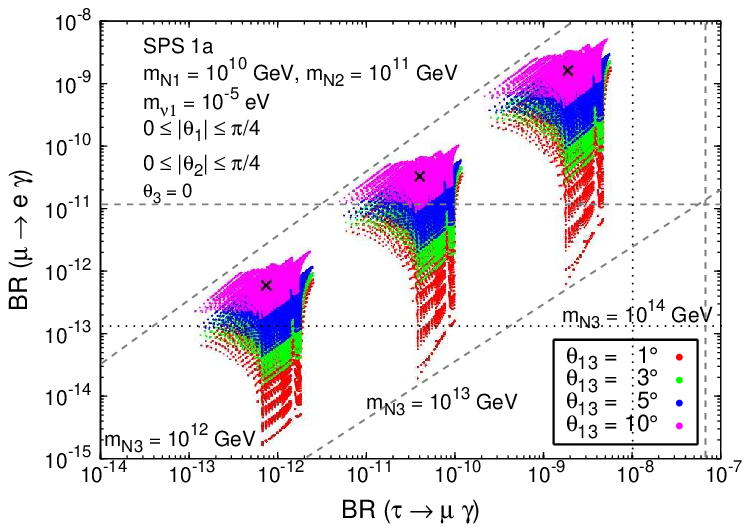}
    \caption{
      Correlation between ${\cal B}(\mu \to e\gamma)$ and 
      ${\cal B}(\tau \to \mu\gamma)$,
      and the dependence on the heaviest right-handed neutrino mass $m_{N_3}$
      and the neutrino mixing angle $\theta_{13}$ in constrained MSSM with three right-handed neutrinos~\cite{Antusch:2006vw}.
      For three values of $m_{N_3}$,
      the range of predicted values for the 
      lepton flavor violating branching fractions
      are illustrated for different values of $\theta_{13}$
      by scanning over other model parameters.
      Horizontal and vertical dashed lines denote experimental bounds,
      with dotted lines showing estimated future sensitivities 
      (note that these are almost an order of magnitude too conservative
      with regard to the SFF sensitivity for ${\cal B}(\tau \to \mu\gamma)$~\cite{Akeroyd:2004mj,Bona:2007qt}).
    } 
    \label{fig:LFV}
  \end{center}
\end{figure}

A working example of MLFV model is for instance the CMSSM
with three right-handed neutrinos~\cite{Antusch:2006vw}. 
The correlations between ${\cal B}(\mu\to e\gamma)$ and 
${\cal B}(\tau\to \mu\gamma)$ are shown in Fig.~\ref{fig:LFV}. 
In this scenario the rate for $\mu \to e\gamma$ decay depends strongly on
the value of the neutrino mixing parameter $\theta_{13}$,
and could be hard to measure if $\theta_{13} < 1^\circ$,
whereas ${\cal B}(\tau \to \mu\gamma)$ is approximately 
independent of this parameter. 
For the choices of parameters used in Fig.~\ref{fig:LFV},
based on the Snowmass point 1~\cite{Allanach:2002nj},
the rates of LFV processes are suppressed -- 
much larger rates for ${\cal B}(\tau \to \mu\gamma)$ 
are possible for other choices of NP parameters.
Large LFV effects in charged lepton decays are found in other examples of 
extending SM with heavy right-handed neutrinos with or without
supersymmetry~\cite{Borzumati:1986qx,Pham:1998fq,Ilakovac:1999md,Masiero:2004js,Hisano:1995cp,Ellis:2002fe,Babu:2002et,Agashe:2006iy}.

Embedding MFV in a GUT setup can lead to qualitatively different conclusions. 
Now the effective weak Hamiltonian for $l_i\to l_j$ processes involves also 
the quark Yukawa couplings $Y_{U,D}$. 
This means that contrary to the MLFV case above, 
the $\mu\to e\gamma$ and $\tau\to \mu\gamma, e\gamma$ rates 
cannot be arbitrarily suppressed by lowering $\Lambda_{LN}$. 
For $\Lambda_{LN}\lesssim 10^{12}$ GeV the GUT induced contribution controlled 
by $Y_{U,D}$ starts to dominate, 
which in turn for NP scale $\Lambda_{LF}\lesssim 10$ TeV gives 
${\cal B}(\mu\to e\gamma)$ above $10^{-13}$ 
within reach of the MEG experiment \cite{Grassi:2005ac}. 
The MLFV and GUT-MFV scenarios can be distinguished by comparing 
different $\tau$ and $\mu$ LFV rates. 
For instance, in the limit where quark-induced terms dominate one has 
${\cal B}(\tau\to \mu\gamma)\propto \lambda^4$ and 
${\cal B}(\mu\to e\gamma)\propto \lambda^{10}$, with $\lambda\simeq 0.22$, 
giving ${\cal B}(\tau\to \mu\gamma)/{\cal B}(\mu\to e\gamma) \sim {\cal O}(10^4)$, 
which allows $\tau\to \mu\gamma$ to be just below the present exclusion bound. 
Further information that distinguishes the two scenarios can be obtained from 
$\tau \to l \pi (l=\mu,e)$, $\pi^0\to \mu^+e^-$, 
$V\to \tau \mu \ (V=J/\psi, \Upsilon)$ and 
$\tau,\mu\to l_1l_2l_3$ decays~\cite{Cirigliano:2006su}. 
Explicit realizations of LFV in supersymmetric GUT models have been discussed in the literature~\cite{Barbieri:1994pv,Barbieri:1995tw,Gomez:1995cv,Calibbi:2006nq}.

Similarly, correlations between different $\tau$ and $\mu$ decays for a 
general 2HDM have been derived~\cite{Paradisi:2005tk,Paradisi:2006jp}. 
The decays $\mu\to e\gamma$ and $\tau\to \mu\gamma$ were found to be the 
most sensitive probes that can be close to present experimental bounds, 
while correlations between different decays are a signature of the theory. 

In supersymmetric extensions of the SM, the $l_i\to l_j\gamma^*$ 
dipole operator typically dominates over the four-lepton operators, 
which leads to a simple prediction \cite{Brignole:2004ah}
\beq
\frac{{\cal B}(l_i\to l_jl_kl_k)}{{\cal B}(l_i\to l_j\gamma)}\simeq \frac{\alpha_{\rm em}}{3\pi}\Big(\log\frac{m_\tau^2}{m_\mu^2}-\frac{11}{4}\Big)={\cal O}(10^{-3})
\eeq
If the off-diagonal slepton mass-matrix element $\delta_{3l}$ and $\tan\beta$ 
are large enough, the Higgs-mediated transitions can alter this conclusion. 
For instance in the decoupling limit~\cite{Paradisi:2005tk}
\beq
\frac{{\cal B}(\tau\to l\mu\mu)}{{\cal B}(\tau\to l \gamma)}\leq \frac{3+5\delta_{l\mu}}{36}\sim {\cal O}(0.1).
\eeq
In Little Higgs Models with $T$-parity on the other hand, 
$Z$ and box-diagram contributions dominate over the radiative operators, 
which then gives distinctly different ratios of decay widths to those in the MSSM,
as shown in Table \ref{tab:ratios}~\cite{Blanke:2007db}. 
In Little Higgs Models with $T$-parity with a NP scale $f\sim 500$ GeV,
the LFV $\tau$ decays can be seen at a SFF.  
In other models $\tau\to e\gamma$, $\tau\to l_1l_2l_3$ or $\tau\to h l$ 
can be enhanced~\cite{Cvetic:2002jy,Saha:2002kt,Black:2002wh,Sher:2002ew,Brignole:2004ah,Li:2005rr,Chen:2006hp,deGouvea:2007xp}. 
Further information on the LFV origin could be provided from Dalitz plot analysis of $\tau\to 3\mu$ with large enough data samples~\cite{Dassinger:2007ru,Matsuzaki:2007hh}.

\begin{table}
% {\renewcommand{\arraystretch}{1.5}
  \begin{center}
    \begin{ruledtabular}
      \caption{Comparison of various ratios of branching ratios in little
        Higgs model with $T$ parity and in the MSSM without and with
        significant Higgs
        contributions~\cite{Blanke:2007db}.}
      \label{tab:ratios}
      \begin{tabular}{cccc}
% \hline\hline
        Ratio & LHT  & MSSM (dipole) & MSSM (Higgs) \\\hline
        $\frac{{\cal B}(\mu\to 3e)}{{\cal B}(\mu\to e\gamma)}$  & 0.4--2.5  & $\sim6\cdot10^{-3}$ &$\sim6\cdot10^{-3}$  \\
        $\frac{{\cal B}(\tau\to 3e)}{{\cal B}(\tau\to e\gamma)}$   & 0.4--2.3     &$\sim1\cdot10^{-2}$ & ${\sim1\cdot10^{-2}}$\\
        $\frac{{\cal B}(\tau\to 3\mu)}{{\cal B}(\tau\to \mu\gamma)}$  &0.4--2.3     &$\sim2\cdot10^{-3}$ & 0.06--0.1 \\
        $\frac{{\cal B}(\tau\to e 2\mu)}{{\cal B}(\tau\to e\gamma)}$  & 0.3--1.6     &$\sim2\cdot10^{-3}$ & 0.02--0.04 \\
        $\frac{{\cal B}(\tau\to \mu 2e)}{{\cal B}(\tau\to \mu\gamma)}$  & 0.3--1.6    &$\sim1\cdot10^{-2}$ & ${\sim1\cdot10^{-2}}$\\
        $\frac{{\cal B}(\tau\to 3e)}{{\cal B}(\tau\to e 2\mu)}$     & 1.3--1.7   &$\sim5$ & 0.3--0.5\\
        $\frac{{\cal B}(\tau\to 3\mu)}{{\cal B}(\tau\to \mu 2e)}$   & 1.2--1.6    &$\sim0.2$ & 5--10 \\
        $\frac{R(\mu\text{Ti}\to e\text{Ti})}{{\cal B}(\mu\to e\gamma)}$  &
        $10^{-2}$--$10^2$     & $\sim 5\cdot 10^{-3}$ & 0.08--0.15 \\
% \hline\hline
      \end{tabular}
    \end{ruledtabular}
  \end{center}
\end{table}

\subsection{Tests of lepton flavor universality in tau decays}
\label{subsec:phys:tau:lu}

A complementary window to NP is provided by precise tests of 
lepton flavor universality in charged current 
$\tau\to \mu\nu\bar \nu$ and $\mu\to e \nu\bar \nu$ decays. 
In the large $\tan \beta$ regime of MSSM the deviations arise from 
Higgs-mediated LFV amplitudes, 
where the effects are generated by LF-conserving but mass dependent couplings.
This is complementary to $K_{l2}$ and $B_{l2}$ decays, 
where deviations are mainly due to LFV couplings \cite{Isidori:2006pk,Masiero:2005wr}. 

It is important to note that, while most of the supersymmetric models discussed above were minimally flavor violating, this is far from being the only possibility
still allowed by the LFV data. To first approximation the rare flavor changing charged lepton decays constrain the following combination of supersymmetric parameters
\beq
\sin 2 \tilde \theta_{ij}^2 \frac{\Delta \tilde m_{ij}^2}{\tilde m^2},
\label{low-eng-LFV}
\eeq
where $\tilde \theta_{ij}$ is the slepton mixing angle with $i,j=1,2,3$ the generation indices, while $\Delta \tilde m_{ij}$ and $\tilde m$
are the difference and the average of $\tilde m_{i,j}$ slepton masses, while for simplicity we suppress the $L,R$ indices for left-handed and right-handed sleptons. Thus the flavor bounds can
be obeyed either if the mixing angles are small or if the sleptons are mass degenerate. Interpolation between the two options exemplifies a set of realistic supersymmetric models discussed by \textcite{Feng:2007ke}, where supersymmetry breaking mechanism was taken to be
a combination of gauge mediated (leading to degeneracy) and gravity mediated supersymmetry breaking supplemented with horizontal symmetries (leading to alignment with split mass spectrum). 

\begin{figure}
\begin{center}
%\hspace*{1.5cm}
\includegraphics[width=\columnwidth]{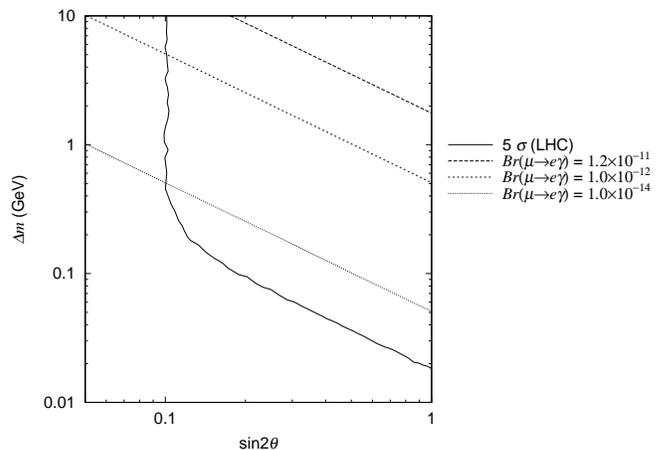} 
\caption{The LHC reach for $196$ fb${}^{-1}$ in the 
  $\tilde \theta_{ij}$--$\Delta \tilde m_{ij}$ plane,
  and the line of the constant 
  ${\cal B}(\mu \to e \gamma)$ in cMSSM with $\tan \beta = 10$,
  $A=0$, $M_0 = 90$ GeV, and $M_{1/2}=250$ GeV~\cite{Hisano:2002iy}.
  \label{osc_sugra}}
\end{center} 
\end{figure}

The high $p_T$ processes at LHC experiments probe a different combination of FV supersymmetric couplings. For degenerate sleptons with large
mixing one may observe oscillations in $\tilde l_i\to l_j\chi^0$ or $\tilde \chi_2^0\to \tilde l_i l_j\to l_i l_j \tilde\chi_1^0$ decay chains. This constrains (taking the limit of both sleptons having the same decay width $\Gamma$ for simplicity) \cite{ArkaniHamed:1996au}
\beq
\sin 2 \tilde \theta_{ij} \frac{(\Delta \tilde m_{ij}/\tilde m)^2}{(\Gamma/\tilde m)^2+(\Delta \tilde m_{ij}/\tilde m)^2},
\eeq
which should be compared with Eq.~\eqref{low-eng-LFV}. 
An example of constraints coming from the LHC and ${\cal B}(\mu\to e \gamma)$ 
based on a preliminary simulation in the cMSSM is shown in Fig~\ref{osc_sugra}.
A qualitatively similar interplay of LHC and SFF constraints 
is expected for $\tau\to \mu \gamma$. 
By having both the LHC high $p_T$ and low energy LFV measurements 
at high enough precision one is able to measure both the mixing angle 
and the mass splitting of the leptons, 
thus probing the nature of the supersymmetry breaking mechanism. 

On the experimental side, a SFF is an ideal experiment to study
lepton flavor violating tau decays due to the large cross-section 
($\sigma(e^+e^- \to \tau^+\tau^-) = (0.919 \pm 0.003) \ {\rm nb}$ at 
$\sqrt{s} = 10.58 \ {\rm GeV}$~\cite{Banerjee:2007is}) and a clean environment.
It has much better sensitivity than the LHC experiments even for the apparently favourable $\tau \to \mu\mu\mu$ channel~\cite{Santinelli:2002ea,Unel:2005fj}. 

The $B$ factories have demonstrated the enormous potential 
for tau physics from an $e^+e^-$ collider running at the $\Upsilon(4{\rm S})$.
The current experimental upper limits for most lepton flavor violating
tau decays are at present in the $10^{-7}$--$10^{-8}$ range~\cite{Hayasaka:2007vc,Abe:2007ev,Aubert:2007pw,Aubert:2006cz,Miyazaki:2007jp,Miyazaki:2006sx,Abe:2007ex},
indicating that a SFF will probe what is phenomenologically a highly 
interesting range, up to two orders of magnitude below the existing bounds.

For many of the LFV $\tau$ channels, the only limitation is due to statistics --
there are no significant backgrounds as the 
$e^+e^- \to \tau^+\tau^-$ process provides a very distinctive signature,
and the neutrinoless final state allows the four-momentum of the
decaying tau lepton to be reconstructed.
In the limit of negligible background, the achievable upper limit
scales with the integrated luminosity.

Special consideration must be given to the radiative decays
$\tau \to \mu \gamma$ and $\tau \to e \gamma$, since for these channels
there is an important background source from SM tau decays 
({\it eg.} $\tau \to \mu \nu_\mu \nu_\tau$) 
combined with a photon from initial state radiation.
This irreducible background is already an important factor
in the current analyses~\cite{Hayasaka:2007vc,Aubert:2005ye,Aubert:2005wa},
and will be dominant at very high luminosities.
Control of these backgrounds and other improvements in the analyses
will have an important effect on the ultimate sensitivity of
a SFF to lepton flavor violating tau decays.

\subsection{CP Violation in the $\tau$ system}
\label{subsec:phys:tau:cpv}

An observation of CP violation in $\tau$ decays would provide an 
incontrovertible NP signal. 
Several NP models allow direct CP violation effects in hadronic $\tau$ decays~\cite{Grossman:1994jb,Kuhn:1996dv,Delepine:2005tw,Davier:2005xq,Delepine:2006fv,Datta:2006kd},
where the only SM background is that from daughter neutral kaons~\cite{Bigi:2005ts,Calderon:2007rg} and is ${\cal O}(10^{-3})$ in $\tau\to \pi K_S^0\nu_\tau$.
Partial rate asymmetries, integrated over the phase space for the decay,
can be measured with subpercent precision at a SFF.
A more comprehensive analysis requires a study of the 
amplitude structure functions~\cite{Kuhn:1991cc,Kuhn:1992nz,Bona:2007qt};
these analyses can also be performed, but benefit from having a polarized beam
to provide a reference axis.

A polarized beam can also be used to make measurements of the $\tau$ 
electric and magnetic dipole moments. 
For the EDM measurement, an improvement of three orders of magnitude on the 
present bounds~\cite{Inami:2002ah} can be achieved~\cite{Bernabeu:2006wf}.
However this range can be saturated only by exotic NP models that can 
avoid stringent bound on the electric dipole moment of the electron. 
For the MDM, the anomalous moment could be measured for the first time at 
a SFF~\cite{Bernabeu:2007rr}.

\section{Comparison of a Super Flavor Factory with LHCb}
\label{sec:comparison}

Since a Super Flavor Factory will take data in the LHC era,
it is reasonable to ask how its physics reach
compares with the flavor physics potential of the LHC experiments,
most notably LHCb~\cite{Nakada:2007zz,Camilleri:2007zz}.
By 2014, the LHCb experiment is expected to have accumulated
$10 \ {\rm fb}^{-1}$ of data
from $pp$ collisions at a luminosity of
$\sim 2 \times 10^{32} \ {\rm cm}^{-2} {\rm s}^{-1}$~\cite{Buchalla:2008jp}.
Moreover, LHCb is planning an upgrade where they would run
at 10 times the initial design luminosity and record a data sample of 
about $100 \ {\rm fb}^{-1}$~\cite{Muheim:2007jk,Dijkstra:2007hm}.

The most striking outcome of any comparison between a SFF and
LHCb is that the strengths of the two experiments 
are largely complementary.
For example, the large boost of the $B$ hadrons produced at LHCb 
allows time-dependent studies of the oscillations of $B_s$ mesons
while many of the measurements that constitute the primary
physics motivation for a SFF cannot be performed
in a high multiplicity
hadronic environment, for example, rare decay modes with missing energy
such as $B^+ \to \ell^+\nu_\ell$ and $B^+ \to K^+\nu\bar{\nu}$. 
Measurements of the CKM matrix elements $|V_{ub}|$ and $|V_{cb}|$
and inclusive analyses of processes such as $b \to s\gamma$ and 
$b\to s\ell^+ \ell^-$ also benefit greatly from 
the clean and relatively simple $e^+e^-$ collider environment.
At LHCb the reconstruction efficiencies are reduced 
for channels containing several neutral particles and 
for studies where the $B$ decay vertex 
must be determined from a $K^0_S$ meson. 
Consequently, a SFF has unique potential to measure the photon polarization
via mixing-induced CP violation in $B^0_d \to K^0_S \pi^0 \gamma$.
Similarly, a SFF is well placed to study possible NP effects in
hadronic $b \to s$ penguin decays as it 
can measure precisely the CP asymmetries in many $B^0_d$ decay modes 
including $\phi K^0$, $\eta^\prime K^0$, $K^0_S K^0_S K^0_S$ and 
$K^0_S \pi^0$. While LHCb will have limited capability for these channels,
it can perform complementary measurements
using decay modes such as $B^0_s \to \phi\gamma$ 
and $B^0_s \to \phi\phi$ for radiative and hadronic 
$b \to s$ transitions, respectively~\cite{Camilleri:2007zz}.

Where there is overlap, 
the strength of the SFF programme in its ability to use multiple 
approaches to reach the objective becomes apparent.
For example, LHCb should be able to measure
$\alpha$ to about $5^\circ$ precision 
using $B\to\rho\pi$~\cite{Nakada:2007zz},
but will not be able to access the full information in the
$\pi\pi$ and $\rho\rho$ channels, which is necessary to reduce
the uncertainty to the $1$--$2^\circ$ level of a SFF.
Similarly, LHCb can certainly measure $\sin(2\beta)$
through mixing-induced CP violation in $B^0_d \to J/\psi K^0_S$ decay
to high accuracy (about 0.01),
but will have less sensitivity to make important complementary measurements
({\it e.g.}, in $J/\psi\,\pi^0$ and $D h^0$).
While LHCb hopes to measure the angle $\gamma$ with a precision
of $2$--$3^\circ$, extrapolations from current B factories show that
a SFF is likely to be able 
to improve this precision to about $1^\circ$. 
LHCb can probably make a precise measurement of the zero of the
forward-backward asymmetry in $B^0 \to K^{*0}\mu^+\mu^-$,
but a SFF can also measure the inclusive channel $b \to s \ell^+\ell^-$,
which, as discussed in Section~\ref{InclXllTheory}
is theoretically a much cleaner and more powerful observable.
The broader programme of a SFF thus provides a 
very comprehensive set of measurements in addition to its
clean experimental environment and superior neutral detection capabilities.
This will be of great importance for the study of
flavor physics in the LHC era.

\section{Summary}
\label{sec:summ}

\begin{table*}[!tb] 
  \caption{ Expected sensitivities at a SFF compared to
    current sensitivities for selected physics quantities.
    This table has been adapted from Table I of \cite{Browder:2007gg}
    and also includes results from the HFAG (Heavy Flavor Averaging
    Group) compilation~\cite{HFAG}. For some unitarity 
    triangle quantities such as $\gamma$ and $\alpha$, due to low
    statistics and non-gaussian behaviour of the uncertainties in
    current measurements
    there is poor agreement on the final uncertainty in the world average.
    For example, for $\gamma$ the CKMfitter group~\cite{Charles:2004jd} 
    obtains $\pm 31^\circ$ while UTfit~\cite{Bona:2006sa} finds 
    $\pm 16^\circ$ due to differences in statistical methodologies. 
    For $|V_{ub}|$ there is considerable debate on the
    treatment of theoretical errors. 
    Representative values from the PDG minireview are
    given as an estimate for the current sensitivity entry below.
    \smallskip}
  \label{tab:sff_sensivite}
  \begin{center}
    \begin{tabular}{ccc}
      \hline \hline
      Observable    & SFF sensitivity & Current sensitivity \\
      \hline
      $\sin(2\beta)$ ($J/\psi K^0$)    & 0.005--0.012  &  0.025  \\ 
      $\gamma$ ($DK$) &    1--2$^\circ$  & $\sim31^\circ$ (CKMfitter)  \\    
      $\alpha$ ($\pi\pi$, $\rho\pi$, $\rho\rho$)  & 1--2$^\circ$ & $\sim 15^\circ$ (CKMfitter) \\
      $|V_{ub}|$(excl) & 3--5\%    &  $\sim 18\%$ (PDG review)  \\
      $|V_{ub}|$(incl) & 3--5\%    &  $\sim 8\%$ (PDG review)  \\
      \hline
      $\bar{\rho}$    &   1.7--3.4\%    &   $^{+20\%}_{-12\%}$  \\
      $\bar{\eta}$    &   0.7--1.7\%    &   $4.6\%$  \\
      \hline  
      $S(\phi K^0)$   & 0.02--0.03  &    0.17     \\
      $S(\eta^{\prime} K^0)$ & 0.01--0.02 &    0.07     \\
      $S(K_S K_S K^0)$ & 0.02--0.03 &    0.20     \\
      \hline   
      ${\cal B}(B\to \tau \nu)$ & 3--4\% & 30\% \\
      ${\cal B}(B\to \mu \nu)$ & 5--6\% & not measured \\
      ${\cal B}(B\to D\tau \nu)$ & 2--2.5\% & 31\% \\  
      \hline
%      ${\cal B}(B\to X_s \nu \bar\nu)$ & &            \\
      $A_{\rm CP}(b\to s\gamma)$ & 0.004--0.005  &  0.037  \\
      $A_{\rm CP}(b\to s\gamma+d\gamma)$ & 0.01  &  0.12  \\
      ${\cal B}(B\to X_d \gamma)$ & 5--10\% & $\sim40\%$   \\ 
      ${\cal B}(B\to \rho\gamma)/{\cal B}(B\to K^*\gamma)$ & 3--4\%& 16\% \\  
      $S(K_S\pi^0\gamma)$ & 0.02--0.03 & 0.24 \\
      $S(\rho^0\gamma)$ & 0.08--0.12 & 0.67 \\
      ${\cal B}(B\to X_s \ell^+ \ell^-)$ & 4--6\% & 23\% \\
      $A^{FB}(B\to X_s \ell^+\ell^-)_{s0}$ & 4--6\% & not measured \\
      ${\cal B}(B\to K\nu \bar{\nu})$ & 16--20\% & not measured \\ 
%      $s_0(b\to s \ell^+\ell^-)$ &  &  \\
%      $s_0(b\to d \ell^+\ell^-)$ &  &  not measured \\
      \hline
      $\phi_D$ &   1--2$^{\circ}$    & $\sim 20^{\circ}$   \\
      \hline
      ${\cal B}(\tau\to\mu\gamma)$ & 2--8$\times 10^{-9}$ & not seen, $<5.0\times 10^{-8}$    \\
      ${\cal B}(\tau\to\mu\mu\mu)$ & 0.2--1$\times 10^{-9}$ &  not seen, $<(2$--$4)\times 10^{-8}$  \\
      ${\cal B}(\tau\to\mu\eta)$ & 0.4--4$\times 10^{-9}$ & not seen, $<5.1\times10^{-8}$   \\
      \hline \hline
    \end{tabular}
  \end{center}
\end{table*}

In this review we have summarized the physics case for a Super Flavor Factory
(SFF); our emphasis has been on searches for New Physics.
Such a high luminosity machine (integrating 50-75 ${\rm ab}^{-1}$) 
will of course be a Super $B$ Factory, but importantly
has enormous potential for exposing New Physics not only in the $B$ sector, 
but also in charm as well as in $\tau$ lepton decays. 

In $B$ physics the range of clean and powerful observables is 
very extensive,  see Table~\ref{tab:sff_sensivite}. 
A quick inspection vividly shows that the SFF will extend
the current reach from the $B$ factories for
many important observables by over an order of magnitude.
Specifically, we should be able to significantly
improve the precision with which we can cleanly measure the
angles ``directly'' and also determine sides of the 
unitarity triangle enhancing our knowledge of these 
fundamental parameters of the SM as
well as checking for new physics effects in $B_d$ mixing and in
$b\to d$ transitions. In addition, there are critically important 
direct searches for New Physics that are also possible.  
For example, we should be able to measure $\sin 2 \beta$
from penguin-dominated $b\to s$ modes with an accuracy of a few
percent. This will either clearly establish the presence of a new CP-odd 
phase in $b\to s$ transitions or allow us to constrain it significantly. 
Improved measurements of direct and time-dependent CP asymmetries 
in a host of modes
and the first results on the zero crossing of the forward-backward 
asymmetries in inclusive radiative $b\to s \ell^+\ell^-$ decays 
will be exciting and extremely informative. 
Furthermore, a large class of null tests will 
either constrain NP or reveal its presence.

While the dramatic increase in luminosity at a SFF will allow significant
improvements in many important existing measurements, the SFF also will
provide an important step change over the $B$ factories in that many new
channels and observables will become accessible for the first time. 
These include $b \to d \gamma$, $b\to d \ell^+\ell^-$
$B \to K^{(*)} \nu \bar \nu$ and semi-inclusive hadronic modes. 
In addition, sensitive probes
of  right-handed currents will become possible through measurements of
time-dependent asymmetries in radiative $b\to s\gamma$ processes 
such as $B \to K_S \pi^0 (\rho^0)\gamma$, 
as well as transverse polarization of the $\tau$ in semitauonic
decays of $B$ mesons.
At the SFF, the high statistics and 
kinematic constraints of production at the $\Upsilon(4S)$
also will allow clean studies of many important inclusive processes
in the recoil of fully reconstructed tagged $B$ mesons.

High luminosity charm studies will also be sensitive to
the effects of new physics; 
the most important of these is a search
for a new CP-odd phase in $D$ mixing ($\phi_D$) with 
a sensitivity of a few degrees.
Improved studies of lepton flavor violation in $\tau$ decays 
with much higher sensitivities could also prove 
to be extremely important in revealing new phenomena
or allowing us to constrain it more effectively.  

A Super Flavor Factory will complement dedicated flavor studies
at the LHC with its sensitivity to decay modes with photons and
multiple neutrinos as well as inclusive processes.
The SFF will extend the reach of the high $p_T$ 
experiments at the LHC in many ways and will help us interpret
whatever type of New Physics is discovered there.

\begin{acknowledgments}
We thank  Rafael Porto for useful discussions and Sebastian Jaeger and Tobias Hurth for comments on the manuscript. Research supported in part by the US Department of Energy, contracts
DE-FG02-04ER41291 (Hawaii) and DE-AC02-98CH10886 (BNL). The work of J. Z. was supported in part by the European Commission RTN network, Contract No.~MRTN-CT-2006-035482 (FLAVIAnet) and by the 
Slovenian Research Agency.
\end{acknowledgments}

%%%%%%%%%%%%%%%%%%%%%%%%%%%%%%%%%
%%%%%%%%%%%%%  bib  %%%%%%%%%%%%%
%%%%%%%%%%%%%%%%%%%%%%%%%%%%%%%%%

\bibliography{RMPrev}

\begin{thebibliography}{594}
\expandafter\ifx\csname natexlab\endcsname\relax\def\natexlab#1{#1}\fi
\expandafter\ifx\csname bibnamefont\endcsname\relax
  \def\bibnamefont#1{#1}\fi
\expandafter\ifx\csname bibfnamefont\endcsname\relax
  \def\bibfnamefont#1{#1}\fi
\expandafter\ifx\csname citenamefont\endcsname\relax
  \def\citenamefont#1{#1}\fi
\expandafter\ifx\csname url\endcsname\relax
  \def\url#1{\texttt{#1}}\fi
\expandafter\ifx\csname urlprefix\endcsname\relax\def\urlprefix{URL }\fi
\providecommand{\bibinfo}[2]{#2}
\providecommand{\eprint}[2][]{\url{#2}}

\bibitem[{Aaltonen \emph{et~al.}(2007{\natexlab{a}})\citenamefont{Aaltonen}
  \emph{et~al.}}]{Aaltonen:2007he}
\bibinfo{author}{\bibnamefont{Aaltonen}, \bibfnamefont{T.}}, \emph{et~al.}
  (\bibinfo{collaboration}{CDF}), \bibinfo{year}{2007}{\natexlab{a}},
  \eprint{arXiv:0712.2397 [hep-ex]}.

\bibitem[{Aaltonen \emph{et~al.}(2007{\natexlab{b}})\citenamefont{Aaltonen}
  \emph{et~al.}}]{Aaltonen:2007kv}
\bibinfo{author}{\bibnamefont{Aaltonen}, \bibfnamefont{T.}}, \emph{et~al.}
  (\bibinfo{collaboration}{CDF}), \bibinfo{year}{2007}{\natexlab{b}},
  \eprint{arXiv:0712.1708 [hep-ex]}.

\bibitem[{Abashian \emph{et~al.}(2002)\citenamefont{Abashian}
  \emph{et~al.}}]{Abashian:2000cg}
\bibinfo{author}{\bibnamefont{Abashian}, \bibfnamefont{A.}}, \emph{et~al.},
  \bibinfo{year}{2002}, \bibinfo{journal}{Nucl. Instrum. Meth.}
  \textbf{\bibinfo{volume}{A479}}, \bibinfo{pages}{117}.

\bibitem[{Abazov \emph{et~al.}(2005)\citenamefont{Abazov}
  \emph{et~al.}}]{Abazov:2005sa}
\bibinfo{author}{\bibnamefont{Abazov}, \bibfnamefont{V.~M.}}, \emph{et~al.}
  (\bibinfo{collaboration}{D0}), \bibinfo{year}{2005}, \bibinfo{journal}{Phys.
  Rev. Lett.} \textbf{\bibinfo{volume}{95}}, \bibinfo{pages}{171801}.

\bibitem[{Abazov \emph{et~al.}(2007)\citenamefont{Abazov}
  \emph{et~al.}}]{Abazov:2007iy}
\bibinfo{author}{\bibnamefont{Abazov}, \bibfnamefont{V.~M.}}, \emph{et~al.}
  (\bibinfo{collaboration}{D0}), \bibinfo{year}{2007}, \bibinfo{journal}{Phys.
  Rev.} \textbf{\bibinfo{volume}{D76}}, \bibinfo{pages}{092001}.

\bibitem[{Abazov \emph{et~al.}(2008)\citenamefont{Abazov}
  \emph{et~al.}}]{Abazov:2008fj}
\bibinfo{author}{\bibnamefont{Abazov}, \bibfnamefont{V.~M.}}, \emph{et~al.}
  (\bibinfo{collaboration}{D0}), \bibinfo{year}{2008}, \eprint{arXiv:0802.2255
  [hep-ex]}.

\bibitem[{Abbiendi \emph{et~al.}(2001)\citenamefont{Abbiendi}
  \emph{et~al.}}]{Abbiendi:2001fi}
\bibinfo{author}{\bibnamefont{Abbiendi}, \bibfnamefont{G.}}, \emph{et~al.}
  (\bibinfo{collaboration}{OPAL}), \bibinfo{year}{2001},
  \bibinfo{journal}{Phys. Lett.} \textbf{\bibinfo{volume}{B520}},
  \bibinfo{pages}{1}.

\bibitem[{Abe \emph{et~al.}(2005)\citenamefont{Abe} \emph{et~al.}}]{Abe:2004mz}
\bibinfo{author}{\bibnamefont{Abe}, \bibfnamefont{K.}}, \emph{et~al.}
  (\bibinfo{collaboration}{Belle}), \bibinfo{year}{2005},
  \bibinfo{journal}{Phys. Rev.} \textbf{\bibinfo{volume}{D71}},
  \bibinfo{pages}{072003}.

\bibitem[{Abe \emph{et~al.}(2006{\natexlab{a}})\citenamefont{Abe}
  \emph{et~al.}}]{Abe:2005rj}
\bibinfo{author}{\bibnamefont{Abe}, \bibfnamefont{K.}}, \emph{et~al.}
  (\bibinfo{collaboration}{Belle}), \bibinfo{year}{2006}{\natexlab{a}},
  \bibinfo{journal}{Phys. Rev. Lett.} \textbf{\bibinfo{volume}{96}},
  \bibinfo{pages}{221601}.

\bibitem[{Abe \emph{et~al.}(2007{\natexlab{a}})\citenamefont{Abe}
  \emph{et~al.}}]{Abe:2007rd}
\bibinfo{author}{\bibnamefont{Abe}, \bibfnamefont{K.}}, \emph{et~al.}
  (\bibinfo{collaboration}{Belle}), \bibinfo{year}{2007}{\natexlab{a}},
  \bibinfo{journal}{Phys. Rev. Lett.} \textbf{\bibinfo{volume}{99}},
  \bibinfo{pages}{131803}.

\bibitem[{Abe \emph{et~al.}(2007{\natexlab{b}})\citenamefont{Abe}
  \emph{et~al.}}]{Abe:2006qx}
\bibinfo{author}{\bibnamefont{Abe}, \bibfnamefont{K.}}, \emph{et~al.}
  (\bibinfo{collaboration}{Belle}), \bibinfo{year}{2007}{\natexlab{b}},
  \bibinfo{journal}{Phys. Rev. Lett.} \textbf{\bibinfo{volume}{99}},
  \bibinfo{pages}{121601}.

\bibitem[{Abe \emph{et~al.}(2007{\natexlab{c}})\citenamefont{Abe}
  \emph{et~al.}}]{Abe:2007xd}
\bibinfo{author}{\bibnamefont{Abe}, \bibfnamefont{K.}}, \emph{et~al.}
  (\bibinfo{collaboration}{Belle}), \bibinfo{year}{2007}{\natexlab{c}},
  \eprint{arXiv:0708.1845 [hep-ex]}.

\bibitem[{Abe \emph{et~al.}(2007{\natexlab{d}})\citenamefont{Abe}
  \emph{et~al.}}]{Abe:2006gy}
\bibinfo{author}{\bibnamefont{Abe}, \bibfnamefont{K.}}, \emph{et~al.}
  (\bibinfo{collaboration}{Belle}), \bibinfo{year}{2007}{\natexlab{d}},
  \bibinfo{journal}{Phys. Rev.} \textbf{\bibinfo{volume}{D76}},
  \bibinfo{pages}{091103}.

\bibitem[{Abe \emph{et~al.}(2007{\natexlab{e}})\citenamefont{Abe}
  \emph{et~al.}}]{Abe:2006xs}
\bibinfo{author}{\bibnamefont{Abe}, \bibfnamefont{K.}}, \emph{et~al.}
  (\bibinfo{collaboration}{Belle}), \bibinfo{year}{2007}{\natexlab{e}},
  \bibinfo{journal}{Phys. Rev. Lett.} \textbf{\bibinfo{volume}{98}},
  \bibinfo{pages}{181804}.

\bibitem[{Abe \emph{et~al.}(2007{\natexlab{f}})\citenamefont{Abe}
  \emph{et~al.}}]{Abe:2007ex}
\bibinfo{author}{\bibnamefont{Abe}, \bibfnamefont{K.}}, \emph{et~al.}
  (\bibinfo{collaboration}{Belle}), \bibinfo{year}{2007}{\natexlab{f}},
  \eprint{arXiv:0708.3276 [hep-ex]}.

\bibitem[{Abe \emph{et~al.}(2008)\citenamefont{Abe} \emph{et~al.}}]{Abe:2007ev}
\bibinfo{author}{\bibnamefont{Abe}, \bibfnamefont{K.}}, \emph{et~al.}
  (\bibinfo{collaboration}{Belle}), \bibinfo{year}{2008},
  \bibinfo{journal}{Phys. Lett.} \textbf{\bibinfo{volume}{B660}},
  \bibinfo{pages}{154}.

\bibitem[{Abe \emph{et~al.}(2004)\citenamefont{Abe} \emph{et~al.}}]{Abe:2004px}
\bibinfo{author}{\bibnamefont{Abe}, \bibfnamefont{M.}}, \emph{et~al.}
  (\bibinfo{collaboration}{KEK-E246}), \bibinfo{year}{2004},
  \bibinfo{journal}{Phys. Rev. Lett.} \textbf{\bibinfo{volume}{93}},
  \bibinfo{pages}{131601}.

\bibitem[{Abe \emph{et~al.}(2006{\natexlab{b}})\citenamefont{Abe}
  \emph{et~al.}}]{Abe:2006de}
\bibinfo{author}{\bibnamefont{Abe}, \bibfnamefont{M.}}, \emph{et~al.},
  \bibinfo{year}{2006}{\natexlab{b}}, \bibinfo{journal}{Phys. Rev.}
  \textbf{\bibinfo{volume}{D73}}, \bibinfo{pages}{072005}.

\bibitem[{Abe \emph{et~al.}(2007{\natexlab{g}})\citenamefont{Abe}
  \emph{et~al.}}]{Abe:2007iz}
\bibinfo{author}{\bibnamefont{Abe}, \bibfnamefont{T.}}, \emph{et~al.},
  \bibinfo{year}{2007}{\natexlab{g}}, \eprint{arXiv:0706.3248
  [physics.ins-det]}.

\bibitem[{Abulencia \emph{et~al.}(2006)\citenamefont{Abulencia}
  \emph{et~al.}}]{Abulencia:2006ze}
\bibinfo{author}{\bibnamefont{Abulencia}, \bibfnamefont{A.}}, \emph{et~al.}
  (\bibinfo{collaboration}{CDF}), \bibinfo{year}{2006}, \bibinfo{journal}{Phys.
  Rev. Lett.} \textbf{\bibinfo{volume}{97}}, \bibinfo{pages}{242003}.

\bibitem[{Acosta \emph{et~al.}(2005)\citenamefont{Acosta}
  \emph{et~al.}}]{Acosta:2004gt}
\bibinfo{author}{\bibnamefont{Acosta}, \bibfnamefont{D.~E.}}, \emph{et~al.}
  (\bibinfo{collaboration}{CDF}), \bibinfo{year}{2005}, \bibinfo{journal}{Phys.
  Rev. Lett.} \textbf{\bibinfo{volume}{94}}, \bibinfo{pages}{101803}.

\bibitem[{\citenamefont{Agashe} \emph{et~al.}(2006)\citenamefont{Agashe,
  Blechman, and Petriello}}]{Agashe:2006iy}
\bibinfo{author}{\bibnamefont{Agashe}, \bibfnamefont{K.}},
  \bibinfo{author}{\bibfnamefont{A.~E.} \bibnamefont{Blechman}}, and
  \bibinfo{author}{\bibfnamefont{F.}~\bibnamefont{Petriello}},
  \bibinfo{year}{2006}, \bibinfo{journal}{Phys. Rev.}
  \textbf{\bibinfo{volume}{D74}}, \bibinfo{pages}{053011}.

\bibitem[{\citenamefont{Agashe} \emph{et~al.}(2003)\citenamefont{Agashe,
  Delgado, May, and Sundrum}}]{Agashe:2003zs}
\bibinfo{author}{\bibnamefont{Agashe}, \bibfnamefont{K.}},
  \bibinfo{author}{\bibfnamefont{A.}~\bibnamefont{Delgado}},
  \bibinfo{author}{\bibfnamefont{M.~J.} \bibnamefont{May}}, and
  \bibinfo{author}{\bibfnamefont{R.}~\bibnamefont{Sundrum}},
  \bibinfo{year}{2003}, \bibinfo{journal}{JHEP} \textbf{\bibinfo{volume}{08}},
  \bibinfo{pages}{050}.

\bibitem[{\citenamefont{Agashe}
  \emph{et~al.}(2005{\natexlab{a}})\citenamefont{Agashe, Papucci, Perez, and
  Pirjol}}]{Agashe:2005hk}
\bibinfo{author}{\bibnamefont{Agashe}, \bibfnamefont{K.}},
  \bibinfo{author}{\bibfnamefont{M.}~\bibnamefont{Papucci}},
  \bibinfo{author}{\bibfnamefont{G.}~\bibnamefont{Perez}}, and
  \bibinfo{author}{\bibfnamefont{D.}~\bibnamefont{Pirjol}},
  \bibinfo{year}{2005}{\natexlab{a}}, \eprint{hep-ph/0509117}.

\bibitem[{\citenamefont{Agashe} \emph{et~al.}(2004)\citenamefont{Agashe, Perez,
  and Soni}}]{Agashe:2004ay}
\bibinfo{author}{\bibnamefont{Agashe}, \bibfnamefont{K.}},
  \bibinfo{author}{\bibfnamefont{G.}~\bibnamefont{Perez}}, and
  \bibinfo{author}{\bibfnamefont{A.}~\bibnamefont{Soni}}, \bibinfo{year}{2004},
  \bibinfo{journal}{Phys. Rev. Lett.} \textbf{\bibinfo{volume}{93}},
  \bibinfo{pages}{201804}.

\bibitem[{\citenamefont{Agashe}
  \emph{et~al.}(2005{\natexlab{b}})\citenamefont{Agashe, Perez, and
  Soni}}]{Agashe:2004cp}
\bibinfo{author}{\bibnamefont{Agashe}, \bibfnamefont{K.}},
  \bibinfo{author}{\bibfnamefont{G.}~\bibnamefont{Perez}}, and
  \bibinfo{author}{\bibfnamefont{A.}~\bibnamefont{Soni}},
  \bibinfo{year}{2005}{\natexlab{b}}, \bibinfo{journal}{Phys. Rev.}
  \textbf{\bibinfo{volume}{D71}}, \bibinfo{pages}{016002}.

\bibitem[{Ahmad \emph{et~al.}(2002)\citenamefont{Ahmad}
  \emph{et~al.}}]{Ahmad:2002jz}
\bibinfo{author}{\bibnamefont{Ahmad}, \bibfnamefont{Q.~R.}}, \emph{et~al.}
  (\bibinfo{collaboration}{SNO}), \bibinfo{year}{2002}, \bibinfo{journal}{Phys.
  Rev. Lett.} \textbf{\bibinfo{volume}{89}}, \bibinfo{pages}{011301}.

\bibitem[{Ahmed \emph{et~al.}(2002)\citenamefont{Ahmed}
  \emph{et~al.}}]{Ahmed:2001eh}
\bibinfo{author}{\bibnamefont{Ahmed}, \bibfnamefont{M.}}, \emph{et~al.}
  (\bibinfo{collaboration}{MEGA}), \bibinfo{year}{2002},
  \bibinfo{journal}{Phys. Rev.} \textbf{\bibinfo{volume}{D65}},
  \bibinfo{pages}{112002}.

\bibitem[{Aihara \emph{et~al.}(2006)\citenamefont{Aihara}
  \emph{et~al.}}]{Aihara:2006dh}
\bibinfo{author}{\bibnamefont{Aihara}, \bibfnamefont{H.}}, \emph{et~al.},
  \bibinfo{year}{2006}, \bibinfo{journal}{Nucl. Instrum. Meth.}
  \textbf{\bibinfo{volume}{A568}}, \bibinfo{pages}{269}.

\bibitem[{\citenamefont{Akai and Morita}(2003)}]{Akai:2004np}
\bibinfo{author}{\bibnamefont{Akai}, \bibfnamefont{K.}}, and
  \bibinfo{author}{\bibfnamefont{Y.}~\bibnamefont{Morita}},
  \bibinfo{year}{2003}, \bibinfo{note}{kEK-PREPRINT-2003-123}.

\bibitem[{\citenamefont{Akeroyd}(2004)}]{Akeroyd:2003jb}
\bibinfo{author}{\bibnamefont{Akeroyd}, \bibfnamefont{A.~G.}},
  \bibinfo{year}{2004}, \bibinfo{journal}{Prog. Theor. Phys.}
  \textbf{\bibinfo{volume}{111}}, \bibinfo{pages}{295}.

\bibitem[{\citenamefont{Akeroyd and Chen}(2007)}]{Akeroyd:2007eh}
\bibinfo{author}{\bibnamefont{Akeroyd}, \bibfnamefont{A.~G.}}, and
  \bibinfo{author}{\bibfnamefont{C.~H.} \bibnamefont{Chen}},
  \bibinfo{year}{2007}, \bibinfo{journal}{Phys. Rev.}
  \textbf{\bibinfo{volume}{D75}}, \bibinfo{pages}{075004}.

\bibitem[{Akeroyd \emph{et~al.}(2004)\citenamefont{Akeroyd}
  \emph{et~al.}}]{Akeroyd:2004mj}
\bibinfo{author}{\bibnamefont{Akeroyd}, \bibfnamefont{A.~G.}}, \emph{et~al.}
  (\bibinfo{collaboration}{SuperKEKB Physics Working Group}),
  \bibinfo{year}{2004}, \eprint{hep-ex/0406071}.

\bibitem[{\citenamefont{Aleksan} \emph{et~al.}(1994)\citenamefont{Aleksan,
  Kayser, and London}}]{Aleksan:1994if}
\bibinfo{author}{\bibnamefont{Aleksan}, \bibfnamefont{R.}},
  \bibinfo{author}{\bibfnamefont{B.}~\bibnamefont{Kayser}}, and
  \bibinfo{author}{\bibfnamefont{D.}~\bibnamefont{London}},
  \bibinfo{year}{1994}, \bibinfo{journal}{Phys. Rev. Lett.}
  \textbf{\bibinfo{volume}{73}}, \bibinfo{pages}{18}.

\bibitem[{\citenamefont{Aleksan} \emph{et~al.}(2003)\citenamefont{Aleksan,
  Petersen, and Soffer}}]{Aleksan:2002mh}
\bibinfo{author}{\bibnamefont{Aleksan}, \bibfnamefont{R.}},
  \bibinfo{author}{\bibfnamefont{T.~C.} \bibnamefont{Petersen}}, and
  \bibinfo{author}{\bibfnamefont{A.}~\bibnamefont{Soffer}},
  \bibinfo{year}{2003}, \bibinfo{journal}{Phys. Rev.}
  \textbf{\bibinfo{volume}{D67}}, \bibinfo{pages}{096002}.

\bibitem[{\citenamefont{Ali and Braun}(1995)}]{Ali:1995uy}
\bibinfo{author}{\bibnamefont{Ali}, \bibfnamefont{A.}}, and
  \bibinfo{author}{\bibfnamefont{V.~M.} \bibnamefont{Braun}},
  \bibinfo{year}{1995}, \bibinfo{journal}{Phys. Lett.}
  \textbf{\bibinfo{volume}{B359}}, \bibinfo{pages}{223}.

\bibitem[{\citenamefont{Ali} \emph{et~al.}(1997)\citenamefont{Ali, Hiller,
  Handoko, and Morozumi}}]{Ali:1996bm}
\bibinfo{author}{\bibnamefont{Ali}, \bibfnamefont{A.}},
  \bibinfo{author}{\bibfnamefont{G.}~\bibnamefont{Hiller}},
  \bibinfo{author}{\bibfnamefont{L.~T.} \bibnamefont{Handoko}}, and
  \bibinfo{author}{\bibfnamefont{T.}~\bibnamefont{Morozumi}},
  \bibinfo{year}{1997}, \bibinfo{journal}{Phys. Rev.}
  \textbf{\bibinfo{volume}{D55}}, \bibinfo{pages}{4105}.

\bibitem[{\citenamefont{Ali} \emph{et~al.}(2006)\citenamefont{Ali, Kramer, and
  Zhu}}]{Ali:2006ew}
\bibinfo{author}{\bibnamefont{Ali}, \bibfnamefont{A.}},
  \bibinfo{author}{\bibfnamefont{G.}~\bibnamefont{Kramer}}, and
  \bibinfo{author}{\bibfnamefont{G.-h.} \bibnamefont{Zhu}},
  \bibinfo{year}{2006}, \bibinfo{journal}{Eur. Phys. J.}
  \textbf{\bibinfo{volume}{C47}}, \bibinfo{pages}{625}.

\bibitem[{\citenamefont{Ali} \emph{et~al.}(2002)\citenamefont{Ali, Lunghi,
  Greub, and Hiller}}]{Ali:2002jg}
\bibinfo{author}{\bibnamefont{Ali}, \bibfnamefont{A.}},
  \bibinfo{author}{\bibfnamefont{E.}~\bibnamefont{Lunghi}},
  \bibinfo{author}{\bibfnamefont{C.}~\bibnamefont{Greub}}, and
  \bibinfo{author}{\bibfnamefont{G.}~\bibnamefont{Hiller}},
  \bibinfo{year}{2002}, \bibinfo{journal}{Phys. Rev.}
  \textbf{\bibinfo{volume}{D66}}, \bibinfo{pages}{034002}.

\bibitem[{\citenamefont{Ali and Parkhomenko}(2002)}]{Ali:2001ez}
\bibinfo{author}{\bibnamefont{Ali}, \bibfnamefont{A.}}, and
  \bibinfo{author}{\bibfnamefont{A.~Y.} \bibnamefont{Parkhomenko}},
  \bibinfo{year}{2002}, \bibinfo{journal}{Eur. Phys. J.}
  \textbf{\bibinfo{volume}{C23}}, \bibinfo{pages}{89}.

\bibitem[{Ali \emph{et~al.}(2007)\citenamefont{Ali} \emph{et~al.}}]{Ali:2007ff}
\bibinfo{author}{\bibnamefont{Ali}, \bibfnamefont{A.}}, \emph{et~al.},
  \bibinfo{year}{2007}, \bibinfo{journal}{Phys. Rev.}
  \textbf{\bibinfo{volume}{D76}}, \bibinfo{pages}{074018}.

\bibitem[{Aliu \emph{et~al.}(2005)\citenamefont{Aliu}
  \emph{et~al.}}]{Aliu:2004sq}
\bibinfo{author}{\bibnamefont{Aliu}, \bibfnamefont{E.}}, \emph{et~al.}
  (\bibinfo{collaboration}{K2K}), \bibinfo{year}{2005}, \bibinfo{journal}{Phys.
  Rev. Lett.} \textbf{\bibinfo{volume}{94}}, \bibinfo{pages}{081802}.

\bibitem[{Allanach \emph{et~al.}(2002)\citenamefont{Allanach}
  \emph{et~al.}}]{Allanach:2002nj}
\bibinfo{author}{\bibnamefont{Allanach}, \bibfnamefont{B.~C.}}, \emph{et~al.},
  \bibinfo{year}{2002}, \eprint{hep-ph/0202233}.

\bibitem[{\citenamefont{Andersen and Gardi}(2006)}]{Andersen:2005mj}
\bibinfo{author}{\bibnamefont{Andersen}, \bibfnamefont{J.~R.}}, and
  \bibinfo{author}{\bibfnamefont{E.}~\bibnamefont{Gardi}},
  \bibinfo{year}{2006}, \bibinfo{journal}{JHEP} \textbf{\bibinfo{volume}{01}},
  \bibinfo{pages}{097}.

\bibitem[{\citenamefont{Antaramian}
  \emph{et~al.}(1992)\citenamefont{Antaramian, Hall, and
  Rasin}}]{Antaramian:1992ya}
\bibinfo{author}{\bibnamefont{Antaramian}, \bibfnamefont{A.}},
  \bibinfo{author}{\bibfnamefont{L.~J.} \bibnamefont{Hall}}, and
  \bibinfo{author}{\bibfnamefont{A.}~\bibnamefont{Rasin}},
  \bibinfo{year}{1992}, \bibinfo{journal}{Phys. Rev. Lett.}
  \textbf{\bibinfo{volume}{69}}, \bibinfo{pages}{1871}.

\bibitem[{Antonio \emph{et~al.}(2008)\citenamefont{Antonio}
  \emph{et~al.}}]{Antonio:2007pb}
\bibinfo{author}{\bibnamefont{Antonio}, \bibfnamefont{D.~J.}}, \emph{et~al.}
  (\bibinfo{collaboration}{RBC}), \bibinfo{year}{2008}, \bibinfo{journal}{Phys.
  Rev. Lett.} \textbf{\bibinfo{volume}{100}}, \bibinfo{pages}{032001}.

\bibitem[{\citenamefont{Antusch} \emph{et~al.}(2006)\citenamefont{Antusch,
  Arganda, Herrero, and Teixeira}}]{Antusch:2006vw}
\bibinfo{author}{\bibnamefont{Antusch}, \bibfnamefont{S.}},
  \bibinfo{author}{\bibfnamefont{E.}~\bibnamefont{Arganda}},
  \bibinfo{author}{\bibfnamefont{M.~J.} \bibnamefont{Herrero}}, and
  \bibinfo{author}{\bibfnamefont{A.~M.} \bibnamefont{Teixeira}},
  \bibinfo{year}{2006}, \bibinfo{journal}{JHEP} \textbf{\bibinfo{volume}{11}},
  \bibinfo{pages}{090}.

\bibitem[{\citenamefont{Arkani-Hamed}
  \emph{et~al.}(1996)\citenamefont{Arkani-Hamed, Cheng, Feng, and
  Hall}}]{ArkaniHamed:1996au}
\bibinfo{author}{\bibnamefont{Arkani-Hamed}, \bibfnamefont{N.}},
  \bibinfo{author}{\bibfnamefont{H.-C.} \bibnamefont{Cheng}},
  \bibinfo{author}{\bibfnamefont{J.~L.} \bibnamefont{Feng}}, and
  \bibinfo{author}{\bibfnamefont{L.~J.} \bibnamefont{Hall}},
  \bibinfo{year}{1996}, \bibinfo{journal}{Phys. Rev. Lett.}
  \textbf{\bibinfo{volume}{77}}, \bibinfo{pages}{1937}.

\bibitem[{\citenamefont{Arnesen} \emph{et~al.}(2006)\citenamefont{Arnesen,
  Ligeti, Rothstein, and Stewart}}]{Arnesen:2006vb}
\bibinfo{author}{\bibnamefont{Arnesen}, \bibfnamefont{C.~M.}},
  \bibinfo{author}{\bibfnamefont{Z.}~\bibnamefont{Ligeti}},
  \bibinfo{author}{\bibfnamefont{I.~Z.} \bibnamefont{Rothstein}}, and
  \bibinfo{author}{\bibfnamefont{I.~W.} \bibnamefont{Stewart}},
  \bibinfo{year}{2006}, \eprint{hep-ph/0607001}.

\bibitem[{\citenamefont{Arnesen} \emph{et~al.}(2005)\citenamefont{Arnesen,
  Grinstein, Rothstein, and Stewart}}]{Arnesen:2005ez}
\bibinfo{author}{\bibnamefont{Arnesen}, \bibfnamefont{M.~C.}},
  \bibinfo{author}{\bibfnamefont{B.}~\bibnamefont{Grinstein}},
  \bibinfo{author}{\bibfnamefont{I.~Z.} \bibnamefont{Rothstein}}, and
  \bibinfo{author}{\bibfnamefont{I.~W.} \bibnamefont{Stewart}},
  \bibinfo{year}{2005}, \bibinfo{journal}{Phys. Rev. Lett.}
  \textbf{\bibinfo{volume}{95}}, \bibinfo{pages}{071802}.

\bibitem[{\citenamefont{Artuso} \emph{et~al.}(2008)\citenamefont{Artuso,
  Meadows, and Petrov}}]{Artuso:2008vf}
\bibinfo{author}{\bibnamefont{Artuso}, \bibfnamefont{M.}},
  \bibinfo{author}{\bibfnamefont{B.}~\bibnamefont{Meadows}}, and
  \bibinfo{author}{\bibfnamefont{A.~A.} \bibnamefont{Petrov}},
  \bibinfo{year}{2008}, \eprint{arXiv:0802.2934 [hep-ph]}.

\bibitem[{\citenamefont{Asatrian} \emph{et~al.}(2002)\citenamefont{Asatrian,
  Bieri, Greub, and Hovhannisyan}}]{Asatrian:2002va}
\bibinfo{author}{\bibnamefont{Asatrian}, \bibfnamefont{H.~M.}},
  \bibinfo{author}{\bibfnamefont{K.}~\bibnamefont{Bieri}},
  \bibinfo{author}{\bibfnamefont{C.}~\bibnamefont{Greub}}, and
  \bibinfo{author}{\bibfnamefont{A.}~\bibnamefont{Hovhannisyan}},
  \bibinfo{year}{2002}, \bibinfo{journal}{Phys. Rev.}
  \textbf{\bibinfo{volume}{D66}}, \bibinfo{pages}{094013}.

\bibitem[{\citenamefont{Asatryan} \emph{et~al.}(2002)\citenamefont{Asatryan,
  Asatrian, Greub, and Walker}}]{Asatryan:2001zw}
\bibinfo{author}{\bibnamefont{Asatryan}, \bibfnamefont{H.~H.}},
  \bibinfo{author}{\bibfnamefont{H.~M.} \bibnamefont{Asatrian}},
  \bibinfo{author}{\bibfnamefont{C.}~\bibnamefont{Greub}}, and
  \bibinfo{author}{\bibfnamefont{M.}~\bibnamefont{Walker}},
  \bibinfo{year}{2002}, \bibinfo{journal}{Phys. Rev.}
  \textbf{\bibinfo{volume}{D65}}, \bibinfo{pages}{074004}.

\bibitem[{Athar \emph{et~al.}(2003)\citenamefont{Athar}
  \emph{et~al.}}]{Athar:2003yg}
\bibinfo{author}{\bibnamefont{Athar}, \bibfnamefont{S.~B.}}, \emph{et~al.}
  (\bibinfo{collaboration}{CLEO}), \bibinfo{year}{2003},
  \bibinfo{journal}{Phys. Rev.} \textbf{\bibinfo{volume}{D68}},
  \bibinfo{pages}{072003}.

\bibitem[{\citenamefont{Atwood}
  \emph{et~al.}(2001{\natexlab{a}})\citenamefont{Atwood, Bar-Shalom, Eilam, and
  Soni}}]{Atwood:2000tu}
\bibinfo{author}{\bibnamefont{Atwood}, \bibfnamefont{D.}},
  \bibinfo{author}{\bibfnamefont{S.}~\bibnamefont{Bar-Shalom}},
  \bibinfo{author}{\bibfnamefont{G.}~\bibnamefont{Eilam}}, and
  \bibinfo{author}{\bibfnamefont{A.}~\bibnamefont{Soni}},
  \bibinfo{year}{2001}{\natexlab{a}}, \bibinfo{journal}{Phys. Rept.}
  \textbf{\bibinfo{volume}{347}}, \bibinfo{pages}{1}.

\bibitem[{\citenamefont{Atwood} \emph{et~al.}(1996)\citenamefont{Atwood, Blok,
  and Soni}}]{Atwood:1994rw}
\bibinfo{author}{\bibnamefont{Atwood}, \bibfnamefont{D.}},
  \bibinfo{author}{\bibfnamefont{B.}~\bibnamefont{Blok}}, and
  \bibinfo{author}{\bibfnamefont{A.}~\bibnamefont{Soni}}, \bibinfo{year}{1996},
  \bibinfo{journal}{Int. J. Mod. Phys.} \textbf{\bibinfo{volume}{A11}},
  \bibinfo{pages}{3743}.

\bibitem[{\citenamefont{Atwood}
  \emph{et~al.}(1997{\natexlab{a}})\citenamefont{Atwood, Dunietz, and
  Soni}}]{Atwood:1996ci}
\bibinfo{author}{\bibnamefont{Atwood}, \bibfnamefont{D.}},
  \bibinfo{author}{\bibfnamefont{I.}~\bibnamefont{Dunietz}}, and
  \bibinfo{author}{\bibfnamefont{A.}~\bibnamefont{Soni}},
  \bibinfo{year}{1997}{\natexlab{a}}, \bibinfo{journal}{Phys. Rev. Lett.}
  \textbf{\bibinfo{volume}{78}}, \bibinfo{pages}{3257}.

\bibitem[{\citenamefont{Atwood}
  \emph{et~al.}(2001{\natexlab{b}})\citenamefont{Atwood, Dunietz, and
  Soni}}]{Atwood:2000ck}
\bibinfo{author}{\bibnamefont{Atwood}, \bibfnamefont{D.}},
  \bibinfo{author}{\bibfnamefont{I.}~\bibnamefont{Dunietz}}, and
  \bibinfo{author}{\bibfnamefont{A.}~\bibnamefont{Soni}},
  \bibinfo{year}{2001}{\natexlab{b}}, \bibinfo{journal}{Phys. Rev.}
  \textbf{\bibinfo{volume}{D63}}, \bibinfo{pages}{036005}.

\bibitem[{\citenamefont{Atwood} \emph{et~al.}(1993)\citenamefont{Atwood, Eilam,
  and Soni}}]{Atwood:1993ka}
\bibinfo{author}{\bibnamefont{Atwood}, \bibfnamefont{D.}},
  \bibinfo{author}{\bibfnamefont{G.}~\bibnamefont{Eilam}}, and
  \bibinfo{author}{\bibfnamefont{A.}~\bibnamefont{Soni}}, \bibinfo{year}{1993},
  \bibinfo{journal}{Phys. Rev. Lett.} \textbf{\bibinfo{volume}{71}},
  \bibinfo{pages}{492}.

\bibitem[{\citenamefont{Atwood} \emph{et~al.}(2005)\citenamefont{Atwood,
  Gershon, Hazumi, and Soni}}]{Atwood:2004jj}
\bibinfo{author}{\bibnamefont{Atwood}, \bibfnamefont{D.}},
  \bibinfo{author}{\bibfnamefont{T.}~\bibnamefont{Gershon}},
  \bibinfo{author}{\bibfnamefont{M.}~\bibnamefont{Hazumi}}, and
  \bibinfo{author}{\bibfnamefont{A.}~\bibnamefont{Soni}}, \bibinfo{year}{2005},
  \bibinfo{journal}{Phys. Rev.} \textbf{\bibinfo{volume}{D71}},
  \bibinfo{pages}{076003}.

\bibitem[{\citenamefont{Atwood} \emph{et~al.}(2007)\citenamefont{Atwood,
  Gershon, Hazumi, and Soni}}]{Atwood:2007qh}
\bibinfo{author}{\bibnamefont{Atwood}, \bibfnamefont{D.}},
  \bibinfo{author}{\bibfnamefont{T.}~\bibnamefont{Gershon}},
  \bibinfo{author}{\bibfnamefont{M.}~\bibnamefont{Hazumi}}, and
  \bibinfo{author}{\bibfnamefont{A.}~\bibnamefont{Soni}}, \bibinfo{year}{2007},
  \eprint{hep-ph/0701021}.

\bibitem[{\citenamefont{Atwood}
  \emph{et~al.}(1997{\natexlab{b}})\citenamefont{Atwood, Gronau, and
  Soni}}]{Atwood:1997zr}
\bibinfo{author}{\bibnamefont{Atwood}, \bibfnamefont{D.}},
  \bibinfo{author}{\bibfnamefont{M.}~\bibnamefont{Gronau}}, and
  \bibinfo{author}{\bibfnamefont{A.}~\bibnamefont{Soni}},
  \bibinfo{year}{1997}{\natexlab{b}}, \bibinfo{journal}{Phys. Rev. Lett.}
  \textbf{\bibinfo{volume}{79}}, \bibinfo{pages}{185}.

\bibitem[{\citenamefont{Atwood and Petrov}(2005)}]{Atwood:2002ak}
\bibinfo{author}{\bibnamefont{Atwood}, \bibfnamefont{D.}}, and
  \bibinfo{author}{\bibfnamefont{A.~A.} \bibnamefont{Petrov}},
  \bibinfo{year}{2005}, \bibinfo{journal}{Phys. Rev.}
  \textbf{\bibinfo{volume}{D71}}, \bibinfo{pages}{054032}.

\bibitem[{\citenamefont{Atwood}
  \emph{et~al.}(1997{\natexlab{c}})\citenamefont{Atwood, Reina, and
  Soni}}]{Atwood:1996vj}
\bibinfo{author}{\bibnamefont{Atwood}, \bibfnamefont{D.}},
  \bibinfo{author}{\bibfnamefont{L.}~\bibnamefont{Reina}}, and
  \bibinfo{author}{\bibfnamefont{A.}~\bibnamefont{Soni}},
  \bibinfo{year}{1997}{\natexlab{c}}, \bibinfo{journal}{Phys. Rev.}
  \textbf{\bibinfo{volume}{D55}}, \bibinfo{pages}{3156}.

\bibitem[{\citenamefont{Atwood and Soni}(1997)}]{Atwood:1997de}
\bibinfo{author}{\bibnamefont{Atwood}, \bibfnamefont{D.}}, and
  \bibinfo{author}{\bibfnamefont{A.}~\bibnamefont{Soni}}, \bibinfo{year}{1997},
  \bibinfo{journal}{Phys. Rev. Lett.} \textbf{\bibinfo{volume}{79}},
  \bibinfo{pages}{5206}.

\bibitem[{\citenamefont{Atwood and Soni}(1998{\natexlab{a}})}]{Atwood:1997iw}
\bibinfo{author}{\bibnamefont{Atwood}, \bibfnamefont{D.}}, and
  \bibinfo{author}{\bibfnamefont{A.}~\bibnamefont{Soni}},
  \bibinfo{year}{1998}{\natexlab{a}}, \bibinfo{journal}{Phys. Rev.}
  \textbf{\bibinfo{volume}{D58}}, \bibinfo{pages}{036005}.

\bibitem[{\citenamefont{Atwood and Soni}(1998{\natexlab{b}})}]{Atwood:1998ib}
\bibinfo{author}{\bibnamefont{Atwood}, \bibfnamefont{D.}}, and
  \bibinfo{author}{\bibfnamefont{A.}~\bibnamefont{Soni}},
  \bibinfo{year}{1998}{\natexlab{b}}, \bibinfo{journal}{Phys. Rev. Lett.}
  \textbf{\bibinfo{volume}{81}}, \bibinfo{pages}{3324}.

\bibitem[{\citenamefont{Atwood and Soni}(2002)}]{Atwood:2001js}
\bibinfo{author}{\bibnamefont{Atwood}, \bibfnamefont{D.}}, and
  \bibinfo{author}{\bibfnamefont{A.}~\bibnamefont{Soni}}, \bibinfo{year}{2002},
  \bibinfo{journal}{Phys. Lett.} \textbf{\bibinfo{volume}{B533}},
  \bibinfo{pages}{37}.

\bibitem[{\citenamefont{Atwood and Soni}(2003{\natexlab{a}})}]{Atwood:2002vw}
\bibinfo{author}{\bibnamefont{Atwood}, \bibfnamefont{D.}}, and
  \bibinfo{author}{\bibfnamefont{A.}~\bibnamefont{Soni}},
  \bibinfo{year}{2003}{\natexlab{a}}, \bibinfo{journal}{Phys. Rev.}
  \textbf{\bibinfo{volume}{D68}}, \bibinfo{pages}{033009}.

\bibitem[{\citenamefont{Atwood and Soni}(2003{\natexlab{b}})}]{Atwood:2003mj}
\bibinfo{author}{\bibnamefont{Atwood}, \bibfnamefont{D.}}, and
  \bibinfo{author}{\bibfnamefont{A.}~\bibnamefont{Soni}},
  \bibinfo{year}{2003}{\natexlab{b}}, \bibinfo{journal}{Phys. Rev.}
  \textbf{\bibinfo{volume}{D68}}, \bibinfo{pages}{033003}.

\bibitem[{\citenamefont{Atwood and Soni}(2005)}]{Atwood:2003jb}
\bibinfo{author}{\bibnamefont{Atwood}, \bibfnamefont{D.}}, and
  \bibinfo{author}{\bibfnamefont{A.}~\bibnamefont{Soni}}, \bibinfo{year}{2005},
  \bibinfo{journal}{Phys. Rev.} \textbf{\bibinfo{volume}{D71}},
  \bibinfo{pages}{013007}.

\bibitem[{Aubert \emph{et~al.}(2002)\citenamefont{Aubert}
  \emph{et~al.}}]{Aubert:2001tu}
\bibinfo{author}{\bibnamefont{Aubert}, \bibfnamefont{B.}}, \emph{et~al.}
  (\bibinfo{collaboration}{\babar}), \bibinfo{year}{2002},
  \bibinfo{journal}{Nucl. Instrum. Meth.} \textbf{\bibinfo{volume}{A479}},
  \bibinfo{pages}{1}.

\bibitem[{Aubert \emph{et~al.}(2003)\citenamefont{Aubert}
  \emph{et~al.}}]{Aubert:2003xz}
\bibinfo{author}{\bibnamefont{Aubert}, \bibfnamefont{B.}}, \emph{et~al.}
  (\bibinfo{collaboration}{\babar}), \bibinfo{year}{2003},
  \bibinfo{journal}{Phys. Rev. Lett.} \textbf{\bibinfo{volume}{91}},
  \bibinfo{pages}{051801}.

\bibitem[{Aubert \emph{et~al.}(2004{\natexlab{a}})\citenamefont{Aubert}
  \emph{et~al.}}]{Aubert:2004te}
\bibinfo{author}{\bibnamefont{Aubert}, \bibfnamefont{B.}}, \emph{et~al.}
  (\bibinfo{collaboration}{\babar}), \bibinfo{year}{2004}{\natexlab{a}},
  \bibinfo{journal}{Phys. Rev.} \textbf{\bibinfo{volume}{D70}},
  \bibinfo{pages}{112006}.

\bibitem[{Aubert \emph{et~al.}(2004{\natexlab{b}})\citenamefont{Aubert}
  \emph{et~al.}}]{Aubert:2004it}
\bibinfo{author}{\bibnamefont{Aubert}, \bibfnamefont{B.}}, \emph{et~al.}
  (\bibinfo{collaboration}{\babar}), \bibinfo{year}{2004}{\natexlab{b}},
  \bibinfo{journal}{Phys. Rev. Lett.} \textbf{\bibinfo{volume}{93}},
  \bibinfo{pages}{081802}.

\bibitem[{Aubert \emph{et~al.}(2004{\natexlab{c}})\citenamefont{Aubert}
  \emph{et~al.}}]{Aubert:2004hq}
\bibinfo{author}{\bibnamefont{Aubert}, \bibfnamefont{B.}}, \emph{et~al.}
  (\bibinfo{collaboration}{\babar}), \bibinfo{year}{2004}{\natexlab{c}},
  \bibinfo{journal}{Phys. Rev. Lett.} \textbf{\bibinfo{volume}{93}},
  \bibinfo{pages}{021804}.

\bibitem[{Aubert \emph{et~al.}(2005{\natexlab{a}})\citenamefont{Aubert}
  \emph{et~al.}}]{Aubert:2004bw}
\bibinfo{author}{\bibnamefont{Aubert}, \bibfnamefont{B.}}, \emph{et~al.}
  (\bibinfo{collaboration}{\babar}), \bibinfo{year}{2005}{\natexlab{a}},
  \bibinfo{journal}{Phys. Rev.} \textbf{\bibinfo{volume}{D71}},
  \bibinfo{pages}{051502}.

\bibitem[{Aubert \emph{et~al.}(2005{\natexlab{b}})\citenamefont{Aubert}
  \emph{et~al.}}]{Aubert:2005cua}
\bibinfo{author}{\bibnamefont{Aubert}, \bibfnamefont{B.}}, \emph{et~al.}
  (\bibinfo{collaboration}{\babar}), \bibinfo{year}{2005}{\natexlab{b}},
  \bibinfo{journal}{Phys. Rev.} \textbf{\bibinfo{volume}{D72}},
  \bibinfo{pages}{052004}.

\bibitem[{Aubert \emph{et~al.}(2005{\natexlab{c}})\citenamefont{Aubert}
  \emph{et~al.}}]{Aubert:2005ye}
\bibinfo{author}{\bibnamefont{Aubert}, \bibfnamefont{B.}}, \emph{et~al.}
  (\bibinfo{collaboration}{\babar}), \bibinfo{year}{2005}{\natexlab{c}},
  \bibinfo{journal}{Phys. Rev. Lett.} \textbf{\bibinfo{volume}{95}},
  \bibinfo{pages}{041802}.

\bibitem[{Aubert \emph{et~al.}(2005{\natexlab{d}})\citenamefont{Aubert}
  \emph{et~al.}}]{Aubert:2004ws}
\bibinfo{author}{\bibnamefont{Aubert}, \bibfnamefont{B.}}, \emph{et~al.}
  (\bibinfo{collaboration}{\babar}), \bibinfo{year}{2005}{\natexlab{d}},
  \bibinfo{journal}{Phys. Rev. Lett.} \textbf{\bibinfo{volume}{94}},
  \bibinfo{pages}{101801}.

\bibitem[{Aubert \emph{et~al.}(2006{\natexlab{a}})\citenamefont{Aubert}
  \emph{et~al.}}]{Aubert:2006qi}
\bibinfo{author}{\bibnamefont{Aubert}, \bibfnamefont{B.}}, \emph{et~al.}
  (\bibinfo{collaboration}{\babar}), \bibinfo{year}{2006}{\natexlab{a}},
  \bibinfo{journal}{Phys. Rev. Lett.} \textbf{\bibinfo{volume}{96}},
  \bibinfo{pages}{221801}.

\bibitem[{Aubert \emph{et~al.}(2006{\natexlab{b}})\citenamefont{Aubert}
  \emph{et~al.}}]{Aubert:2006gg}
\bibinfo{author}{\bibnamefont{Aubert}, \bibfnamefont{B.}}, \emph{et~al.}
  (\bibinfo{collaboration}{\babar}), \bibinfo{year}{2006}{\natexlab{b}},
  \bibinfo{journal}{Phys. Rev. Lett.} \textbf{\bibinfo{volume}{97}},
  \bibinfo{pages}{171803}.

\bibitem[{Aubert \emph{et~al.}(2006{\natexlab{c}})\citenamefont{Aubert}
  \emph{et~al.}}]{Aubert:2006am}
\bibinfo{author}{\bibnamefont{Aubert}, \bibfnamefont{B.}}, \emph{et~al.}
  (\bibinfo{collaboration}{\babar}), \bibinfo{year}{2006}{\natexlab{c}},
  \eprint{hep-ex/0607104}.

\bibitem[{Aubert \emph{et~al.}(2006{\natexlab{d}})\citenamefont{Aubert}
  \emph{et~al.}}]{Aubert:2005kf}
\bibinfo{author}{\bibnamefont{Aubert}, \bibfnamefont{B.}}, \emph{et~al.}
  (\bibinfo{collaboration}{\babar}), \bibinfo{year}{2006}{\natexlab{d}},
  \bibinfo{journal}{Phys. Rev.} \textbf{\bibinfo{volume}{D73}},
  \bibinfo{pages}{012004}.

\bibitem[{Aubert \emph{et~al.}(2006{\natexlab{e}})\citenamefont{Aubert}
  \emph{et~al.}}]{Aubert:2006vb}
\bibinfo{author}{\bibnamefont{Aubert}, \bibfnamefont{B.}}, \emph{et~al.}
  (\bibinfo{collaboration}{\babar}), \bibinfo{year}{2006}{\natexlab{e}},
  \bibinfo{journal}{Phys. Rev.} \textbf{\bibinfo{volume}{D73}},
  \bibinfo{pages}{092001}.

\bibitem[{Aubert \emph{et~al.}(2006{\natexlab{f}})\citenamefont{Aubert}
  \emph{et~al.}}]{Aubert:2006ai}
\bibinfo{author}{\bibnamefont{Aubert}, \bibfnamefont{B.}}, \emph{et~al.}
  (\bibinfo{collaboration}{\babar}), \bibinfo{year}{2006}{\natexlab{f}},
  \eprint{hep-ex/0607101}.

\bibitem[{Aubert \emph{et~al.}(2006{\natexlab{g}})\citenamefont{Aubert}
  \emph{et~al.}}]{Aubert:2006cx}
\bibinfo{author}{\bibnamefont{Aubert}, \bibfnamefont{B.}}, \emph{et~al.}
  (\bibinfo{collaboration}{\babar}), \bibinfo{year}{2006}{\natexlab{g}},
  \bibinfo{journal}{Phys. Rev.} \textbf{\bibinfo{volume}{D74}},
  \bibinfo{pages}{092004}.

\bibitem[{Aubert \emph{et~al.}(2006{\natexlab{h}})\citenamefont{Aubert}
  \emph{et~al.}}]{Aubert:2006gm}
\bibinfo{author}{\bibnamefont{Aubert}, \bibfnamefont{B.}}, \emph{et~al.}
  (\bibinfo{collaboration}{\babar}), \bibinfo{year}{2006}{\natexlab{h}},
  \bibinfo{journal}{Phys. Rev. Lett.} \textbf{\bibinfo{volume}{97}},
  \bibinfo{pages}{171805}.

\bibitem[{Aubert \emph{et~al.}(2006{\natexlab{i}})\citenamefont{Aubert}
  \emph{et~al.}}]{Aubert:2006ana}
\bibinfo{author}{\bibnamefont{Aubert}, \bibfnamefont{B.}}, \emph{et~al.}
  (\bibinfo{collaboration}{\babar}), \bibinfo{year}{2006}{\natexlab{i}},
  \eprint{hep-ex/0607053}.

\bibitem[{Aubert \emph{et~al.}(2006{\natexlab{j}})\citenamefont{Aubert}
  \emph{et~al.}}]{Aubert:2005wa}
\bibinfo{author}{\bibnamefont{Aubert}, \bibfnamefont{B.}}, \emph{et~al.}
  (\bibinfo{collaboration}{\babar}), \bibinfo{year}{2006}{\natexlab{j}},
  \bibinfo{journal}{Phys. Rev. Lett.} \textbf{\bibinfo{volume}{96}},
  \bibinfo{pages}{041801}.

\bibitem[{Aubert \emph{et~al.}(2007{\natexlab{a}})\citenamefont{Aubert}
  \emph{et~al.}}]{Aubert:2007bx}
\bibinfo{author}{\bibnamefont{Aubert}, \bibfnamefont{B.}}, \emph{et~al.}
  (\bibinfo{collaboration}{\babar}), \bibinfo{year}{2007}{\natexlab{a}},
  \bibinfo{journal}{Phys. Rev.} \textbf{\bibinfo{volume}{D76}},
  \bibinfo{pages}{052002}.

\bibitem[{Aubert \emph{et~al.}(2007{\natexlab{b}})\citenamefont{Aubert}
  \emph{et~al.}}]{Aubert:2007si}
\bibinfo{author}{\bibnamefont{Aubert}, \bibfnamefont{B.}}, \emph{et~al.}
  (\bibinfo{collaboration}{\babar}), \bibinfo{year}{2007}{\natexlab{b}},
  \bibinfo{journal}{Phys. Rev.} \textbf{\bibinfo{volume}{D76}},
  \bibinfo{pages}{031103}.

\bibitem[{Aubert \emph{et~al.}(2007{\natexlab{c}})\citenamefont{Aubert}
  \emph{et~al.}}]{Aubert:2006pu}
\bibinfo{author}{\bibnamefont{Aubert}, \bibfnamefont{B.}}, \emph{et~al.}
  (\bibinfo{collaboration}{\babar}), \bibinfo{year}{2007}{\natexlab{c}},
  \bibinfo{journal}{Phys. Rev. Lett.} \textbf{\bibinfo{volume}{98}},
  \bibinfo{pages}{151802}.

\bibitem[{Aubert \emph{et~al.}(2007{\natexlab{d}})\citenamefont{Aubert}
  \emph{et~al.}}]{Aubert:2007wf}
\bibinfo{author}{\bibnamefont{Aubert}, \bibfnamefont{B.}}, \emph{et~al.}
  (\bibinfo{collaboration}{\babar}), \bibinfo{year}{2007}{\natexlab{d}},
  \bibinfo{journal}{Phys. Rev. Lett.} \textbf{\bibinfo{volume}{98}},
  \bibinfo{pages}{211802}.

\bibitem[{Aubert \emph{et~al.}(2007{\natexlab{e}})\citenamefont{Aubert}
  \emph{et~al.}}]{Aubert:2007pw}
\bibinfo{author}{\bibnamefont{Aubert}, \bibfnamefont{B.}}, \emph{et~al.}
  (\bibinfo{collaboration}{\babar}), \bibinfo{year}{2007}{\natexlab{e}},
  \bibinfo{journal}{Phys. Rev. Lett.} \textbf{\bibinfo{volume}{99}},
  \bibinfo{pages}{251803}.

\bibitem[{Aubert \emph{et~al.}(2007{\natexlab{f}})\citenamefont{Aubert}
  \emph{et~al.}}]{Aubert:2007hm}
\bibinfo{author}{\bibnamefont{Aubert}, \bibfnamefont{B.}}, \emph{et~al.}
  (\bibinfo{collaboration}{\babar}), \bibinfo{year}{2007}{\natexlab{f}},
  \bibinfo{journal}{Phys. Rev. Lett.} \textbf{\bibinfo{volume}{99}},
  \bibinfo{pages}{171803}.

\bibitem[{Aubert \emph{et~al.}(2007{\natexlab{g}})\citenamefont{Aubert}
  \emph{et~al.}}]{Aubert:2006fha}
\bibinfo{author}{\bibnamefont{Aubert}, \bibfnamefont{B.}}, \emph{et~al.}
  (\bibinfo{collaboration}{\babar}), \bibinfo{year}{2007}{\natexlab{g}},
  \bibinfo{journal}{Phys. Rev.} \textbf{\bibinfo{volume}{D75}},
  \bibinfo{pages}{012008}.

\bibitem[{Aubert \emph{et~al.}(2007{\natexlab{h}})\citenamefont{Aubert}
  \emph{et~al.}}]{Aubert:2005xk}
\bibinfo{author}{\bibnamefont{Aubert}, \bibfnamefont{B.}}, \emph{et~al.}
  (\bibinfo{collaboration}{\babar}), \bibinfo{year}{2007}{\natexlab{h}},
  \bibinfo{journal}{Phys. Rev. Lett.} \textbf{\bibinfo{volume}{98}},
  \bibinfo{pages}{211804}.

\bibitem[{Aubert \emph{et~al.}(2007{\natexlab{i}})\citenamefont{Aubert}
  \emph{et~al.}}]{Aubert:2007rp}
\bibinfo{author}{\bibnamefont{Aubert}, \bibfnamefont{B.}}, \emph{et~al.}
  (\bibinfo{collaboration}{\babar}), \bibinfo{year}{2007}{\natexlab{i}},
  \bibinfo{journal}{Phys. Rev. Lett.} \textbf{\bibinfo{volume}{99}},
  \bibinfo{pages}{231802}.

\bibitem[{Aubert \emph{et~al.}(2007{\natexlab{j}})\citenamefont{Aubert}
  \emph{et~al.}}]{Aubert:2007me}
\bibinfo{author}{\bibnamefont{Aubert}, \bibfnamefont{B.}}, \emph{et~al.}
  (\bibinfo{collaboration}{\babar}), \bibinfo{year}{2007}{\natexlab{j}},
  \bibinfo{journal}{Phys. Rev.} \textbf{\bibinfo{volume}{D76}},
  \bibinfo{pages}{091101}.

\bibitem[{Aubert \emph{et~al.}(2007{\natexlab{k}})\citenamefont{Aubert}
  \emph{et~al.}}]{Aubert:2007ub}
\bibinfo{author}{\bibnamefont{Aubert}, \bibfnamefont{B.}}, \emph{et~al.}
  (\bibinfo{collaboration}{\babar}), \bibinfo{year}{2007}{\natexlab{k}},
  \bibinfo{journal}{Phys. Rev.} \textbf{\bibinfo{volume}{D76}},
  \bibinfo{pages}{071101}.

\bibitem[{Aubert \emph{et~al.}(2007{\natexlab{l}})\citenamefont{Aubert}
  \emph{et~al.}}]{Aubert:2007jn}
\bibinfo{author}{\bibnamefont{Aubert}, \bibfnamefont{B.}}, \emph{et~al.}
  (\bibinfo{collaboration}{\babar}), \bibinfo{year}{2007}{\natexlab{l}},
  \bibinfo{journal}{Phys. Rev.} \textbf{\bibinfo{volume}{D76}},
  \bibinfo{pages}{012004}.

\bibitem[{Aubert \emph{et~al.}(2007{\natexlab{m}})\citenamefont{Aubert}
  \emph{et~al.}}]{Aubert:2006px}
\bibinfo{author}{\bibnamefont{Aubert}, \bibfnamefont{B.}}, \emph{et~al.}
  (\bibinfo{collaboration}{\babar}), \bibinfo{year}{2007}{\natexlab{m}},
  \bibinfo{journal}{Phys. Rev. Lett.} \textbf{\bibinfo{volume}{98}},
  \bibinfo{pages}{091801}.

\bibitem[{Aubert \emph{et~al.}(2007{\natexlab{n}})\citenamefont{Aubert}
  \emph{et~al.}}]{Aubert:2007qj}
\bibinfo{author}{\bibnamefont{Aubert}, \bibfnamefont{B.}}, \emph{et~al.}
  (\bibinfo{collaboration}{\babar}), \bibinfo{year}{2007}{\natexlab{n}},
  \eprint{arXiv:0708.1614 [hep-ex]}.

\bibitem[{Aubert \emph{et~al.}(2007{\natexlab{o}})\citenamefont{Aubert}
  \emph{et~al.}}]{Aubert:2007sd}
\bibinfo{author}{\bibnamefont{Aubert}, \bibfnamefont{B.}}, \emph{et~al.}
  (\bibinfo{collaboration}{\babar}), \bibinfo{year}{2007}{\natexlab{o}},
  \bibinfo{journal}{Phys. Rev. Lett.} \textbf{\bibinfo{volume}{99}},
  \bibinfo{pages}{161802}.

\bibitem[{Aubert \emph{et~al.}(2007{\natexlab{p}})\citenamefont{Aubert}
  \emph{et~al.}}]{Aubert:2007rb}
\bibinfo{author}{\bibnamefont{Aubert}, \bibfnamefont{B.}}, \emph{et~al.}
  (\bibinfo{collaboration}{\babar}), \bibinfo{year}{2007}{\natexlab{p}},
  \eprint{arXiv:0708.3702 [hep-ex]}.

\bibitem[{Aubert \emph{et~al.}(2007{\natexlab{q}})\citenamefont{Aubert}
  \emph{et~al.}}]{Aubert:2006wv}
\bibinfo{author}{\bibnamefont{Aubert}, \bibfnamefont{B.}}, \emph{et~al.}
  (\bibinfo{collaboration}{\babar}), \bibinfo{year}{2007}{\natexlab{q}},
  \bibinfo{journal}{Phys. Rev. Lett.} \textbf{\bibinfo{volume}{98}},
  \bibinfo{pages}{031801}.

\bibitem[{Aubert \emph{et~al.}(2007{\natexlab{r}})\citenamefont{Aubert}
  \emph{et~al.}}]{Aubert:2007mj}
\bibinfo{author}{\bibnamefont{Aubert}, \bibfnamefont{B.}}, \emph{et~al.}
  (\bibinfo{collaboration}{\babar}), \bibinfo{year}{2007}{\natexlab{r}},
  \bibinfo{journal}{Phys. Rev. Lett.} \textbf{\bibinfo{volume}{99}},
  \bibinfo{pages}{021603}.

\bibitem[{Aubert \emph{et~al.}(2007{\natexlab{s}})\citenamefont{Aubert}
  \emph{et~al.}}]{Aubert:2007ds}
\bibinfo{author}{\bibnamefont{Aubert}, \bibfnamefont{B.}}, \emph{et~al.}
  (\bibinfo{collaboration}{\babar}), \bibinfo{year}{2007}{\natexlab{s}},
  \eprint{arXiv:0709.1698 [hep-ex]}.

\bibitem[{Aubert \emph{et~al.}(2007{\natexlab{t}})\citenamefont{Aubert}
  \emph{et~al.}}]{Aubert:2006cz}
\bibinfo{author}{\bibnamefont{Aubert}, \bibfnamefont{B.}}, \emph{et~al.}
  (\bibinfo{collaboration}{\babar}), \bibinfo{year}{2007}{\natexlab{t}},
  \bibinfo{journal}{Phys. Rev. Lett.} \textbf{\bibinfo{volume}{98}},
  \bibinfo{pages}{061803}.

\bibitem[{Aubert \emph{et~al.}(2007{\natexlab{u}})\citenamefont{Aubert}
  \emph{et~al.}}]{Aubert:2007hh}
\bibinfo{author}{\bibnamefont{Aubert}, \bibfnamefont{B.}}, \emph{et~al.}
  (\bibinfo{collaboration}{\babar}), \bibinfo{year}{2007}{\natexlab{u}},
  \bibinfo{journal}{Phys. Rev.} \textbf{\bibinfo{volume}{D76}},
  \bibinfo{pages}{091102}.

\bibitem[{Aubert \emph{et~al.}(2007{\natexlab{v}})\citenamefont{Aubert}
  \emph{et~al.}}]{Aubert:2007qs}
\bibinfo{author}{\bibnamefont{Aubert}, \bibfnamefont{B.}}, \emph{et~al.}
  (\bibinfo{collaboration}{\babar}), \bibinfo{year}{2007}{\natexlab{v}},
  \eprint{arXiv:0708.1630 [hep-ex]}.

\bibitem[{Aubert \emph{et~al.}(2007{\natexlab{w}})\citenamefont{Aubert}
  \emph{et~al.}}]{Aubert:2007vi}
\bibinfo{author}{\bibnamefont{Aubert}, \bibfnamefont{B.}}, \emph{et~al.}
  (\bibinfo{collaboration}{\babar}), \bibinfo{year}{2007}{\natexlab{w}},
  \eprint{arXiv:0708.2097 [hep-ex]}.

\bibitem[{Aubert \emph{et~al.}(2008{\natexlab{a}})\citenamefont{Aubert}
  \emph{et~al.}}]{Aubert:2007xj}
\bibinfo{author}{\bibnamefont{Aubert}, \bibfnamefont{B.}}, \emph{et~al.}
  (\bibinfo{collaboration}{\babar}), \bibinfo{year}{2008}{\natexlab{a}},
  \bibinfo{journal}{Phys. Rev.} \textbf{\bibinfo{volume}{D77}},
  \bibinfo{pages}{011107}.

\bibitem[{Aubert \emph{et~al.}(2008{\natexlab{b}})\citenamefont{Aubert}
  \emph{et~al.}}]{Aubert:2007mgb}
\bibinfo{author}{\bibnamefont{Aubert}, \bibfnamefont{B.}}, \emph{et~al.}
  (\bibinfo{collaboration}{\babar}), \bibinfo{year}{2008}{\natexlab{b}},
  \bibinfo{journal}{Phys. Rev.} \textbf{\bibinfo{volume}{D77}},
  \bibinfo{pages}{012003}.

\bibitem[{\citenamefont{Babu and Kolda}(2002)}]{Babu:2002et}
\bibinfo{author}{\bibnamefont{Babu}, \bibfnamefont{K.~S.}}, and
  \bibinfo{author}{\bibfnamefont{C.}~\bibnamefont{Kolda}},
  \bibinfo{year}{2002}, \bibinfo{journal}{Phys. Rev. Lett.}
  \textbf{\bibinfo{volume}{89}}, \bibinfo{pages}{241802}.

\bibitem[{\citenamefont{Babu and Kolda}(2000)}]{Babu:1999hn}
\bibinfo{author}{\bibnamefont{Babu}, \bibfnamefont{K.~S.}}, and
  \bibinfo{author}{\bibfnamefont{C.~F.} \bibnamefont{Kolda}},
  \bibinfo{year}{2000}, \bibinfo{journal}{Phys. Rev. Lett.}
  \textbf{\bibinfo{volume}{84}}, \bibinfo{pages}{228}.

\bibitem[{\citenamefont{Baek} \emph{et~al.}(2005)\citenamefont{Baek, Hamel,
  London, Datta, and Suprun}}]{Baek:2004rp}
\bibinfo{author}{\bibnamefont{Baek}, \bibfnamefont{S.}},
  \bibinfo{author}{\bibfnamefont{P.}~\bibnamefont{Hamel}},
  \bibinfo{author}{\bibfnamefont{D.}~\bibnamefont{London}},
  \bibinfo{author}{\bibfnamefont{A.}~\bibnamefont{Datta}}, and
  \bibinfo{author}{\bibfnamefont{D.~A.} \bibnamefont{Suprun}},
  \bibinfo{year}{2005}, \bibinfo{journal}{Phys. Rev.}
  \textbf{\bibinfo{volume}{D71}}, \bibinfo{pages}{057502}.

\bibitem[{\citenamefont{Ball}(2006)}]{Ball:2006yj}
\bibinfo{author}{\bibnamefont{Ball}, \bibfnamefont{P.}}, \bibinfo{year}{2006},
  \eprint{hep-ph/0612190}.

\bibitem[{\citenamefont{Ball} \emph{et~al.}(2007)\citenamefont{Ball, Jones, and
  Zwicky}}]{Ball:2006eu}
\bibinfo{author}{\bibnamefont{Ball}, \bibfnamefont{P.}},
  \bibinfo{author}{\bibfnamefont{G.~W.} \bibnamefont{Jones}}, and
  \bibinfo{author}{\bibfnamefont{R.}~\bibnamefont{Zwicky}},
  \bibinfo{year}{2007}, \bibinfo{journal}{Phys. Rev.}
  \textbf{\bibinfo{volume}{D75}}, \bibinfo{pages}{054004}.

\bibitem[{\citenamefont{Ball and Zwicky}(2005)}]{Ball:2004rg}
\bibinfo{author}{\bibnamefont{Ball}, \bibfnamefont{P.}}, and
  \bibinfo{author}{\bibfnamefont{R.}~\bibnamefont{Zwicky}},
  \bibinfo{year}{2005}, \bibinfo{journal}{Phys. Rev.}
  \textbf{\bibinfo{volume}{D71}}, \bibinfo{pages}{014029}.

\bibitem[{\citenamefont{Ball and Zwicky}(2006{\natexlab{a}})}]{Ball:2006cva}
\bibinfo{author}{\bibnamefont{Ball}, \bibfnamefont{P.}}, and
  \bibinfo{author}{\bibfnamefont{R.}~\bibnamefont{Zwicky}},
  \bibinfo{year}{2006}{\natexlab{a}}, \bibinfo{journal}{Phys. Lett.}
  \textbf{\bibinfo{volume}{B642}}, \bibinfo{pages}{478}.

\bibitem[{\citenamefont{Ball and Zwicky}(2006{\natexlab{b}})}]{Ball:2006nr}
\bibinfo{author}{\bibnamefont{Ball}, \bibfnamefont{P.}}, and
  \bibinfo{author}{\bibfnamefont{R.}~\bibnamefont{Zwicky}},
  \bibinfo{year}{2006}{\natexlab{b}}, \bibinfo{journal}{JHEP}
  \textbf{\bibinfo{volume}{04}}, \bibinfo{pages}{046}.

\bibitem[{\citenamefont{Bander} \emph{et~al.}(1979)\citenamefont{Bander,
  Silverman, and Soni}}]{Bander:1979px}
\bibinfo{author}{\bibnamefont{Bander}, \bibfnamefont{M.}},
  \bibinfo{author}{\bibfnamefont{D.}~\bibnamefont{Silverman}}, and
  \bibinfo{author}{\bibfnamefont{A.}~\bibnamefont{Soni}}, \bibinfo{year}{1979},
  \bibinfo{journal}{Phys. Rev. Lett.} \textbf{\bibinfo{volume}{43}},
  \bibinfo{pages}{242}.

\bibitem[{\citenamefont{Bander} \emph{et~al.}(1980)\citenamefont{Bander,
  Silverman, and Soni}}]{Bander:1979jx}
\bibinfo{author}{\bibnamefont{Bander}, \bibfnamefont{M.}},
  \bibinfo{author}{\bibfnamefont{D.}~\bibnamefont{Silverman}}, and
  \bibinfo{author}{\bibfnamefont{A.}~\bibnamefont{Soni}}, \bibinfo{year}{1980},
  \bibinfo{journal}{Phys. Rev. Lett.} \textbf{\bibinfo{volume}{44}},
  \bibinfo{pages}{7}, \bibinfo{note}{[Erratum-ibid.\ {\bf 44}, 962 (1980)]}.

\bibitem[{\citenamefont{Banerjee} \emph{et~al.}(2007)\citenamefont{Banerjee,
  Pietrzyk, Roney, and Was}}]{Banerjee:2007is}
\bibinfo{author}{\bibnamefont{Banerjee}, \bibfnamefont{S.}},
  \bibinfo{author}{\bibfnamefont{B.}~\bibnamefont{Pietrzyk}},
  \bibinfo{author}{\bibfnamefont{J.~M.} \bibnamefont{Roney}}, and
  \bibinfo{author}{\bibfnamefont{Z.}~\bibnamefont{Was}}, \bibinfo{year}{2007},
  \eprint{arXiv:0706.3235 [hep-ph]}.

\bibitem[{Baracchini \emph{et~al.}(2007)\citenamefont{Baracchini}
  \emph{et~al.}}]{Baracchini:2007ei}
\bibinfo{author}{\bibnamefont{Baracchini}, \bibfnamefont{E.}}, \emph{et~al.},
  \bibinfo{year}{2007}, \bibinfo{journal}{JHEP} \textbf{\bibinfo{volume}{08}},
  \bibinfo{pages}{005}.

\bibitem[{\citenamefont{Baranowski and Misiak}(2000)}]{Baranowski:1999tq}
\bibinfo{author}{\bibnamefont{Baranowski}, \bibfnamefont{K.}}, and
  \bibinfo{author}{\bibfnamefont{M.}~\bibnamefont{Misiak}},
  \bibinfo{year}{2000}, \bibinfo{journal}{Phys. Lett.}
  \textbf{\bibinfo{volume}{B483}}, \bibinfo{pages}{410}.

\bibitem[{Barate \emph{et~al.}(2001)\citenamefont{Barate}
  \emph{et~al.}}]{Barate:2000rc}
\bibinfo{author}{\bibnamefont{Barate}, \bibfnamefont{R.}}, \emph{et~al.}
  (\bibinfo{collaboration}{ALEPH}), \bibinfo{year}{2001},
  \bibinfo{journal}{Eur. Phys. J.} \textbf{\bibinfo{volume}{C19}},
  \bibinfo{pages}{213}.

\bibitem[{Barberio \emph{et~al.}(2007)\citenamefont{Barberio}
  \emph{et~al.}}]{HFAG}
\bibinfo{author}{\bibnamefont{Barberio}, \bibfnamefont{E.}}, \emph{et~al.}
  (\bibinfo{collaboration}{Heavy Flavor Averaging Group (HFAG)}),
  \bibinfo{year}{2007}, \eprint{arXiv:0704.3575 [hep-ex]}.

\bibitem[{\citenamefont{Barbieri} \emph{et~al.}(1996)\citenamefont{Barbieri,
  Dvali, and Hall}}]{Barbieri:1995uv}
\bibinfo{author}{\bibnamefont{Barbieri}, \bibfnamefont{R.}},
  \bibinfo{author}{\bibfnamefont{G.~R.} \bibnamefont{Dvali}}, and
  \bibinfo{author}{\bibfnamefont{L.~J.} \bibnamefont{Hall}},
  \bibinfo{year}{1996}, \bibinfo{journal}{Phys. Lett.}
  \textbf{\bibinfo{volume}{B377}}, \bibinfo{pages}{76}.

\bibitem[{\citenamefont{Barbieri and Hall}(1994)}]{Barbieri:1994pv}
\bibinfo{author}{\bibnamefont{Barbieri}, \bibfnamefont{R.}}, and
  \bibinfo{author}{\bibfnamefont{L.~J.} \bibnamefont{Hall}},
  \bibinfo{year}{1994}, \bibinfo{journal}{Phys. Lett.}
  \textbf{\bibinfo{volume}{B338}}, \bibinfo{pages}{212}.

\bibitem[{\citenamefont{Barbieri} \emph{et~al.}(1995)\citenamefont{Barbieri,
  Hall, and Strumia}}]{Barbieri:1995tw}
\bibinfo{author}{\bibnamefont{Barbieri}, \bibfnamefont{R.}},
  \bibinfo{author}{\bibfnamefont{L.~J.} \bibnamefont{Hall}}, and
  \bibinfo{author}{\bibfnamefont{A.}~\bibnamefont{Strumia}},
  \bibinfo{year}{1995}, \bibinfo{journal}{Nucl. Phys.}
  \textbf{\bibinfo{volume}{B445}}, \bibinfo{pages}{219}.

\bibitem[{\citenamefont{Barenboim} \emph{et~al.}(2007)\citenamefont{Barenboim,
  Paradisi, Vives, Lunghi, and Porod}}]{Barenboim:2007sk}
\bibinfo{author}{\bibnamefont{Barenboim}, \bibfnamefont{G.}},
  \bibinfo{author}{\bibfnamefont{P.}~\bibnamefont{Paradisi}},
  \bibinfo{author}{\bibfnamefont{O.}~\bibnamefont{Vives}},
  \bibinfo{author}{\bibfnamefont{E.}~\bibnamefont{Lunghi}}, and
  \bibinfo{author}{\bibfnamefont{W.}~\bibnamefont{Porod}},
  \bibinfo{year}{2007}, \eprint{arXiv:0712.3559 [hep-ph]}.

\bibitem[{\citenamefont{Barger} \emph{et~al.}(2006)\citenamefont{Barger,
  Langacker, Lee, and Shaughnessy}}]{Barger:2006dh}
\bibinfo{author}{\bibnamefont{Barger}, \bibfnamefont{V.}},
  \bibinfo{author}{\bibfnamefont{P.}~\bibnamefont{Langacker}},
  \bibinfo{author}{\bibfnamefont{H.-S.} \bibnamefont{Lee}}, and
  \bibinfo{author}{\bibfnamefont{G.}~\bibnamefont{Shaughnessy}},
  \bibinfo{year}{2006}, \bibinfo{journal}{Phys. Rev.}
  \textbf{\bibinfo{volume}{D73}}, \bibinfo{pages}{115010}.

\bibitem[{Bartl \emph{et~al.}(2001)\citenamefont{Bartl}
  \emph{et~al.}}]{Bartl:2001wc}
\bibinfo{author}{\bibnamefont{Bartl}, \bibfnamefont{A.}}, \emph{et~al.},
  \bibinfo{year}{2001}, \bibinfo{journal}{Phys. Rev.}
  \textbf{\bibinfo{volume}{D64}}, \bibinfo{pages}{076009}.

\bibitem[{\citenamefont{Bauer} \emph{et~al.}(2000)\citenamefont{Bauer, Ligeti,
  and Luke}}]{Bauer:2000xf}
\bibinfo{author}{\bibnamefont{Bauer}, \bibfnamefont{C.~W.}},
  \bibinfo{author}{\bibfnamefont{Z.}~\bibnamefont{Ligeti}}, and
  \bibinfo{author}{\bibfnamefont{M.~E.} \bibnamefont{Luke}},
  \bibinfo{year}{2000}, \bibinfo{journal}{Phys. Lett.}
  \textbf{\bibinfo{volume}{B479}}, \bibinfo{pages}{395}.

\bibitem[{\citenamefont{Bauer} \emph{et~al.}(2001)\citenamefont{Bauer, Ligeti,
  and Luke}}]{Bauer:2001rc}
\bibinfo{author}{\bibnamefont{Bauer}, \bibfnamefont{C.~W.}},
  \bibinfo{author}{\bibfnamefont{Z.}~\bibnamefont{Ligeti}}, and
  \bibinfo{author}{\bibfnamefont{M.~E.} \bibnamefont{Luke}},
  \bibinfo{year}{2001}, \bibinfo{journal}{Phys. Rev.}
  \textbf{\bibinfo{volume}{D64}}, \bibinfo{pages}{113004}.

\bibitem[{\citenamefont{Bauer and Manohar}(2004)}]{Bauer:2003pi}
\bibinfo{author}{\bibnamefont{Bauer}, \bibfnamefont{C.~W.}}, and
  \bibinfo{author}{\bibfnamefont{A.~V.} \bibnamefont{Manohar}},
  \bibinfo{year}{2004}, \bibinfo{journal}{Phys. Rev.}
  \textbf{\bibinfo{volume}{D70}}, \bibinfo{pages}{034024}.

\bibitem[{\citenamefont{Bauer} \emph{et~al.}(2005)\citenamefont{Bauer, Pirjol,
  Rothstein, and Stewart}}]{Bauer:2005wb}
\bibinfo{author}{\bibnamefont{Bauer}, \bibfnamefont{C.~W.}},
  \bibinfo{author}{\bibfnamefont{D.}~\bibnamefont{Pirjol}},
  \bibinfo{author}{\bibfnamefont{I.~Z.} \bibnamefont{Rothstein}}, and
  \bibinfo{author}{\bibfnamefont{I.~W.} \bibnamefont{Stewart}},
  \bibinfo{year}{2005}, \bibinfo{journal}{Phys. Rev.}
  \textbf{\bibinfo{volume}{D72}}, \bibinfo{pages}{098502}.

\bibitem[{\citenamefont{Bauer} \emph{et~al.}(2002)\citenamefont{Bauer, Pirjol,
  and Stewart}}]{Bauer:2001yt}
\bibinfo{author}{\bibnamefont{Bauer}, \bibfnamefont{C.~W.}},
  \bibinfo{author}{\bibfnamefont{D.}~\bibnamefont{Pirjol}}, and
  \bibinfo{author}{\bibfnamefont{I.~W.} \bibnamefont{Stewart}},
  \bibinfo{year}{2002}, \bibinfo{journal}{Phys. Rev.}
  \textbf{\bibinfo{volume}{D65}}, \bibinfo{pages}{054022}.

\bibitem[{\citenamefont{Bauer} \emph{et~al.}(2003)\citenamefont{Bauer, Pirjol,
  and Stewart}}]{Bauer:2002aj}
\bibinfo{author}{\bibnamefont{Bauer}, \bibfnamefont{C.~W.}},
  \bibinfo{author}{\bibfnamefont{D.}~\bibnamefont{Pirjol}}, and
  \bibinfo{author}{\bibfnamefont{I.~W.} \bibnamefont{Stewart}},
  \bibinfo{year}{2003}, \bibinfo{journal}{Phys. Rev.}
  \textbf{\bibinfo{volume}{D67}}, \bibinfo{pages}{071502}.

\bibitem[{\citenamefont{Beall} \emph{et~al.}(1982)\citenamefont{Beall, Bander,
  and Soni}}]{Beall:1981ze}
\bibinfo{author}{\bibnamefont{Beall}, \bibfnamefont{G.}},
  \bibinfo{author}{\bibfnamefont{M.}~\bibnamefont{Bander}}, and
  \bibinfo{author}{\bibfnamefont{A.}~\bibnamefont{Soni}}, \bibinfo{year}{1982},
  \bibinfo{journal}{Phys. Rev. Lett.} \textbf{\bibinfo{volume}{48}},
  \bibinfo{pages}{848}.

\bibitem[{\citenamefont{Becher and Hill}(2006)}]{Becher:2005bg}
\bibinfo{author}{\bibnamefont{Becher}, \bibfnamefont{T.}}, and
  \bibinfo{author}{\bibfnamefont{R.~J.} \bibnamefont{Hill}},
  \bibinfo{year}{2006}, \bibinfo{journal}{Phys. Lett.}
  \textbf{\bibinfo{volume}{B633}}, \bibinfo{pages}{61}.

\bibitem[{\citenamefont{Becher and Neubert}(2006)}]{Becher:2006qw}
\bibinfo{author}{\bibnamefont{Becher}, \bibfnamefont{T.}}, and
  \bibinfo{author}{\bibfnamefont{M.}~\bibnamefont{Neubert}},
  \bibinfo{year}{2006}, \bibinfo{journal}{Phys. Lett.}
  \textbf{\bibinfo{volume}{B637}}, \bibinfo{pages}{251}.

\bibitem[{\citenamefont{Becher and Neubert}(2007)}]{Becher:2006pu}
\bibinfo{author}{\bibnamefont{Becher}, \bibfnamefont{T.}}, and
  \bibinfo{author}{\bibfnamefont{M.}~\bibnamefont{Neubert}},
  \bibinfo{year}{2007}, \bibinfo{journal}{Phys. Rev. Lett.}
  \textbf{\bibinfo{volume}{98}}, \bibinfo{pages}{022003}.

\bibitem[{\citenamefont{Becirevic and Kaidalov}(2000)}]{Becirevic:1999kt}
\bibinfo{author}{\bibnamefont{Becirevic}, \bibfnamefont{D.}}, and
  \bibinfo{author}{\bibfnamefont{A.~B.} \bibnamefont{Kaidalov}},
  \bibinfo{year}{2000}, \bibinfo{journal}{Phys. Lett.}
  \textbf{\bibinfo{volume}{B478}}, \bibinfo{pages}{417}.

\bibitem[{\citenamefont{Becirevic} \emph{et~al.}(2007)\citenamefont{Becirevic,
  Lubicz, and Mescia}}]{Becirevic:2006nm}
\bibinfo{author}{\bibnamefont{Becirevic}, \bibfnamefont{D.}},
  \bibinfo{author}{\bibfnamefont{V.}~\bibnamefont{Lubicz}}, and
  \bibinfo{author}{\bibfnamefont{F.}~\bibnamefont{Mescia}},
  \bibinfo{year}{2007}, \bibinfo{journal}{Nucl. Phys.}
  \textbf{\bibinfo{volume}{B769}}, \bibinfo{pages}{31}.

\bibitem[{Bellgardt \emph{et~al.}(1988)\citenamefont{Bellgardt}
  \emph{et~al.}}]{Bellgardt:1987du}
\bibinfo{author}{\bibnamefont{Bellgardt}, \bibfnamefont{U.}}, \emph{et~al.}
  (\bibinfo{collaboration}{SINDRUM}), \bibinfo{year}{1988},
  \bibinfo{journal}{Nucl. Phys.} \textbf{\bibinfo{volume}{B299}},
  \bibinfo{pages}{1}.

\bibitem[{\citenamefont{Beneke}(2005)}]{Beneke:2005pu}
\bibinfo{author}{\bibnamefont{Beneke}, \bibfnamefont{M.}},
  \bibinfo{year}{2005}, \bibinfo{journal}{Phys. Lett.}
  \textbf{\bibinfo{volume}{B620}}, \bibinfo{pages}{143}.

\bibitem[{\citenamefont{Beneke}
  \emph{et~al.}(2005{\natexlab{a}})\citenamefont{Beneke, Buchalla, Neubert, and
  Sachrajda}}]{Beneke:2004bn}
\bibinfo{author}{\bibnamefont{Beneke}, \bibfnamefont{M.}},
  \bibinfo{author}{\bibfnamefont{G.}~\bibnamefont{Buchalla}},
  \bibinfo{author}{\bibfnamefont{M.}~\bibnamefont{Neubert}}, and
  \bibinfo{author}{\bibfnamefont{C.~T.} \bibnamefont{Sachrajda}},
  \bibinfo{year}{2005}{\natexlab{a}}, \bibinfo{journal}{Phys. Rev.}
  \textbf{\bibinfo{volume}{D72}}, \bibinfo{pages}{098501}.

\bibitem[{\citenamefont{Beneke}
  \emph{et~al.}(2005{\natexlab{b}})\citenamefont{Beneke, Campanario, Mannel,
  and Pecjak}}]{Beneke:2004in}
\bibinfo{author}{\bibnamefont{Beneke}, \bibfnamefont{M.}},
  \bibinfo{author}{\bibfnamefont{F.}~\bibnamefont{Campanario}},
  \bibinfo{author}{\bibfnamefont{T.}~\bibnamefont{Mannel}}, and
  \bibinfo{author}{\bibfnamefont{B.~D.} \bibnamefont{Pecjak}},
  \bibinfo{year}{2005}{\natexlab{b}}, \bibinfo{journal}{JHEP}
  \textbf{\bibinfo{volume}{06}}, \bibinfo{pages}{071}.

\bibitem[{\citenamefont{Beneke and Feldmann}(2001)}]{Beneke:2000wa}
\bibinfo{author}{\bibnamefont{Beneke}, \bibfnamefont{M.}}, and
  \bibinfo{author}{\bibfnamefont{T.}~\bibnamefont{Feldmann}},
  \bibinfo{year}{2001}, \bibinfo{journal}{Nucl. Phys.}
  \textbf{\bibinfo{volume}{B592}}, \bibinfo{pages}{3}.

\bibitem[{\citenamefont{Beneke and Feldmann}(2004)}]{Beneke:2003pa}
\bibinfo{author}{\bibnamefont{Beneke}, \bibfnamefont{M.}}, and
  \bibinfo{author}{\bibfnamefont{T.}~\bibnamefont{Feldmann}},
  \bibinfo{year}{2004}, \bibinfo{journal}{Nucl. Phys.}
  \textbf{\bibinfo{volume}{B685}}, \bibinfo{pages}{249}.

\bibitem[{\citenamefont{Beneke} \emph{et~al.}(2001)\citenamefont{Beneke,
  Feldmann, and Seidel}}]{Beneke:2001at}
\bibinfo{author}{\bibnamefont{Beneke}, \bibfnamefont{M.}},
  \bibinfo{author}{\bibfnamefont{T.}~\bibnamefont{Feldmann}}, and
  \bibinfo{author}{\bibfnamefont{D.}~\bibnamefont{Seidel}},
  \bibinfo{year}{2001}, \bibinfo{journal}{Nucl. Phys.}
  \textbf{\bibinfo{volume}{B612}}, \bibinfo{pages}{25}.

\bibitem[{\citenamefont{Beneke}
  \emph{et~al.}(2005{\natexlab{c}})\citenamefont{Beneke, Feldmann, and
  Seidel}}]{Beneke:2004dp}
\bibinfo{author}{\bibnamefont{Beneke}, \bibfnamefont{M.}},
  \bibinfo{author}{\bibfnamefont{T.}~\bibnamefont{Feldmann}}, and
  \bibinfo{author}{\bibfnamefont{D.}~\bibnamefont{Seidel}},
  \bibinfo{year}{2005}{\natexlab{c}}, \bibinfo{journal}{Eur. Phys. J.}
  \textbf{\bibinfo{volume}{C41}}, \bibinfo{pages}{173}.

\bibitem[{\citenamefont{Beneke} \emph{et~al.}(2006)\citenamefont{Beneke,
  Gronau, Rohrer, and Spranger}}]{Beneke:2006rb}
\bibinfo{author}{\bibnamefont{Beneke}, \bibfnamefont{M.}},
  \bibinfo{author}{\bibfnamefont{M.}~\bibnamefont{Gronau}},
  \bibinfo{author}{\bibfnamefont{J.}~\bibnamefont{Rohrer}}, and
  \bibinfo{author}{\bibfnamefont{M.}~\bibnamefont{Spranger}},
  \bibinfo{year}{2006}, \bibinfo{journal}{Phys. Lett.}
  \textbf{\bibinfo{volume}{B638}}, \bibinfo{pages}{68}.

\bibitem[{\citenamefont{Beneke and Neubert}(2003)}]{Beneke:2003zv}
\bibinfo{author}{\bibnamefont{Beneke}, \bibfnamefont{M.}}, and
  \bibinfo{author}{\bibfnamefont{M.}~\bibnamefont{Neubert}},
  \bibinfo{year}{2003}, \bibinfo{journal}{Nucl. Phys.}
  \textbf{\bibinfo{volume}{B675}}, \bibinfo{pages}{333}.

\bibitem[{\citenamefont{Beneke and Yang}(2006)}]{Beneke:2005gs}
\bibinfo{author}{\bibnamefont{Beneke}, \bibfnamefont{M.}}, and
  \bibinfo{author}{\bibfnamefont{D.}~\bibnamefont{Yang}}, \bibinfo{year}{2006},
  \bibinfo{journal}{Nucl. Phys.} \textbf{\bibinfo{volume}{B736}},
  \bibinfo{pages}{34}.

\bibitem[{Bennett \emph{et~al.}(2006)\citenamefont{Bennett}
  \emph{et~al.}}]{Bennett:2006fi}
\bibinfo{author}{\bibnamefont{Bennett}, \bibfnamefont{G.~W.}}, \emph{et~al.}
  (\bibinfo{collaboration}{Muon G-2}), \bibinfo{year}{2006},
  \bibinfo{journal}{Phys. Rev.} \textbf{\bibinfo{volume}{D73}},
  \bibinfo{pages}{072003}.

\bibitem[{\citenamefont{Bergmann and Perez}(2001)}]{Bergmann:2001pm}
\bibinfo{author}{\bibnamefont{Bergmann}, \bibfnamefont{S.}}, and
  \bibinfo{author}{\bibfnamefont{G.}~\bibnamefont{Perez}},
  \bibinfo{year}{2001}, \bibinfo{journal}{Phys. Rev.}
  \textbf{\bibinfo{volume}{D64}}, \bibinfo{pages}{115009}.

\bibitem[{\citenamefont{Bernabeu} \emph{et~al.}(2008)\citenamefont{Bernabeu,
  Gonzalez-Sprinberg, Papavassiliou, and Vidal}}]{Bernabeu:2007rr}
\bibinfo{author}{\bibnamefont{Bernabeu}, \bibfnamefont{J.}},
  \bibinfo{author}{\bibfnamefont{G.~A.} \bibnamefont{Gonzalez-Sprinberg}},
  \bibinfo{author}{\bibfnamefont{J.}~\bibnamefont{Papavassiliou}}, and
  \bibinfo{author}{\bibfnamefont{J.}~\bibnamefont{Vidal}},
  \bibinfo{year}{2008}, \bibinfo{journal}{Nucl. Phys.}
  \textbf{\bibinfo{volume}{B790}}, \bibinfo{pages}{160}.

\bibitem[{\citenamefont{Bernabeu} \emph{et~al.}(2007)\citenamefont{Bernabeu,
  Gonzalez-Sprinberg, and Vidal}}]{Bernabeu:2006wf}
\bibinfo{author}{\bibnamefont{Bernabeu}, \bibfnamefont{J.}},
  \bibinfo{author}{\bibfnamefont{G.~A.} \bibnamefont{Gonzalez-Sprinberg}}, and
  \bibinfo{author}{\bibfnamefont{J.}~\bibnamefont{Vidal}},
  \bibinfo{year}{2007}, \bibinfo{journal}{Nucl. Phys.}
  \textbf{\bibinfo{volume}{B763}}, \bibinfo{pages}{283}.

\bibitem[{\citenamefont{Bernard} \emph{et~al.}(1998)\citenamefont{Bernard,
  Blum, and Soni}}]{Bernard:1998dg}
\bibinfo{author}{\bibnamefont{Bernard}, \bibfnamefont{C.~W.}},
  \bibinfo{author}{\bibfnamefont{T.}~\bibnamefont{Blum}}, and
  \bibinfo{author}{\bibfnamefont{A.}~\bibnamefont{Soni}}, \bibinfo{year}{1998},
  \bibinfo{journal}{Phys. Rev.} \textbf{\bibinfo{volume}{D58}},
  \bibinfo{pages}{014501}.

\bibitem[{\citenamefont{Bernard} \emph{et~al.}(1994)\citenamefont{Bernard,
  Hsieh, and Soni}}]{Bernard:1993yt}
\bibinfo{author}{\bibnamefont{Bernard}, \bibfnamefont{C.~W.}},
  \bibinfo{author}{\bibfnamefont{P.}~\bibnamefont{Hsieh}}, and
  \bibinfo{author}{\bibfnamefont{A.}~\bibnamefont{Soni}}, \bibinfo{year}{1994},
  \bibinfo{journal}{Phys. Rev. Lett.} \textbf{\bibinfo{volume}{72}},
  \bibinfo{pages}{1402}.

\bibitem[{Besson \emph{et~al.}(1985)\citenamefont{Besson}
  \emph{et~al.}}]{Besson:1984bd}
\bibinfo{author}{\bibnamefont{Besson}, \bibfnamefont{D.}}, \emph{et~al.}
  (\bibinfo{collaboration}{CLEO}), \bibinfo{year}{1985},
  \bibinfo{journal}{Phys. Rev. Lett.} \textbf{\bibinfo{volume}{54}},
  \bibinfo{pages}{381}.

\bibitem[{\citenamefont{Bianco} \emph{et~al.}(2003)\citenamefont{Bianco,
  Fabbri, Benson, and Bigi}}]{Bianco:2003vb}
\bibinfo{author}{\bibnamefont{Bianco}, \bibfnamefont{S.}},
  \bibinfo{author}{\bibfnamefont{F.~L.} \bibnamefont{Fabbri}},
  \bibinfo{author}{\bibfnamefont{D.}~\bibnamefont{Benson}}, and
  \bibinfo{author}{\bibfnamefont{I.}~\bibnamefont{Bigi}}, \bibinfo{year}{2003},
  \bibinfo{journal}{Riv. Nuovo Cim.} \textbf{\bibinfo{volume}{26N7}},
  \bibinfo{pages}{1}.

\bibitem[{\citenamefont{Bigi}(2007)}]{Bigi:2007vs}
\bibinfo{author}{\bibnamefont{Bigi}, \bibfnamefont{I.~I.}},
  \bibinfo{year}{2007}, \eprint{arXiv:0710.2714 [hep-ph]}.

\bibitem[{\citenamefont{Bigi and Sanda}(2005)}]{Bigi:2005ts}
\bibinfo{author}{\bibnamefont{Bigi}, \bibfnamefont{I.~I.}}, and
  \bibinfo{author}{\bibfnamefont{A.~I.} \bibnamefont{Sanda}},
  \bibinfo{year}{2005}, \bibinfo{journal}{Phys. Lett.}
  \textbf{\bibinfo{volume}{B625}}, \bibinfo{pages}{47}.

\bibitem[{\citenamefont{Bigi}
  \emph{et~al.}(1994{\natexlab{a}})\citenamefont{Bigi, Blok, Shifman, and
  Vainshtein}}]{Bigi:1993fm}
\bibinfo{author}{\bibnamefont{Bigi}, \bibfnamefont{I.~I.~Y.}},
  \bibinfo{author}{\bibfnamefont{B.}~\bibnamefont{Blok}},
  \bibinfo{author}{\bibfnamefont{M.~A.} \bibnamefont{Shifman}}, and
  \bibinfo{author}{\bibfnamefont{A.~I.} \bibnamefont{Vainshtein}},
  \bibinfo{year}{1994}{\natexlab{a}}, \bibinfo{journal}{Phys. Lett.}
  \textbf{\bibinfo{volume}{B323}}, \bibinfo{pages}{408}.

\bibitem[{\citenamefont{Bigi and Sanda}(1981)}]{Bigi:1981qs}
\bibinfo{author}{\bibnamefont{Bigi}, \bibfnamefont{I.~I.~Y.}}, and
  \bibinfo{author}{\bibfnamefont{A.~I.} \bibnamefont{Sanda}},
  \bibinfo{year}{1981}, \bibinfo{journal}{Nucl. Phys.}
  \textbf{\bibinfo{volume}{B193}}, \bibinfo{pages}{85}.

\bibitem[{\citenamefont{Bigi} \emph{et~al.}(1993)\citenamefont{Bigi, Shifman,
  Uraltsev, and Vainshtein}}]{Bigi:1993fe}
\bibinfo{author}{\bibnamefont{Bigi}, \bibfnamefont{I.~I.~Y.}},
  \bibinfo{author}{\bibfnamefont{M.~A.} \bibnamefont{Shifman}},
  \bibinfo{author}{\bibfnamefont{N.~G.} \bibnamefont{Uraltsev}}, and
  \bibinfo{author}{\bibfnamefont{A.~I.} \bibnamefont{Vainshtein}},
  \bibinfo{year}{1993}, \bibinfo{journal}{Phys. Rev. Lett.}
  \textbf{\bibinfo{volume}{71}}, \bibinfo{pages}{496}.

\bibitem[{\citenamefont{Bigi}
  \emph{et~al.}(1994{\natexlab{b}})\citenamefont{Bigi, Shifman, Uraltsev, and
  Vainshtein}}]{Bigi:1993ex}
\bibinfo{author}{\bibnamefont{Bigi}, \bibfnamefont{I.~I.~Y.}},
  \bibinfo{author}{\bibfnamefont{M.~A.} \bibnamefont{Shifman}},
  \bibinfo{author}{\bibfnamefont{N.~G.} \bibnamefont{Uraltsev}}, and
  \bibinfo{author}{\bibfnamefont{A.~I.} \bibnamefont{Vainshtein}},
  \bibinfo{year}{1994}{\natexlab{b}}, \bibinfo{journal}{Int. J. Mod. Phys.}
  \textbf{\bibinfo{volume}{A9}}, \bibinfo{pages}{2467}.

\bibitem[{\citenamefont{Bigi and Uraltsev}(1994)}]{Bigi:1993bh}
\bibinfo{author}{\bibnamefont{Bigi}, \bibfnamefont{I.~I.~Y.}}, and
  \bibinfo{author}{\bibfnamefont{N.~G.} \bibnamefont{Uraltsev}},
  \bibinfo{year}{1994}, \bibinfo{journal}{Nucl. Phys.}
  \textbf{\bibinfo{volume}{B423}}, \bibinfo{pages}{33}.

\bibitem[{\citenamefont{Bird} \emph{et~al.}(2004)\citenamefont{Bird, Jackson,
  Kowalewski, and Pospelov}}]{Bird:2004ts}
\bibinfo{author}{\bibnamefont{Bird}, \bibfnamefont{C.}},
  \bibinfo{author}{\bibfnamefont{P.}~\bibnamefont{Jackson}},
  \bibinfo{author}{\bibfnamefont{R.}~\bibnamefont{Kowalewski}}, and
  \bibinfo{author}{\bibfnamefont{M.}~\bibnamefont{Pospelov}},
  \bibinfo{year}{2004}, \bibinfo{journal}{Phys. Rev. Lett.}
  \textbf{\bibinfo{volume}{93}}, \bibinfo{pages}{201803}.

\bibitem[{Bizjak \emph{et~al.}(2005)\citenamefont{Bizjak}
  \emph{et~al.}}]{Bizjak:2005hn}
\bibinfo{author}{\bibnamefont{Bizjak}, \bibfnamefont{I.}}, \emph{et~al.}
  (\bibinfo{collaboration}{Belle}), \bibinfo{year}{2005},
  \bibinfo{journal}{Phys. Rev. Lett.} \textbf{\bibinfo{volume}{95}},
  \bibinfo{pages}{241801}.

\bibitem[{\citenamefont{Black} \emph{et~al.}(2002)\citenamefont{Black, Han, He,
  and Sher}}]{Black:2002wh}
\bibinfo{author}{\bibnamefont{Black}, \bibfnamefont{D.}},
  \bibinfo{author}{\bibfnamefont{T.}~\bibnamefont{Han}},
  \bibinfo{author}{\bibfnamefont{H.-J.} \bibnamefont{He}}, and
  \bibinfo{author}{\bibfnamefont{M.}~\bibnamefont{Sher}}, \bibinfo{year}{2002},
  \bibinfo{journal}{Phys. Rev.} \textbf{\bibinfo{volume}{D66}},
  \bibinfo{pages}{053002}.

\bibitem[{\citenamefont{Blanke}
  \emph{et~al.}(2007{\natexlab{a}})\citenamefont{Blanke, Buras, Duling,
  Poschenrieder, and Tarantino}}]{Blanke:2007db}
\bibinfo{author}{\bibnamefont{Blanke}, \bibfnamefont{M.}},
  \bibinfo{author}{\bibfnamefont{A.~J.} \bibnamefont{Buras}},
  \bibinfo{author}{\bibfnamefont{B.}~\bibnamefont{Duling}},
  \bibinfo{author}{\bibfnamefont{A.}~\bibnamefont{Poschenrieder}}, and
  \bibinfo{author}{\bibfnamefont{C.}~\bibnamefont{Tarantino}},
  \bibinfo{year}{2007}{\natexlab{a}}, \bibinfo{journal}{JHEP}
  \textbf{\bibinfo{volume}{05}}, \bibinfo{pages}{013}.

\bibitem[{\citenamefont{Blanke}
  \emph{et~al.}(2007{\natexlab{b}})\citenamefont{Blanke, Buras, Duling,
  Poschenrieder, and Tarantino}}]{Blanke:2006eb}
\bibinfo{author}{\bibnamefont{Blanke}, \bibfnamefont{M.}},
  \bibinfo{author}{\bibfnamefont{A.~J.} \bibnamefont{Buras}},
  \bibinfo{author}{\bibfnamefont{B.}~\bibnamefont{Duling}},
  \bibinfo{author}{\bibfnamefont{A.}~\bibnamefont{Poschenrieder}}, and
  \bibinfo{author}{\bibfnamefont{C.}~\bibnamefont{Tarantino}},
  \bibinfo{year}{2007}{\natexlab{b}}, \bibinfo{journal}{JHEP}
  \textbf{\bibinfo{volume}{01}}, \bibinfo{pages}{066}.

\bibitem[{\citenamefont{Blanke} \emph{et~al.}(2006)\citenamefont{Blanke, Buras,
  Guadagnoli, and Tarantino}}]{Blanke:2006ig}
\bibinfo{author}{\bibnamefont{Blanke}, \bibfnamefont{M.}},
  \bibinfo{author}{\bibfnamefont{A.~J.} \bibnamefont{Buras}},
  \bibinfo{author}{\bibfnamefont{D.}~\bibnamefont{Guadagnoli}}, and
  \bibinfo{author}{\bibfnamefont{C.}~\bibnamefont{Tarantino}},
  \bibinfo{year}{2006}, \bibinfo{journal}{JHEP} \textbf{\bibinfo{volume}{10}},
  \bibinfo{pages}{003}.

\bibitem[{\citenamefont{Blok} \emph{et~al.}(1994)\citenamefont{Blok, Koyrakh,
  Shifman, and Vainshtein}}]{Blok:1993va}
\bibinfo{author}{\bibnamefont{Blok}, \bibfnamefont{B.}},
  \bibinfo{author}{\bibfnamefont{L.}~\bibnamefont{Koyrakh}},
  \bibinfo{author}{\bibfnamefont{M.~A.} \bibnamefont{Shifman}}, and
  \bibinfo{author}{\bibfnamefont{A.~I.} \bibnamefont{Vainshtein}},
  \bibinfo{year}{1994}, \bibinfo{journal}{Phys. Rev.}
  \textbf{\bibinfo{volume}{D49}}, \bibinfo{pages}{3356},
  \bibinfo{note}{[Erratum-ibid.\ D {\bf 50}, 3572 (1994)]}.

\bibitem[{\citenamefont{Bobeth} \emph{et~al.}(2001)\citenamefont{Bobeth,
  Ewerth, Kruger, and Urban}}]{Bobeth:2001sq}
\bibinfo{author}{\bibnamefont{Bobeth}, \bibfnamefont{C.}},
  \bibinfo{author}{\bibfnamefont{T.}~\bibnamefont{Ewerth}},
  \bibinfo{author}{\bibfnamefont{F.}~\bibnamefont{Kruger}}, and
  \bibinfo{author}{\bibfnamefont{J.}~\bibnamefont{Urban}},
  \bibinfo{year}{2001}, \bibinfo{journal}{Phys. Rev.}
  \textbf{\bibinfo{volume}{D64}}, \bibinfo{pages}{074014}.

\bibitem[{\citenamefont{Bobeth} \emph{et~al.}(2002)\citenamefont{Bobeth,
  Ewerth, Kruger, and Urban}}]{Bobeth:2002ch}
\bibinfo{author}{\bibnamefont{Bobeth}, \bibfnamefont{C.}},
  \bibinfo{author}{\bibfnamefont{T.}~\bibnamefont{Ewerth}},
  \bibinfo{author}{\bibfnamefont{F.}~\bibnamefont{Kruger}}, and
  \bibinfo{author}{\bibfnamefont{J.}~\bibnamefont{Urban}},
  \bibinfo{year}{2002}, \bibinfo{journal}{Phys. Rev.}
  \textbf{\bibinfo{volume}{D66}}, \bibinfo{pages}{074021}.

\bibitem[{\citenamefont{Bobeth} \emph{et~al.}(2004)\citenamefont{Bobeth,
  Gambino, Gorbahn, and Haisch}}]{Bobeth:2003at}
\bibinfo{author}{\bibnamefont{Bobeth}, \bibfnamefont{C.}},
  \bibinfo{author}{\bibfnamefont{P.}~\bibnamefont{Gambino}},
  \bibinfo{author}{\bibfnamefont{M.}~\bibnamefont{Gorbahn}}, and
  \bibinfo{author}{\bibfnamefont{U.}~\bibnamefont{Haisch}},
  \bibinfo{year}{2004}, \bibinfo{journal}{JHEP} \textbf{\bibinfo{volume}{04}},
  \bibinfo{pages}{071}.

\bibitem[{\citenamefont{Bobeth} \emph{et~al.}(2007)\citenamefont{Bobeth,
  Hiller, and Piranishvili}}]{Bobeth:2007dw}
\bibinfo{author}{\bibnamefont{Bobeth}, \bibfnamefont{C.}},
  \bibinfo{author}{\bibfnamefont{G.}~\bibnamefont{Hiller}}, and
  \bibinfo{author}{\bibfnamefont{G.}~\bibnamefont{Piranishvili}},
  \bibinfo{year}{2007}, \bibinfo{journal}{JHEP} \textbf{\bibinfo{volume}{12}},
  \bibinfo{pages}{040}.

\bibitem[{\citenamefont{Bobeth} \emph{et~al.}(2000)\citenamefont{Bobeth,
  Misiak, and Urban}}]{Bobeth:1999mk}
\bibinfo{author}{\bibnamefont{Bobeth}, \bibfnamefont{C.}},
  \bibinfo{author}{\bibfnamefont{M.}~\bibnamefont{Misiak}}, and
  \bibinfo{author}{\bibfnamefont{J.}~\bibnamefont{Urban}},
  \bibinfo{year}{2000}, \bibinfo{journal}{Nucl. Phys.}
  \textbf{\bibinfo{volume}{B574}}, \bibinfo{pages}{291}.

\bibitem[{Bona \emph{et~al.}(2008)\citenamefont{Bona}
  \emph{et~al.}}]{Bona:2008jn}
\bibinfo{author}{\bibnamefont{Bona}, \bibfnamefont{.~M.}}, \emph{et~al.}
  (\bibinfo{collaboration}{UTfit}), \bibinfo{year}{2008},
  \eprint{arXiv:0803.0659 [hep-ph]}.

\bibitem[{Bona \emph{et~al.}(2006{\natexlab{a}})\citenamefont{Bona}
  \emph{et~al.}}]{Bona:2005eu}
\bibinfo{author}{\bibnamefont{Bona}, \bibfnamefont{M.}}, \emph{et~al.}
  (\bibinfo{collaboration}{UTfit}), \bibinfo{year}{2006}{\natexlab{a}},
  \bibinfo{journal}{JHEP} \textbf{\bibinfo{volume}{03}}, \bibinfo{pages}{080}.

\bibitem[{Bona \emph{et~al.}(2006{\natexlab{b}})\citenamefont{Bona}
  \emph{et~al.}}]{Bona:2006sa}
\bibinfo{author}{\bibnamefont{Bona}, \bibfnamefont{M.}}, \emph{et~al.}
  (\bibinfo{collaboration}{UTfit}), \bibinfo{year}{2006}{\natexlab{b}},
  \bibinfo{journal}{Phys. Rev. Lett.} \textbf{\bibinfo{volume}{97}},
  \bibinfo{pages}{151803}.

\bibitem[{Bona \emph{et~al.}(2007{\natexlab{a}})\citenamefont{Bona}
  \emph{et~al.}}]{Bona:2007qta}
\bibinfo{author}{\bibnamefont{Bona}, \bibfnamefont{M.}}, \emph{et~al.}
  (\bibinfo{collaboration}{UTfit}), \bibinfo{year}{2007}{\natexlab{a}},
  \bibinfo{journal}{Phys. Rev.} \textbf{\bibinfo{volume}{D76}},
  \bibinfo{pages}{014015}.

\bibitem[{Bona \emph{et~al.}(2007{\natexlab{b}})\citenamefont{Bona}
  \emph{et~al.}}]{Bona:2007vi}
\bibinfo{author}{\bibnamefont{Bona}, \bibfnamefont{M.}}, \emph{et~al.}
  (\bibinfo{collaboration}{UTfit}), \bibinfo{year}{2007}{\natexlab{b}},
  \eprint{arXiv:0707.0636 [hep-ph]}.

\bibitem[{Bona \emph{et~al.}(2007{\natexlab{c}})\citenamefont{Bona}
  \emph{et~al.}}]{Bona:2007qt}
\bibinfo{author}{\bibnamefont{Bona}, \bibfnamefont{M.}}, \emph{et~al.},
  \bibinfo{year}{2007}{\natexlab{c}}, \eprint{arXiv:0709.0451 [hep-ex]}.

\bibitem[{\citenamefont{Bondar and Gershon}(2004)}]{Bondar:2004bi}
\bibinfo{author}{\bibnamefont{Bondar}, \bibfnamefont{A.}}, and
  \bibinfo{author}{\bibfnamefont{T.}~\bibnamefont{Gershon}},
  \bibinfo{year}{2004}, \bibinfo{journal}{Phys. Rev.}
  \textbf{\bibinfo{volume}{D70}}, \bibinfo{pages}{091503}.

\bibitem[{\citenamefont{Bondar} \emph{et~al.}(2005)\citenamefont{Bondar,
  Gershon, and Krokovny}}]{Bondar:2005gk}
\bibinfo{author}{\bibnamefont{Bondar}, \bibfnamefont{A.}},
  \bibinfo{author}{\bibfnamefont{T.}~\bibnamefont{Gershon}}, and
  \bibinfo{author}{\bibfnamefont{P.}~\bibnamefont{Krokovny}},
  \bibinfo{year}{2005}, \bibinfo{journal}{Phys. Lett.}
  \textbf{\bibinfo{volume}{B624}}, \bibinfo{pages}{1}.

\bibitem[{\citenamefont{Bondar and Poluektov}(2006)}]{Bondar:2005ki}
\bibinfo{author}{\bibnamefont{Bondar}, \bibfnamefont{A.}}, and
  \bibinfo{author}{\bibfnamefont{A.}~\bibnamefont{Poluektov}},
  \bibinfo{year}{2006}, \bibinfo{journal}{Eur. Phys. J.}
  \textbf{\bibinfo{volume}{C47}}, \bibinfo{pages}{347}.

\bibitem[{\citenamefont{Bondar and Poluektov}(2008)}]{Bondar:2008hh}
\bibinfo{author}{\bibnamefont{Bondar}, \bibfnamefont{A.}}, and
  \bibinfo{author}{\bibfnamefont{A.}~\bibnamefont{Poluektov}},
  \bibinfo{year}{2008}, \eprint{arXiv:0801.0840 [hep-ex]}.

\bibitem[{\citenamefont{Boos} \emph{et~al.}(2004)\citenamefont{Boos, Mannel,
  and Reuter}}]{Boos:2004xp}
\bibinfo{author}{\bibnamefont{Boos}, \bibfnamefont{H.}},
  \bibinfo{author}{\bibfnamefont{T.}~\bibnamefont{Mannel}}, and
  \bibinfo{author}{\bibfnamefont{J.}~\bibnamefont{Reuter}},
  \bibinfo{year}{2004}, \bibinfo{journal}{Phys. Rev.}
  \textbf{\bibinfo{volume}{D70}}, \bibinfo{pages}{036006}.

\bibitem[{\citenamefont{Borzumati and Greub}(1998)}]{Borzumati:1998tg}
\bibinfo{author}{\bibnamefont{Borzumati}, \bibfnamefont{F.}}, and
  \bibinfo{author}{\bibfnamefont{C.}~\bibnamefont{Greub}},
  \bibinfo{year}{1998}, \bibinfo{journal}{Phys. Rev.}
  \textbf{\bibinfo{volume}{D58}}, \bibinfo{pages}{074004}.

\bibitem[{\citenamefont{Borzumati and Masiero}(1986)}]{Borzumati:1986qx}
\bibinfo{author}{\bibnamefont{Borzumati}, \bibfnamefont{F.}}, and
  \bibinfo{author}{\bibfnamefont{A.}~\bibnamefont{Masiero}},
  \bibinfo{year}{1986}, \bibinfo{journal}{Phys. Rev. Lett.}
  \textbf{\bibinfo{volume}{57}}, \bibinfo{pages}{961}.

\bibitem[{\citenamefont{Bosch and Buchalla}(2002)}]{Bosch:2001gv}
\bibinfo{author}{\bibnamefont{Bosch}, \bibfnamefont{S.~W.}}, and
  \bibinfo{author}{\bibfnamefont{G.}~\bibnamefont{Buchalla}},
  \bibinfo{year}{2002}, \bibinfo{journal}{Nucl. Phys.}
  \textbf{\bibinfo{volume}{B621}}, \bibinfo{pages}{459}.

\bibitem[{\citenamefont{Bosch and Buchalla}(2005)}]{Bosch:2004nd}
\bibinfo{author}{\bibnamefont{Bosch}, \bibfnamefont{S.~W.}}, and
  \bibinfo{author}{\bibfnamefont{G.}~\bibnamefont{Buchalla}},
  \bibinfo{year}{2005}, \bibinfo{journal}{JHEP} \textbf{\bibinfo{volume}{01}},
  \bibinfo{pages}{035}.

\bibitem[{\citenamefont{Bosch}
  \emph{et~al.}(2004{\natexlab{a}})\citenamefont{Bosch, Lange, Neubert, and
  Paz}}]{Bosch:2004th}
\bibinfo{author}{\bibnamefont{Bosch}, \bibfnamefont{S.~W.}},
  \bibinfo{author}{\bibfnamefont{B.~O.} \bibnamefont{Lange}},
  \bibinfo{author}{\bibfnamefont{M.}~\bibnamefont{Neubert}}, and
  \bibinfo{author}{\bibfnamefont{G.}~\bibnamefont{Paz}},
  \bibinfo{year}{2004}{\natexlab{a}}, \bibinfo{journal}{Nucl. Phys.}
  \textbf{\bibinfo{volume}{B699}}, \bibinfo{pages}{335}.

\bibitem[{\citenamefont{Bosch}
  \emph{et~al.}(2004{\natexlab{b}})\citenamefont{Bosch, Neubert, and
  Paz}}]{Bosch:2004cb}
\bibinfo{author}{\bibnamefont{Bosch}, \bibfnamefont{S.~W.}},
  \bibinfo{author}{\bibfnamefont{M.}~\bibnamefont{Neubert}}, and
  \bibinfo{author}{\bibfnamefont{G.}~\bibnamefont{Paz}},
  \bibinfo{year}{2004}{\natexlab{b}}, \bibinfo{journal}{JHEP}
  \textbf{\bibinfo{volume}{11}}, \bibinfo{pages}{073}.

\bibitem[{\citenamefont{Boyd} \emph{et~al.}(1995)\citenamefont{Boyd, Grinstein,
  and Lebed}}]{Boyd:1994tt}
\bibinfo{author}{\bibnamefont{Boyd}, \bibfnamefont{C.~G.}},
  \bibinfo{author}{\bibfnamefont{B.}~\bibnamefont{Grinstein}}, and
  \bibinfo{author}{\bibfnamefont{R.~F.} \bibnamefont{Lebed}},
  \bibinfo{year}{1995}, \bibinfo{journal}{Phys. Rev. Lett.}
  \textbf{\bibinfo{volume}{74}}, \bibinfo{pages}{4603}.

\bibitem[{\citenamefont{Boyd and Savage}(1997)}]{Boyd:1997qw}
\bibinfo{author}{\bibnamefont{Boyd}, \bibfnamefont{C.~G.}}, and
  \bibinfo{author}{\bibfnamefont{M.~J.} \bibnamefont{Savage}},
  \bibinfo{year}{1997}, \bibinfo{journal}{Phys. Rev.}
  \textbf{\bibinfo{volume}{D56}}, \bibinfo{pages}{303}.

\bibitem[{\citenamefont{Brignole} \emph{et~al.}(1994)\citenamefont{Brignole,
  Ibanez, and Munoz}}]{Brignole:1993dj}
\bibinfo{author}{\bibnamefont{Brignole}, \bibfnamefont{A.}},
  \bibinfo{author}{\bibfnamefont{L.~E.} \bibnamefont{Ibanez}}, and
  \bibinfo{author}{\bibfnamefont{C.}~\bibnamefont{Munoz}},
  \bibinfo{year}{1994}, \bibinfo{journal}{Nucl. Phys.}
  \textbf{\bibinfo{volume}{B422}}, \bibinfo{pages}{125},
  \bibinfo{note}{[Erratum-ibid.\ B {\bf 436}, 747 (1995)]}.

\bibitem[{\citenamefont{Brignole and Rossi}(2004)}]{Brignole:2004ah}
\bibinfo{author}{\bibnamefont{Brignole}, \bibfnamefont{A.}}, and
  \bibinfo{author}{\bibfnamefont{A.}~\bibnamefont{Rossi}},
  \bibinfo{year}{2004}, \bibinfo{journal}{Nucl. Phys.}
  \textbf{\bibinfo{volume}{B701}}, \bibinfo{pages}{3}.

\bibitem[{Brooks \emph{et~al.}(1999)\citenamefont{Brooks}
  \emph{et~al.}}]{Brooks:1999pu}
\bibinfo{author}{\bibnamefont{Brooks}, \bibfnamefont{M.~L.}}, \emph{et~al.}
  (\bibinfo{collaboration}{MEGA}), \bibinfo{year}{1999},
  \bibinfo{journal}{Phys. Rev. Lett.} \textbf{\bibinfo{volume}{83}},
  \bibinfo{pages}{1521}.

\bibitem[{Browder \emph{et~al.}(2007)\citenamefont{Browder}
  \emph{et~al.}}]{Browder:2007gg}
\bibinfo{author}{\bibnamefont{Browder}, \bibfnamefont{T.}}, \emph{et~al.},
  \bibinfo{year}{2007}, \eprint{arXiv:0710.3799 [hep-ph]}.

\bibitem[{\citenamefont{Buchalla} \emph{et~al.}(1996)\citenamefont{Buchalla,
  Buras, and Lautenbacher}}]{Buchalla:1995vs}
\bibinfo{author}{\bibnamefont{Buchalla}, \bibfnamefont{G.}},
  \bibinfo{author}{\bibfnamefont{A.~J.} \bibnamefont{Buras}}, and
  \bibinfo{author}{\bibfnamefont{M.~E.} \bibnamefont{Lautenbacher}},
  \bibinfo{year}{1996}, \bibinfo{journal}{Rev. Mod. Phys.}
  \textbf{\bibinfo{volume}{68}}, \bibinfo{pages}{1125}.

\bibitem[{\citenamefont{Buchalla} \emph{et~al.}(2005)\citenamefont{Buchalla,
  Hiller, Nir, and Raz}}]{Buchalla:2005us}
\bibinfo{author}{\bibnamefont{Buchalla}, \bibfnamefont{G.}},
  \bibinfo{author}{\bibfnamefont{G.}~\bibnamefont{Hiller}},
  \bibinfo{author}{\bibfnamefont{Y.}~\bibnamefont{Nir}}, and
  \bibinfo{author}{\bibfnamefont{G.}~\bibnamefont{Raz}}, \bibinfo{year}{2005},
  \bibinfo{journal}{JHEP} \textbf{\bibinfo{volume}{09}}, \bibinfo{pages}{074}.

\bibitem[{Buchalla \emph{et~al.}(2008)\citenamefont{Buchalla}
  \emph{et~al.}}]{Buchalla:2008jp}
\bibinfo{author}{\bibnamefont{Buchalla}, \bibfnamefont{G.}}, \emph{et~al.},
  \bibinfo{year}{2008}, \eprint{arXiv:0801.1833 [hep-ph]}.

\bibitem[{\citenamefont{Buras}(2003)}]{Buras:2003jf}
\bibinfo{author}{\bibnamefont{Buras}, \bibfnamefont{A.~J.}},
  \bibinfo{year}{2003}, \bibinfo{journal}{Acta Phys. Polon.}
  \textbf{\bibinfo{volume}{B34}}, \bibinfo{pages}{5615}.

\bibitem[{\citenamefont{Buras}
  \emph{et~al.}(2001{\natexlab{a}})\citenamefont{Buras, Chankowski, Rosiek, and
  Slawianowska}}]{Buras:2001mb}
\bibinfo{author}{\bibnamefont{Buras}, \bibfnamefont{A.~J.}},
  \bibinfo{author}{\bibfnamefont{P.~H.} \bibnamefont{Chankowski}},
  \bibinfo{author}{\bibfnamefont{J.}~\bibnamefont{Rosiek}}, and
  \bibinfo{author}{\bibfnamefont{L.}~\bibnamefont{Slawianowska}},
  \bibinfo{year}{2001}{\natexlab{a}}, \bibinfo{journal}{Nucl. Phys.}
  \textbf{\bibinfo{volume}{B619}}, \bibinfo{pages}{434}.

\bibitem[{\citenamefont{Buras and Fleischer}(1999)}]{Buras:1998rb}
\bibinfo{author}{\bibnamefont{Buras}, \bibfnamefont{A.~J.}}, and
  \bibinfo{author}{\bibfnamefont{R.}~\bibnamefont{Fleischer}},
  \bibinfo{year}{1999}, \bibinfo{journal}{Eur. Phys. J.}
  \textbf{\bibinfo{volume}{C11}}, \bibinfo{pages}{93}.

\bibitem[{\citenamefont{Buras} \emph{et~al.}(2003)\citenamefont{Buras,
  Fleischer, Recksiegel, and Schwab}}]{Buras:2003yc}
\bibinfo{author}{\bibnamefont{Buras}, \bibfnamefont{A.~J.}},
  \bibinfo{author}{\bibfnamefont{R.}~\bibnamefont{Fleischer}},
  \bibinfo{author}{\bibfnamefont{S.}~\bibnamefont{Recksiegel}}, and
  \bibinfo{author}{\bibfnamefont{F.}~\bibnamefont{Schwab}},
  \bibinfo{year}{2003}, \bibinfo{journal}{Eur. Phys. J.}
  \textbf{\bibinfo{volume}{C32}}, \bibinfo{pages}{45}.

\bibitem[{\citenamefont{Buras}
  \emph{et~al.}(2004{\natexlab{a}})\citenamefont{Buras, Fleischer, Recksiegel,
  and Schwab}}]{Buras:2004ub}
\bibinfo{author}{\bibnamefont{Buras}, \bibfnamefont{A.~J.}},
  \bibinfo{author}{\bibfnamefont{R.}~\bibnamefont{Fleischer}},
  \bibinfo{author}{\bibfnamefont{S.}~\bibnamefont{Recksiegel}}, and
  \bibinfo{author}{\bibfnamefont{F.}~\bibnamefont{Schwab}},
  \bibinfo{year}{2004}{\natexlab{a}}, \bibinfo{journal}{Nucl. Phys.}
  \textbf{\bibinfo{volume}{B697}}, \bibinfo{pages}{133}.

\bibitem[{\citenamefont{Buras}
  \emph{et~al.}(2004{\natexlab{b}})\citenamefont{Buras, Fleischer, Recksiegel,
  and Schwab}}]{Buras:2003dj}
\bibinfo{author}{\bibnamefont{Buras}, \bibfnamefont{A.~J.}},
  \bibinfo{author}{\bibfnamefont{R.}~\bibnamefont{Fleischer}},
  \bibinfo{author}{\bibfnamefont{S.}~\bibnamefont{Recksiegel}}, and
  \bibinfo{author}{\bibfnamefont{F.}~\bibnamefont{Schwab}},
  \bibinfo{year}{2004}{\natexlab{b}}, \bibinfo{journal}{Phys. Rev. Lett.}
  \textbf{\bibinfo{volume}{92}}, \bibinfo{pages}{101804}.

\bibitem[{\citenamefont{Buras} \emph{et~al.}(2005)\citenamefont{Buras,
  Fleischer, Recksiegel, and Schwab}}]{Buras:2004th}
\bibinfo{author}{\bibnamefont{Buras}, \bibfnamefont{A.~J.}},
  \bibinfo{author}{\bibfnamefont{R.}~\bibnamefont{Fleischer}},
  \bibinfo{author}{\bibfnamefont{S.}~\bibnamefont{Recksiegel}}, and
  \bibinfo{author}{\bibfnamefont{F.}~\bibnamefont{Schwab}},
  \bibinfo{year}{2005}, \bibinfo{journal}{Acta Phys. Polon.}
  \textbf{\bibinfo{volume}{B36}}, \bibinfo{pages}{2015}.

\bibitem[{\citenamefont{Buras} \emph{et~al.}(2006)\citenamefont{Buras,
  Fleischer, Recksiegel, and Schwab}}]{Buras:2005cv}
\bibinfo{author}{\bibnamefont{Buras}, \bibfnamefont{A.~J.}},
  \bibinfo{author}{\bibfnamefont{R.}~\bibnamefont{Fleischer}},
  \bibinfo{author}{\bibfnamefont{S.}~\bibnamefont{Recksiegel}}, and
  \bibinfo{author}{\bibfnamefont{F.}~\bibnamefont{Schwab}},
  \bibinfo{year}{2006}, \bibinfo{journal}{Eur. Phys. J.}
  \textbf{\bibinfo{volume}{C45}}, \bibinfo{pages}{701}.

\bibitem[{\citenamefont{Buras}
  \emph{et~al.}(2001{\natexlab{b}})\citenamefont{Buras, Gambino, Gorbahn,
  Jager, and Silvestrini}}]{Buras:2000dm}
\bibinfo{author}{\bibnamefont{Buras}, \bibfnamefont{A.~J.}},
  \bibinfo{author}{\bibfnamefont{P.}~\bibnamefont{Gambino}},
  \bibinfo{author}{\bibfnamefont{M.}~\bibnamefont{Gorbahn}},
  \bibinfo{author}{\bibfnamefont{S.}~\bibnamefont{Jager}}, and
  \bibinfo{author}{\bibfnamefont{L.}~\bibnamefont{Silvestrini}},
  \bibinfo{year}{2001}{\natexlab{b}}, \bibinfo{journal}{Phys. Lett.}
  \textbf{\bibinfo{volume}{B500}}, \bibinfo{pages}{161}.

\bibitem[{\citenamefont{Buras} \emph{et~al.}(1990)\citenamefont{Buras, Jamin,
  and Weisz}}]{Buras:1990fn}
\bibinfo{author}{\bibnamefont{Buras}, \bibfnamefont{A.~J.}},
  \bibinfo{author}{\bibfnamefont{M.}~\bibnamefont{Jamin}}, and
  \bibinfo{author}{\bibfnamefont{P.~H.} \bibnamefont{Weisz}},
  \bibinfo{year}{1990}, \bibinfo{journal}{Nucl. Phys.}
  \textbf{\bibinfo{volume}{B347}}, \bibinfo{pages}{491}.

\bibitem[{\citenamefont{Burdman}(1998)}]{Burdman:1998mk}
\bibinfo{author}{\bibnamefont{Burdman}, \bibfnamefont{G.}},
  \bibinfo{year}{1998}, \bibinfo{journal}{Phys. Rev.}
  \textbf{\bibinfo{volume}{D57}}, \bibinfo{pages}{4254}.

\bibitem[{\citenamefont{Burdman and Nomura}(2004)}]{Burdman:2003ya}
\bibinfo{author}{\bibnamefont{Burdman}, \bibfnamefont{G.}}, and
  \bibinfo{author}{\bibfnamefont{Y.}~\bibnamefont{Nomura}},
  \bibinfo{year}{2004}, \bibinfo{journal}{Phys. Rev.}
  \textbf{\bibinfo{volume}{D69}}, \bibinfo{pages}{115013}.

\bibitem[{\citenamefont{Burdman and Shipsey}(2003)}]{Burdman:2003rs}
\bibinfo{author}{\bibnamefont{Burdman}, \bibfnamefont{G.}}, and
  \bibinfo{author}{\bibfnamefont{I.}~\bibnamefont{Shipsey}},
  \bibinfo{year}{2003}, \bibinfo{journal}{Ann. Rev. Nucl. Part. Sci.}
  \textbf{\bibinfo{volume}{53}}, \bibinfo{pages}{431}.

\bibitem[{\citenamefont{Cabibbo}(1963)}]{Cabibbo:1963yz}
\bibinfo{author}{\bibnamefont{Cabibbo}, \bibfnamefont{N.}},
  \bibinfo{year}{1963}, \bibinfo{journal}{Phys. Rev. Lett.}
  \textbf{\bibinfo{volume}{10}}, \bibinfo{pages}{531}.

\bibitem[{Cacciapaglia \emph{et~al.}(2007)\citenamefont{Cacciapaglia}
  \emph{et~al.}}]{Cacciapaglia:2007fw}
\bibinfo{author}{\bibnamefont{Cacciapaglia}, \bibfnamefont{G.}}, \emph{et~al.},
  \bibinfo{year}{2007}, \eprint{arXiv:0709.1714 [hep-ph]}.

\bibitem[{\citenamefont{Calderon} \emph{et~al.}(2007)\citenamefont{Calderon,
  Delepine, and Castro}}]{Calderon:2007rg}
\bibinfo{author}{\bibnamefont{Calderon}, \bibfnamefont{G.}},
  \bibinfo{author}{\bibfnamefont{D.}~\bibnamefont{Delepine}}, and
  \bibinfo{author}{\bibfnamefont{G.~L.} \bibnamefont{Castro}},
  \bibinfo{year}{2007}, \bibinfo{journal}{Phys. Rev.}
  \textbf{\bibinfo{volume}{D75}}, \bibinfo{pages}{076001}.

\bibitem[{\citenamefont{Calibbi} \emph{et~al.}(2006)\citenamefont{Calibbi,
  Faccia, Masiero, and Vempati}}]{Calibbi:2006nq}
\bibinfo{author}{\bibnamefont{Calibbi}, \bibfnamefont{L.}},
  \bibinfo{author}{\bibfnamefont{A.}~\bibnamefont{Faccia}},
  \bibinfo{author}{\bibfnamefont{A.}~\bibnamefont{Masiero}}, and
  \bibinfo{author}{\bibfnamefont{S.~K.} \bibnamefont{Vempati}},
  \bibinfo{year}{2006}, \bibinfo{journal}{Phys. Rev.}
  \textbf{\bibinfo{volume}{D74}}, \bibinfo{pages}{116002}.

\bibitem[{\citenamefont{Camilleri}(2007)}]{Camilleri:2007zz}
\bibinfo{author}{\bibnamefont{Camilleri}, \bibfnamefont{L.}}
  (\bibinfo{collaboration}{LHCb}), \bibinfo{year}{2007},
  \bibinfo{note}{cERN-LHCB-2007-096}.

\bibitem[{\citenamefont{Carena} \emph{et~al.}(2001)\citenamefont{Carena,
  Garcia, Nierste, and Wagner}}]{Carena:2000uj}
\bibinfo{author}{\bibnamefont{Carena}, \bibfnamefont{M.}},
  \bibinfo{author}{\bibfnamefont{D.}~\bibnamefont{Garcia}},
  \bibinfo{author}{\bibfnamefont{U.}~\bibnamefont{Nierste}}, and
  \bibinfo{author}{\bibfnamefont{C.~E.~M.} \bibnamefont{Wagner}},
  \bibinfo{year}{2001}, \bibinfo{journal}{Phys. Lett.}
  \textbf{\bibinfo{volume}{B499}}, \bibinfo{pages}{141}.

\bibitem[{\citenamefont{Carena} \emph{et~al.}(2006)\citenamefont{Carena, Menon,
  Noriega-Papaqui, Szynkman, and Wagner}}]{Carena:2006ai}
\bibinfo{author}{\bibnamefont{Carena}, \bibfnamefont{M.~S.}},
  \bibinfo{author}{\bibfnamefont{A.}~\bibnamefont{Menon}},
  \bibinfo{author}{\bibfnamefont{R.}~\bibnamefont{Noriega-Papaqui}},
  \bibinfo{author}{\bibfnamefont{A.}~\bibnamefont{Szynkman}}, and
  \bibinfo{author}{\bibfnamefont{C.~E.~M.} \bibnamefont{Wagner}},
  \bibinfo{year}{2006}, \bibinfo{journal}{Phys. Rev.}
  \textbf{\bibinfo{volume}{D74}}, \bibinfo{pages}{015009}.

\bibitem[{\citenamefont{Carter and Sanda}(1981)}]{Carter:1980tk}
\bibinfo{author}{\bibnamefont{Carter}, \bibfnamefont{A.~B.}}, and
  \bibinfo{author}{\bibfnamefont{A.~I.} \bibnamefont{Sanda}},
  \bibinfo{year}{1981}, \bibinfo{journal}{Phys. Rev.}
  \textbf{\bibinfo{volume}{D23}}, \bibinfo{pages}{1567}.

\bibitem[{\citenamefont{Cavoto} \emph{et~al.}(2007)\citenamefont{Cavoto,
  Fleischer, Trabelsi, and Zupan}}]{Cavoto:2007fp}
\bibinfo{author}{\bibnamefont{Cavoto}, \bibfnamefont{G.}},
  \bibinfo{author}{\bibfnamefont{R.}~\bibnamefont{Fleischer}},
  \bibinfo{author}{\bibfnamefont{K.}~\bibnamefont{Trabelsi}}, and
  \bibinfo{author}{\bibfnamefont{J.}~\bibnamefont{Zupan}},
  \bibinfo{year}{2007}, \eprint{arXiv:0706.4227 [hep-ph]}.

\bibitem[{\citenamefont{Chankowski and Slawianowska}(2001)}]{Chankowski:2000ng}
\bibinfo{author}{\bibnamefont{Chankowski}, \bibfnamefont{P.~H.}}, and
  \bibinfo{author}{\bibfnamefont{L.}~\bibnamefont{Slawianowska}},
  \bibinfo{year}{2001}, \bibinfo{journal}{Phys. Rev.}
  \textbf{\bibinfo{volume}{D63}}, \bibinfo{pages}{054012}.

\bibitem[{Chao \emph{et~al.}(2004)\citenamefont{Chao}
  \emph{et~al.}}]{Chao:2004mn}
\bibinfo{author}{\bibnamefont{Chao}, \bibfnamefont{Y.}}, \emph{et~al.}
  (\bibinfo{collaboration}{Belle}), \bibinfo{year}{2004},
  \bibinfo{journal}{Phys. Rev. Lett.} \textbf{\bibinfo{volume}{93}},
  \bibinfo{pages}{191802}.

\bibitem[{\citenamefont{Charles} \emph{et~al.}(1998)\citenamefont{Charles,
  Le~Yaouanc, Oliver, Pene, and Raynal}}]{Charles:1998vf}
\bibinfo{author}{\bibnamefont{Charles}, \bibfnamefont{J.}},
  \bibinfo{author}{\bibfnamefont{A.}~\bibnamefont{Le~Yaouanc}},
  \bibinfo{author}{\bibfnamefont{L.}~\bibnamefont{Oliver}},
  \bibinfo{author}{\bibfnamefont{O.}~\bibnamefont{Pene}}, and
  \bibinfo{author}{\bibfnamefont{J.~C.} \bibnamefont{Raynal}},
  \bibinfo{year}{1998}, \bibinfo{journal}{Phys. Lett.}
  \textbf{\bibinfo{volume}{B425}}, \bibinfo{pages}{375},
  \bibinfo{note}{[Erratum-ibid.\ B {\bf 433}, 441 (1998)]}.

\bibitem[{Charles \emph{et~al.}(2005)\citenamefont{Charles}
  \emph{et~al.}}]{Charles:2004jd}
\bibinfo{author}{\bibnamefont{Charles}, \bibfnamefont{J.}}, \emph{et~al.}
  (\bibinfo{collaboration}{CKMfitter Group}), \bibinfo{year}{2005},
  \bibinfo{journal}{Eur. Phys. J.} \textbf{\bibinfo{volume}{C41}},
  \bibinfo{pages}{1}, \bibinfo{note}{updated in
  www.slac.stanford.edu/xorg/ckmfitter}.

\bibitem[{\citenamefont{Chay} \emph{et~al.}(1990)\citenamefont{Chay, Georgi,
  and Grinstein}}]{Chay:1990da}
\bibinfo{author}{\bibnamefont{Chay}, \bibfnamefont{J.}},
  \bibinfo{author}{\bibfnamefont{H.}~\bibnamefont{Georgi}}, and
  \bibinfo{author}{\bibfnamefont{B.}~\bibnamefont{Grinstein}},
  \bibinfo{year}{1990}, \bibinfo{journal}{Phys. Lett.}
  \textbf{\bibinfo{volume}{B247}}, \bibinfo{pages}{399}.

\bibitem[{\citenamefont{Chay} \emph{et~al.}(2006)\citenamefont{Chay, Kim,
  Leibovich, and Zupan}}]{Chay:2006ve}
\bibinfo{author}{\bibnamefont{Chay}, \bibfnamefont{J.}},
  \bibinfo{author}{\bibfnamefont{C.}~\bibnamefont{Kim}},
  \bibinfo{author}{\bibfnamefont{A.~K.} \bibnamefont{Leibovich}}, and
  \bibinfo{author}{\bibfnamefont{J.}~\bibnamefont{Zupan}},
  \bibinfo{year}{2006}, \bibinfo{journal}{Phys. Rev.}
  \textbf{\bibinfo{volume}{D74}}, \bibinfo{pages}{074022}.

\bibitem[{\citenamefont{Chay} \emph{et~al.}(2007)\citenamefont{Chay, Kim,
  Leibovich, and Zupan}}]{Chay:2007ej}
\bibinfo{author}{\bibnamefont{Chay}, \bibfnamefont{J.}},
  \bibinfo{author}{\bibfnamefont{C.}~\bibnamefont{Kim}},
  \bibinfo{author}{\bibfnamefont{A.~K.} \bibnamefont{Leibovich}}, and
  \bibinfo{author}{\bibfnamefont{J.}~\bibnamefont{Zupan}},
  \bibinfo{year}{2007}, \bibinfo{journal}{Phys. Rev.}
  \textbf{\bibinfo{volume}{D76}}, \bibinfo{pages}{094031}.

\bibitem[{\citenamefont{Chen and Geng}(2006{\natexlab{a}})}]{Chen:2006nua}
\bibinfo{author}{\bibnamefont{Chen}, \bibfnamefont{C.-H.}}, and
  \bibinfo{author}{\bibfnamefont{C.-Q.} \bibnamefont{Geng}},
  \bibinfo{year}{2006}{\natexlab{a}}, \bibinfo{journal}{JHEP}
  \textbf{\bibinfo{volume}{10}}, \bibinfo{pages}{053}.

\bibitem[{\citenamefont{Chen and Geng}(2006{\natexlab{b}})}]{Chen:2006hp}
\bibinfo{author}{\bibnamefont{Chen}, \bibfnamefont{C.-H.}}, and
  \bibinfo{author}{\bibfnamefont{C.-Q.} \bibnamefont{Geng}},
  \bibinfo{year}{2006}{\natexlab{b}}, \bibinfo{journal}{Phys. Rev.}
  \textbf{\bibinfo{volume}{D74}}, \bibinfo{pages}{035010}.

\bibitem[{\citenamefont{Chen and Geng}(2007)}]{Chen:2007pua}
\bibinfo{author}{\bibnamefont{Chen}, \bibfnamefont{C.-H.}}, and
  \bibinfo{author}{\bibfnamefont{C.-Q.} \bibnamefont{Geng}},
  \bibinfo{year}{2007}, \eprint{arXiv:0709.0235 [hep-ph]}.

\bibitem[{Chen \emph{et~al.}(2007{\natexlab{a}})\citenamefont{Chen}
  \emph{et~al.}}]{Chen:2006nk}
\bibinfo{author}{\bibnamefont{Chen}, \bibfnamefont{K.~F.}}, \emph{et~al.}
  (\bibinfo{collaboration}{Belle}), \bibinfo{year}{2007}{\natexlab{a}},
  \bibinfo{journal}{Phys. Rev. Lett.} \textbf{\bibinfo{volume}{98}},
  \bibinfo{pages}{031802}.

\bibitem[{Chen \emph{et~al.}(2007{\natexlab{b}})\citenamefont{Chen}
  \emph{et~al.}}]{Chen:2007zk}
\bibinfo{author}{\bibnamefont{Chen}, \bibfnamefont{K.~F.}}, \emph{et~al.}
  (\bibinfo{collaboration}{Belle}), \bibinfo{year}{2007}{\natexlab{b}},
  \bibinfo{journal}{Phys. Rev. Lett.} \textbf{\bibinfo{volume}{99}},
  \bibinfo{pages}{221802}.

\bibitem[{Chen \emph{et~al.}(2001)\citenamefont{Chen}
  \emph{et~al.}}]{Chen:2001fja}
\bibinfo{author}{\bibnamefont{Chen}, \bibfnamefont{S.}}, \emph{et~al.}
  (\bibinfo{collaboration}{CLEO}), \bibinfo{year}{2001},
  \bibinfo{journal}{Phys. Rev. Lett.} \textbf{\bibinfo{volume}{87}},
  \bibinfo{pages}{251807}.

\bibitem[{\citenamefont{Cheng and Low}(2003)}]{Cheng:2003ju}
\bibinfo{author}{\bibnamefont{Cheng}, \bibfnamefont{H.-C.}}, and
  \bibinfo{author}{\bibfnamefont{I.}~\bibnamefont{Low}}, \bibinfo{year}{2003},
  \bibinfo{journal}{JHEP} \textbf{\bibinfo{volume}{09}}, \bibinfo{pages}{051}.

\bibitem[{\citenamefont{Cheng}
  \emph{et~al.}(2005{\natexlab{a}})\citenamefont{Cheng, Chua, and
  Soni}}]{Cheng:2005ug}
\bibinfo{author}{\bibnamefont{Cheng}, \bibfnamefont{H.-Y.}},
  \bibinfo{author}{\bibfnamefont{C.-K.} \bibnamefont{Chua}}, and
  \bibinfo{author}{\bibfnamefont{A.}~\bibnamefont{Soni}},
  \bibinfo{year}{2005}{\natexlab{a}}, \bibinfo{journal}{Phys. Rev.}
  \textbf{\bibinfo{volume}{D72}}, \bibinfo{pages}{094003}.

\bibitem[{\citenamefont{Cheng}
  \emph{et~al.}(2005{\natexlab{b}})\citenamefont{Cheng, Chua, and
  Soni}}]{Cheng:2005bg}
\bibinfo{author}{\bibnamefont{Cheng}, \bibfnamefont{H.-Y.}},
  \bibinfo{author}{\bibfnamefont{C.-K.} \bibnamefont{Chua}}, and
  \bibinfo{author}{\bibfnamefont{A.}~\bibnamefont{Soni}},
  \bibinfo{year}{2005}{\natexlab{b}}, \bibinfo{journal}{Phys. Rev.}
  \textbf{\bibinfo{volume}{D72}}, \bibinfo{pages}{014006}.

\bibitem[{\citenamefont{Cheng}
  \emph{et~al.}(2005{\natexlab{c}})\citenamefont{Cheng, Chua, and
  Soni}}]{Cheng:2004ru}
\bibinfo{author}{\bibnamefont{Cheng}, \bibfnamefont{H.-Y.}},
  \bibinfo{author}{\bibfnamefont{C.-K.} \bibnamefont{Chua}}, and
  \bibinfo{author}{\bibfnamefont{A.}~\bibnamefont{Soni}},
  \bibinfo{year}{2005}{\natexlab{c}}, \bibinfo{journal}{Phys. Rev.}
  \textbf{\bibinfo{volume}{D71}}, \bibinfo{pages}{014030}.

\bibitem[{\citenamefont{Cheng and Sher}(1987)}]{Cheng:1987rs}
\bibinfo{author}{\bibnamefont{Cheng}, \bibfnamefont{T.~P.}}, and
  \bibinfo{author}{\bibfnamefont{M.}~\bibnamefont{Sher}}, \bibinfo{year}{1987},
  \bibinfo{journal}{Phys. Rev.} \textbf{\bibinfo{volume}{D35}},
  \bibinfo{pages}{3484}.

\bibitem[{\citenamefont{Chetyrkin} \emph{et~al.}(1997)\citenamefont{Chetyrkin,
  Misiak, and Munz}}]{Chetyrkin:1996vx}
\bibinfo{author}{\bibnamefont{Chetyrkin}, \bibfnamefont{K.~G.}},
  \bibinfo{author}{\bibfnamefont{M.}~\bibnamefont{Misiak}}, and
  \bibinfo{author}{\bibfnamefont{M.}~\bibnamefont{Munz}}, \bibinfo{year}{1997},
  \bibinfo{journal}{Phys. Lett.} \textbf{\bibinfo{volume}{B400}},
  \bibinfo{pages}{206}, \bibinfo{note}{[Erratum-ibid.\ B {\bf 425}, 414
  (1998)]}.

\bibitem[{\citenamefont{Chivukula and Georgi}(1987)}]{Chivukula:1987py}
\bibinfo{author}{\bibnamefont{Chivukula}, \bibfnamefont{R.~S.}}, and
  \bibinfo{author}{\bibfnamefont{H.}~\bibnamefont{Georgi}},
  \bibinfo{year}{1987}, \bibinfo{journal}{Phys. Lett.}
  \textbf{\bibinfo{volume}{B188}}, \bibinfo{pages}{99}.

\bibitem[{\citenamefont{Chun and Lee}(2003)}]{Chun:2003rg}
\bibinfo{author}{\bibnamefont{Chun}, \bibfnamefont{E.~J.}}, and
  \bibinfo{author}{\bibfnamefont{J.~S.} \bibnamefont{Lee}},
  \bibinfo{year}{2003}, \eprint{hep-ph/0307108}.

\bibitem[{Chung \emph{et~al.}(2005)\citenamefont{Chung}
  \emph{et~al.}}]{Chung:2003fi}
\bibinfo{author}{\bibnamefont{Chung}, \bibfnamefont{D.~J.~H.}}, \emph{et~al.},
  \bibinfo{year}{2005}, \bibinfo{journal}{Phys. Rept.}
  \textbf{\bibinfo{volume}{407}}, \bibinfo{pages}{1}.

\bibitem[{\citenamefont{Cirigliano and Grinstein}(2006)}]{Cirigliano:2006su}
\bibinfo{author}{\bibnamefont{Cirigliano}, \bibfnamefont{V.}}, and
  \bibinfo{author}{\bibfnamefont{B.}~\bibnamefont{Grinstein}},
  \bibinfo{year}{2006}, \bibinfo{journal}{Nucl. Phys.}
  \textbf{\bibinfo{volume}{B752}}, \bibinfo{pages}{18}.

\bibitem[{\citenamefont{Cirigliano}
  \emph{et~al.}(2005)\citenamefont{Cirigliano, Grinstein, Isidori, and
  Wise}}]{Cirigliano:2005ck}
\bibinfo{author}{\bibnamefont{Cirigliano}, \bibfnamefont{V.}},
  \bibinfo{author}{\bibfnamefont{B.}~\bibnamefont{Grinstein}},
  \bibinfo{author}{\bibfnamefont{G.}~\bibnamefont{Isidori}}, and
  \bibinfo{author}{\bibfnamefont{M.~B.} \bibnamefont{Wise}},
  \bibinfo{year}{2005}, \bibinfo{journal}{Nucl. Phys.}
  \textbf{\bibinfo{volume}{B728}}, \bibinfo{pages}{121}.

\bibitem[{\citenamefont{Ciuchini}
  \emph{et~al.}(1998{\natexlab{a}})\citenamefont{Ciuchini, Degrassi, Gambino,
  and Giudice}}]{Ciuchini:1998xy}
\bibinfo{author}{\bibnamefont{Ciuchini}, \bibfnamefont{M.}},
  \bibinfo{author}{\bibfnamefont{G.}~\bibnamefont{Degrassi}},
  \bibinfo{author}{\bibfnamefont{P.}~\bibnamefont{Gambino}}, and
  \bibinfo{author}{\bibfnamefont{G.~F.} \bibnamefont{Giudice}},
  \bibinfo{year}{1998}{\natexlab{a}}, \bibinfo{journal}{Nucl. Phys.}
  \textbf{\bibinfo{volume}{B534}}, \bibinfo{pages}{3}.

\bibitem[{\citenamefont{Ciuchini}
  \emph{et~al.}(1998{\natexlab{b}})\citenamefont{Ciuchini, Degrassi, Gambino,
  and Giudice}}]{Ciuchini:1997xe}
\bibinfo{author}{\bibnamefont{Ciuchini}, \bibfnamefont{M.}},
  \bibinfo{author}{\bibfnamefont{G.}~\bibnamefont{Degrassi}},
  \bibinfo{author}{\bibfnamefont{P.}~\bibnamefont{Gambino}}, and
  \bibinfo{author}{\bibfnamefont{G.~F.} \bibnamefont{Giudice}},
  \bibinfo{year}{1998}{\natexlab{b}}, \bibinfo{journal}{Nucl. Phys.}
  \textbf{\bibinfo{volume}{B527}}, \bibinfo{pages}{21}.

\bibitem[{\citenamefont{Ciuchini}
  \emph{et~al.}(1997{\natexlab{a}})\citenamefont{Ciuchini, Franco, Martinelli,
  Masiero, and Silvestrini}}]{Ciuchini:1997zp}
\bibinfo{author}{\bibnamefont{Ciuchini}, \bibfnamefont{M.}},
  \bibinfo{author}{\bibfnamefont{E.}~\bibnamefont{Franco}},
  \bibinfo{author}{\bibfnamefont{G.}~\bibnamefont{Martinelli}},
  \bibinfo{author}{\bibfnamefont{A.}~\bibnamefont{Masiero}}, and
  \bibinfo{author}{\bibfnamefont{L.}~\bibnamefont{Silvestrini}},
  \bibinfo{year}{1997}{\natexlab{a}}, \bibinfo{journal}{Phys. Rev. Lett.}
  \textbf{\bibinfo{volume}{79}}, \bibinfo{pages}{978}.

\bibitem[{\citenamefont{Ciuchini} \emph{et~al.}(2001)\citenamefont{Ciuchini,
  Franco, Martinelli, Pierini, and Silvestrini}}]{Ciuchini:2001gv}
\bibinfo{author}{\bibnamefont{Ciuchini}, \bibfnamefont{M.}},
  \bibinfo{author}{\bibfnamefont{E.}~\bibnamefont{Franco}},
  \bibinfo{author}{\bibfnamefont{G.}~\bibnamefont{Martinelli}},
  \bibinfo{author}{\bibfnamefont{M.}~\bibnamefont{Pierini}}, and
  \bibinfo{author}{\bibfnamefont{L.}~\bibnamefont{Silvestrini}},
  \bibinfo{year}{2001}, \bibinfo{journal}{Phys. Lett.}
  \textbf{\bibinfo{volume}{B515}}, \bibinfo{pages}{33}.

\bibitem[{\citenamefont{Ciuchini}
  \emph{et~al.}(1997{\natexlab{b}})\citenamefont{Ciuchini, Franco, Martinelli,
  and Silvestrini}}]{Ciuchini:1997hb}
\bibinfo{author}{\bibnamefont{Ciuchini}, \bibfnamefont{M.}},
  \bibinfo{author}{\bibfnamefont{E.}~\bibnamefont{Franco}},
  \bibinfo{author}{\bibfnamefont{G.}~\bibnamefont{Martinelli}}, and
  \bibinfo{author}{\bibfnamefont{L.}~\bibnamefont{Silvestrini}},
  \bibinfo{year}{1997}{\natexlab{b}}, \bibinfo{journal}{Nucl. Phys.}
  \textbf{\bibinfo{volume}{B501}}, \bibinfo{pages}{271}.

\bibitem[{\citenamefont{Ciuchini} \emph{et~al.}(2005)\citenamefont{Ciuchini,
  Pierini, and Silvestrini}}]{Ciuchini:2005mg}
\bibinfo{author}{\bibnamefont{Ciuchini}, \bibfnamefont{M.}},
  \bibinfo{author}{\bibfnamefont{M.}~\bibnamefont{Pierini}}, and
  \bibinfo{author}{\bibfnamefont{L.}~\bibnamefont{Silvestrini}},
  \bibinfo{year}{2005}, \bibinfo{journal}{Phys. Rev. Lett.}
  \textbf{\bibinfo{volume}{95}}, \bibinfo{pages}{221804}.

\bibitem[{Ciuchini \emph{et~al.}(2007{\natexlab{a}})\citenamefont{Ciuchini}
  \emph{et~al.}}]{Ciuchini:2007cw}
\bibinfo{author}{\bibnamefont{Ciuchini}, \bibfnamefont{M.}}, \emph{et~al.},
  \bibinfo{year}{2007}{\natexlab{a}}, \bibinfo{journal}{Phys. Lett.}
  \textbf{\bibinfo{volume}{B655}}, \bibinfo{pages}{162}.

\bibitem[{Ciuchini \emph{et~al.}(2007{\natexlab{b}})\citenamefont{Ciuchini}
  \emph{et~al.}}]{Ciuchini:2007ha}
\bibinfo{author}{\bibnamefont{Ciuchini}, \bibfnamefont{M.}}, \emph{et~al.},
  \bibinfo{year}{2007}{\natexlab{b}}, \bibinfo{journal}{Nucl. Phys.}
  \textbf{\bibinfo{volume}{B783}}, \bibinfo{pages}{112}.

\bibitem[{Coan \emph{et~al.}(2001)\citenamefont{Coan}
  \emph{et~al.}}]{Coan:2000pu}
\bibinfo{author}{\bibnamefont{Coan}, \bibfnamefont{T.~E.}}, \emph{et~al.}
  (\bibinfo{collaboration}{CLEO}), \bibinfo{year}{2001},
  \bibinfo{journal}{Phys. Rev. Lett.} \textbf{\bibinfo{volume}{86}},
  \bibinfo{pages}{5661}.

\bibitem[{\citenamefont{Cvetic} \emph{et~al.}(2002)\citenamefont{Cvetic, Dib,
  Kim, and Kim}}]{Cvetic:2002jy}
\bibinfo{author}{\bibnamefont{Cvetic}, \bibfnamefont{G.}},
  \bibinfo{author}{\bibfnamefont{C.}~\bibnamefont{Dib}},
  \bibinfo{author}{\bibfnamefont{C.~S.} \bibnamefont{Kim}}, and
  \bibinfo{author}{\bibfnamefont{J.~D.} \bibnamefont{Kim}},
  \bibinfo{year}{2002}, \bibinfo{journal}{Phys. Rev.}
  \textbf{\bibinfo{volume}{D66}}, \bibinfo{pages}{034008}.

\bibitem[{\citenamefont{Czarnecki}(1996)}]{Czarnecki:1996gu}
\bibinfo{author}{\bibnamefont{Czarnecki}, \bibfnamefont{A.}},
  \bibinfo{year}{1996}, \bibinfo{journal}{Phys. Rev. Lett.}
  \textbf{\bibinfo{volume}{76}}, \bibinfo{pages}{4124}.

\bibitem[{\citenamefont{Czarnecki and Marciano}(1998)}]{Czarnecki:1998tn}
\bibinfo{author}{\bibnamefont{Czarnecki}, \bibfnamefont{A.}}, and
  \bibinfo{author}{\bibfnamefont{W.~J.} \bibnamefont{Marciano}},
  \bibinfo{year}{1998}, \bibinfo{journal}{Phys. Rev. Lett.}
  \textbf{\bibinfo{volume}{81}}, \bibinfo{pages}{277}.

\bibitem[{\citenamefont{Czarnecki and Melnikov}(1997)}]{Czarnecki:1997cf}
\bibinfo{author}{\bibnamefont{Czarnecki}, \bibfnamefont{A.}}, and
  \bibinfo{author}{\bibfnamefont{K.}~\bibnamefont{Melnikov}},
  \bibinfo{year}{1997}, \bibinfo{journal}{Nucl. Phys.}
  \textbf{\bibinfo{volume}{B505}}, \bibinfo{pages}{65}.

\bibitem[{\citenamefont{Dai} \emph{et~al.}(1997)\citenamefont{Dai, Huang, and
  Huang}}]{Dai:1996vg}
\bibinfo{author}{\bibnamefont{Dai}, \bibfnamefont{Y.-B.}},
  \bibinfo{author}{\bibfnamefont{C.-S.} \bibnamefont{Huang}}, and
  \bibinfo{author}{\bibfnamefont{H.-W.} \bibnamefont{Huang}},
  \bibinfo{year}{1997}, \bibinfo{journal}{Phys. Lett.}
  \textbf{\bibinfo{volume}{B390}}, \bibinfo{pages}{257}.

\bibitem[{Dalgic \emph{et~al.}(2006)\citenamefont{Dalgic}
  \emph{et~al.}}]{Dalgic:2006dt}
\bibinfo{author}{\bibnamefont{Dalgic}, \bibfnamefont{E.}}, \emph{et~al.},
  \bibinfo{year}{2006}, \bibinfo{journal}{Phys. Rev.}
  \textbf{\bibinfo{volume}{D73}}, \bibinfo{pages}{074502}.

\bibitem[{Dalgic \emph{et~al.}(2007)\citenamefont{Dalgic}
  \emph{et~al.}}]{Dalgic:2006gp}
\bibinfo{author}{\bibnamefont{Dalgic}, \bibfnamefont{E.}}, \emph{et~al.},
  \bibinfo{year}{2007}, \bibinfo{journal}{Phys. Rev.}
  \textbf{\bibinfo{volume}{D76}}, \bibinfo{pages}{011501}.

\bibitem[{\citenamefont{D'Ambrosio}
  \emph{et~al.}(2002)\citenamefont{D'Ambrosio, Giudice, Isidori, and
  Strumia}}]{D'Ambrosio:2002ex}
\bibinfo{author}{\bibnamefont{D'Ambrosio}, \bibfnamefont{G.}},
  \bibinfo{author}{\bibfnamefont{G.~F.} \bibnamefont{Giudice}},
  \bibinfo{author}{\bibfnamefont{G.}~\bibnamefont{Isidori}}, and
  \bibinfo{author}{\bibfnamefont{A.}~\bibnamefont{Strumia}},
  \bibinfo{year}{2002}, \bibinfo{journal}{Nucl. Phys.}
  \textbf{\bibinfo{volume}{B645}}, \bibinfo{pages}{155}.

\bibitem[{\citenamefont{Das and Kao}(1996)}]{Das:1995df}
\bibinfo{author}{\bibnamefont{Das}, \bibfnamefont{A.~K.}}, and
  \bibinfo{author}{\bibfnamefont{C.}~\bibnamefont{Kao}}, \bibinfo{year}{1996},
  \bibinfo{journal}{Phys. Lett.} \textbf{\bibinfo{volume}{B372}},
  \bibinfo{pages}{106}.

\bibitem[{\citenamefont{Dassinger} \emph{et~al.}(2007)\citenamefont{Dassinger,
  Feldmann, Mannel, and Turczyk}}]{Dassinger:2007ru}
\bibinfo{author}{\bibnamefont{Dassinger}, \bibfnamefont{B.~M.}},
  \bibinfo{author}{\bibfnamefont{T.}~\bibnamefont{Feldmann}},
  \bibinfo{author}{\bibfnamefont{T.}~\bibnamefont{Mannel}}, and
  \bibinfo{author}{\bibfnamefont{S.}~\bibnamefont{Turczyk}},
  \bibinfo{year}{2007}, \bibinfo{journal}{JHEP} \textbf{\bibinfo{volume}{10}},
  \bibinfo{pages}{039}.

\bibitem[{\citenamefont{Datta} \emph{et~al.}(2007)\citenamefont{Datta, Kiers,
  London, O'Donnell, and Szynkman}}]{Datta:2006kd}
\bibinfo{author}{\bibnamefont{Datta}, \bibfnamefont{A.}},
  \bibinfo{author}{\bibfnamefont{K.}~\bibnamefont{Kiers}},
  \bibinfo{author}{\bibfnamefont{D.}~\bibnamefont{London}},
  \bibinfo{author}{\bibfnamefont{P.~J.} \bibnamefont{O'Donnell}}, and
  \bibinfo{author}{\bibfnamefont{A.}~\bibnamefont{Szynkman}},
  \bibinfo{year}{2007}, \bibinfo{journal}{Phys. Rev.}
  \textbf{\bibinfo{volume}{D75}}, \bibinfo{pages}{074007}.

\bibitem[{\citenamefont{Davidson and Palorini}(2006)}]{Davidson:2006bd}
\bibinfo{author}{\bibnamefont{Davidson}, \bibfnamefont{S.}}, and
  \bibinfo{author}{\bibfnamefont{F.}~\bibnamefont{Palorini}},
  \bibinfo{year}{2006}, \bibinfo{journal}{Phys. Lett.}
  \textbf{\bibinfo{volume}{B642}}, \bibinfo{pages}{72}.

\bibitem[{\citenamefont{Davier} \emph{et~al.}(2006)\citenamefont{Davier,
  Hocker, and Zhang}}]{Davier:2005xq}
\bibinfo{author}{\bibnamefont{Davier}, \bibfnamefont{M.}},
  \bibinfo{author}{\bibfnamefont{A.}~\bibnamefont{Hocker}}, and
  \bibinfo{author}{\bibfnamefont{Z.}~\bibnamefont{Zhang}},
  \bibinfo{year}{2006}, \bibinfo{journal}{Rev. Mod. Phys.}
  \textbf{\bibinfo{volume}{78}}, \bibinfo{pages}{1043}.

\bibitem[{\citenamefont{Davis} \emph{et~al.}(1968)\citenamefont{Davis, Harmer,
  and Hoffman}}]{Davis:1968cp}
\bibinfo{author}{\bibnamefont{Davis}, \bibfnamefont{J., Raymond}},
  \bibinfo{author}{\bibfnamefont{D.~S.} \bibnamefont{Harmer}}, and
  \bibinfo{author}{\bibfnamefont{K.~C.} \bibnamefont{Hoffman}},
  \bibinfo{year}{1968}, \bibinfo{journal}{Phys. Rev. Lett.}
  \textbf{\bibinfo{volume}{20}}, \bibinfo{pages}{1205}.

\bibitem[{\citenamefont{Davoudiasl}
  \emph{et~al.}(2000)\citenamefont{Davoudiasl, Hewett, and
  Rizzo}}]{Davoudiasl:1999jd}
\bibinfo{author}{\bibnamefont{Davoudiasl}, \bibfnamefont{H.}},
  \bibinfo{author}{\bibfnamefont{J.~L.} \bibnamefont{Hewett}}, and
  \bibinfo{author}{\bibfnamefont{T.~G.} \bibnamefont{Rizzo}},
  \bibinfo{year}{2000}, \bibinfo{journal}{Phys. Rev. Lett.}
  \textbf{\bibinfo{volume}{84}}, \bibinfo{pages}{2080}.

\bibitem[{\citenamefont{Davoudiasl and Soni}(2007)}]{Davoudiasl:2007zx}
\bibinfo{author}{\bibnamefont{Davoudiasl}, \bibfnamefont{H.}}, and
  \bibinfo{author}{\bibfnamefont{A.}~\bibnamefont{Soni}}, \bibinfo{year}{2007},
  \bibinfo{journal}{Phys. Rev.} \textbf{\bibinfo{volume}{D76}},
  \bibinfo{pages}{095015}.

\bibitem[{\citenamefont{Dedes and Pilaftsis}(2003)}]{Dedes:2002er}
\bibinfo{author}{\bibnamefont{Dedes}, \bibfnamefont{A.}}, and
  \bibinfo{author}{\bibfnamefont{A.}~\bibnamefont{Pilaftsis}},
  \bibinfo{year}{2003}, \bibinfo{journal}{Phys. Rev.}
  \textbf{\bibinfo{volume}{D67}}, \bibinfo{pages}{015012}.

\bibitem[{\citenamefont{Delepine} \emph{et~al.}(2006)\citenamefont{Delepine,
  Faisl, Khalil, and Castro}}]{Delepine:2006fv}
\bibinfo{author}{\bibnamefont{Delepine}, \bibfnamefont{D.}},
  \bibinfo{author}{\bibfnamefont{G.}~\bibnamefont{Faisl}},
  \bibinfo{author}{\bibfnamefont{S.}~\bibnamefont{Khalil}}, and
  \bibinfo{author}{\bibfnamefont{G.~L.} \bibnamefont{Castro}},
  \bibinfo{year}{2006}, \bibinfo{journal}{Phys. Rev.}
  \textbf{\bibinfo{volume}{D74}}, \bibinfo{pages}{056004}.

\bibitem[{\citenamefont{Delepine} \emph{et~al.}(2005)\citenamefont{Delepine,
  Lopez~Castro, and Lopez~Lozano}}]{Delepine:2005tw}
\bibinfo{author}{\bibnamefont{Delepine}, \bibfnamefont{D.}},
  \bibinfo{author}{\bibfnamefont{G.}~\bibnamefont{Lopez~Castro}}, and
  \bibinfo{author}{\bibfnamefont{L.~T.} \bibnamefont{Lopez~Lozano}},
  \bibinfo{year}{2005}, \bibinfo{journal}{Phys. Rev.}
  \textbf{\bibinfo{volume}{D72}}, \bibinfo{pages}{033009}.

\bibitem[{\citenamefont{Dermisek} \emph{et~al.}(2007)\citenamefont{Dermisek,
  Gunion, and McElrath}}]{Dermisek:2006py}
\bibinfo{author}{\bibnamefont{Dermisek}, \bibfnamefont{R.}},
  \bibinfo{author}{\bibfnamefont{J.~F.} \bibnamefont{Gunion}}, and
  \bibinfo{author}{\bibfnamefont{B.}~\bibnamefont{McElrath}},
  \bibinfo{year}{2007}, \bibinfo{journal}{Phys. Rev.}
  \textbf{\bibinfo{volume}{D76}}, \bibinfo{pages}{051105}.

\bibitem[{\citenamefont{Descotes-Genon and
  Sachrajda}(2004)}]{Descotes-Genon:2004hd}
\bibinfo{author}{\bibnamefont{Descotes-Genon}, \bibfnamefont{S.}}, and
  \bibinfo{author}{\bibfnamefont{C.~T.} \bibnamefont{Sachrajda}},
  \bibinfo{year}{2004}, \bibinfo{journal}{Nucl. Phys.}
  \textbf{\bibinfo{volume}{B693}}, \bibinfo{pages}{103}.

\bibitem[{\citenamefont{Diehl and Hiller}(2001)}]{Diehl:2001ey}
\bibinfo{author}{\bibnamefont{Diehl}, \bibfnamefont{M.}}, and
  \bibinfo{author}{\bibfnamefont{G.}~\bibnamefont{Hiller}},
  \bibinfo{year}{2001}, \bibinfo{journal}{Phys. Lett.}
  \textbf{\bibinfo{volume}{B517}}, \bibinfo{pages}{125}.

\bibitem[{\citenamefont{Dijkstra}(2007)}]{Dijkstra:2007hm}
\bibinfo{author}{\bibnamefont{Dijkstra}, \bibfnamefont{H.}},
  \bibinfo{year}{2007}, \eprint{arXiv:0708.2665 [hep-ex]}.

\bibitem[{\citenamefont{Dimopoulos and Sutter}(1995)}]{Dimopoulos:1995ju}
\bibinfo{author}{\bibnamefont{Dimopoulos}, \bibfnamefont{S.}}, and
  \bibinfo{author}{\bibfnamefont{D.~W.} \bibnamefont{Sutter}},
  \bibinfo{year}{1995}, \bibinfo{journal}{Nucl. Phys.}
  \textbf{\bibinfo{volume}{B452}}, \bibinfo{pages}{496}.

\bibitem[{\citenamefont{Dine} \emph{et~al.}(1993)\citenamefont{Dine, Leigh, and
  Kagan}}]{Dine:1993np}
\bibinfo{author}{\bibnamefont{Dine}, \bibfnamefont{M.}},
  \bibinfo{author}{\bibfnamefont{R.~G.} \bibnamefont{Leigh}}, and
  \bibinfo{author}{\bibfnamefont{A.}~\bibnamefont{Kagan}},
  \bibinfo{year}{1993}, \bibinfo{journal}{Phys. Rev.}
  \textbf{\bibinfo{volume}{D48}}, \bibinfo{pages}{4269}.

\bibitem[{Dohmen \emph{et~al.}(1993)\citenamefont{Dohmen}
  \emph{et~al.}}]{Dohmen:1993mp}
\bibinfo{author}{\bibnamefont{Dohmen}, \bibfnamefont{C.}}, \emph{et~al.}
  (\bibinfo{collaboration}{SINDRUM II.}), \bibinfo{year}{1993},
  \bibinfo{journal}{Phys. Lett.} \textbf{\bibinfo{volume}{B317}},
  \bibinfo{pages}{631}.

\bibitem[{\citenamefont{Drutskoy}(2006)}]{Drutskoy:2006dw}
\bibinfo{author}{\bibnamefont{Drutskoy}, \bibfnamefont{A.}},
  \bibinfo{year}{2006}, \eprint{hep-ex/0605110}.

\bibitem[{\citenamefont{Dunietz}(1995)}]{Dunietz:1995cp}
\bibinfo{author}{\bibnamefont{Dunietz}, \bibfnamefont{I.}},
  \bibinfo{year}{1995}, \bibinfo{journal}{Phys. Rev.}
  \textbf{\bibinfo{volume}{D52}}, \bibinfo{pages}{3048}.

\bibitem[{\citenamefont{Dunietz}(1998)}]{Dunietz:1997in}
\bibinfo{author}{\bibnamefont{Dunietz}, \bibfnamefont{I.}},
  \bibinfo{year}{1998}, \bibinfo{journal}{Phys. Lett.}
  \textbf{\bibinfo{volume}{B427}}, \bibinfo{pages}{179}.

\bibitem[{\citenamefont{Dunietz} \emph{et~al.}(2001)\citenamefont{Dunietz,
  Fleischer, and Nierste}}]{Dunietz:2000cr}
\bibinfo{author}{\bibnamefont{Dunietz}, \bibfnamefont{I.}},
  \bibinfo{author}{\bibfnamefont{R.}~\bibnamefont{Fleischer}}, and
  \bibinfo{author}{\bibfnamefont{U.}~\bibnamefont{Nierste}},
  \bibinfo{year}{2001}, \bibinfo{journal}{Phys. Rev.}
  \textbf{\bibinfo{volume}{D63}}, \bibinfo{pages}{114015}.

\bibitem[{Eguchi \emph{et~al.}(2003)\citenamefont{Eguchi}
  \emph{et~al.}}]{Eguchi:2002dm}
\bibinfo{author}{\bibnamefont{Eguchi}, \bibfnamefont{K.}}, \emph{et~al.}
  (\bibinfo{collaboration}{KamLAND}), \bibinfo{year}{2003},
  \bibinfo{journal}{Phys. Rev. Lett.} \textbf{\bibinfo{volume}{90}},
  \bibinfo{pages}{021802}.

\bibitem[{\citenamefont{Ellis}
  \emph{et~al.}(2007{\natexlab{a}})\citenamefont{Ellis, Lee, and
  Pilaftsis}}]{Ellis:2007kb}
\bibinfo{author}{\bibnamefont{Ellis}, \bibfnamefont{J.}},
  \bibinfo{author}{\bibfnamefont{J.~S.} \bibnamefont{Lee}}, and
  \bibinfo{author}{\bibfnamefont{A.}~\bibnamefont{Pilaftsis}},
  \bibinfo{year}{2007}{\natexlab{a}}, \bibinfo{journal}{Phys. Rev.}
  \textbf{\bibinfo{volume}{D76}}, \bibinfo{pages}{115011}.

\bibitem[{\citenamefont{Ellis}
  \emph{et~al.}(2007{\natexlab{b}})\citenamefont{Ellis, Heinemeyer, Olive,
  Weber, and Weiglein}}]{Ellis:2007fu}
\bibinfo{author}{\bibnamefont{Ellis}, \bibfnamefont{J.~R.}},
  \bibinfo{author}{\bibfnamefont{S.}~\bibnamefont{Heinemeyer}},
  \bibinfo{author}{\bibfnamefont{K.~A.} \bibnamefont{Olive}},
  \bibinfo{author}{\bibfnamefont{A.~M.} \bibnamefont{Weber}}, and
  \bibinfo{author}{\bibfnamefont{G.}~\bibnamefont{Weiglein}},
  \bibinfo{year}{2007}{\natexlab{b}}, \bibinfo{journal}{JHEP}
  \textbf{\bibinfo{volume}{08}}, \bibinfo{pages}{083}.

\bibitem[{\citenamefont{Ellis} \emph{et~al.}(2002)\citenamefont{Ellis, Hisano,
  Raidal, and Shimizu}}]{Ellis:2002fe}
\bibinfo{author}{\bibnamefont{Ellis}, \bibfnamefont{J.~R.}},
  \bibinfo{author}{\bibfnamefont{J.}~\bibnamefont{Hisano}},
  \bibinfo{author}{\bibfnamefont{M.}~\bibnamefont{Raidal}}, and
  \bibinfo{author}{\bibfnamefont{Y.}~\bibnamefont{Shimizu}},
  \bibinfo{year}{2002}, \bibinfo{journal}{Phys. Rev.}
  \textbf{\bibinfo{volume}{D66}}, \bibinfo{pages}{115013}.

\bibitem[{\citenamefont{Ellis} \emph{et~al.}(1986)\citenamefont{Ellis, Joshi,
  and Matsuda}}]{Ellis:1985xy}
\bibinfo{author}{\bibnamefont{Ellis}, \bibfnamefont{R.~G.}},
  \bibinfo{author}{\bibfnamefont{G.~C.} \bibnamefont{Joshi}}, and
  \bibinfo{author}{\bibfnamefont{M.}~\bibnamefont{Matsuda}},
  \bibinfo{year}{1986}, \bibinfo{journal}{Phys. Lett.}
  \textbf{\bibinfo{volume}{B179}}, \bibinfo{pages}{119}.

\bibitem[{\citenamefont{Engelhard} \emph{et~al.}(2005)\citenamefont{Engelhard,
  Nir, and Raz}}]{Engelhard:2005hu}
\bibinfo{author}{\bibnamefont{Engelhard}, \bibfnamefont{G.}},
  \bibinfo{author}{\bibfnamefont{Y.}~\bibnamefont{Nir}}, and
  \bibinfo{author}{\bibfnamefont{G.}~\bibnamefont{Raz}}, \bibinfo{year}{2005},
  \bibinfo{journal}{Phys. Rev.} \textbf{\bibinfo{volume}{D72}},
  \bibinfo{pages}{075013}.

\bibitem[{\citenamefont{Engelhard and Raz}(2005)}]{Engelhard:2005ky}
\bibinfo{author}{\bibnamefont{Engelhard}, \bibfnamefont{G.}}, and
  \bibinfo{author}{\bibfnamefont{G.}~\bibnamefont{Raz}}, \bibinfo{year}{2005},
  \bibinfo{journal}{Phys. Rev.} \textbf{\bibinfo{volume}{D72}},
  \bibinfo{pages}{114017}.

\bibitem[{\citenamefont{Fajfer} \emph{et~al.}(2006)\citenamefont{Fajfer,
  Kamenik, and Kosnik}}]{Fajfer:2006av}
\bibinfo{author}{\bibnamefont{Fajfer}, \bibfnamefont{S.}},
  \bibinfo{author}{\bibfnamefont{J.}~\bibnamefont{Kamenik}}, and
  \bibinfo{author}{\bibfnamefont{N.}~\bibnamefont{Kosnik}},
  \bibinfo{year}{2006}, \bibinfo{journal}{Phys. Rev.}
  \textbf{\bibinfo{volume}{D74}}, \bibinfo{pages}{034027}.

\bibitem[{\citenamefont{Fajfer and Singer}(2000)}]{Fajfer:2000ny}
\bibinfo{author}{\bibnamefont{Fajfer}, \bibfnamefont{S.}}, and
  \bibinfo{author}{\bibfnamefont{P.}~\bibnamefont{Singer}},
  \bibinfo{year}{2000}, \bibinfo{journal}{Phys. Rev.}
  \textbf{\bibinfo{volume}{D62}}, \bibinfo{pages}{117702}.

\bibitem[{\citenamefont{Falk} \emph{et~al.}(2004)\citenamefont{Falk, Ligeti,
  Nir, and Quinn}}]{Falk:2003uq}
\bibinfo{author}{\bibnamefont{Falk}, \bibfnamefont{A.~F.}},
  \bibinfo{author}{\bibfnamefont{Z.}~\bibnamefont{Ligeti}},
  \bibinfo{author}{\bibfnamefont{Y.}~\bibnamefont{Nir}}, and
  \bibinfo{author}{\bibfnamefont{H.}~\bibnamefont{Quinn}},
  \bibinfo{year}{2004}, \bibinfo{journal}{Phys. Rev.}
  \textbf{\bibinfo{volume}{D69}}, \bibinfo{pages}{011502}.

\bibitem[{\citenamefont{Falk} \emph{et~al.}(1994)\citenamefont{Falk, Luke, and
  Savage}}]{Falk:1993dh}
\bibinfo{author}{\bibnamefont{Falk}, \bibfnamefont{A.~F.}},
  \bibinfo{author}{\bibfnamefont{M.~E.} \bibnamefont{Luke}}, and
  \bibinfo{author}{\bibfnamefont{M.~J.} \bibnamefont{Savage}},
  \bibinfo{year}{1994}, \bibinfo{journal}{Phys. Rev.}
  \textbf{\bibinfo{volume}{D49}}, \bibinfo{pages}{3367}.

\bibitem[{\citenamefont{Falk and Petrov}(2000)}]{Falk:2000ga}
\bibinfo{author}{\bibnamefont{Falk}, \bibfnamefont{A.~F.}}, and
  \bibinfo{author}{\bibfnamefont{A.~A.} \bibnamefont{Petrov}},
  \bibinfo{year}{2000}, \bibinfo{journal}{Phys. Rev. Lett.}
  \textbf{\bibinfo{volume}{85}}, \bibinfo{pages}{252}.

\bibitem[{\citenamefont{Feldmann and Mannel}(2007)}]{Feldmann:2006jk}
\bibinfo{author}{\bibnamefont{Feldmann}, \bibfnamefont{T.}}, and
  \bibinfo{author}{\bibfnamefont{T.}~\bibnamefont{Mannel}},
  \bibinfo{year}{2007}, \bibinfo{journal}{JHEP} \textbf{\bibinfo{volume}{02}},
  \bibinfo{pages}{067}.

\bibitem[{\citenamefont{Feng} \emph{et~al.}(2007)\citenamefont{Feng, Lester,
  Nir, and Shadmi}}]{Feng:2007ke}
\bibinfo{author}{\bibnamefont{Feng}, \bibfnamefont{J.~L.}},
  \bibinfo{author}{\bibfnamefont{C.~G.} \bibnamefont{Lester}},
  \bibinfo{author}{\bibfnamefont{Y.}~\bibnamefont{Nir}}, and
  \bibinfo{author}{\bibfnamefont{Y.}~\bibnamefont{Shadmi}},
  \bibinfo{year}{2007}, \eprint{arXiv:0712.0674 [hep-ph]}.

\bibitem[{\citenamefont{Fitzpatrick}
  \emph{et~al.}(2007)\citenamefont{Fitzpatrick, Perez, and
  Randall}}]{Fitzpatrick:2007sa}
\bibinfo{author}{\bibnamefont{Fitzpatrick}, \bibfnamefont{A.~L.}},
  \bibinfo{author}{\bibfnamefont{G.}~\bibnamefont{Perez}}, and
  \bibinfo{author}{\bibfnamefont{L.}~\bibnamefont{Randall}},
  \bibinfo{year}{2007}, \eprint{arXiv:0710.1869 [hep-ph]}.

\bibitem[{\citenamefont{Fleischer}(1997)}]{Fleischer:1996bv}
\bibinfo{author}{\bibnamefont{Fleischer}, \bibfnamefont{R.}},
  \bibinfo{year}{1997}, \bibinfo{journal}{Int. J. Mod. Phys.}
  \textbf{\bibinfo{volume}{A12}}, \bibinfo{pages}{2459}.

\bibitem[{\citenamefont{Fleischer}(2003)}]{Fleischer:2003ai}
\bibinfo{author}{\bibnamefont{Fleischer}, \bibfnamefont{R.}},
  \bibinfo{year}{2003}, \bibinfo{journal}{Phys. Lett.}
  \textbf{\bibinfo{volume}{B562}}, \bibinfo{pages}{234}.

\bibitem[{\citenamefont{Fleischer}(2004)}]{Fleischer:2003rx}
\bibinfo{author}{\bibnamefont{Fleischer}, \bibfnamefont{R.}},
  \bibinfo{year}{2004}, \bibinfo{journal}{Eur. Phys. J.}
  \textbf{\bibinfo{volume}{C33}}, \bibinfo{pages}{s268}.

\bibitem[{\citenamefont{Fleischer} \emph{et~al.}(2007)\citenamefont{Fleischer,
  Recksiegel, and Schwab}}]{Fleischer:2007mq}
\bibinfo{author}{\bibnamefont{Fleischer}, \bibfnamefont{R.}},
  \bibinfo{author}{\bibfnamefont{S.}~\bibnamefont{Recksiegel}}, and
  \bibinfo{author}{\bibfnamefont{F.}~\bibnamefont{Schwab}},
  \bibinfo{year}{2007}, \bibinfo{journal}{Eur. Phys. J.}
  \textbf{\bibinfo{volume}{C51}}, \bibinfo{pages}{55}.

\bibitem[{\citenamefont{Fritzsch}(2008)}]{Fritzsch:story}
\bibinfo{author}{\bibnamefont{Fritzsch}, \bibfnamefont{H.}},
  \bibinfo{year}{2008}, \bibinfo{note}{personal communication}.

\bibitem[{Fukuda \emph{et~al.}(1998)\citenamefont{Fukuda}
  \emph{et~al.}}]{Fukuda:1998mi}
\bibinfo{author}{\bibnamefont{Fukuda}, \bibfnamefont{Y.}}, \emph{et~al.}
  (\bibinfo{collaboration}{Super-Kamiokande}), \bibinfo{year}{1998},
  \bibinfo{journal}{Phys. Rev. Lett.} \textbf{\bibinfo{volume}{81}},
  \bibinfo{pages}{1562}.

\bibitem[{\citenamefont{Fullana and Sanchis-Lozano}(2007)}]{Fullana:2007uq}
\bibinfo{author}{\bibnamefont{Fullana}, \bibfnamefont{E.}}, and
  \bibinfo{author}{\bibfnamefont{M.-A.} \bibnamefont{Sanchis-Lozano}},
  \bibinfo{year}{2007}, \bibinfo{journal}{Phys. Lett.}
  \textbf{\bibinfo{volume}{B653}}, \bibinfo{pages}{67}.

\bibitem[{\citenamefont{Gabbiani} \emph{et~al.}(1996)\citenamefont{Gabbiani,
  Gabrielli, Masiero, and Silvestrini}}]{Gabbiani:1996hi}
\bibinfo{author}{\bibnamefont{Gabbiani}, \bibfnamefont{F.}},
  \bibinfo{author}{\bibfnamefont{E.}~\bibnamefont{Gabrielli}},
  \bibinfo{author}{\bibfnamefont{A.}~\bibnamefont{Masiero}}, and
  \bibinfo{author}{\bibfnamefont{L.}~\bibnamefont{Silvestrini}},
  \bibinfo{year}{1996}, \bibinfo{journal}{Nucl. Phys.}
  \textbf{\bibinfo{volume}{B477}}, \bibinfo{pages}{321}.

\bibitem[{\citenamefont{Gaillard and Lee}(1974)}]{Gaillard:1974hs}
\bibinfo{author}{\bibnamefont{Gaillard}, \bibfnamefont{M.~K.}}, and
  \bibinfo{author}{\bibfnamefont{B.~W.} \bibnamefont{Lee}},
  \bibinfo{year}{1974}, \bibinfo{journal}{Phys. Rev.}
  \textbf{\bibinfo{volume}{D10}}, \bibinfo{pages}{897}.

\bibitem[{\citenamefont{Gambino} \emph{et~al.}(2003)\citenamefont{Gambino,
  Gorbahn, and Haisch}}]{Gambino:2003zm}
\bibinfo{author}{\bibnamefont{Gambino}, \bibfnamefont{P.}},
  \bibinfo{author}{\bibfnamefont{M.}~\bibnamefont{Gorbahn}}, and
  \bibinfo{author}{\bibfnamefont{U.}~\bibnamefont{Haisch}},
  \bibinfo{year}{2003}, \bibinfo{journal}{Nucl. Phys.}
  \textbf{\bibinfo{volume}{B673}}, \bibinfo{pages}{238}.

\bibitem[{\citenamefont{Gambino and Haisch}(2000)}]{Gambino:2000fz}
\bibinfo{author}{\bibnamefont{Gambino}, \bibfnamefont{P.}}, and
  \bibinfo{author}{\bibfnamefont{U.}~\bibnamefont{Haisch}},
  \bibinfo{year}{2000}, \bibinfo{journal}{JHEP} \textbf{\bibinfo{volume}{09}},
  \bibinfo{pages}{001}.

\bibitem[{\citenamefont{Gambino and Haisch}(2001)}]{Gambino:2001au}
\bibinfo{author}{\bibnamefont{Gambino}, \bibfnamefont{P.}}, and
  \bibinfo{author}{\bibfnamefont{U.}~\bibnamefont{Haisch}},
  \bibinfo{year}{2001}, \bibinfo{journal}{JHEP} \textbf{\bibinfo{volume}{10}},
  \bibinfo{pages}{020}.

\bibitem[{\citenamefont{Gambino} \emph{et~al.}(2005)\citenamefont{Gambino,
  Haisch, and Misiak}}]{Gambino:2004mv}
\bibinfo{author}{\bibnamefont{Gambino}, \bibfnamefont{P.}},
  \bibinfo{author}{\bibfnamefont{U.}~\bibnamefont{Haisch}}, and
  \bibinfo{author}{\bibfnamefont{M.}~\bibnamefont{Misiak}},
  \bibinfo{year}{2005}, \bibinfo{journal}{Phys. Rev. Lett.}
  \textbf{\bibinfo{volume}{94}}, \bibinfo{pages}{061803}.

\bibitem[{\citenamefont{Gardner}(1999)}]{Gardner:1998gz}
\bibinfo{author}{\bibnamefont{Gardner}, \bibfnamefont{S.}},
  \bibinfo{year}{1999}, \bibinfo{journal}{Phys. Rev.}
  \textbf{\bibinfo{volume}{D59}}, \bibinfo{pages}{077502}.

\bibitem[{\citenamefont{Garisto}(1995)}]{Garisto:1994vz}
\bibinfo{author}{\bibnamefont{Garisto}, \bibfnamefont{R.}},
  \bibinfo{year}{1995}, \bibinfo{journal}{Phys. Rev.}
  \textbf{\bibinfo{volume}{D51}}, \bibinfo{pages}{1107}.

\bibitem[{\citenamefont{Gavela} \emph{et~al.}(1994)\citenamefont{Gavela,
  Hernandez, Orloff, Pene, and Quimbay}}]{Gavela:1994dt}
\bibinfo{author}{\bibnamefont{Gavela}, \bibfnamefont{M.~B.}},
  \bibinfo{author}{\bibfnamefont{P.}~\bibnamefont{Hernandez}},
  \bibinfo{author}{\bibfnamefont{J.}~\bibnamefont{Orloff}},
  \bibinfo{author}{\bibfnamefont{O.}~\bibnamefont{Pene}}, and
  \bibinfo{author}{\bibfnamefont{C.}~\bibnamefont{Quimbay}},
  \bibinfo{year}{1994}, \bibinfo{journal}{Nucl. Phys.}
  \textbf{\bibinfo{volume}{B430}}, \bibinfo{pages}{382}.

\bibitem[{\citenamefont{Gemintern} \emph{et~al.}(2004)\citenamefont{Gemintern,
  Bar-Shalom, and Eilam}}]{Gemintern:2004bw}
\bibinfo{author}{\bibnamefont{Gemintern}, \bibfnamefont{A.}},
  \bibinfo{author}{\bibfnamefont{S.}~\bibnamefont{Bar-Shalom}}, and
  \bibinfo{author}{\bibfnamefont{G.}~\bibnamefont{Eilam}},
  \bibinfo{year}{2004}, \bibinfo{journal}{Phys. Rev.}
  \textbf{\bibinfo{volume}{D70}}, \bibinfo{pages}{035008}.

\bibitem[{\citenamefont{Georgi}(2007)}]{Georgi:2007ek}
\bibinfo{author}{\bibnamefont{Georgi}, \bibfnamefont{H.}},
  \bibinfo{year}{2007}, \bibinfo{journal}{Phys. Rev. Lett.}
  \textbf{\bibinfo{volume}{98}}, \bibinfo{pages}{221601}.

\bibitem[{\citenamefont{Gershon and Hazumi}(2004)}]{Gershon:2004tk}
\bibinfo{author}{\bibnamefont{Gershon}, \bibfnamefont{T.}}, and
  \bibinfo{author}{\bibfnamefont{M.}~\bibnamefont{Hazumi}},
  \bibinfo{year}{2004}, \bibinfo{journal}{Phys. Lett.}
  \textbf{\bibinfo{volume}{B596}}, \bibinfo{pages}{163}.

\bibitem[{\citenamefont{Gershon and Soni}(2007)}]{Gershon:2006mt}
\bibinfo{author}{\bibnamefont{Gershon}, \bibfnamefont{T.}}, and
  \bibinfo{author}{\bibfnamefont{A.}~\bibnamefont{Soni}}, \bibinfo{year}{2007},
  \bibinfo{journal}{J. Phys.} \textbf{\bibinfo{volume}{G33}},
  \bibinfo{pages}{479}.

\bibitem[{\citenamefont{Gherghetta and Pomarol}(2000)}]{Gherghetta:2000qt}
\bibinfo{author}{\bibnamefont{Gherghetta}, \bibfnamefont{T.}}, and
  \bibinfo{author}{\bibfnamefont{A.}~\bibnamefont{Pomarol}},
  \bibinfo{year}{2000}, \bibinfo{journal}{Nucl. Phys.}
  \textbf{\bibinfo{volume}{B586}}, \bibinfo{pages}{141}.

\bibitem[{\citenamefont{Ghinculov} \emph{et~al.}(2003)\citenamefont{Ghinculov,
  Hurth, Isidori, and Yao}}]{Ghinculov:2002pe}
\bibinfo{author}{\bibnamefont{Ghinculov}, \bibfnamefont{A.}},
  \bibinfo{author}{\bibfnamefont{T.}~\bibnamefont{Hurth}},
  \bibinfo{author}{\bibfnamefont{G.}~\bibnamefont{Isidori}}, and
  \bibinfo{author}{\bibfnamefont{Y.~P.} \bibnamefont{Yao}},
  \bibinfo{year}{2003}, \bibinfo{journal}{Nucl. Phys.}
  \textbf{\bibinfo{volume}{B648}}, \bibinfo{pages}{254}.

\bibitem[{\citenamefont{Ghinculov} \emph{et~al.}(2004)\citenamefont{Ghinculov,
  Hurth, Isidori, and Yao}}]{Ghinculov:2003qd}
\bibinfo{author}{\bibnamefont{Ghinculov}, \bibfnamefont{A.}},
  \bibinfo{author}{\bibfnamefont{T.}~\bibnamefont{Hurth}},
  \bibinfo{author}{\bibfnamefont{G.}~\bibnamefont{Isidori}}, and
  \bibinfo{author}{\bibfnamefont{Y.~P.} \bibnamefont{Yao}},
  \bibinfo{year}{2004}, \bibinfo{journal}{Nucl. Phys.}
  \textbf{\bibinfo{volume}{B685}}, \bibinfo{pages}{351}.

\bibitem[{\citenamefont{Girardello and Grisaru}(1982)}]{Girardello:1981wz}
\bibinfo{author}{\bibnamefont{Girardello}, \bibfnamefont{L.}}, and
  \bibinfo{author}{\bibfnamefont{M.~T.} \bibnamefont{Grisaru}},
  \bibinfo{year}{1982}, \bibinfo{journal}{Nucl. Phys.}
  \textbf{\bibinfo{volume}{B194}}, \bibinfo{pages}{65}.

\bibitem[{\citenamefont{Giri} \emph{et~al.}(2003)\citenamefont{Giri, Grossman,
  Soffer, and Zupan}}]{Giri:2003ty}
\bibinfo{author}{\bibnamefont{Giri}, \bibfnamefont{A.}},
  \bibinfo{author}{\bibfnamefont{Y.}~\bibnamefont{Grossman}},
  \bibinfo{author}{\bibfnamefont{A.}~\bibnamefont{Soffer}}, and
  \bibinfo{author}{\bibfnamefont{J.}~\bibnamefont{Zupan}},
  \bibinfo{year}{2003}, \bibinfo{journal}{Phys. Rev.}
  \textbf{\bibinfo{volume}{D68}}, \bibinfo{pages}{054018}.

\bibitem[{\citenamefont{Giudice and Rattazzi}(1999)}]{Giudice:1998bp}
\bibinfo{author}{\bibnamefont{Giudice}, \bibfnamefont{G.~F.}}, and
  \bibinfo{author}{\bibfnamefont{R.}~\bibnamefont{Rattazzi}},
  \bibinfo{year}{1999}, \bibinfo{journal}{Phys. Rept.}
  \textbf{\bibinfo{volume}{322}}, \bibinfo{pages}{419}.

\bibitem[{\citenamefont{Glashow} \emph{et~al.}(1970)\citenamefont{Glashow,
  Iliopoulos, and Maiani}}]{Glashow:1970gm}
\bibinfo{author}{\bibnamefont{Glashow}, \bibfnamefont{S.~L.}},
  \bibinfo{author}{\bibfnamefont{J.}~\bibnamefont{Iliopoulos}}, and
  \bibinfo{author}{\bibfnamefont{L.}~\bibnamefont{Maiani}},
  \bibinfo{year}{1970}, \bibinfo{journal}{Phys. Rev.}
  \textbf{\bibinfo{volume}{D2}}, \bibinfo{pages}{1285}.

\bibitem[{\citenamefont{Glashow and Weinberg}(1977)}]{Glashow:1976nt}
\bibinfo{author}{\bibnamefont{Glashow}, \bibfnamefont{S.~L.}}, and
  \bibinfo{author}{\bibfnamefont{S.}~\bibnamefont{Weinberg}},
  \bibinfo{year}{1977}, \bibinfo{journal}{Phys. Rev.}
  \textbf{\bibinfo{volume}{D15}}, \bibinfo{pages}{1958}.

\bibitem[{\citenamefont{Golowich} \emph{et~al.}(2007)\citenamefont{Golowich,
  Hewett, Pakvasa, and Petrov}}]{Golowich:2007ka}
\bibinfo{author}{\bibnamefont{Golowich}, \bibfnamefont{E.}},
  \bibinfo{author}{\bibfnamefont{J.}~\bibnamefont{Hewett}},
  \bibinfo{author}{\bibfnamefont{S.}~\bibnamefont{Pakvasa}}, and
  \bibinfo{author}{\bibfnamefont{A.~A.} \bibnamefont{Petrov}},
  \bibinfo{year}{2007}, \bibinfo{journal}{Phys. Rev.}
  \textbf{\bibinfo{volume}{D76}}, \bibinfo{pages}{095009}.

\bibitem[{\citenamefont{Gomez and Goldberg}(1996)}]{Gomez:1995cv}
\bibinfo{author}{\bibnamefont{Gomez}, \bibfnamefont{M.~E.}}, and
  \bibinfo{author}{\bibfnamefont{H.}~\bibnamefont{Goldberg}},
  \bibinfo{year}{1996}, \bibinfo{journal}{Phys. Rev.}
  \textbf{\bibinfo{volume}{D53}}, \bibinfo{pages}{5244}.

\bibitem[{\citenamefont{Gorbahn and Haisch}(2005)}]{Gorbahn:2004my}
\bibinfo{author}{\bibnamefont{Gorbahn}, \bibfnamefont{M.}}, and
  \bibinfo{author}{\bibfnamefont{U.}~\bibnamefont{Haisch}},
  \bibinfo{year}{2005}, \bibinfo{journal}{Nucl. Phys.}
  \textbf{\bibinfo{volume}{B713}}, \bibinfo{pages}{291}.

\bibitem[{\citenamefont{Gorbahn} \emph{et~al.}(2005)\citenamefont{Gorbahn,
  Haisch, and Misiak}}]{Gorbahn:2005sa}
\bibinfo{author}{\bibnamefont{Gorbahn}, \bibfnamefont{M.}},
  \bibinfo{author}{\bibfnamefont{U.}~\bibnamefont{Haisch}}, and
  \bibinfo{author}{\bibfnamefont{M.}~\bibnamefont{Misiak}},
  \bibinfo{year}{2005}, \bibinfo{journal}{Phys. Rev. Lett.}
  \textbf{\bibinfo{volume}{95}}, \bibinfo{pages}{102004}.

\bibitem[{\citenamefont{Goto} \emph{et~al.}(2002)\citenamefont{Goto, Okada,
  Shimizu, Shindou, and Tanaka}}]{Goto:2002xt}
\bibinfo{author}{\bibnamefont{Goto}, \bibfnamefont{T.}},
  \bibinfo{author}{\bibfnamefont{Y.}~\bibnamefont{Okada}},
  \bibinfo{author}{\bibfnamefont{Y.}~\bibnamefont{Shimizu}},
  \bibinfo{author}{\bibfnamefont{T.}~\bibnamefont{Shindou}}, and
  \bibinfo{author}{\bibfnamefont{M.}~\bibnamefont{Tanaka}},
  \bibinfo{year}{2002}, \bibinfo{journal}{Phys. Rev.}
  \textbf{\bibinfo{volume}{D66}}, \bibinfo{pages}{035009}.

\bibitem[{\citenamefont{Goto} \emph{et~al.}(2004)\citenamefont{Goto, Okada,
  Shimizu, Shindou, and Tanaka}}]{Goto:2003iu}
\bibinfo{author}{\bibnamefont{Goto}, \bibfnamefont{T.}},
  \bibinfo{author}{\bibfnamefont{Y.}~\bibnamefont{Okada}},
  \bibinfo{author}{\bibfnamefont{Y.}~\bibnamefont{Shimizu}},
  \bibinfo{author}{\bibfnamefont{T.}~\bibnamefont{Shindou}}, and
  \bibinfo{author}{\bibfnamefont{M.}~\bibnamefont{Tanaka}},
  \bibinfo{year}{2004}, \bibinfo{journal}{Phys. Rev.}
  \textbf{\bibinfo{volume}{D70}}, \bibinfo{pages}{035012}.

\bibitem[{\citenamefont{Goto} \emph{et~al.}(2007)\citenamefont{Goto, Okada,
  Shindou, and Tanaka}}]{Goto:2007ee}
\bibinfo{author}{\bibnamefont{Goto}, \bibfnamefont{T.}},
  \bibinfo{author}{\bibfnamefont{Y.}~\bibnamefont{Okada}},
  \bibinfo{author}{\bibfnamefont{T.}~\bibnamefont{Shindou}}, and
  \bibinfo{author}{\bibfnamefont{M.}~\bibnamefont{Tanaka}},
  \bibinfo{year}{2007}, \eprint{arXiv:0711.2935 [hep-ph]}.

\bibitem[{\citenamefont{de~Gouvea and Jenkins}(2007)}]{deGouvea:2007xp}
\bibinfo{author}{\bibnamefont{de~Gouvea}, \bibfnamefont{A.}}, and
  \bibinfo{author}{\bibfnamefont{J.}~\bibnamefont{Jenkins}},
  \bibinfo{year}{2007}, \eprint{arXiv:0708.1344 [hep-ph]}.

\bibitem[{\citenamefont{Grassi}(2005)}]{Grassi:2005ac}
\bibinfo{author}{\bibnamefont{Grassi}, \bibfnamefont{M.}}
  (\bibinfo{collaboration}{MEG}), \bibinfo{year}{2005}, \bibinfo{journal}{Nucl.
  Phys. Proc. Suppl.} \textbf{\bibinfo{volume}{149}}, \bibinfo{pages}{369}.

\bibitem[{\citenamefont{Gremm} \emph{et~al.}(1995)\citenamefont{Gremm, Kruger,
  and Sehgal}}]{Gremm:1995nx}
\bibinfo{author}{\bibnamefont{Gremm}, \bibfnamefont{M.}},
  \bibinfo{author}{\bibfnamefont{F.}~\bibnamefont{Kruger}}, and
  \bibinfo{author}{\bibfnamefont{L.~M.} \bibnamefont{Sehgal}},
  \bibinfo{year}{1995}, \bibinfo{journal}{Phys. Lett.}
  \textbf{\bibinfo{volume}{B355}}, \bibinfo{pages}{579}.

\bibitem[{\citenamefont{Grinstein} \emph{et~al.}(2007)\citenamefont{Grinstein,
  Cirigliano, Isidori, and Wise}}]{Grinstein:2006cg}
\bibinfo{author}{\bibnamefont{Grinstein}, \bibfnamefont{B.}},
  \bibinfo{author}{\bibfnamefont{V.}~\bibnamefont{Cirigliano}},
  \bibinfo{author}{\bibfnamefont{G.}~\bibnamefont{Isidori}}, and
  \bibinfo{author}{\bibfnamefont{M.~B.} \bibnamefont{Wise}},
  \bibinfo{year}{2007}, \bibinfo{journal}{Nucl. Phys.}
  \textbf{\bibinfo{volume}{B763}}, \bibinfo{pages}{35}.

\bibitem[{\citenamefont{Grinstein} \emph{et~al.}(2005)\citenamefont{Grinstein,
  Grossman, Ligeti, and Pirjol}}]{Grinstein:2004uu}
\bibinfo{author}{\bibnamefont{Grinstein}, \bibfnamefont{B.}},
  \bibinfo{author}{\bibfnamefont{Y.}~\bibnamefont{Grossman}},
  \bibinfo{author}{\bibfnamefont{Z.}~\bibnamefont{Ligeti}}, and
  \bibinfo{author}{\bibfnamefont{D.}~\bibnamefont{Pirjol}},
  \bibinfo{year}{2005}, \bibinfo{journal}{Phys. Rev.}
  \textbf{\bibinfo{volume}{D71}}, \bibinfo{pages}{011504}.

\bibitem[{\citenamefont{Grinstein} \emph{et~al.}(2008)\citenamefont{Grinstein,
  Intriligator, and Rothstein}}]{Grinstein:2008qk}
\bibinfo{author}{\bibnamefont{Grinstein}, \bibfnamefont{B.}},
  \bibinfo{author}{\bibfnamefont{K.}~\bibnamefont{Intriligator}}, and
  \bibinfo{author}{\bibfnamefont{I.~Z.} \bibnamefont{Rothstein}},
  \bibinfo{year}{2008}, \eprint{arXiv:0801.1140 [hep-ph]}.

\bibitem[{\citenamefont{Grinstein and Pirjol}(2000)}]{Grinstein:2000pc}
\bibinfo{author}{\bibnamefont{Grinstein}, \bibfnamefont{B.}}, and
  \bibinfo{author}{\bibfnamefont{D.}~\bibnamefont{Pirjol}},
  \bibinfo{year}{2000}, \bibinfo{journal}{Phys. Rev.}
  \textbf{\bibinfo{volume}{D62}}, \bibinfo{pages}{093002}.

\bibitem[{\citenamefont{Grinstein and Pirjol}(2004)}]{Grinstein:2004vb}
\bibinfo{author}{\bibnamefont{Grinstein}, \bibfnamefont{B.}}, and
  \bibinfo{author}{\bibfnamefont{D.}~\bibnamefont{Pirjol}},
  \bibinfo{year}{2004}, \bibinfo{journal}{Phys. Rev.}
  \textbf{\bibinfo{volume}{D70}}, \bibinfo{pages}{114005}.

\bibitem[{\citenamefont{Grinstein and Pirjol}(2005)}]{Grinstein:2005ry}
\bibinfo{author}{\bibnamefont{Grinstein}, \bibfnamefont{B.}}, and
  \bibinfo{author}{\bibfnamefont{D.}~\bibnamefont{Pirjol}},
  \bibinfo{year}{2005}, \bibinfo{journal}{Phys. Lett.}
  \textbf{\bibinfo{volume}{B615}}, \bibinfo{pages}{213}.

\bibitem[{\citenamefont{Grinstein and
  Pirjol}(2006{\natexlab{a}})}]{Grinstein:2005nu}
\bibinfo{author}{\bibnamefont{Grinstein}, \bibfnamefont{B.}}, and
  \bibinfo{author}{\bibfnamefont{D.}~\bibnamefont{Pirjol}},
  \bibinfo{year}{2006}{\natexlab{a}}, \bibinfo{journal}{Phys. Rev.}
  \textbf{\bibinfo{volume}{D73}}, \bibinfo{pages}{014013}.

\bibitem[{\citenamefont{Grinstein and
  Pirjol}(2006{\natexlab{b}})}]{Grinstein:2005ud}
\bibinfo{author}{\bibnamefont{Grinstein}, \bibfnamefont{B.}}, and
  \bibinfo{author}{\bibfnamefont{D.}~\bibnamefont{Pirjol}},
  \bibinfo{year}{2006}{\natexlab{b}}, \bibinfo{journal}{Phys. Rev.}
  \textbf{\bibinfo{volume}{D73}}, \bibinfo{pages}{094027}.

\bibitem[{\citenamefont{Grinstein} \emph{et~al.}(1989)\citenamefont{Grinstein,
  Savage, and Wise}}]{Grinstein:1988me}
\bibinfo{author}{\bibnamefont{Grinstein}, \bibfnamefont{B.}},
  \bibinfo{author}{\bibfnamefont{M.~J.} \bibnamefont{Savage}}, and
  \bibinfo{author}{\bibfnamefont{M.~B.} \bibnamefont{Wise}},
  \bibinfo{year}{1989}, \bibinfo{journal}{Nucl. Phys.}
  \textbf{\bibinfo{volume}{B319}}, \bibinfo{pages}{271}.

\bibitem[{\citenamefont{Gronau}(1989)}]{Gronau:1989ia}
\bibinfo{author}{\bibnamefont{Gronau}, \bibfnamefont{M.}},
  \bibinfo{year}{1989}, \bibinfo{journal}{Phys. Rev. Lett.}
  \textbf{\bibinfo{volume}{63}}, \bibinfo{pages}{1451}.

\bibitem[{\citenamefont{Gronau}(2000)}]{Gronau:2000zy}
\bibinfo{author}{\bibnamefont{Gronau}, \bibfnamefont{M.}},
  \bibinfo{year}{2000}, \bibinfo{journal}{Phys. Lett.}
  \textbf{\bibinfo{volume}{B492}}, \bibinfo{pages}{297}.

\bibitem[{\citenamefont{Gronau}(2003)}]{Gronau:2002mu}
\bibinfo{author}{\bibnamefont{Gronau}, \bibfnamefont{M.}},
  \bibinfo{year}{2003}, \bibinfo{journal}{Phys. Lett.}
  \textbf{\bibinfo{volume}{B557}}, \bibinfo{pages}{198}.

\bibitem[{\citenamefont{Gronau}(2005)}]{Gronau:2005kz}
\bibinfo{author}{\bibnamefont{Gronau}, \bibfnamefont{M.}},
  \bibinfo{year}{2005}, \bibinfo{journal}{Phys. Lett.}
  \textbf{\bibinfo{volume}{B627}}, \bibinfo{pages}{82}.

\bibitem[{\citenamefont{Gronau} \emph{et~al.}(2002)\citenamefont{Gronau,
  Grossman, Pirjol, and Ryd}}]{Gronau:2001ng}
\bibinfo{author}{\bibnamefont{Gronau}, \bibfnamefont{M.}},
  \bibinfo{author}{\bibfnamefont{Y.}~\bibnamefont{Grossman}},
  \bibinfo{author}{\bibfnamefont{D.}~\bibnamefont{Pirjol}}, and
  \bibinfo{author}{\bibfnamefont{A.}~\bibnamefont{Ryd}}, \bibinfo{year}{2002},
  \bibinfo{journal}{Phys. Rev. Lett.} \textbf{\bibinfo{volume}{88}},
  \bibinfo{pages}{051802}.

\bibitem[{\citenamefont{Gronau}
  \emph{et~al.}(2006{\natexlab{a}})\citenamefont{Gronau, Grossman, Raz, and
  Rosner}}]{Gronau:2006eb}
\bibinfo{author}{\bibnamefont{Gronau}, \bibfnamefont{M.}},
  \bibinfo{author}{\bibfnamefont{Y.}~\bibnamefont{Grossman}},
  \bibinfo{author}{\bibfnamefont{G.}~\bibnamefont{Raz}}, and
  \bibinfo{author}{\bibfnamefont{J.~L.} \bibnamefont{Rosner}},
  \bibinfo{year}{2006}{\natexlab{a}}, \bibinfo{journal}{Phys. Lett.}
  \textbf{\bibinfo{volume}{B635}}, \bibinfo{pages}{207}.

\bibitem[{\citenamefont{Gronau} \emph{et~al.}(2001)\citenamefont{Gronau,
  Grossman, and Rosner}}]{Gronau:2001nr}
\bibinfo{author}{\bibnamefont{Gronau}, \bibfnamefont{M.}},
  \bibinfo{author}{\bibfnamefont{Y.}~\bibnamefont{Grossman}}, and
  \bibinfo{author}{\bibfnamefont{J.~L.} \bibnamefont{Rosner}},
  \bibinfo{year}{2001}, \bibinfo{journal}{Phys. Lett.}
  \textbf{\bibinfo{volume}{B508}}, \bibinfo{pages}{37}.

\bibitem[{\citenamefont{Gronau}
  \emph{et~al.}(2004{\natexlab{a}})\citenamefont{Gronau, Grossman, and
  Rosner}}]{Gronau:2003kx}
\bibinfo{author}{\bibnamefont{Gronau}, \bibfnamefont{M.}},
  \bibinfo{author}{\bibfnamefont{Y.}~\bibnamefont{Grossman}}, and
  \bibinfo{author}{\bibfnamefont{J.~L.} \bibnamefont{Rosner}},
  \bibinfo{year}{2004}{\natexlab{a}}, \bibinfo{journal}{Phys. Lett.}
  \textbf{\bibinfo{volume}{B579}}, \bibinfo{pages}{331}.

\bibitem[{\citenamefont{Gronau}
  \emph{et~al.}(2004{\natexlab{b}})\citenamefont{Gronau, Grossman, Shuhmaher,
  Soffer, and Zupan}}]{Gronau:2004gt}
\bibinfo{author}{\bibnamefont{Gronau}, \bibfnamefont{M.}},
  \bibinfo{author}{\bibfnamefont{Y.}~\bibnamefont{Grossman}},
  \bibinfo{author}{\bibfnamefont{N.}~\bibnamefont{Shuhmaher}},
  \bibinfo{author}{\bibfnamefont{A.}~\bibnamefont{Soffer}}, and
  \bibinfo{author}{\bibfnamefont{J.}~\bibnamefont{Zupan}},
  \bibinfo{year}{2004}{\natexlab{b}}, \bibinfo{journal}{Phys. Rev.}
  \textbf{\bibinfo{volume}{D69}}, \bibinfo{pages}{113003}.

\bibitem[{\citenamefont{Gronau} \emph{et~al.}(2007)\citenamefont{Gronau,
  Grossman, Surujon, and Zupan}}]{Gronau:2007bh}
\bibinfo{author}{\bibnamefont{Gronau}, \bibfnamefont{M.}},
  \bibinfo{author}{\bibfnamefont{Y.}~\bibnamefont{Grossman}},
  \bibinfo{author}{\bibfnamefont{Z.}~\bibnamefont{Surujon}}, and
  \bibinfo{author}{\bibfnamefont{J.}~\bibnamefont{Zupan}},
  \bibinfo{year}{2007}, \bibinfo{journal}{Phys. Lett.}
  \textbf{\bibinfo{volume}{B649}}, \bibinfo{pages}{61}.

\bibitem[{\citenamefont{Gronau and London}(1990)}]{Gronau:1990ka}
\bibinfo{author}{\bibnamefont{Gronau}, \bibfnamefont{M.}}, and
  \bibinfo{author}{\bibfnamefont{D.}~\bibnamefont{London}},
  \bibinfo{year}{1990}, \bibinfo{journal}{Phys. Rev. Lett.}
  \textbf{\bibinfo{volume}{65}}, \bibinfo{pages}{3381}.

\bibitem[{\citenamefont{Gronau and London.}(1991)}]{Gronau:1990ra}
\bibinfo{author}{\bibnamefont{Gronau}, \bibfnamefont{M.}}, and
  \bibinfo{author}{\bibfnamefont{D.}~\bibnamefont{London.}},
  \bibinfo{year}{1991}, \bibinfo{journal}{Phys. Lett.}
  \textbf{\bibinfo{volume}{B253}}, \bibinfo{pages}{483}.

\bibitem[{\citenamefont{Gronau and Pirjol}(2002)}]{Gronau:2002rz}
\bibinfo{author}{\bibnamefont{Gronau}, \bibfnamefont{M.}}, and
  \bibinfo{author}{\bibfnamefont{D.}~\bibnamefont{Pirjol}},
  \bibinfo{year}{2002}, \bibinfo{journal}{Phys. Rev.}
  \textbf{\bibinfo{volume}{D66}}, \bibinfo{pages}{054008}.

\bibitem[{\citenamefont{Gronau} \emph{et~al.}(2003)\citenamefont{Gronau,
  Pirjol, and Wyler}}]{Gronau:2002nk}
\bibinfo{author}{\bibnamefont{Gronau}, \bibfnamefont{M.}},
  \bibinfo{author}{\bibfnamefont{D.}~\bibnamefont{Pirjol}}, and
  \bibinfo{author}{\bibfnamefont{D.}~\bibnamefont{Wyler}},
  \bibinfo{year}{2003}, \bibinfo{journal}{Phys. Rev. Lett.}
  \textbf{\bibinfo{volume}{90}}, \bibinfo{pages}{051801}.

\bibitem[{\citenamefont{Gronau} \emph{et~al.}(1999)\citenamefont{Gronau,
  Pirjol, and Yan}}]{Gronau:1998fn}
\bibinfo{author}{\bibnamefont{Gronau}, \bibfnamefont{M.}},
  \bibinfo{author}{\bibfnamefont{D.}~\bibnamefont{Pirjol}}, and
  \bibinfo{author}{\bibfnamefont{T.-M.} \bibnamefont{Yan}},
  \bibinfo{year}{1999}, \bibinfo{journal}{Phys. Rev.}
  \textbf{\bibinfo{volume}{D60}}, \bibinfo{pages}{034021},
  \bibinfo{note}{[Erratum-ibid.\ D {\bf 69}, 119901 (2004)]}.

\bibitem[{\citenamefont{Gronau and Rosner}(1999)}]{Gronau:1998ep}
\bibinfo{author}{\bibnamefont{Gronau}, \bibfnamefont{M.}}, and
  \bibinfo{author}{\bibfnamefont{J.~L.} \bibnamefont{Rosner}},
  \bibinfo{year}{1999}, \bibinfo{journal}{Phys. Rev.}
  \textbf{\bibinfo{volume}{D59}}, \bibinfo{pages}{113002}.

\bibitem[{\citenamefont{Gronau and Rosner}(2005)}]{Gronau:2005gz}
\bibinfo{author}{\bibnamefont{Gronau}, \bibfnamefont{M.}}, and
  \bibinfo{author}{\bibfnamefont{J.~L.} \bibnamefont{Rosner}},
  \bibinfo{year}{2005}, \bibinfo{journal}{Phys. Rev.}
  \textbf{\bibinfo{volume}{D71}}, \bibinfo{pages}{074019}.

\bibitem[{\citenamefont{Gronau}
  \emph{et~al.}(2004{\natexlab{c}})\citenamefont{Gronau, Rosner, and
  Zupan}}]{Gronau:2004hp}
\bibinfo{author}{\bibnamefont{Gronau}, \bibfnamefont{M.}},
  \bibinfo{author}{\bibfnamefont{J.~L.} \bibnamefont{Rosner}}, and
  \bibinfo{author}{\bibfnamefont{J.}~\bibnamefont{Zupan}},
  \bibinfo{year}{2004}{\natexlab{c}}, \bibinfo{journal}{Phys. Lett.}
  \textbf{\bibinfo{volume}{B596}}, \bibinfo{pages}{107}.

\bibitem[{\citenamefont{Gronau}
  \emph{et~al.}(2006{\natexlab{b}})\citenamefont{Gronau, Rosner, and
  Zupan}}]{Gronau:2006qh}
\bibinfo{author}{\bibnamefont{Gronau}, \bibfnamefont{M.}},
  \bibinfo{author}{\bibfnamefont{J.~L.} \bibnamefont{Rosner}}, and
  \bibinfo{author}{\bibfnamefont{J.}~\bibnamefont{Zupan}},
  \bibinfo{year}{2006}{\natexlab{b}}, \bibinfo{journal}{Phys. Rev.}
  \textbf{\bibinfo{volume}{D74}}, \bibinfo{pages}{093003}.

\bibitem[{\citenamefont{Gronau and Wyler}(1991)}]{Gronau:1991dp}
\bibinfo{author}{\bibnamefont{Gronau}, \bibfnamefont{M.}}, and
  \bibinfo{author}{\bibfnamefont{D.}~\bibnamefont{Wyler}},
  \bibinfo{year}{1991}, \bibinfo{journal}{Phys. Lett.}
  \textbf{\bibinfo{volume}{B265}}, \bibinfo{pages}{172}.

\bibitem[{\citenamefont{Gronau and Zupan}(2004)}]{Gronau:2004tm}
\bibinfo{author}{\bibnamefont{Gronau}, \bibfnamefont{M.}}, and
  \bibinfo{author}{\bibfnamefont{J.}~\bibnamefont{Zupan}},
  \bibinfo{year}{2004}, \bibinfo{journal}{Phys. Rev.}
  \textbf{\bibinfo{volume}{D70}}, \bibinfo{pages}{074031}.

\bibitem[{\citenamefont{Gronau and Zupan}(2005)}]{Gronau:2005pq}
\bibinfo{author}{\bibnamefont{Gronau}, \bibfnamefont{M.}}, and
  \bibinfo{author}{\bibfnamefont{J.}~\bibnamefont{Zupan}},
  \bibinfo{year}{2005}, \bibinfo{journal}{Phys. Rev.}
  \textbf{\bibinfo{volume}{D71}}, \bibinfo{pages}{074017}.

\bibitem[{\citenamefont{Grossman}(1994)}]{Grossman:1994jb}
\bibinfo{author}{\bibnamefont{Grossman}, \bibfnamefont{Y.}},
  \bibinfo{year}{1994}, \bibinfo{journal}{Nucl. Phys.}
  \textbf{\bibinfo{volume}{B426}}, \bibinfo{pages}{355}.

\bibitem[{\citenamefont{Grossman} \emph{et~al.}(2002)\citenamefont{Grossman,
  Kagan, and Ligeti}}]{Grossman:2002bu}
\bibinfo{author}{\bibnamefont{Grossman}, \bibfnamefont{Y.}},
  \bibinfo{author}{\bibfnamefont{A.~L.} \bibnamefont{Kagan}}, and
  \bibinfo{author}{\bibfnamefont{Z.}~\bibnamefont{Ligeti}},
  \bibinfo{year}{2002}, \bibinfo{journal}{Phys. Lett.}
  \textbf{\bibinfo{volume}{B538}}, \bibinfo{pages}{327}.

\bibitem[{\citenamefont{Grossman}
  \emph{et~al.}(2007{\natexlab{a}})\citenamefont{Grossman, Kagan, and
  Nir}}]{Grossman:2006jg}
\bibinfo{author}{\bibnamefont{Grossman}, \bibfnamefont{Y.}},
  \bibinfo{author}{\bibfnamefont{A.~L.} \bibnamefont{Kagan}}, and
  \bibinfo{author}{\bibfnamefont{Y.}~\bibnamefont{Nir}},
  \bibinfo{year}{2007}{\natexlab{a}}, \bibinfo{journal}{Phys. Rev.}
  \textbf{\bibinfo{volume}{D75}}, \bibinfo{pages}{036008}.

\bibitem[{\citenamefont{Grossman and Ligeti}(1994)}]{Grossman:1994ax}
\bibinfo{author}{\bibnamefont{Grossman}, \bibfnamefont{Y.}}, and
  \bibinfo{author}{\bibfnamefont{Z.}~\bibnamefont{Ligeti}},
  \bibinfo{year}{1994}, \bibinfo{journal}{Phys. Lett.}
  \textbf{\bibinfo{volume}{B332}}, \bibinfo{pages}{373}.

\bibitem[{\citenamefont{Grossman and Ligeti}(1995)}]{Grossman:1994eb}
\bibinfo{author}{\bibnamefont{Grossman}, \bibfnamefont{Y.}}, and
  \bibinfo{author}{\bibfnamefont{Z.}~\bibnamefont{Ligeti}},
  \bibinfo{year}{1995}, \bibinfo{journal}{Phys. Lett.}
  \textbf{\bibinfo{volume}{B347}}, \bibinfo{pages}{399}.

\bibitem[{\citenamefont{Grossman} \emph{et~al.}(1996)\citenamefont{Grossman,
  Ligeti, and Nardi}}]{Grossman:1995gt}
\bibinfo{author}{\bibnamefont{Grossman}, \bibfnamefont{Y.}},
  \bibinfo{author}{\bibfnamefont{Z.}~\bibnamefont{Ligeti}}, and
  \bibinfo{author}{\bibfnamefont{E.}~\bibnamefont{Nardi}},
  \bibinfo{year}{1996}, \bibinfo{journal}{Nucl. Phys.}
  \textbf{\bibinfo{volume}{B465}}, \bibinfo{pages}{369}.

\bibitem[{\citenamefont{Grossman}
  \emph{et~al.}(2003{\natexlab{a}})\citenamefont{Grossman, Ligeti, Nir, and
  Quinn}}]{Grossman:2003qp}
\bibinfo{author}{\bibnamefont{Grossman}, \bibfnamefont{Y.}},
  \bibinfo{author}{\bibfnamefont{Z.}~\bibnamefont{Ligeti}},
  \bibinfo{author}{\bibfnamefont{Y.}~\bibnamefont{Nir}}, and
  \bibinfo{author}{\bibfnamefont{H.}~\bibnamefont{Quinn}},
  \bibinfo{year}{2003}{\natexlab{a}}, \bibinfo{journal}{Phys. Rev.}
  \textbf{\bibinfo{volume}{D68}}, \bibinfo{pages}{015004}.

\bibitem[{\citenamefont{Grossman}
  \emph{et~al.}(2003{\natexlab{b}})\citenamefont{Grossman, Ligeti, and
  Soffer}}]{Grossman:2002aq}
\bibinfo{author}{\bibnamefont{Grossman}, \bibfnamefont{Y.}},
  \bibinfo{author}{\bibfnamefont{Z.}~\bibnamefont{Ligeti}}, and
  \bibinfo{author}{\bibfnamefont{A.}~\bibnamefont{Soffer}},
  \bibinfo{year}{2003}{\natexlab{b}}, \bibinfo{journal}{Phys. Rev.}
  \textbf{\bibinfo{volume}{D67}}, \bibinfo{pages}{071301}.

\bibitem[{\citenamefont{Grossman and Neubert}(2000)}]{Grossman:1999ra}
\bibinfo{author}{\bibnamefont{Grossman}, \bibfnamefont{Y.}}, and
  \bibinfo{author}{\bibfnamefont{M.}~\bibnamefont{Neubert}},
  \bibinfo{year}{2000}, \bibinfo{journal}{Phys. Lett.}
  \textbf{\bibinfo{volume}{B474}}, \bibinfo{pages}{361}.

\bibitem[{\citenamefont{Grossman} \emph{et~al.}(1998)\citenamefont{Grossman,
  Nir, and Rattazzi}}]{Grossman:1997pa}
\bibinfo{author}{\bibnamefont{Grossman}, \bibfnamefont{Y.}},
  \bibinfo{author}{\bibfnamefont{Y.}~\bibnamefont{Nir}}, and
  \bibinfo{author}{\bibfnamefont{R.}~\bibnamefont{Rattazzi}},
  \bibinfo{year}{1998}, \bibinfo{journal}{Adv. Ser. Direct. High Energy Phys.}
  \textbf{\bibinfo{volume}{15}}, \bibinfo{pages}{755}.

\bibitem[{\citenamefont{Grossman}
  \emph{et~al.}(2007{\natexlab{b}})\citenamefont{Grossman, Nir, Thaler,
  Volansky, and Zupan}}]{Grossman:2007bd}
\bibinfo{author}{\bibnamefont{Grossman}, \bibfnamefont{Y.}},
  \bibinfo{author}{\bibfnamefont{Y.}~\bibnamefont{Nir}},
  \bibinfo{author}{\bibfnamefont{J.}~\bibnamefont{Thaler}},
  \bibinfo{author}{\bibfnamefont{T.}~\bibnamefont{Volansky}}, and
  \bibinfo{author}{\bibfnamefont{J.}~\bibnamefont{Zupan}},
  \bibinfo{year}{2007}{\natexlab{b}}, \bibinfo{journal}{Phys. Rev.}
  \textbf{\bibinfo{volume}{D76}}, \bibinfo{pages}{096006}.

\bibitem[{\citenamefont{Grossman} \emph{et~al.}(2005)\citenamefont{Grossman,
  Soffer, and Zupan}}]{Grossman:2005rp}
\bibinfo{author}{\bibnamefont{Grossman}, \bibfnamefont{Y.}},
  \bibinfo{author}{\bibfnamefont{A.}~\bibnamefont{Soffer}}, and
  \bibinfo{author}{\bibfnamefont{J.}~\bibnamefont{Zupan}},
  \bibinfo{year}{2005}, \bibinfo{journal}{Phys. Rev.}
  \textbf{\bibinfo{volume}{D72}}, \bibinfo{pages}{031501}.

\bibitem[{\citenamefont{Grossman and Worah}(1997)}]{Grossman:1996ke}
\bibinfo{author}{\bibnamefont{Grossman}, \bibfnamefont{Y.}}, and
  \bibinfo{author}{\bibfnamefont{M.~P.} \bibnamefont{Worah}},
  \bibinfo{year}{1997}, \bibinfo{journal}{Phys. Lett.}
  \textbf{\bibinfo{volume}{B395}}, \bibinfo{pages}{241}.

\bibitem[{\citenamefont{Grzadkowski and Hou}(1992)}]{Grzadkowski:1992qj}
\bibinfo{author}{\bibnamefont{Grzadkowski}, \bibfnamefont{B.}}, and
  \bibinfo{author}{\bibfnamefont{W.-S.} \bibnamefont{Hou}},
  \bibinfo{year}{1992}, \bibinfo{journal}{Phys. Lett.}
  \textbf{\bibinfo{volume}{B283}}, \bibinfo{pages}{427}.

\bibitem[{\citenamefont{Gunion} \emph{et~al.}(2006)\citenamefont{Gunion,
  Hooper, and McElrath}}]{Gunion:2005rw}
\bibinfo{author}{\bibnamefont{Gunion}, \bibfnamefont{J.~F.}},
  \bibinfo{author}{\bibfnamefont{D.}~\bibnamefont{Hooper}}, and
  \bibinfo{author}{\bibfnamefont{B.}~\bibnamefont{McElrath}},
  \bibinfo{year}{2006}, \bibinfo{journal}{Phys. Rev.}
  \textbf{\bibinfo{volume}{D73}}, \bibinfo{pages}{015011}.

\bibitem[{\citenamefont{Haber}(1998)}]{Haber:1997if}
\bibinfo{author}{\bibnamefont{Haber}, \bibfnamefont{H.~E.}},
  \bibinfo{year}{1998}, \bibinfo{journal}{Nucl. Phys. Proc. Suppl.}
  \textbf{\bibinfo{volume}{62}}, \bibinfo{pages}{469}.

\bibitem[{\citenamefont{Haber and Kane}(1985)}]{Haber:1984rc}
\bibinfo{author}{\bibnamefont{Haber}, \bibfnamefont{H.~E.}}, and
  \bibinfo{author}{\bibfnamefont{G.~L.} \bibnamefont{Kane}},
  \bibinfo{year}{1985}, \bibinfo{journal}{Phys. Rept.}
  \textbf{\bibinfo{volume}{117}}, \bibinfo{pages}{75}.

\bibitem[{\citenamefont{Haber} \emph{et~al.}(1979)\citenamefont{Haber, Kane,
  and Sterling}}]{Haber:1978jt}
\bibinfo{author}{\bibnamefont{Haber}, \bibfnamefont{H.~E.}},
  \bibinfo{author}{\bibfnamefont{G.~L.} \bibnamefont{Kane}}, and
  \bibinfo{author}{\bibfnamefont{T.}~\bibnamefont{Sterling}},
  \bibinfo{year}{1979}, \bibinfo{journal}{Nucl. Phys.}
  \textbf{\bibinfo{volume}{B161}}, \bibinfo{pages}{493}.

\bibitem[{\citenamefont{Hagelin}(1981)}]{Hagelin:1981zk}
\bibinfo{author}{\bibnamefont{Hagelin}, \bibfnamefont{J.~S.}},
  \bibinfo{year}{1981}, \bibinfo{journal}{Nucl. Phys.}
  \textbf{\bibinfo{volume}{B193}}, \bibinfo{pages}{123}.

\bibitem[{\citenamefont{Hall} \emph{et~al.}(1986)\citenamefont{Hall,
  Kostelecky, and Raby}}]{Hall:1985dx}
\bibinfo{author}{\bibnamefont{Hall}, \bibfnamefont{L.~J.}},
  \bibinfo{author}{\bibfnamefont{V.~A.} \bibnamefont{Kostelecky}}, and
  \bibinfo{author}{\bibfnamefont{S.}~\bibnamefont{Raby}}, \bibinfo{year}{1986},
  \bibinfo{journal}{Nucl. Phys.} \textbf{\bibinfo{volume}{B267}},
  \bibinfo{pages}{415}.

\bibitem[{\citenamefont{Hall and Randall}(1990)}]{Hall:1990ac}
\bibinfo{author}{\bibnamefont{Hall}, \bibfnamefont{L.~J.}}, and
  \bibinfo{author}{\bibfnamefont{L.}~\bibnamefont{Randall}},
  \bibinfo{year}{1990}, \bibinfo{journal}{Phys. Rev. Lett.}
  \textbf{\bibinfo{volume}{65}}, \bibinfo{pages}{2939}.

\bibitem[{\citenamefont{Han} \emph{et~al.}(2004)\citenamefont{Han, Langacker,
  and McElrath}}]{Han:2004yd}
\bibinfo{author}{\bibnamefont{Han}, \bibfnamefont{T.}},
  \bibinfo{author}{\bibfnamefont{P.}~\bibnamefont{Langacker}}, and
  \bibinfo{author}{\bibfnamefont{B.}~\bibnamefont{McElrath}},
  \bibinfo{year}{2004}, \bibinfo{journal}{Phys. Rev.}
  \textbf{\bibinfo{volume}{D70}}, \bibinfo{pages}{115006}.

\bibitem[{Hashimoto \emph{et~al.}(2004)\citenamefont{Hashimoto}
  \emph{et~al.}}]{Hashimoto:2004sm}
\bibinfo{author}{\bibnamefont{Hashimoto}, \bibfnamefont{e.~., S.}},
  \emph{et~al.}, \bibinfo{year}{2004}, \bibinfo{note}{{\rm KEK-REPORT-2004-4}}.

\bibitem[{\citenamefont{Hashimoto} \emph{et~al.}(2002)\citenamefont{Hashimoto,
  Kronfeld, Mackenzie, Ryan, and Simone}}]{Hashimoto:2001nb}
\bibinfo{author}{\bibnamefont{Hashimoto}, \bibfnamefont{S.}},
  \bibinfo{author}{\bibfnamefont{A.~S.} \bibnamefont{Kronfeld}},
  \bibinfo{author}{\bibfnamefont{P.~B.} \bibnamefont{Mackenzie}},
  \bibinfo{author}{\bibfnamefont{S.~M.} \bibnamefont{Ryan}}, and
  \bibinfo{author}{\bibfnamefont{J.~N.} \bibnamefont{Simone}},
  \bibinfo{year}{2002}, \bibinfo{journal}{Phys. Rev.}
  \textbf{\bibinfo{volume}{D66}}, \bibinfo{pages}{014503}.

\bibitem[{Hashimoto \emph{et~al.}(2000)\citenamefont{Hashimoto}
  \emph{et~al.}}]{Hashimoto:1999yp}
\bibinfo{author}{\bibnamefont{Hashimoto}, \bibfnamefont{S.}}, \emph{et~al.},
  \bibinfo{year}{2000}, \bibinfo{journal}{Phys. Rev.}
  \textbf{\bibinfo{volume}{D61}}, \bibinfo{pages}{014502}.

\bibitem[{Hayasaka \emph{et~al.}(2007)\citenamefont{Hayasaka}
  \emph{et~al.}}]{Hayasaka:2007vc}
\bibinfo{author}{\bibnamefont{Hayasaka}, \bibfnamefont{K.}}, \emph{et~al.}
  (\bibinfo{collaboration}{Belle}), \bibinfo{year}{2007},
  \eprint{arXiv:0705.0650 [hep-ex]}.

\bibitem[{\citenamefont{Heinemeyer}
  \emph{et~al.}(2006)\citenamefont{Heinemeyer, Hollik, and
  Weiglein}}]{Heinemeyer:2004gx}
\bibinfo{author}{\bibnamefont{Heinemeyer}, \bibfnamefont{S.}},
  \bibinfo{author}{\bibfnamefont{W.}~\bibnamefont{Hollik}}, and
  \bibinfo{author}{\bibfnamefont{G.}~\bibnamefont{Weiglein}},
  \bibinfo{year}{2006}, \bibinfo{journal}{Phys. Rept.}
  \textbf{\bibinfo{volume}{425}}, \bibinfo{pages}{265}.

\bibitem[{\citenamefont{Hewett}(1996)}]{Hewett:1995dk}
\bibinfo{author}{\bibnamefont{Hewett}, \bibfnamefont{J.~L.}},
  \bibinfo{year}{1996}, \bibinfo{journal}{Phys. Rev.}
  \textbf{\bibinfo{volume}{D53}}, \bibinfo{pages}{4964}.

\bibitem[{Hewett \emph{et~al.}(2004)\citenamefont{Hewett}
  \emph{et~al.}}]{Hewett:2004tv}
\bibinfo{author}{\bibnamefont{Hewett}, \bibfnamefont{J.~L.}}, \emph{et~al.},
  \bibinfo{year}{2004}, \eprint{hep-ph/0503261}.

\bibitem[{\citenamefont{Hill}(2006)}]{Hill:2005ju}
\bibinfo{author}{\bibnamefont{Hill}, \bibfnamefont{R.~J.}},
  \bibinfo{year}{2006}, \bibinfo{journal}{Phys. Rev.}
  \textbf{\bibinfo{volume}{D73}}, \bibinfo{pages}{014012}.

\bibitem[{\citenamefont{Hill} \emph{et~al.}(2004)\citenamefont{Hill, Becher,
  Lee, and Neubert}}]{Hill:2004if}
\bibinfo{author}{\bibnamefont{Hill}, \bibfnamefont{R.~J.}},
  \bibinfo{author}{\bibfnamefont{T.}~\bibnamefont{Becher}},
  \bibinfo{author}{\bibfnamefont{S.~J.} \bibnamefont{Lee}}, and
  \bibinfo{author}{\bibfnamefont{M.}~\bibnamefont{Neubert}},
  \bibinfo{year}{2004}, \bibinfo{journal}{JHEP} \textbf{\bibinfo{volume}{07}},
  \bibinfo{pages}{081}.

\bibitem[{\citenamefont{Hiller}(2004)}]{Hiller:2004ii}
\bibinfo{author}{\bibnamefont{Hiller}, \bibfnamefont{G.}},
  \bibinfo{year}{2004}, \bibinfo{journal}{Phys. Rev.}
  \textbf{\bibinfo{volume}{D70}}, \bibinfo{pages}{034018}.

\bibitem[{\citenamefont{Hiller and Kagan}(2002)}]{Hiller:2001zj}
\bibinfo{author}{\bibnamefont{Hiller}, \bibfnamefont{G.}}, and
  \bibinfo{author}{\bibfnamefont{A.}~\bibnamefont{Kagan}},
  \bibinfo{year}{2002}, \bibinfo{journal}{Phys. Rev.}
  \textbf{\bibinfo{volume}{D65}}, \bibinfo{pages}{074038}.

\bibitem[{\citenamefont{Hiller} \emph{et~al.}(2007)\citenamefont{Hiller,
  Knecht, Legger, and Schietinger}}]{Hiller:2007ur}
\bibinfo{author}{\bibnamefont{Hiller}, \bibfnamefont{G.}},
  \bibinfo{author}{\bibfnamefont{M.}~\bibnamefont{Knecht}},
  \bibinfo{author}{\bibfnamefont{F.}~\bibnamefont{Legger}}, and
  \bibinfo{author}{\bibfnamefont{T.}~\bibnamefont{Schietinger}},
  \bibinfo{year}{2007}, \bibinfo{journal}{Phys. Lett.}
  \textbf{\bibinfo{volume}{B649}}, \bibinfo{pages}{152}.

\bibitem[{\citenamefont{Hiller and Kruger}(2004)}]{Hiller:2003js}
\bibinfo{author}{\bibnamefont{Hiller}, \bibfnamefont{G.}}, and
  \bibinfo{author}{\bibfnamefont{F.}~\bibnamefont{Kruger}},
  \bibinfo{year}{2004}, \bibinfo{journal}{Phys. Rev.}
  \textbf{\bibinfo{volume}{D69}}, \bibinfo{pages}{074020}.

\bibitem[{\citenamefont{Hirata}(1995)}]{Hirata:1994jn}
\bibinfo{author}{\bibnamefont{Hirata}, \bibfnamefont{K.}},
  \bibinfo{year}{1995}, \bibinfo{journal}{Phys. Rev. Lett.}
  \textbf{\bibinfo{volume}{74}}, \bibinfo{pages}{2228}.

\bibitem[{\citenamefont{Hisano} \emph{et~al.}(2002)\citenamefont{Hisano,
  Kitano, and Nojiri}}]{Hisano:2002iy}
\bibinfo{author}{\bibnamefont{Hisano}, \bibfnamefont{J.}},
  \bibinfo{author}{\bibfnamefont{R.}~\bibnamefont{Kitano}}, and
  \bibinfo{author}{\bibfnamefont{M.~M.} \bibnamefont{Nojiri}},
  \bibinfo{year}{2002}, \bibinfo{journal}{Phys. Rev.}
  \textbf{\bibinfo{volume}{D65}}, \bibinfo{pages}{116002}.

\bibitem[{\citenamefont{Hisano} \emph{et~al.}(1996)\citenamefont{Hisano, Moroi,
  Tobe, and Yamaguchi}}]{Hisano:1995cp}
\bibinfo{author}{\bibnamefont{Hisano}, \bibfnamefont{J.}},
  \bibinfo{author}{\bibfnamefont{T.}~\bibnamefont{Moroi}},
  \bibinfo{author}{\bibfnamefont{K.}~\bibnamefont{Tobe}}, and
  \bibinfo{author}{\bibfnamefont{M.}~\bibnamefont{Yamaguchi}},
  \bibinfo{year}{1996}, \bibinfo{journal}{Phys. Rev.}
  \textbf{\bibinfo{volume}{D53}}, \bibinfo{pages}{2442}.

\bibitem[{Hokuue \emph{et~al.}(2007)\citenamefont{Hokuue}
  \emph{et~al.}}]{Hokuue:2006nr}
\bibinfo{author}{\bibnamefont{Hokuue}, \bibfnamefont{T.}}, \emph{et~al.}
  (\bibinfo{collaboration}{Belle}), \bibinfo{year}{2007},
  \bibinfo{journal}{Phys. Lett.} \textbf{\bibinfo{volume}{B648}},
  \bibinfo{pages}{139}.

\bibitem[{\citenamefont{Hooper and Profumo}(2007)}]{Hooper:2007qk}
\bibinfo{author}{\bibnamefont{Hooper}, \bibfnamefont{D.}}, and
  \bibinfo{author}{\bibfnamefont{S.}~\bibnamefont{Profumo}},
  \bibinfo{year}{2007}, \bibinfo{journal}{Phys. Rept.}
  \textbf{\bibinfo{volume}{453}}, \bibinfo{pages}{29}.

\bibitem[{\citenamefont{Hou}(1992)}]{Hou:1991un}
\bibinfo{author}{\bibnamefont{Hou}, \bibfnamefont{W.-S.}},
  \bibinfo{year}{1992}, \bibinfo{journal}{Phys. Lett.}
  \textbf{\bibinfo{volume}{B296}}, \bibinfo{pages}{179}.

\bibitem[{\citenamefont{Hou}(1993)}]{Hou:1992sy}
\bibinfo{author}{\bibnamefont{Hou}, \bibfnamefont{W.-S.}},
  \bibinfo{year}{1993}, \bibinfo{journal}{Phys. Rev.}
  \textbf{\bibinfo{volume}{D48}}, \bibinfo{pages}{2342}.

\bibitem[{\citenamefont{Hou} \emph{et~al.}(2006)\citenamefont{Hou, Nagashima,
  and Soddu}}]{Hou:2006du}
\bibinfo{author}{\bibnamefont{Hou}, \bibfnamefont{W.-S.}},
  \bibinfo{author}{\bibfnamefont{M.}~\bibnamefont{Nagashima}}, and
  \bibinfo{author}{\bibfnamefont{A.}~\bibnamefont{Soddu}},
  \bibinfo{year}{2006}, \eprint{hep-ph/0605080}.

\bibitem[{\citenamefont{Hou and Tseng}(1998)}]{Hou:1997wy}
\bibinfo{author}{\bibnamefont{Hou}, \bibfnamefont{W.-S.}}, and
  \bibinfo{author}{\bibfnamefont{B.}~\bibnamefont{Tseng}},
  \bibinfo{year}{1998}, \bibinfo{journal}{Phys. Rev. Lett.}
  \textbf{\bibinfo{volume}{80}}, \bibinfo{pages}{434}.

\bibitem[{\citenamefont{Hou and Willey}(1988)}]{Hou:1987kf}
\bibinfo{author}{\bibnamefont{Hou}, \bibfnamefont{W.-S.}}, and
  \bibinfo{author}{\bibfnamefont{R.~S.} \bibnamefont{Willey}},
  \bibinfo{year}{1988}, \bibinfo{journal}{Phys. Lett.}
  \textbf{\bibinfo{volume}{B202}}, \bibinfo{pages}{591}.

\bibitem[{\citenamefont{Huang and Wu}(2007)}]{Huang:2007ax}
\bibinfo{author}{\bibnamefont{Huang}, \bibfnamefont{C.-S.}}, and
  \bibinfo{author}{\bibfnamefont{X.-H.} \bibnamefont{Wu}},
  \bibinfo{year}{2007}, \eprint{arXiv:0707.1268 [hep-ph]}.

\bibitem[{\citenamefont{Huber} \emph{et~al.}(2007)\citenamefont{Huber, Hurth,
  and Lunghi}}]{Huber:2007vv}
\bibinfo{author}{\bibnamefont{Huber}, \bibfnamefont{T.}},
  \bibinfo{author}{\bibfnamefont{T.}~\bibnamefont{Hurth}}, and
  \bibinfo{author}{\bibfnamefont{E.}~\bibnamefont{Lunghi}},
  \bibinfo{year}{2007}, \eprint{arXiv:0712.3009 [hep-ph]}.

\bibitem[{\citenamefont{Huber} \emph{et~al.}(2006)\citenamefont{Huber, Lunghi,
  Misiak, and Wyler}}]{Huber:2005ig}
\bibinfo{author}{\bibnamefont{Huber}, \bibfnamefont{T.}},
  \bibinfo{author}{\bibfnamefont{E.}~\bibnamefont{Lunghi}},
  \bibinfo{author}{\bibfnamefont{M.}~\bibnamefont{Misiak}}, and
  \bibinfo{author}{\bibfnamefont{D.}~\bibnamefont{Wyler}},
  \bibinfo{year}{2006}, \bibinfo{journal}{Nucl. Phys.}
  \textbf{\bibinfo{volume}{B740}}, \bibinfo{pages}{105}.

\bibitem[{\citenamefont{Huitu} \emph{et~al.}(1998)\citenamefont{Huitu, Zhang,
  Lu, and Singer}}]{Huitu:1998vn}
\bibinfo{author}{\bibnamefont{Huitu}, \bibfnamefont{K.}},
  \bibinfo{author}{\bibfnamefont{D.~X.} \bibnamefont{Zhang}},
  \bibinfo{author}{\bibfnamefont{C.~D.} \bibnamefont{Lu}}, and
  \bibinfo{author}{\bibfnamefont{P.}~\bibnamefont{Singer}},
  \bibinfo{year}{1998}, \bibinfo{journal}{Phys. Rev. Lett.}
  \textbf{\bibinfo{volume}{81}}, \bibinfo{pages}{4313}.

\bibitem[{\citenamefont{Hurth} \emph{et~al.}(2005)\citenamefont{Hurth, Lunghi,
  and Porod}}]{Hurth:2003dk}
\bibinfo{author}{\bibnamefont{Hurth}, \bibfnamefont{T.}},
  \bibinfo{author}{\bibfnamefont{E.}~\bibnamefont{Lunghi}}, and
  \bibinfo{author}{\bibfnamefont{W.}~\bibnamefont{Porod}},
  \bibinfo{year}{2005}, \bibinfo{journal}{Nucl. Phys.}
  \textbf{\bibinfo{volume}{B704}}, \bibinfo{pages}{56}.

\bibitem[{\citenamefont{Hurth and Mannel}(2001)}]{Hurth:2001yb}
\bibinfo{author}{\bibnamefont{Hurth}, \bibfnamefont{T.}}, and
  \bibinfo{author}{\bibfnamefont{T.}~\bibnamefont{Mannel}},
  \bibinfo{year}{2001}, \bibinfo{journal}{Phys. Lett.}
  \textbf{\bibinfo{volume}{B511}}, \bibinfo{pages}{196}.

\bibitem[{Ikado \emph{et~al.}(2006)\citenamefont{Ikado}
  \emph{et~al.}}]{Ikado:2006un}
\bibinfo{author}{\bibnamefont{Ikado}, \bibfnamefont{K.}}, \emph{et~al.},
  \bibinfo{year}{2006}, \bibinfo{journal}{Phys. Rev. Lett.}
  \textbf{\bibinfo{volume}{97}}, \bibinfo{pages}{251802}.

\bibitem[{\citenamefont{Ilakovac}(2000)}]{Ilakovac:1999md}
\bibinfo{author}{\bibnamefont{Ilakovac}, \bibfnamefont{A.}},
  \bibinfo{year}{2000}, \bibinfo{journal}{Phys. Rev.}
  \textbf{\bibinfo{volume}{D62}}, \bibinfo{pages}{036010}.

\bibitem[{Inami \emph{et~al.}(2003)\citenamefont{Inami}
  \emph{et~al.}}]{Inami:2002ah}
\bibinfo{author}{\bibnamefont{Inami}, \bibfnamefont{K.}}, \emph{et~al.}
  (\bibinfo{collaboration}{Belle}), \bibinfo{year}{2003},
  \bibinfo{journal}{Phys. Lett.} \textbf{\bibinfo{volume}{B551}},
  \bibinfo{pages}{16}.

\bibitem[{\citenamefont{Ishino} \emph{et~al.}(2007)\citenamefont{Ishino,
  Hazumi, Nakao, and Yoshikawa}}]{Ishino:2007pt}
\bibinfo{author}{\bibnamefont{Ishino}, \bibfnamefont{H.}},
  \bibinfo{author}{\bibfnamefont{M.}~\bibnamefont{Hazumi}},
  \bibinfo{author}{\bibfnamefont{M.}~\bibnamefont{Nakao}}, and
  \bibinfo{author}{\bibfnamefont{T.}~\bibnamefont{Yoshikawa}},
  \bibinfo{year}{2007}, \eprint{hep-ex/0703039}.

\bibitem[{Ishino \emph{et~al.}(2007)\citenamefont{Ishino}
  \emph{et~al.}}]{Ishino:2006if}
\bibinfo{author}{\bibnamefont{Ishino}, \bibfnamefont{H.}}, \emph{et~al.}
  (\bibinfo{collaboration}{Belle}), \bibinfo{year}{2007},
  \bibinfo{journal}{Phys. Rev. Lett.} \textbf{\bibinfo{volume}{98}},
  \bibinfo{pages}{211801}.

\bibitem[{\citenamefont{Isidori} \emph{et~al.}(2007)\citenamefont{Isidori,
  Mescia, Paradisi, and Temes}}]{Isidori:2007jw}
\bibinfo{author}{\bibnamefont{Isidori}, \bibfnamefont{G.}},
  \bibinfo{author}{\bibfnamefont{F.}~\bibnamefont{Mescia}},
  \bibinfo{author}{\bibfnamefont{P.}~\bibnamefont{Paradisi}}, and
  \bibinfo{author}{\bibfnamefont{D.}~\bibnamefont{Temes}},
  \bibinfo{year}{2007}, \bibinfo{journal}{Phys. Rev.}
  \textbf{\bibinfo{volume}{D75}}, \bibinfo{pages}{115019}.

\bibitem[{\citenamefont{Isidori and Paradisi}(2006)}]{Isidori:2006pk}
\bibinfo{author}{\bibnamefont{Isidori}, \bibfnamefont{G.}}, and
  \bibinfo{author}{\bibfnamefont{P.}~\bibnamefont{Paradisi}},
  \bibinfo{year}{2006}, \bibinfo{journal}{Phys. Lett.}
  \textbf{\bibinfo{volume}{B639}}, \bibinfo{pages}{499}.

\bibitem[{\citenamefont{Isidori and Retico}(2001)}]{Isidori:2001fv}
\bibinfo{author}{\bibnamefont{Isidori}, \bibfnamefont{G.}}, and
  \bibinfo{author}{\bibfnamefont{A.}~\bibnamefont{Retico}},
  \bibinfo{year}{2001}, \bibinfo{journal}{JHEP} \textbf{\bibinfo{volume}{11}},
  \bibinfo{pages}{001}.

\bibitem[{Iwasaki \emph{et~al.}(2005)\citenamefont{Iwasaki}
  \emph{et~al.}}]{Iwasaki:2005sy}
\bibinfo{author}{\bibnamefont{Iwasaki}, \bibfnamefont{M.}}, \emph{et~al.}
  (\bibinfo{collaboration}{Belle}), \bibinfo{year}{2005},
  \bibinfo{journal}{Phys. Rev.} \textbf{\bibinfo{volume}{D72}},
  \bibinfo{pages}{092005}.

\bibitem[{\citenamefont{Kagan and Neubert}(1998)}]{Kagan:1998bh}
\bibinfo{author}{\bibnamefont{Kagan}, \bibfnamefont{A.~L.}}, and
  \bibinfo{author}{\bibfnamefont{M.}~\bibnamefont{Neubert}},
  \bibinfo{year}{1998}, \bibinfo{journal}{Phys. Rev.}
  \textbf{\bibinfo{volume}{D58}}, \bibinfo{pages}{094012}.

\bibitem[{\citenamefont{Kagan and Neubert}(1999)}]{Kagan:1998ym}
\bibinfo{author}{\bibnamefont{Kagan}, \bibfnamefont{A.~L.}}, and
  \bibinfo{author}{\bibfnamefont{M.}~\bibnamefont{Neubert}},
  \bibinfo{year}{1999}, \bibinfo{journal}{Eur. Phys. J.}
  \textbf{\bibinfo{volume}{C7}}, \bibinfo{pages}{5}.

\bibitem[{\citenamefont{Kajita}(2006)}]{Kajita:2006cy}
\bibinfo{author}{\bibnamefont{Kajita}, \bibfnamefont{T.}},
  \bibinfo{year}{2006}, \bibinfo{journal}{Rept. Prog. Phys.}
  \textbf{\bibinfo{volume}{69}}, \bibinfo{pages}{1607}.

\bibitem[{\citenamefont{Kamenik and Mescia}(2008)}]{Kamenik:2008tj}
\bibinfo{author}{\bibnamefont{Kamenik}, \bibfnamefont{J.~F.}}, and
  \bibinfo{author}{\bibfnamefont{F.}~\bibnamefont{Mescia}},
  \bibinfo{year}{2008}, \eprint{arXiv:0802.3790 [hep-ph]}.

\bibitem[{\citenamefont{Kane} \emph{et~al.}(1994)\citenamefont{Kane, Kolda,
  Roszkowski, and Wells}}]{Kane:1993td}
\bibinfo{author}{\bibnamefont{Kane}, \bibfnamefont{G.~L.}},
  \bibinfo{author}{\bibfnamefont{C.~F.} \bibnamefont{Kolda}},
  \bibinfo{author}{\bibfnamefont{L.}~\bibnamefont{Roszkowski}}, and
  \bibinfo{author}{\bibfnamefont{J.~D.} \bibnamefont{Wells}},
  \bibinfo{year}{1994}, \bibinfo{journal}{Phys. Rev.}
  \textbf{\bibinfo{volume}{D49}}, \bibinfo{pages}{6173}.

\bibitem[{\citenamefont{Kaplunovsky and Louis}(1993)}]{Kaplunovsky:1993rd}
\bibinfo{author}{\bibnamefont{Kaplunovsky}, \bibfnamefont{V.~S.}}, and
  \bibinfo{author}{\bibfnamefont{J.}~\bibnamefont{Louis}},
  \bibinfo{year}{1993}, \bibinfo{journal}{Phys. Lett.}
  \textbf{\bibinfo{volume}{B306}}, \bibinfo{pages}{269}.

\bibitem[{\citenamefont{Kayser and London}(2000)}]{Kayser:1999bu}
\bibinfo{author}{\bibnamefont{Kayser}, \bibfnamefont{B.}}, and
  \bibinfo{author}{\bibfnamefont{D.}~\bibnamefont{London}},
  \bibinfo{year}{2000}, \bibinfo{journal}{Phys. Rev.}
  \textbf{\bibinfo{volume}{D61}}, \bibinfo{pages}{116013}.

\bibitem[{\citenamefont{Keum} \emph{et~al.}(2001)\citenamefont{Keum, Li, and
  Sanda}}]{Keum:2000wi}
\bibinfo{author}{\bibnamefont{Keum}, \bibfnamefont{Y.~Y.}},
  \bibinfo{author}{\bibfnamefont{H.-N.} \bibnamefont{Li}}, and
  \bibinfo{author}{\bibfnamefont{A.~I.} \bibnamefont{Sanda}},
  \bibinfo{year}{2001}, \bibinfo{journal}{Phys. Rev.}
  \textbf{\bibinfo{volume}{D63}}, \bibinfo{pages}{054008}.

\bibitem[{\citenamefont{Khodjamirian}
  \emph{et~al.}(1997)\citenamefont{Khodjamirian, Ruckl, Stoll, and
  Wyler}}]{Khodjamirian:1997tg}
\bibinfo{author}{\bibnamefont{Khodjamirian}, \bibfnamefont{A.}},
  \bibinfo{author}{\bibfnamefont{R.}~\bibnamefont{Ruckl}},
  \bibinfo{author}{\bibfnamefont{G.}~\bibnamefont{Stoll}}, and
  \bibinfo{author}{\bibfnamefont{D.}~\bibnamefont{Wyler}},
  \bibinfo{year}{1997}, \bibinfo{journal}{Phys. Lett.}
  \textbf{\bibinfo{volume}{B402}}, \bibinfo{pages}{167}.

\bibitem[{\citenamefont{Khodjamirian}
  \emph{et~al.}(1995)\citenamefont{Khodjamirian, Stoll, and
  Wyler}}]{Khodjamirian:1995uc}
\bibinfo{author}{\bibnamefont{Khodjamirian}, \bibfnamefont{A.}},
  \bibinfo{author}{\bibfnamefont{G.}~\bibnamefont{Stoll}}, and
  \bibinfo{author}{\bibfnamefont{D.}~\bibnamefont{Wyler}},
  \bibinfo{year}{1995}, \bibinfo{journal}{Phys. Lett.}
  \textbf{\bibinfo{volume}{B358}}, \bibinfo{pages}{129}.

\bibitem[{\citenamefont{Kiers} \emph{et~al.}(2002)\citenamefont{Kiers, Kolb,
  Lee, Soni, and Wu}}]{Kiers:2002cz}
\bibinfo{author}{\bibnamefont{Kiers}, \bibfnamefont{K.}},
  \bibinfo{author}{\bibfnamefont{J.}~\bibnamefont{Kolb}},
  \bibinfo{author}{\bibfnamefont{J.}~\bibnamefont{Lee}},
  \bibinfo{author}{\bibfnamefont{A.}~\bibnamefont{Soni}}, and
  \bibinfo{author}{\bibfnamefont{G.-H.} \bibnamefont{Wu}},
  \bibinfo{year}{2002}, \bibinfo{journal}{Phys. Rev.}
  \textbf{\bibinfo{volume}{D66}}, \bibinfo{pages}{095002}.

\bibitem[{\citenamefont{Kiers and Soni}(1997)}]{Kiers:1997zt}
\bibinfo{author}{\bibnamefont{Kiers}, \bibfnamefont{K.}}, and
  \bibinfo{author}{\bibfnamefont{A.}~\bibnamefont{Soni}}, \bibinfo{year}{1997},
  \bibinfo{journal}{Phys. Rev.} \textbf{\bibinfo{volume}{D56}},
  \bibinfo{pages}{5786}.

\bibitem[{\citenamefont{Kiers} \emph{et~al.}(1999)\citenamefont{Kiers, Soni,
  and Wu}}]{Kiers:1998ry}
\bibinfo{author}{\bibnamefont{Kiers}, \bibfnamefont{K.}},
  \bibinfo{author}{\bibfnamefont{A.}~\bibnamefont{Soni}}, and
  \bibinfo{author}{\bibfnamefont{G.-H.} \bibnamefont{Wu}},
  \bibinfo{year}{1999}, \bibinfo{journal}{Phys. Rev.}
  \textbf{\bibinfo{volume}{D59}}, \bibinfo{pages}{096001}.

\bibitem[{\citenamefont{Kiers} \emph{et~al.}(2000)\citenamefont{Kiers, Soni,
  and Wu}}]{Kiers:2000xy}
\bibinfo{author}{\bibnamefont{Kiers}, \bibfnamefont{K.}},
  \bibinfo{author}{\bibfnamefont{A.}~\bibnamefont{Soni}}, and
  \bibinfo{author}{\bibfnamefont{G.-H.} \bibnamefont{Wu}},
  \bibinfo{year}{2000}, \bibinfo{journal}{Phys. Rev.}
  \textbf{\bibinfo{volume}{D62}}, \bibinfo{pages}{116004}.

\bibitem[{\citenamefont{Kim and Yoshikawa}(2007)}]{Kim:2007fx}
\bibinfo{author}{\bibnamefont{Kim}, \bibfnamefont{C.~S.}}, and
  \bibinfo{author}{\bibfnamefont{T.}~\bibnamefont{Yoshikawa}},
  \bibinfo{year}{2007}, \eprint{arXiv:0711.3880 [hep-ph]}.

\bibitem[{\citenamefont{Kobayashi and Maskawa}(1973)}]{Kobayashi:1973fv}
\bibinfo{author}{\bibnamefont{Kobayashi}, \bibfnamefont{M.}}, and
  \bibinfo{author}{\bibfnamefont{T.}~\bibnamefont{Maskawa}},
  \bibinfo{year}{1973}, \bibinfo{journal}{Prog. Theor. Phys.}
  \textbf{\bibinfo{volume}{49}}, \bibinfo{pages}{652}.

\bibitem[{Koppenburg \emph{et~al.}(2004)\citenamefont{Koppenburg}
  \emph{et~al.}}]{Koppenburg:2004fz}
\bibinfo{author}{\bibnamefont{Koppenburg}, \bibfnamefont{P.}}, \emph{et~al.}
  (\bibinfo{collaboration}{Belle}), \bibinfo{year}{2004},
  \bibinfo{journal}{Phys. Rev. Lett.} \textbf{\bibinfo{volume}{93}},
  \bibinfo{pages}{061803}.

\bibitem[{\citenamefont{Korchemsky and Sterman}(1994)}]{Korchemsky:1994jb}
\bibinfo{author}{\bibnamefont{Korchemsky}, \bibfnamefont{G.~P.}}, and
  \bibinfo{author}{\bibfnamefont{G.}~\bibnamefont{Sterman}},
  \bibinfo{year}{1994}, \bibinfo{journal}{Phys. Lett.}
  \textbf{\bibinfo{volume}{B340}}, \bibinfo{pages}{96}.

\bibitem[{Krokovny \emph{et~al.}(2006)\citenamefont{Krokovny}
  \emph{et~al.}}]{Krokovny:2006sv}
\bibinfo{author}{\bibnamefont{Krokovny}, \bibfnamefont{P.}}, \emph{et~al.}
  (\bibinfo{collaboration}{Belle}), \bibinfo{year}{2006},
  \bibinfo{journal}{Phys. Rev. Lett.} \textbf{\bibinfo{volume}{97}},
  \bibinfo{pages}{081801}.

\bibitem[{\citenamefont{Kruger and Matias}(2005)}]{Kruger:2005ep}
\bibinfo{author}{\bibnamefont{Kruger}, \bibfnamefont{F.}}, and
  \bibinfo{author}{\bibfnamefont{J.}~\bibnamefont{Matias}},
  \bibinfo{year}{2005}, \bibinfo{journal}{Phys. Rev.}
  \textbf{\bibinfo{volume}{D71}}, \bibinfo{pages}{094009}.

\bibitem[{\citenamefont{Kruger and Sehgal}(1996)}]{Kruger:1996cv}
\bibinfo{author}{\bibnamefont{Kruger}, \bibfnamefont{F.}}, and
  \bibinfo{author}{\bibfnamefont{L.~M.} \bibnamefont{Sehgal}},
  \bibinfo{year}{1996}, \bibinfo{journal}{Phys. Lett.}
  \textbf{\bibinfo{volume}{B380}}, \bibinfo{pages}{199}.

\bibitem[{\citenamefont{Kuhn and Mirkes}(1992{\natexlab{a}})}]{Kuhn:1991cc}
\bibinfo{author}{\bibnamefont{Kuhn}, \bibfnamefont{J.~H.}}, and
  \bibinfo{author}{\bibfnamefont{E.}~\bibnamefont{Mirkes}},
  \bibinfo{year}{1992}{\natexlab{a}}, \bibinfo{journal}{Phys. Lett.}
  \textbf{\bibinfo{volume}{B286}}, \bibinfo{pages}{381}.

\bibitem[{\citenamefont{Kuhn and Mirkes}(1992{\natexlab{b}})}]{Kuhn:1992nz}
\bibinfo{author}{\bibnamefont{Kuhn}, \bibfnamefont{J.~H.}}, and
  \bibinfo{author}{\bibfnamefont{E.}~\bibnamefont{Mirkes}},
  \bibinfo{year}{1992}{\natexlab{b}}, \bibinfo{journal}{Z. Phys.}
  \textbf{\bibinfo{volume}{C56}}, \bibinfo{pages}{661}.

\bibitem[{\citenamefont{Kuhn and Mirkes}(1997)}]{Kuhn:1996dv}
\bibinfo{author}{\bibnamefont{Kuhn}, \bibfnamefont{J.~H.}}, and
  \bibinfo{author}{\bibfnamefont{E.}~\bibnamefont{Mirkes}},
  \bibinfo{year}{1997}, \bibinfo{journal}{Phys. Lett.}
  \textbf{\bibinfo{volume}{B398}}, \bibinfo{pages}{407}.

\bibitem[{\citenamefont{Kuno}(2005)}]{Kuno:2005mm}
\bibinfo{author}{\bibnamefont{Kuno}, \bibfnamefont{Y.}}, \bibinfo{year}{2005},
  \bibinfo{journal}{Nucl. Phys. Proc. Suppl.} \textbf{\bibinfo{volume}{149}},
  \bibinfo{pages}{376}.

\bibitem[{Kusaka \emph{et~al.}(2007{\natexlab{a}})\citenamefont{Kusaka}
  \emph{et~al.}}]{Kusaka:2007mj}
\bibinfo{author}{\bibnamefont{Kusaka}, \bibfnamefont{A.}}, \emph{et~al.}
  (\bibinfo{collaboration}{Belle}), \bibinfo{year}{2007}{\natexlab{a}},
  \eprint{arXiv:0710.4974 [hep-ex]}.

\bibitem[{Kusaka \emph{et~al.}(2007{\natexlab{b}})\citenamefont{Kusaka}
  \emph{et~al.}}]{Kusaka:2007dv}
\bibinfo{author}{\bibnamefont{Kusaka}, \bibfnamefont{A.}}, \emph{et~al.}
  (\bibinfo{collaboration}{Belle}), \bibinfo{year}{2007}{\natexlab{b}},
  \bibinfo{journal}{Phys. Rev. Lett.} \textbf{\bibinfo{volume}{98}},
  \bibinfo{pages}{221602}.

\bibitem[{\citenamefont{Laiho}(2007)}]{Laiho:2007pn}
\bibinfo{author}{\bibnamefont{Laiho}, \bibfnamefont{J.}}
  (\bibinfo{collaboration}{Fermilab Lattice and MILC}), \bibinfo{year}{2007},
  \eprint{arXiv:0710.1111 [hep-lat]}.

\bibitem[{\citenamefont{Lange}(2006)}]{Lange:2005xz}
\bibinfo{author}{\bibnamefont{Lange}, \bibfnamefont{B.~O.}},
  \bibinfo{year}{2006}, \bibinfo{journal}{JHEP} \textbf{\bibinfo{volume}{01}},
  \bibinfo{pages}{104}.

\bibitem[{\citenamefont{Lange}
  \emph{et~al.}(2005{\natexlab{a}})\citenamefont{Lange, Neubert, and
  Paz}}]{Lange:2005yw}
\bibinfo{author}{\bibnamefont{Lange}, \bibfnamefont{B.~O.}},
  \bibinfo{author}{\bibfnamefont{M.}~\bibnamefont{Neubert}}, and
  \bibinfo{author}{\bibfnamefont{G.}~\bibnamefont{Paz}},
  \bibinfo{year}{2005}{\natexlab{a}}, \bibinfo{journal}{Phys. Rev.}
  \textbf{\bibinfo{volume}{D72}}, \bibinfo{pages}{073006}.

\bibitem[{\citenamefont{Lange}
  \emph{et~al.}(2005{\natexlab{b}})\citenamefont{Lange, Neubert, and
  Paz}}]{Lange:2005qn}
\bibinfo{author}{\bibnamefont{Lange}, \bibfnamefont{B.~O.}},
  \bibinfo{author}{\bibfnamefont{M.}~\bibnamefont{Neubert}}, and
  \bibinfo{author}{\bibfnamefont{G.}~\bibnamefont{Paz}},
  \bibinfo{year}{2005}{\natexlab{b}}, \bibinfo{journal}{JHEP}
  \textbf{\bibinfo{volume}{10}}, \bibinfo{pages}{084}.

\bibitem[{\citenamefont{Lee}(2008)}]{Lee:2008vs}
\bibinfo{author}{\bibnamefont{Lee}, \bibfnamefont{K.~S.~M.}},
  \bibinfo{year}{2008}, \eprint{arXiv:0802.0873 [hep-ph]}.

\bibitem[{\citenamefont{Lee} \emph{et~al.}(2006)\citenamefont{Lee, Ligeti,
  Stewart, and Tackmann}}]{Lee:2005pwa}
\bibinfo{author}{\bibnamefont{Lee}, \bibfnamefont{K.~S.~M.}},
  \bibinfo{author}{\bibfnamefont{Z.}~\bibnamefont{Ligeti}},
  \bibinfo{author}{\bibfnamefont{I.~W.} \bibnamefont{Stewart}}, and
  \bibinfo{author}{\bibfnamefont{F.~J.} \bibnamefont{Tackmann}},
  \bibinfo{year}{2006}, \bibinfo{journal}{Phys. Rev.}
  \textbf{\bibinfo{volume}{D74}}, \bibinfo{pages}{011501}.

\bibitem[{\citenamefont{Lee} \emph{et~al.}(2007)\citenamefont{Lee, Ligeti,
  Stewart, and Tackmann}}]{Lee:2006gs}
\bibinfo{author}{\bibnamefont{Lee}, \bibfnamefont{K.~S.~M.}},
  \bibinfo{author}{\bibfnamefont{Z.}~\bibnamefont{Ligeti}},
  \bibinfo{author}{\bibfnamefont{I.~W.} \bibnamefont{Stewart}}, and
  \bibinfo{author}{\bibfnamefont{F.~J.} \bibnamefont{Tackmann}},
  \bibinfo{year}{2007}, \bibinfo{journal}{Phys. Rev.}
  \textbf{\bibinfo{volume}{D75}}, \bibinfo{pages}{034016}.

\bibitem[{\citenamefont{Lee and Stewart}(2005)}]{Lee:2004ja}
\bibinfo{author}{\bibnamefont{Lee}, \bibfnamefont{K.~S.~M.}}, and
  \bibinfo{author}{\bibfnamefont{I.~W.} \bibnamefont{Stewart}},
  \bibinfo{year}{2005}, \bibinfo{journal}{Nucl. Phys.}
  \textbf{\bibinfo{volume}{B721}}, \bibinfo{pages}{325}.

\bibitem[{\citenamefont{Lee}(1973)}]{Lee:1973iz}
\bibinfo{author}{\bibnamefont{Lee}, \bibfnamefont{T.~D.}},
  \bibinfo{year}{1973}, \bibinfo{journal}{Phys. Rev.}
  \textbf{\bibinfo{volume}{D8}}, \bibinfo{pages}{1226}.

\bibitem[{\citenamefont{Legger and Schietinger}(2007)}]{Legger:2006cq}
\bibinfo{author}{\bibnamefont{Legger}, \bibfnamefont{F.}}, and
  \bibinfo{author}{\bibfnamefont{T.}~\bibnamefont{Schietinger}},
  \bibinfo{year}{2007}, \bibinfo{journal}{Phys. Lett.}
  \textbf{\bibinfo{volume}{B645}}, \bibinfo{pages}{204}.

\bibitem[{\citenamefont{Leibovich} \emph{et~al.}(2000)\citenamefont{Leibovich,
  Low, and Rothstein}}]{Leibovich:2000ey}
\bibinfo{author}{\bibnamefont{Leibovich}, \bibfnamefont{A.~K.}},
  \bibinfo{author}{\bibfnamefont{I.}~\bibnamefont{Low}}, and
  \bibinfo{author}{\bibfnamefont{I.~Z.} \bibnamefont{Rothstein}},
  \bibinfo{year}{2000}, \bibinfo{journal}{Phys. Lett.}
  \textbf{\bibinfo{volume}{B486}}, \bibinfo{pages}{86}.

\bibitem[{\citenamefont{Lenz}(2007)}]{Lenz:2007nj}
\bibinfo{author}{\bibnamefont{Lenz}, \bibfnamefont{A.}}, \bibinfo{year}{2007},
  \bibinfo{journal}{Phys. Rev.} \textbf{\bibinfo{volume}{D76}},
  \bibinfo{pages}{065006}.

\bibitem[{\citenamefont{Lenz and Nierste}(2007)}]{Lenz:2006hd}
\bibinfo{author}{\bibnamefont{Lenz}, \bibfnamefont{A.}}, and
  \bibinfo{author}{\bibfnamefont{U.}~\bibnamefont{Nierste}},
  \bibinfo{year}{2007}, \bibinfo{journal}{JHEP} \textbf{\bibinfo{volume}{06}},
  \bibinfo{pages}{072}.

\bibitem[{\citenamefont{Leurer} \emph{et~al.}(1994)\citenamefont{Leurer, Nir,
  and Seiberg}}]{Leurer:1993gy}
\bibinfo{author}{\bibnamefont{Leurer}, \bibfnamefont{M.}},
  \bibinfo{author}{\bibfnamefont{Y.}~\bibnamefont{Nir}}, and
  \bibinfo{author}{\bibfnamefont{N.}~\bibnamefont{Seiberg}},
  \bibinfo{year}{1994}, \bibinfo{journal}{Nucl. Phys.}
  \textbf{\bibinfo{volume}{B420}}, \bibinfo{pages}{468}.

\bibitem[{\citenamefont{Li and Mishima}(2006)}]{Li:2006jv}
\bibinfo{author}{\bibnamefont{Li}, \bibfnamefont{H.-n.}}, and
  \bibinfo{author}{\bibfnamefont{S.}~\bibnamefont{Mishima}},
  \bibinfo{year}{2006}, \bibinfo{journal}{Phys. Rev.}
  \textbf{\bibinfo{volume}{D74}}, \bibinfo{pages}{094020}.

\bibitem[{\citenamefont{Li and Mishima}(2007)}]{Li:2006vq}
\bibinfo{author}{\bibnamefont{Li}, \bibfnamefont{H.-n.}}, and
  \bibinfo{author}{\bibfnamefont{S.}~\bibnamefont{Mishima}},
  \bibinfo{year}{2007}, \bibinfo{journal}{JHEP} \textbf{\bibinfo{volume}{03}},
  \bibinfo{pages}{009}.

\bibitem[{\citenamefont{Li} \emph{et~al.}(2006)\citenamefont{Li, Yang, and
  Zhang}}]{Li:2005rr}
\bibinfo{author}{\bibnamefont{Li}, \bibfnamefont{W.-j.}},
  \bibinfo{author}{\bibfnamefont{Y.-d.} \bibnamefont{Yang}}, and
  \bibinfo{author}{\bibfnamefont{X.-d.} \bibnamefont{Zhang}},
  \bibinfo{year}{2006}, \bibinfo{journal}{Phys. Rev.}
  \textbf{\bibinfo{volume}{D73}}, \bibinfo{pages}{073005}.

\bibitem[{\citenamefont{Ligeti} \emph{et~al.}(1997)\citenamefont{Ligeti,
  Randall, and Wise}}]{Ligeti:1997tc}
\bibinfo{author}{\bibnamefont{Ligeti}, \bibfnamefont{Z.}},
  \bibinfo{author}{\bibfnamefont{L.}~\bibnamefont{Randall}}, and
  \bibinfo{author}{\bibfnamefont{M.~B.} \bibnamefont{Wise}},
  \bibinfo{year}{1997}, \bibinfo{journal}{Phys. Lett.}
  \textbf{\bibinfo{volume}{B402}}, \bibinfo{pages}{178}.

\bibitem[{\citenamefont{Ligeti and Tackmann}(2007)}]{Ligeti:2007sn}
\bibinfo{author}{\bibnamefont{Ligeti}, \bibfnamefont{Z.}}, and
  \bibinfo{author}{\bibfnamefont{F.~J.} \bibnamefont{Tackmann}},
  \bibinfo{year}{2007}, \bibinfo{journal}{Phys. Lett.}
  \textbf{\bibinfo{volume}{B653}}, \bibinfo{pages}{404}.

\bibitem[{\citenamefont{Ligeti and Wise}(1996)}]{Ligeti:1995yz}
\bibinfo{author}{\bibnamefont{Ligeti}, \bibfnamefont{Z.}}, and
  \bibinfo{author}{\bibfnamefont{M.~B.} \bibnamefont{Wise}},
  \bibinfo{year}{1996}, \bibinfo{journal}{Phys. Rev.}
  \textbf{\bibinfo{volume}{D53}}, \bibinfo{pages}{4937}.

\bibitem[{Link \emph{et~al.}(2005)\citenamefont{Link}
  \emph{et~al.}}]{Link:2005th}
\bibinfo{author}{\bibnamefont{Link}, \bibfnamefont{J.~M.}}, \emph{et~al.}
  (\bibinfo{collaboration}{FOCUS}), \bibinfo{year}{2005},
  \bibinfo{journal}{Phys. Lett.} \textbf{\bibinfo{volume}{B622}},
  \bibinfo{pages}{239}.

\bibitem[{\citenamefont{Lipkin}(1999)}]{Lipkin:1998ie}
\bibinfo{author}{\bibnamefont{Lipkin}, \bibfnamefont{H.~J.}},
  \bibinfo{year}{1999}, \bibinfo{journal}{Phys. Lett.}
  \textbf{\bibinfo{volume}{B445}}, \bibinfo{pages}{403}.

\bibitem[{\citenamefont{Lipkin} \emph{et~al.}(1991)\citenamefont{Lipkin, Nir,
  Quinn, and Snyder}}]{Lipkin:1991st}
\bibinfo{author}{\bibnamefont{Lipkin}, \bibfnamefont{H.~J.}},
  \bibinfo{author}{\bibfnamefont{Y.}~\bibnamefont{Nir}},
  \bibinfo{author}{\bibfnamefont{H.~R.} \bibnamefont{Quinn}}, and
  \bibinfo{author}{\bibfnamefont{A.}~\bibnamefont{Snyder}},
  \bibinfo{year}{1991}, \bibinfo{journal}{Phys. Rev.}
  \textbf{\bibinfo{volume}{D44}}, \bibinfo{pages}{1454}.

\bibitem[{\citenamefont{London} \emph{et~al.}(2000)\citenamefont{London, Sinha,
  and Sinha}}]{London:2000zi}
\bibinfo{author}{\bibnamefont{London}, \bibfnamefont{D.}},
  \bibinfo{author}{\bibfnamefont{N.}~\bibnamefont{Sinha}}, and
  \bibinfo{author}{\bibfnamefont{R.}~\bibnamefont{Sinha}},
  \bibinfo{year}{2000}, \bibinfo{journal}{Phys. Rev. Lett.}
  \textbf{\bibinfo{volume}{85}}, \bibinfo{pages}{1807}.

\bibitem[{\citenamefont{London and Soni}(1997)}]{London:1997zk}
\bibinfo{author}{\bibnamefont{London}, \bibfnamefont{D.}}, and
  \bibinfo{author}{\bibfnamefont{A.}~\bibnamefont{Soni}}, \bibinfo{year}{1997},
  \bibinfo{journal}{Phys. Lett.} \textbf{\bibinfo{volume}{B407}},
  \bibinfo{pages}{61}.

\bibitem[{Lovelock \emph{et~al.}(1985)\citenamefont{Lovelock}
  \emph{et~al.}}]{Lovelock:1985nb}
\bibinfo{author}{\bibnamefont{Lovelock}, \bibfnamefont{D.~M.~J.}},
  \emph{et~al.}, \bibinfo{year}{1985}, \bibinfo{journal}{Phys. Rev. Lett.}
  \textbf{\bibinfo{volume}{54}}, \bibinfo{pages}{377}.

\bibitem[{\citenamefont{Luke}(1990)}]{Luke:1990eg}
\bibinfo{author}{\bibnamefont{Luke}, \bibfnamefont{M.~E.}},
  \bibinfo{year}{1990}, \bibinfo{journal}{Phys. Lett.}
  \textbf{\bibinfo{volume}{B252}}, \bibinfo{pages}{447}.

\bibitem[{\citenamefont{Lunghi and Matias}(2007)}]{Lunghi:2006hc}
\bibinfo{author}{\bibnamefont{Lunghi}, \bibfnamefont{E.}}, and
  \bibinfo{author}{\bibfnamefont{J.}~\bibnamefont{Matias}},
  \bibinfo{year}{2007}, \bibinfo{journal}{JHEP} \textbf{\bibinfo{volume}{04}},
  \bibinfo{pages}{058}.

\bibitem[{\citenamefont{Lunghi} \emph{et~al.}(2006)\citenamefont{Lunghi, Porod,
  and Vives}}]{Lunghi:2006uf}
\bibinfo{author}{\bibnamefont{Lunghi}, \bibfnamefont{E.}},
  \bibinfo{author}{\bibfnamefont{W.}~\bibnamefont{Porod}}, and
  \bibinfo{author}{\bibfnamefont{O.}~\bibnamefont{Vives}},
  \bibinfo{year}{2006}, \bibinfo{journal}{Phys. Rev.}
  \textbf{\bibinfo{volume}{D74}}, \bibinfo{pages}{075003}.

\bibitem[{\citenamefont{Lunghi and Soni}(2007)}]{Lunghi:2007ak}
\bibinfo{author}{\bibnamefont{Lunghi}, \bibfnamefont{E.}}, and
  \bibinfo{author}{\bibfnamefont{A.}~\bibnamefont{Soni}}, \bibinfo{year}{2007},
  \bibinfo{journal}{JHEP} \textbf{\bibinfo{volume}{09}}, \bibinfo{pages}{053}.

\bibitem[{\citenamefont{Mannel and Neubert}(1994)}]{Mannel:1994pm}
\bibinfo{author}{\bibnamefont{Mannel}, \bibfnamefont{T.}}, and
  \bibinfo{author}{\bibfnamefont{M.}~\bibnamefont{Neubert}},
  \bibinfo{year}{1994}, \bibinfo{journal}{Phys. Rev.}
  \textbf{\bibinfo{volume}{D50}}, \bibinfo{pages}{2037}.

\bibitem[{\citenamefont{Mannel and Recksiegel}(1997)}]{Mannel:1997pc}
\bibinfo{author}{\bibnamefont{Mannel}, \bibfnamefont{T.}}, and
  \bibinfo{author}{\bibfnamefont{S.}~\bibnamefont{Recksiegel}},
  \bibinfo{year}{1997}, \bibinfo{journal}{Acta Phys. Polon.}
  \textbf{\bibinfo{volume}{B28}}, \bibinfo{pages}{2489}.

\bibitem[{\citenamefont{Manohar and Wise}(1994)}]{Manohar:1993qn}
\bibinfo{author}{\bibnamefont{Manohar}, \bibfnamefont{A.~V.}}, and
  \bibinfo{author}{\bibfnamefont{M.~B.} \bibnamefont{Wise}},
  \bibinfo{year}{1994}, \bibinfo{journal}{Phys. Rev.}
  \textbf{\bibinfo{volume}{D49}}, \bibinfo{pages}{1310}.

\bibitem[{\citenamefont{Martin}(1997)}]{Martin:1997ns}
\bibinfo{author}{\bibnamefont{Martin}, \bibfnamefont{S.~P.}},
  \bibinfo{year}{1997}, \eprint{hep-ph/9709356}.

\bibitem[{\citenamefont{Masiero} \emph{et~al.}(2006)\citenamefont{Masiero,
  Paradisi, and Petronzio}}]{Masiero:2005wr}
\bibinfo{author}{\bibnamefont{Masiero}, \bibfnamefont{A.}},
  \bibinfo{author}{\bibfnamefont{P.}~\bibnamefont{Paradisi}}, and
  \bibinfo{author}{\bibfnamefont{R.}~\bibnamefont{Petronzio}},
  \bibinfo{year}{2006}, \bibinfo{journal}{Phys. Rev.}
  \textbf{\bibinfo{volume}{D74}}, \bibinfo{pages}{011701}.

\bibitem[{\citenamefont{Masiero} \emph{et~al.}(2004)\citenamefont{Masiero,
  Vempati, and Vives}}]{Masiero:2004js}
\bibinfo{author}{\bibnamefont{Masiero}, \bibfnamefont{A.}},
  \bibinfo{author}{\bibfnamefont{S.~K.} \bibnamefont{Vempati}}, and
  \bibinfo{author}{\bibfnamefont{O.}~\bibnamefont{Vives}},
  \bibinfo{year}{2004}, \bibinfo{journal}{New J. Phys.}
  \textbf{\bibinfo{volume}{6}}, \bibinfo{pages}{202}.

\bibitem[{\citenamefont{Matsumori and Sanda}(2006)}]{Matsumori:2005ax}
\bibinfo{author}{\bibnamefont{Matsumori}, \bibfnamefont{M.}}, and
  \bibinfo{author}{\bibfnamefont{A.~I.} \bibnamefont{Sanda}},
  \bibinfo{year}{2006}, \bibinfo{journal}{Phys. Rev.}
  \textbf{\bibinfo{volume}{D73}}, \bibinfo{pages}{114022}.

\bibitem[{\citenamefont{Matsuzaki and Sanda}(2007)}]{Matsuzaki:2007hh}
\bibinfo{author}{\bibnamefont{Matsuzaki}, \bibfnamefont{A.}}, and
  \bibinfo{author}{\bibfnamefont{A.~I.} \bibnamefont{Sanda}},
  \bibinfo{year}{2007}, \eprint{arXiv:0711.0792 [hep-ph]}.

\bibitem[{Matyja \emph{et~al.}(2007)\citenamefont{Matyja}
  \emph{et~al.}}]{Matyja:2007kt}
\bibinfo{author}{\bibnamefont{Matyja}, \bibfnamefont{A.}}, \emph{et~al.}
  (\bibinfo{collaboration}{Belle}), \bibinfo{year}{2007},
  \bibinfo{journal}{Phys. Rev. Lett.} \textbf{\bibinfo{volume}{99}},
  \bibinfo{pages}{191807}.

\bibitem[{\citenamefont{McElrath}(2005)}]{McElrath:2005bp}
\bibinfo{author}{\bibnamefont{McElrath}, \bibfnamefont{B.}},
  \bibinfo{year}{2005}, \bibinfo{journal}{Phys. Rev.}
  \textbf{\bibinfo{volume}{D72}}, \bibinfo{pages}{103508}.

\bibitem[{\citenamefont{Miki} \emph{et~al.}(2002)\citenamefont{Miki, Miura, and
  Tanaka}}]{Miki:2002nz}
\bibinfo{author}{\bibnamefont{Miki}, \bibfnamefont{T.}},
  \bibinfo{author}{\bibfnamefont{T.}~\bibnamefont{Miura}}, and
  \bibinfo{author}{\bibfnamefont{M.}~\bibnamefont{Tanaka}},
  \bibinfo{year}{2002}, \eprint{hep-ph/0210051}.

\bibitem[{\citenamefont{Miller} \emph{et~al.}(2007)\citenamefont{Miller,
  de~Rafael, and Roberts}}]{Miller:2007kk}
\bibinfo{author}{\bibnamefont{Miller}, \bibfnamefont{J.~P.}},
  \bibinfo{author}{\bibfnamefont{E.}~\bibnamefont{de~Rafael}}, and
  \bibinfo{author}{\bibfnamefont{B.~L.} \bibnamefont{Roberts}},
  \bibinfo{year}{2007}, \bibinfo{journal}{Rept. Prog. Phys.}
  \textbf{\bibinfo{volume}{70}}, \bibinfo{pages}{795}.

\bibitem[{\citenamefont{Misiak} \emph{et~al.}(1998)\citenamefont{Misiak,
  Pokorski, and Rosiek}}]{Misiak:1997ei}
\bibinfo{author}{\bibnamefont{Misiak}, \bibfnamefont{M.}},
  \bibinfo{author}{\bibfnamefont{S.}~\bibnamefont{Pokorski}}, and
  \bibinfo{author}{\bibfnamefont{J.}~\bibnamefont{Rosiek}},
  \bibinfo{year}{1998}, \bibinfo{journal}{Adv. Ser. Direct. High Energy Phys.}
  \textbf{\bibinfo{volume}{15}}, \bibinfo{pages}{795}.

\bibitem[{\citenamefont{Misiak and Steinhauser}(2004)}]{Misiak:2004ew}
\bibinfo{author}{\bibnamefont{Misiak}, \bibfnamefont{M.}}, and
  \bibinfo{author}{\bibfnamefont{M.}~\bibnamefont{Steinhauser}},
  \bibinfo{year}{2004}, \bibinfo{journal}{Nucl. Phys.}
  \textbf{\bibinfo{volume}{B683}}, \bibinfo{pages}{277}.

\bibitem[{\citenamefont{Misiak and Steinhauser}(2007)}]{Misiak:2006ab}
\bibinfo{author}{\bibnamefont{Misiak}, \bibfnamefont{M.}}, and
  \bibinfo{author}{\bibfnamefont{M.}~\bibnamefont{Steinhauser}},
  \bibinfo{year}{2007}, \bibinfo{journal}{Nucl. Phys.}
  \textbf{\bibinfo{volume}{B764}}, \bibinfo{pages}{62}.

\bibitem[{Misiak \emph{et~al.}(2007)\citenamefont{Misiak}
  \emph{et~al.}}]{Misiak:2006zs}
\bibinfo{author}{\bibnamefont{Misiak}, \bibfnamefont{M.}}, \emph{et~al.},
  \bibinfo{year}{2007}, \bibinfo{journal}{Phys. Rev. Lett.}
  \textbf{\bibinfo{volume}{98}}, \bibinfo{pages}{022002}.

\bibitem[{Miyazaki \emph{et~al.}(2006)\citenamefont{Miyazaki}
  \emph{et~al.}}]{Miyazaki:2006sx}
\bibinfo{author}{\bibnamefont{Miyazaki}, \bibfnamefont{Y.}}, \emph{et~al.}
  (\bibinfo{collaboration}{Belle}), \bibinfo{year}{2006},
  \bibinfo{journal}{Phys. Lett.} \textbf{\bibinfo{volume}{B639}},
  \bibinfo{pages}{159}.

\bibitem[{Miyazaki \emph{et~al.}(2007{\natexlab{a}})\citenamefont{Miyazaki}
  \emph{et~al.}}]{Miyazaki:2007jp}
\bibinfo{author}{\bibnamefont{Miyazaki}, \bibfnamefont{Y.}}, \emph{et~al.}
  (\bibinfo{collaboration}{Belle}), \bibinfo{year}{2007}{\natexlab{a}},
  \bibinfo{journal}{Phys. Lett.} \textbf{\bibinfo{volume}{B648}},
  \bibinfo{pages}{341}.

\bibitem[{Miyazaki \emph{et~al.}(2007{\natexlab{b}})\citenamefont{Miyazaki}
  \emph{et~al.}}]{Miyazaki:2007zw}
\bibinfo{author}{\bibnamefont{Miyazaki}, \bibfnamefont{Y.}}, \emph{et~al.}
  (\bibinfo{collaboration}{Belle}), \bibinfo{year}{2007}{\natexlab{b}},
  \eprint{arXiv:0711.2189 [hep-ex]}.

\bibitem[{\citenamefont{Mohanta and Giri}(2007)}]{Mohanta:2007ad}
\bibinfo{author}{\bibnamefont{Mohanta}, \bibfnamefont{R.}}, and
  \bibinfo{author}{\bibfnamefont{A.~K.} \bibnamefont{Giri}},
  \bibinfo{year}{2007}, \bibinfo{journal}{Phys. Rev.}
  \textbf{\bibinfo{volume}{D76}}, \bibinfo{pages}{075015}.

\bibitem[{\citenamefont{Mohapatra and Pati}(1975)}]{Mohapatra:1974hk}
\bibinfo{author}{\bibnamefont{Mohapatra}, \bibfnamefont{R.~N.}}, and
  \bibinfo{author}{\bibfnamefont{J.~C.} \bibnamefont{Pati}},
  \bibinfo{year}{1975}, \bibinfo{journal}{Phys. Rev.}
  \textbf{\bibinfo{volume}{D11}}, \bibinfo{pages}{566}.

\bibitem[{\citenamefont{Muheim}(2007)}]{Muheim:2007jk}
\bibinfo{author}{\bibnamefont{Muheim}, \bibfnamefont{F.}},
  \bibinfo{year}{2007}, \bibinfo{journal}{Nucl. Phys. Proc. Suppl.}
  \textbf{\bibinfo{volume}{170}}, \bibinfo{pages}{317}.

\bibitem[{\citenamefont{Nakada}(2007)}]{Nakada:2007zz}
\bibinfo{author}{\bibnamefont{Nakada}, \bibfnamefont{T.}}
  (\bibinfo{collaboration}{LHCb}), \bibinfo{year}{2007}, \bibinfo{journal}{Acta
  Phys. Polon.} \textbf{\bibinfo{volume}{B38}}, \bibinfo{pages}{299}.

\bibitem[{Nakao \emph{et~al.}(2004)\citenamefont{Nakao}
  \emph{et~al.}}]{Nakao:2004th}
\bibinfo{author}{\bibnamefont{Nakao}, \bibfnamefont{M.}}, \emph{et~al.}
  (\bibinfo{collaboration}{Belle}), \bibinfo{year}{2004},
  \bibinfo{journal}{Phys. Rev.} \textbf{\bibinfo{volume}{D69}},
  \bibinfo{pages}{112001}.

\bibitem[{\citenamefont{Neubert}(1994)}]{Neubert:1993um}
\bibinfo{author}{\bibnamefont{Neubert}, \bibfnamefont{M.}},
  \bibinfo{year}{1994}, \bibinfo{journal}{Phys. Rev.}
  \textbf{\bibinfo{volume}{D49}}, \bibinfo{pages}{4623}.

\bibitem[{\citenamefont{Neubert}(1999)}]{Neubert:1998re}
\bibinfo{author}{\bibnamefont{Neubert}, \bibfnamefont{M.}},
  \bibinfo{year}{1999}, \bibinfo{journal}{JHEP} \textbf{\bibinfo{volume}{02}},
  \bibinfo{pages}{014}.

\bibitem[{\citenamefont{Neubert}(2005)}]{Neubert:2005nt}
\bibinfo{author}{\bibnamefont{Neubert}, \bibfnamefont{M.}},
  \bibinfo{year}{2005}, \bibinfo{journal}{Phys. Rev.}
  \textbf{\bibinfo{volume}{D72}}, \bibinfo{pages}{074025}.

\bibitem[{\citenamefont{Neubert}(2008)}]{Neubert:2008cp}
\bibinfo{author}{\bibnamefont{Neubert}, \bibfnamefont{M.}},
  \bibinfo{year}{2008}, \eprint{arXiv:0801.0675 [hep-ph]}.

\bibitem[{\citenamefont{Neubert and
  Rosner}(1998{\natexlab{a}})}]{Neubert:1998jq}
\bibinfo{author}{\bibnamefont{Neubert}, \bibfnamefont{M.}}, and
  \bibinfo{author}{\bibfnamefont{J.~L.} \bibnamefont{Rosner}},
  \bibinfo{year}{1998}{\natexlab{a}}, \bibinfo{journal}{Phys. Rev. Lett.}
  \textbf{\bibinfo{volume}{81}}, \bibinfo{pages}{5076}.

\bibitem[{\citenamefont{Neubert and
  Rosner}(1998{\natexlab{b}})}]{Neubert:1998pt}
\bibinfo{author}{\bibnamefont{Neubert}, \bibfnamefont{M.}}, and
  \bibinfo{author}{\bibfnamefont{J.~L.} \bibnamefont{Rosner}},
  \bibinfo{year}{1998}{\natexlab{b}}, \bibinfo{journal}{Phys. Lett.}
  \textbf{\bibinfo{volume}{B441}}, \bibinfo{pages}{403}.

\bibitem[{\citenamefont{Nierste} \emph{et~al.}(2008)\citenamefont{Nierste,
  Trine, and Westhoff}}]{Nierste:2008qe}
\bibinfo{author}{\bibnamefont{Nierste}, \bibfnamefont{U.}},
  \bibinfo{author}{\bibfnamefont{S.}~\bibnamefont{Trine}}, and
  \bibinfo{author}{\bibfnamefont{S.}~\bibnamefont{Westhoff}},
  \bibinfo{year}{2008}, \eprint{arXiv:0801.4938 [hep-ph]}.

\bibitem[{\citenamefont{Nilles}(1984)}]{Nilles:1983ge}
\bibinfo{author}{\bibnamefont{Nilles}, \bibfnamefont{H.~P.}},
  \bibinfo{year}{1984}, \bibinfo{journal}{Phys. Rept.}
  \textbf{\bibinfo{volume}{110}}, \bibinfo{pages}{1}.

\bibitem[{\citenamefont{Nir}(2007)}]{Nir:2007ac}
\bibinfo{author}{\bibnamefont{Nir}, \bibfnamefont{Y.}}, \bibinfo{year}{2007},
  \bibinfo{journal}{JHEP} \textbf{\bibinfo{volume}{05}}, \bibinfo{pages}{102}.

\bibitem[{\citenamefont{Nir and Seiberg}(1993)}]{Nir:1993mx}
\bibinfo{author}{\bibnamefont{Nir}, \bibfnamefont{Y.}}, and
  \bibinfo{author}{\bibfnamefont{N.}~\bibnamefont{Seiberg}},
  \bibinfo{year}{1993}, \bibinfo{journal}{Phys. Lett.}
  \textbf{\bibinfo{volume}{B309}}, \bibinfo{pages}{337}.

\bibitem[{Nishida \emph{et~al.}(2004)\citenamefont{Nishida}
  \emph{et~al.}}]{Nishida:2003yw}
\bibinfo{author}{\bibnamefont{Nishida}, \bibfnamefont{S.}}, \emph{et~al.}
  (\bibinfo{collaboration}{Belle}), \bibinfo{year}{2004},
  \bibinfo{journal}{Phys. Rev. Lett.} \textbf{\bibinfo{volume}{93}},
  \bibinfo{pages}{031803}.

\bibitem[{\citenamefont{Nobes and Trottier}(2004)}]{Nobes:2003nc}
\bibinfo{author}{\bibnamefont{Nobes}, \bibfnamefont{M.~A.}}, and
  \bibinfo{author}{\bibfnamefont{H.~D.} \bibnamefont{Trottier}},
  \bibinfo{year}{2004}, \bibinfo{journal}{Nucl. Phys. Proc. Suppl.}
  \textbf{\bibinfo{volume}{129}}, \bibinfo{pages}{355}.

\bibitem[{\citenamefont{Oide and Yokoya}(1989)}]{Oide:1989qz}
\bibinfo{author}{\bibnamefont{Oide}, \bibfnamefont{K.}}, and
  \bibinfo{author}{\bibfnamefont{K.}~\bibnamefont{Yokoya}},
  \bibinfo{year}{1989}, \bibinfo{journal}{Phys. Rev.}
  \textbf{\bibinfo{volume}{A40}}, \bibinfo{pages}{315}.

\bibitem[{\citenamefont{Okamoto}(2006)}]{Okamoto:2005zg}
\bibinfo{author}{\bibnamefont{Okamoto}, \bibfnamefont{M.}},
  \bibinfo{year}{2006}, \bibinfo{journal}{PoS}
  \textbf{\bibinfo{volume}{LAT2005}}, \bibinfo{pages}{013}.

\bibitem[{Okamoto \emph{et~al.}(2005)\citenamefont{Okamoto}
  \emph{et~al.}}]{Okamoto:2004xg}
\bibinfo{author}{\bibnamefont{Okamoto}, \bibfnamefont{M.}}, \emph{et~al.},
  \bibinfo{year}{2005}, \bibinfo{journal}{Nucl. Phys. Proc. Suppl.}
  \textbf{\bibinfo{volume}{140}}, \bibinfo{pages}{461}.

\bibitem[{\citenamefont{Oktay} \emph{et~al.}(2004)\citenamefont{Oktay,
  El-Khadra, Kronfeld, and Mackenzie}}]{Oktay:2003gk}
\bibinfo{author}{\bibnamefont{Oktay}, \bibfnamefont{M.~B.}},
  \bibinfo{author}{\bibfnamefont{A.~X.} \bibnamefont{El-Khadra}},
  \bibinfo{author}{\bibfnamefont{A.~S.} \bibnamefont{Kronfeld}}, and
  \bibinfo{author}{\bibfnamefont{P.~B.} \bibnamefont{Mackenzie}},
  \bibinfo{year}{2004}, \bibinfo{journal}{Nucl. Phys. Proc. Suppl.}
  \textbf{\bibinfo{volume}{129}}, \bibinfo{pages}{349}.

\bibitem[{\citenamefont{Orlovsky and Shevchenko}(2007)}]{Orlovsky:2007hv}
\bibinfo{author}{\bibnamefont{Orlovsky}, \bibfnamefont{V.~D.}}, and
  \bibinfo{author}{\bibfnamefont{V.~I.} \bibnamefont{Shevchenko}},
  \bibinfo{year}{2007}, \eprint{arXiv:0708.4302 [hep-ph]}.

\bibitem[{\citenamefont{Paradisi}(2006{\natexlab{a}})}]{Paradisi:2006jp}
\bibinfo{author}{\bibnamefont{Paradisi}, \bibfnamefont{P.}},
  \bibinfo{year}{2006}{\natexlab{a}}, \bibinfo{journal}{JHEP}
  \textbf{\bibinfo{volume}{08}}, \bibinfo{pages}{047}.

\bibitem[{\citenamefont{Paradisi}(2006{\natexlab{b}})}]{Paradisi:2005tk}
\bibinfo{author}{\bibnamefont{Paradisi}, \bibfnamefont{P.}},
  \bibinfo{year}{2006}{\natexlab{b}}, \bibinfo{journal}{JHEP}
  \textbf{\bibinfo{volume}{02}}, \bibinfo{pages}{050}.

\bibitem[{\citenamefont{Paz}(2006)}]{Paz:2006me}
\bibinfo{author}{\bibnamefont{Paz}, \bibfnamefont{G.}}, \bibinfo{year}{2006},
  \eprint{hep-ph/0607217}.

\bibitem[{\citenamefont{Peccei and Quinn}(1977{\natexlab{a}})}]{Peccei:1977ur}
\bibinfo{author}{\bibnamefont{Peccei}, \bibfnamefont{R.~D.}}, and
  \bibinfo{author}{\bibfnamefont{H.~R.} \bibnamefont{Quinn}},
  \bibinfo{year}{1977}{\natexlab{a}}, \bibinfo{journal}{Phys. Rev.}
  \textbf{\bibinfo{volume}{D16}}, \bibinfo{pages}{1791}.

\bibitem[{\citenamefont{Peccei and Quinn}(1977{\natexlab{b}})}]{Peccei:1977hh}
\bibinfo{author}{\bibnamefont{Peccei}, \bibfnamefont{R.~D.}}, and
  \bibinfo{author}{\bibfnamefont{H.~R.} \bibnamefont{Quinn}},
  \bibinfo{year}{1977}{\natexlab{b}}, \bibinfo{journal}{Phys. Rev. Lett.}
  \textbf{\bibinfo{volume}{38}}, \bibinfo{pages}{1440}.

\bibitem[{\citenamefont{Pham}(1999)}]{Pham:1998fq}
\bibinfo{author}{\bibnamefont{Pham}, \bibfnamefont{X.-Y.}},
  \bibinfo{year}{1999}, \bibinfo{journal}{Eur. Phys. J.}
  \textbf{\bibinfo{volume}{C8}}, \bibinfo{pages}{513}.

\bibitem[{\citenamefont{Piwinski}(1977)}]{Piwinski:1977ts}
\bibinfo{author}{\bibnamefont{Piwinski}, \bibfnamefont{A.}},
  \bibinfo{year}{1977}, \bibinfo{note}{{DESY} 77/18}.

\bibitem[{\citenamefont{Polci} \emph{et~al.}(2006)\citenamefont{Polci, Schune,
  and Stocchi}}]{Polci:2006aw}
\bibinfo{author}{\bibnamefont{Polci}, \bibfnamefont{F.}},
  \bibinfo{author}{\bibfnamefont{M.~H.} \bibnamefont{Schune}}, and
  \bibinfo{author}{\bibfnamefont{A.}~\bibnamefont{Stocchi}},
  \bibinfo{year}{2006}, \eprint{hep-ph/0605129}.

\bibitem[{Poluektov \emph{et~al.}(2004)\citenamefont{Poluektov}
  \emph{et~al.}}]{Poluektov:2004mf}
\bibinfo{author}{\bibnamefont{Poluektov}, \bibfnamefont{A.}}, \emph{et~al.}
  (\bibinfo{collaboration}{Belle}), \bibinfo{year}{2004},
  \bibinfo{journal}{Phys. Rev.} \textbf{\bibinfo{volume}{D70}},
  \bibinfo{pages}{072003}.

\bibitem[{Poluektov \emph{et~al.}(2006)\citenamefont{Poluektov}
  \emph{et~al.}}]{Poluektov:2006ia}
\bibinfo{author}{\bibnamefont{Poluektov}, \bibfnamefont{A.}}, \emph{et~al.}
  (\bibinfo{collaboration}{Belle}), \bibinfo{year}{2006},
  \bibinfo{journal}{Phys. Rev.} \textbf{\bibinfo{volume}{D73}},
  \bibinfo{pages}{112009}.

\bibitem[{Raidal \emph{et~al.}(2008)\citenamefont{Raidal}
  \emph{et~al.}}]{Raidal:2008jk}
\bibinfo{author}{\bibnamefont{Raidal}, \bibfnamefont{M.}}, \emph{et~al.},
  \bibinfo{year}{2008}, \eprint{arXiv:0801.1826 [hep-ph]}.

\bibitem[{\citenamefont{Raimondi} \emph{et~al.}(2007)\citenamefont{Raimondi,
  Shatilov, and Zobov}}]{Raimondi:2007vi}
\bibinfo{author}{\bibnamefont{Raimondi}, \bibfnamefont{P.}},
  \bibinfo{author}{\bibfnamefont{D.~N.} \bibnamefont{Shatilov}}, and
  \bibinfo{author}{\bibfnamefont{M.}~\bibnamefont{Zobov}},
  \bibinfo{year}{2007}, \eprint{physics/0702033}.

\bibitem[{\citenamefont{Randall and
  Sundrum}(1999{\natexlab{a}})}]{Randall:1999ee}
\bibinfo{author}{\bibnamefont{Randall}, \bibfnamefont{L.}}, and
  \bibinfo{author}{\bibfnamefont{R.}~\bibnamefont{Sundrum}},
  \bibinfo{year}{1999}{\natexlab{a}}, \bibinfo{journal}{Phys. Rev. Lett.}
  \textbf{\bibinfo{volume}{83}}, \bibinfo{pages}{3370}.

\bibitem[{\citenamefont{Randall and
  Sundrum}(1999{\natexlab{b}})}]{Randall:1998uk}
\bibinfo{author}{\bibnamefont{Randall}, \bibfnamefont{L.}}, and
  \bibinfo{author}{\bibfnamefont{R.}~\bibnamefont{Sundrum}},
  \bibinfo{year}{1999}{\natexlab{b}}, \bibinfo{journal}{Nucl. Phys.}
  \textbf{\bibinfo{volume}{B557}}, \bibinfo{pages}{79}.

\bibitem[{Re \emph{et~al.}(2006)\citenamefont{Re} \emph{et~al.}}]{Re:2006nk}
\bibinfo{author}{\bibnamefont{Re}, \bibfnamefont{V.}}, \emph{et~al.},
  \bibinfo{year}{2006}, \bibinfo{journal}{Nucl. Instrum. Meth.}
  \textbf{\bibinfo{volume}{A569}}, \bibinfo{pages}{1}.

\bibitem[{\citenamefont{Regan} \emph{et~al.}(2002)\citenamefont{Regan, Commins,
  Schmidt, and DeMille}}]{Regan:2002ta}
\bibinfo{author}{\bibnamefont{Regan}, \bibfnamefont{B.~C.}},
  \bibinfo{author}{\bibfnamefont{E.~D.} \bibnamefont{Commins}},
  \bibinfo{author}{\bibfnamefont{C.~J.} \bibnamefont{Schmidt}}, and
  \bibinfo{author}{\bibfnamefont{D.}~\bibnamefont{DeMille}},
  \bibinfo{year}{2002}, \bibinfo{journal}{Phys. Rev. Lett.}
  \textbf{\bibinfo{volume}{88}}, \bibinfo{pages}{071805}.

\bibitem[{\citenamefont{Reina} \emph{et~al.}(1997)\citenamefont{Reina,
  Ricciardi, and Soni}}]{Reina:1997my}
\bibinfo{author}{\bibnamefont{Reina}, \bibfnamefont{L.}},
  \bibinfo{author}{\bibfnamefont{G.}~\bibnamefont{Ricciardi}}, and
  \bibinfo{author}{\bibfnamefont{A.}~\bibnamefont{Soni}}, \bibinfo{year}{1997},
  \bibinfo{journal}{Phys. Rev.} \textbf{\bibinfo{volume}{D56}},
  \bibinfo{pages}{5805}.

\bibitem[{\citenamefont{Ritt}(2006)}]{Ritt:2006cg}
\bibinfo{author}{\bibnamefont{Ritt}, \bibfnamefont{S.}}
  (\bibinfo{collaboration}{MEG}), \bibinfo{year}{2006}, \bibinfo{journal}{Nucl.
  Phys. Proc. Suppl.} \textbf{\bibinfo{volume}{162}}, \bibinfo{pages}{279}.

\bibitem[{\citenamefont{Rosner and Stone}(2008)}]{Rosner:2008yu}
\bibinfo{author}{\bibnamefont{Rosner}, \bibfnamefont{J.~L.}}, and
  \bibinfo{author}{\bibfnamefont{S.}~\bibnamefont{Stone}},
  \bibinfo{year}{2008}, \eprint{arXiv:0802.1043 [hep-ex]}.

\bibitem[{\citenamefont{Saha and Kundu}(2002)}]{Saha:2002kt}
\bibinfo{author}{\bibnamefont{Saha}, \bibfnamefont{J.~P.}}, and
  \bibinfo{author}{\bibfnamefont{A.}~\bibnamefont{Kundu}},
  \bibinfo{year}{2002}, \bibinfo{journal}{Phys. Rev.}
  \textbf{\bibinfo{volume}{D66}}, \bibinfo{pages}{054021}.

\bibitem[{\citenamefont{Santinelli}(2002)}]{Santinelli:2002ea}
\bibinfo{author}{\bibnamefont{Santinelli}, \bibfnamefont{R.}},
  \bibinfo{year}{2002}, \bibinfo{journal}{eConf}
  \textbf{\bibinfo{volume}{C0209101}}, \bibinfo{pages}{WE14}.

\bibitem[{Sato \emph{et~al.}(2006)\citenamefont{Sato}
  \emph{et~al.}}]{Sato:2006zza}
\bibinfo{author}{\bibnamefont{Sato}, \bibfnamefont{A.}}, \emph{et~al.},
  \bibinfo{year}{2006}, \bibinfo{note}{prepared for European Particle
  Accelerator Conference (EPAC 06), Edinburgh, Scotland, 26-30 Jun 2006}.

\bibitem[{Schumann \emph{et~al.}(2006)\citenamefont{Schumann}
  \emph{et~al.}}]{Schumann:2006bg}
\bibinfo{author}{\bibnamefont{Schumann}, \bibfnamefont{J.}}, \emph{et~al.}
  (\bibinfo{collaboration}{Belle}), \bibinfo{year}{2006},
  \bibinfo{journal}{Phys. Rev. Lett.} \textbf{\bibinfo{volume}{97}},
  \bibinfo{pages}{061802}.

\bibitem[{Schwanda \emph{et~al.}(2007)\citenamefont{Schwanda}
  \emph{et~al.}}]{Schwanda:2006nf}
\bibinfo{author}{\bibnamefont{Schwanda}, \bibfnamefont{C.}}, \emph{et~al.}
  (\bibinfo{collaboration}{Belle}), \bibinfo{year}{2007},
  \bibinfo{journal}{Phys. Rev.} \textbf{\bibinfo{volume}{D75}},
  \bibinfo{pages}{032005}.

\bibitem[{\citenamefont{Sher}(2002)}]{Sher:2002ew}
\bibinfo{author}{\bibnamefont{Sher}, \bibfnamefont{M.}}, \bibinfo{year}{2002},
  \bibinfo{journal}{Phys. Rev.} \textbf{\bibinfo{volume}{D66}},
  \bibinfo{pages}{057301}.

\bibitem[{\citenamefont{Silva and Soffer}(2000)}]{Silva:1999bd}
\bibinfo{author}{\bibnamefont{Silva}, \bibfnamefont{J.~P.}}, and
  \bibinfo{author}{\bibfnamefont{A.}~\bibnamefont{Soffer}},
  \bibinfo{year}{2000}, \bibinfo{journal}{Phys. Rev.}
  \textbf{\bibinfo{volume}{D61}}, \bibinfo{pages}{112001}.

\bibitem[{\citenamefont{Silvestrini}(2007)}]{Silvestrini:2007yf}
\bibinfo{author}{\bibnamefont{Silvestrini}, \bibfnamefont{L.}},
  \bibinfo{year}{2007}, \bibinfo{journal}{Ann. Rev. Nucl. Part. Sci.}
  \textbf{\bibinfo{volume}{57}}, \bibinfo{pages}{405}.

\bibitem[{\citenamefont{Sinha}(2004)}]{Sinha:2004ct}
\bibinfo{author}{\bibnamefont{Sinha}, \bibfnamefont{N.}}, \bibinfo{year}{2004},
  \bibinfo{journal}{Phys. Rev.} \textbf{\bibinfo{volume}{D70}},
  \bibinfo{pages}{097501}.

\bibitem[{\citenamefont{Sinha} \emph{et~al.}(2007)\citenamefont{Sinha, Sinha,
  Browder, Deshpande, and Pakvasa}}]{Sinha:2007ck}
\bibinfo{author}{\bibnamefont{Sinha}, \bibfnamefont{N.}},
  \bibinfo{author}{\bibfnamefont{R.}~\bibnamefont{Sinha}},
  \bibinfo{author}{\bibfnamefont{T.~E.} \bibnamefont{Browder}},
  \bibinfo{author}{\bibfnamefont{N.~G.} \bibnamefont{Deshpande}}, and
  \bibinfo{author}{\bibfnamefont{S.}~\bibnamefont{Pakvasa}},
  \bibinfo{year}{2007}, \eprint{arXiv:0708.0454 [hep-ph]}.

\bibitem[{\citenamefont{Snyder and Quinn}(1993)}]{Snyder:1993mx}
\bibinfo{author}{\bibnamefont{Snyder}, \bibfnamefont{A.~E.}}, and
  \bibinfo{author}{\bibfnamefont{H.~R.} \bibnamefont{Quinn}},
  \bibinfo{year}{1993}, \bibinfo{journal}{Phys. Rev.}
  \textbf{\bibinfo{volume}{D48}}, \bibinfo{pages}{2139}.

\bibitem[{\citenamefont{Soares}(1991)}]{Soares:1991te}
\bibinfo{author}{\bibnamefont{Soares}, \bibfnamefont{J.~M.}},
  \bibinfo{year}{1991}, \bibinfo{journal}{Nucl. Phys.}
  \textbf{\bibinfo{volume}{B367}}, \bibinfo{pages}{575}.

\bibitem[{\citenamefont{Soffer}(1998)}]{Soffer:1998un}
\bibinfo{author}{\bibnamefont{Soffer}, \bibfnamefont{A.}},
  \bibinfo{year}{1998}, \eprint{hep-ex/9801018}.

\bibitem[{\citenamefont{Soni and Zupan}(2007)}]{Soni:2005jj}
\bibinfo{author}{\bibnamefont{Soni}, \bibfnamefont{A.}}, and
  \bibinfo{author}{\bibfnamefont{J.}~\bibnamefont{Zupan}},
  \bibinfo{year}{2007}, \bibinfo{journal}{Phys. Rev.}
  \textbf{\bibinfo{volume}{D75}}, \bibinfo{pages}{014024}.

\bibitem[{Staric \emph{et~al.}(2007)\citenamefont{Staric}
  \emph{et~al.}}]{Staric:2007dt}
\bibinfo{author}{\bibnamefont{Staric}, \bibfnamefont{M.}}, \emph{et~al.}
  (\bibinfo{collaboration}{Belle}), \bibinfo{year}{2007},
  \bibinfo{journal}{Phys. Rev. Lett.} \textbf{\bibinfo{volume}{98}},
  \bibinfo{pages}{211803}.

\bibitem[{\citenamefont{Suprun} \emph{et~al.}(2002)\citenamefont{Suprun,
  Chiang, and Rosner}}]{Suprun:2001ms}
\bibinfo{author}{\bibnamefont{Suprun}, \bibfnamefont{D.~A.}},
  \bibinfo{author}{\bibfnamefont{C.-W.} \bibnamefont{Chiang}}, and
  \bibinfo{author}{\bibfnamefont{J.~L.} \bibnamefont{Rosner}},
  \bibinfo{year}{2002}, \bibinfo{journal}{Phys. Rev.}
  \textbf{\bibinfo{volume}{D65}}, \bibinfo{pages}{054025}.

\bibitem[{Tajima \emph{et~al.}(2007)\citenamefont{Tajima}
  \emph{et~al.}}]{Tajima:2006nc}
\bibinfo{author}{\bibnamefont{Tajima}, \bibfnamefont{O.}}, \emph{et~al.}
  (\bibinfo{collaboration}{Belle}), \bibinfo{year}{2007},
  \bibinfo{journal}{Phys. Rev. Lett.} \textbf{\bibinfo{volume}{98}},
  \bibinfo{pages}{132001}.

\bibitem[{\citenamefont{Tanaka}(1995)}]{Tanaka:1994ay}
\bibinfo{author}{\bibnamefont{Tanaka}, \bibfnamefont{M.}},
  \bibinfo{year}{1995}, \bibinfo{journal}{Z. Phys.}
  \textbf{\bibinfo{volume}{C67}}, \bibinfo{pages}{321}.

\bibitem[{\citenamefont{Tantalo}(2007)}]{Tantalo:2007ai}
\bibinfo{author}{\bibnamefont{Tantalo}, \bibfnamefont{N.}},
  \bibinfo{year}{2007}, \eprint{hep-ph/0703241}.

\bibitem[{\citenamefont{Uhlig}(2007)}]{Uhlig:2006xf}
\bibinfo{author}{\bibnamefont{Uhlig}, \bibfnamefont{S.}}, \bibinfo{year}{2007},
  \bibinfo{journal}{JHEP} \textbf{\bibinfo{volume}{11}}, \bibinfo{pages}{066}.

\bibitem[{\citenamefont{Unel}(2005)}]{Unel:2005fj}
\bibinfo{author}{\bibnamefont{Unel}, \bibfnamefont{N.~G.}},
  \bibinfo{year}{2005}, \eprint{hep-ex/0505030}.

\bibitem[{Urquijo \emph{et~al.}(2007)\citenamefont{Urquijo}
  \emph{et~al.}}]{Urquijo:2006wd}
\bibinfo{author}{\bibnamefont{Urquijo}, \bibfnamefont{P.}}, \emph{et~al.}
  (\bibinfo{collaboration}{Belle}), \bibinfo{year}{2007},
  \bibinfo{journal}{Phys. Rev.} \textbf{\bibinfo{volume}{D75}},
  \bibinfo{pages}{032001}.

\bibitem[{Ushiroda \emph{et~al.}(2006)\citenamefont{Ushiroda}
  \emph{et~al.}}]{Ushiroda:2006fi}
\bibinfo{author}{\bibnamefont{Ushiroda}, \bibfnamefont{Y.}}, \emph{et~al.}
  (\bibinfo{collaboration}{Belle}), \bibinfo{year}{2006},
  \bibinfo{journal}{Phys. Rev.} \textbf{\bibinfo{volume}{D74}},
  \bibinfo{pages}{111104}.

\bibitem[{Ushiroda \emph{et~al.}(2007)\citenamefont{Ushiroda}
  \emph{et~al.}}]{Ushiroda:2007jf}
\bibinfo{author}{\bibnamefont{Ushiroda}, \bibfnamefont{Y.}}, \emph{et~al.}
  (\bibinfo{collaboration}{Belle}), \bibinfo{year}{2007},
  \eprint{arXiv:0709.2769 [hep-ex]}.

\bibitem[{\citenamefont{Voloshin}(2001)}]{Voloshin:2001xi}
\bibinfo{author}{\bibnamefont{Voloshin}, \bibfnamefont{M.~B.}},
  \bibinfo{year}{2001}, \bibinfo{journal}{Phys. Lett.}
  \textbf{\bibinfo{volume}{B515}}, \bibinfo{pages}{74}.

\bibitem[{\citenamefont{Weinberg}(1976)}]{Weinberg:1976hu}
\bibinfo{author}{\bibnamefont{Weinberg}, \bibfnamefont{S.}},
  \bibinfo{year}{1976}, \bibinfo{journal}{Phys. Rev. Lett.}
  \textbf{\bibinfo{volume}{37}}, \bibinfo{pages}{657}.

\bibitem[{Wicht \emph{et~al.}(2007)\citenamefont{Wicht}
  \emph{et~al.}}]{Wicht:2007ni}
\bibinfo{author}{\bibnamefont{Wicht}, \bibfnamefont{J.}}, \emph{et~al.}
  (\bibinfo{collaboration}{Belle}), \bibinfo{year}{2007},
  \eprint{arXiv:0712.2659 [hep-ex]}.

\bibitem[{\citenamefont{Wilczek}(1977)}]{Wilczek:1977zn}
\bibinfo{author}{\bibnamefont{Wilczek}, \bibfnamefont{F.}},
  \bibinfo{year}{1977}, \bibinfo{journal}{Phys. Rev. Lett.}
  \textbf{\bibinfo{volume}{39}}, \bibinfo{pages}{1304}.

\bibitem[{\citenamefont{Williamson and Zupan}(2006)}]{Williamson:2006hb}
\bibinfo{author}{\bibnamefont{Williamson}, \bibfnamefont{A.~R.}}, and
  \bibinfo{author}{\bibfnamefont{J.}~\bibnamefont{Zupan}},
  \bibinfo{year}{2006}, \bibinfo{journal}{Phys. Rev.}
  \textbf{\bibinfo{volume}{D74}}, \bibinfo{pages}{014003}.

\bibitem[{\citenamefont{Wu and Soni}(2000)}]{Wu:1999nc}
\bibinfo{author}{\bibnamefont{Wu}, \bibfnamefont{G.-H.}}, and
  \bibinfo{author}{\bibfnamefont{A.}~\bibnamefont{Soni}}, \bibinfo{year}{2000},
  \bibinfo{journal}{Phys. Rev.} \textbf{\bibinfo{volume}{D62}},
  \bibinfo{pages}{056005}.

\bibitem[{Yang \emph{et~al.}(2005)\citenamefont{Yang}
  \emph{et~al.}}]{Yang:2004as}
\bibinfo{author}{\bibnamefont{Yang}, \bibfnamefont{H.}}, \emph{et~al.},
  \bibinfo{year}{2005}, \bibinfo{journal}{Phys. Rev. Lett.}
  \textbf{\bibinfo{volume}{94}}, \bibinfo{pages}{111802}.

\bibitem[{Yao \emph{et~al.}(2006)\citenamefont{Yao}
  \emph{et~al.}}]{VcbYao:2006px}
\bibinfo{author}{\bibnamefont{Yao}, \bibfnamefont{W.~M.}}, \emph{et~al.}
  (\bibinfo{collaboration}{Particle Data Group}), \bibinfo{year}{2006},
  \bibinfo{journal}{J. Phys.} \textbf{\bibinfo{volume}{G33}},
  \bibinfo{pages}{1}, \bibinfo{note}{{R. Kowalewski and T. Mannel}, {\it
  Determination of $V_{cb}$ and $V_{ub}$}}.

\bibitem[{\citenamefont{Zupan}(2007{\natexlab{a}})}]{Zupan:2007fq}
\bibinfo{author}{\bibnamefont{Zupan}, \bibfnamefont{J.}},
  \bibinfo{year}{2007}{\natexlab{a}}, \bibinfo{journal}{Nucl. Phys. Proc.
  Suppl.} \textbf{\bibinfo{volume}{170}}, \bibinfo{pages}{33}.

\bibitem[{\citenamefont{Zupan}(2007{\natexlab{b}})}]{Zupan:2007ca}
\bibinfo{author}{\bibnamefont{Zupan}, \bibfnamefont{J.}},
  \bibinfo{year}{2007}{\natexlab{b}}, \eprint{arXiv:0707.1323 [hep-ph]}.

\bibitem[{\citenamefont{Zupan}(2007{\natexlab{c}})}]{Zupan:2007zz}
\bibinfo{author}{\bibnamefont{Zupan}, \bibfnamefont{J.}},
  \bibinfo{year}{2007}{\natexlab{c}}, \bibinfo{journal}{Nucl. Phys. Proc.
  Suppl.} \textbf{\bibinfo{volume}{170}}, \bibinfo{pages}{65}.

\bibitem[{\citenamefont{Zwicky}(2007)}]{Zwicky:2007vv}
\bibinfo{author}{\bibnamefont{Zwicky}, \bibfnamefont{R.}},
  \bibinfo{year}{2007}, \eprint{arXiv:0707.0677 [hep-ph]}.

\end{thebibliography}

\end{document}